\documentclass[prb,aps, twocolumn, floats,floatfix, amsmath,amssymb]{revtex4-1}

\pdfoutput=1
\usepackage[dvips]{epsfig}
\usepackage{float}
\usepackage{amsmath}
\usepackage{amssymb}
\usepackage{amsfonts}
\usepackage{euscript}
\usepackage{enumerate}
\usepackage{hhline}
\usepackage{threeparttable}
\usepackage{multirow}
\usepackage{supertabular}
\usepackage{pslatex}
\usepackage{tabularx}
\usepackage{color}

\usepackage{graphicx}
\usepackage{dcolumn}
\usepackage{bm}

\begin{document}

\title{Modulated longitudinal gates on encoded spin-qubits via curvature couplings to a superconducting cavity}

\author{Rusko Ruskov and Charles Tahan}

\affiliation{Laboratory for Physical Sciences, 8050 Greenmead Dr., College Park, MD 20740}

\email{charlie@tahan.com, ruskovr@lps.umd.edu}

\date{\today}

\begin{abstract}

We propose entangling operations
based on
the energy curvature couplings
of encoded spin qubits to a superconducting cavity, exploring the non-linear qubit response
to a gate voltage variation.
For a two-qubit ($n$-qubit) entangling gate we explore acquired
geometric phases via a time-modulated longitudinal $\sigma_z$-coupling, offering gate times of 10s of ns
even when the qubits and the cavity are far detuned.
No dipole moment is necessary: the qubit transverse $\sigma_x$-coupling to the resonator is zero at the full sweet spot
of the encoded spin qubit of interest
(a triple quantum dot three-electron exchange-only qubit or
a double quantum dot singlet-triplet qubit).
This approach allows always-on, exchange-only qubits, for example, to stay on their ``sweet spots''
during gate operations, minimizing the charge noise and eliminating an always-on static longitudinal
qubit-qubit coupling.
We calculate the main gate errors due to
the (1) diffusion (Johnson) noise and (2) damping of the resonator,
the (3)
$1/f$-charge noise qubit gate dephasing and
$1/f$-noise on the longitudinal coupling,
(4) qubit dephasing and ac-Stark frequency shifts
via photon fluctuations in the resonator,
and (5) spin-dependent resonator frequency shifts
(via a ``dispersive-like''
static
curvature coupling),
most of them
associated with the non-zero qubit energy curvature (quantum capacitance).
Using spin-echo-like error suppression
at optimal regimes, gate infidelities of $10^{-2}-10^{-3}$ can be achieved with experimentally existing parameters.
The proposed schemes seem suitable for remote spin-to-spin entanglement
of two spin-qubits or
a cluster of spin-qubits:
an important resource of quantum computing.

\end{abstract}

\maketitle

\section{Introduction}
Electron spin qubits in semiconductors have made steady progress in coherence times,
gate operations, and quantum measurements, towards the goal of a spin-based quantum computer
\cite{Petta2005S,Maune2012N,Medford2013PRL,Veldhorst2015N,Kawakami2016PNAS,WardQuantInf2016,Pla2012N,Muhonen2014NN,
Petta2017Science,MiPetta2018N,Landig-Wallraff-Ensslin-Ihn2018N}.
%
Despite the inherent protection of single QD spin qubits  by the
nature of the single electron spin \cite{Nowack2007S,Yang2013NC,VeldhorstRuskov2015PRB},
spin-qubits in multi-electron multi-QDs provide further advantages such as:
(i) encoding the spin-qubit in decoherence free subspaces \cite{DiVincenzo2000N,FongQuInfComp2011} (DFS),
where qubit states can be
partially
protected against charge (electric) and spin (magnetic) global noises:
e.g. in triple quantum dot (TQD) three-electron qubits
\cite{Gaudreau2006PRL,Gaudreau2009APL,LairdMarcus2010PRB,Gaudreau2011NP,Medford2013PRL,
Medford2013NatNano,EngHRLs2015SAdv,Landig-Wallraff-Ensslin-Ihn2018N};
(ii) the potential to choose gate parameter regimes, including the
so-called sweet spots
\cite{TaylorSrinivasa2013PRL,RussBurkard2015PRB,RussGinzelBurkard2016PRB,AEON2016,RuskovVeldhorst2018PRB}
(partial or full sweet spots) where, e.g. the charge noise can be
further minimized, which was recently experimentally confirmed for
singlet-triplet (S-T) DQD qubits
\cite{ReedHunter2016PRL,MartinsKuemethMarcus2016PRL};
(iii) fast single-qubit  gates, based only on exchange  interaction
\cite{LairdMarcus2010PRB,Gaudreau2011NP,Medford2013PRL,Medford2013NatNano,EngHRLs2015SAdv}
(TQD qubits with electric field control and no need of magnetic field gradient),
or using inter-dot magnetic field gradients \cite{Pioro-Ladriere2008NP,Kawakami2014Nn} (S-T qubits).

Exchange based two-qubit gates \cite{DialYacoby2013PRL,Doherty2013PRL,Veldhorst2015N,AEON2016},
however, are locally operated  and do not allow remote coupling of
qubits, e.g. at mm distances that are much larger than the qubit size (of tens of nm).
A solution would be to couple spins via a superconducting (SC) GHz resonator
using a transverse dipole coupling
\cite{BurkardIamoglu2006PRB,FreyWallraff2012PRL,Petersson2012N,TosiMorello2014AIP,Viennot2015Sci,MiPetta2018N}
(already  a resource for SC qubits
\cite{Blais2004PRA,Majer2007N,DiCarlo2010N,KorotkovSiddiqi2012N,RisteDiCarlo2013N,KorotkovSiddiqi2014PRL}).
Recently, studies of the resonant exchange (RX) ``always on'' qubit \cite{TaylorSrinivasa2013PRL},
based on a triple quantum dot (TQD) 3-electron system,
offered strong spin-cavity coupling,
and showed that it can maintain
a partial sweet spot
to gate detuning fluctuations (see also Refs.\onlinecite{RussBurkard2015PRB,Srinivasa2016PRB}).

The transverse ($\sigma_x$) coupling to a spin-qubit
\cite{Childress2004,BurkardIamoglu2006PRB,ShnirmanSchon2012PRL,HuLiuNori2012PRB,
FreyWallraff2012PRL,Petersson2012N,Viennot2015Sci,MiPetta2018N},
requires a  non-zero transition  electric dipole moment (e.d.m.),
$\bm{d}_{\perp} \neq 0$,   
leading to a Jaynes-Cummings (J-C)  interaction with the cavity \cite{BurkardIamoglu2006PRB,TaylorSrinivasa2013PRL}:
${\cal H}_{\rm tr} \propto g_{\perp} \sigma_x\, (\hat{a} + \hat{a}^{\dagger})
\simeq g_{\perp} (\sigma_{-}\hat{a}^{\dagger} + \sigma_{+}\hat{a} )$,
where $\hat{a}$ is the electric field cavity mode annihilation operator
and $\sigma_{-} \equiv |-\rangle \langle +|$ is the qubit lowering operator.
%
It was predicted \cite{Childress2004,BurkardIamoglu2006PRB,ShnirmanSchon2012PRL,HuLiuNori2012PRB}
and recently measured \cite{MiPetta2018N,Landig-Wallraff-Ensslin-Ihn2018N,Samkharadze-Vandersypen2017preprint}
to be at the range from one to tens of MHz.
In a dispersive regime (i.e., avoiding direct excitations),
where the qubit-resonator detuning is large ($\Delta \gg g_{\perp}$),
the leading dispersive Hamiltonian,
${\cal H}_{\rm tr} \propto \frac{g_{\perp}^2}{\Delta}\, \hat{a}^{\dagger} \hat{a} \sigma_z$,
cannot couple directly the qubits via photon exchange since it commutes with the qubit Hamiltonian  ${\cal H}_q$.
(It can entangle the qubits, however, via geometric accumulated phases\cite{CrossGambetta2015PRA,PaikChow2016PRL},
without a net exchange, though the expected entangling rate would be relatively slow,
$\sim \frac{g_{\perp}^4}{\Delta^2}$,
see below).
Higher order terms in the J-C interaction  can couple two qubits
($\sim \sigma^{(1)}_{-} \sigma^{(2)}_{+}$) via virtual photon exchange
with a   rate \cite{BurkardIamoglu2006PRB,Srinivasa2016PRB},
$\sim \frac{g^{(1)}_{\perp} g^{(2)}_{\perp}}{\Delta}$,
suppressed in the dispersive limit.
Reaching faster entangling gates, with a rate $\sim g_{\perp}$, is possible
using sideband transitions
via strong resonant driving of the qubit \cite{Blais2007PRA,Wallraff2007PRL,Leek2009PRB,Srinivasa2016PRB}.

These approaches, however, come with several caveats for encoded spin qubits, namely:
(i) the necessity of strong transverse coupling, $g_{\perp}$, will also imply
much stronger sensitivity to charge noise
via gate voltage fluctuations;
(ii) a strong transverse dipole coupling also would imply an increased coupling
to spurious TLS charge fluctuators
(that is another source of charge noise,
see e.g. Ref.~\cite{YouHuAshhabNori2007PRB,Kerman2010PRL,KhezriDresselKorotkov2015PRA});
(iii) the effective entangling interaction will be generally worsened
by higher-order transitions in the qubit-resonator system \cite{SankKorotkov2016PRL}
since $[{\cal H}_{\rm tr},{\cal H}_q] \neq 0$
in higher orders of $g_{\perp}/\Delta$;
(iv) Also, turning off this
coupling via a larger detuning $\Delta$
may be difficult for both spin and superconducting qubits (see, e.g. Ref.~\onlinecite{Kerman2013}).

\begin{figure} [t!] 
    \centering
        \includegraphics[width=0.37\textwidth]{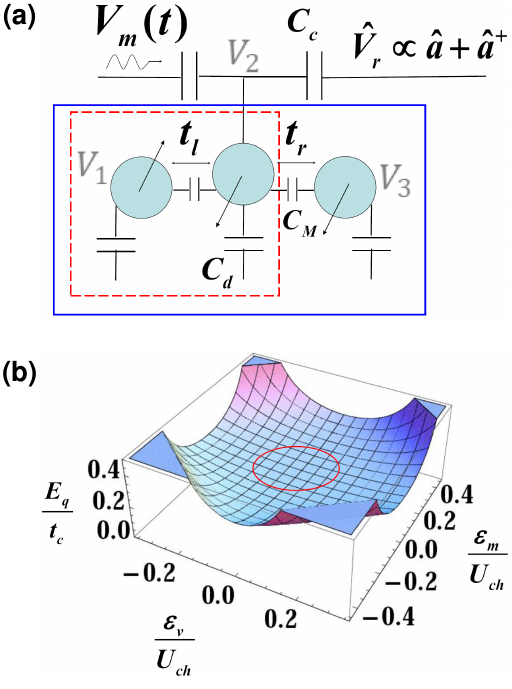}
        \caption{(a) A TQD exchange only qubit (solid, blue box) or a DQD S-T qubit (dashed, red box)
        capacitively coupled to a SC resonator.
        The   
        curvature couplings to a SC resonator   
        arise via the resonator quantized voltage drop $\hat{V}_r = V_{\rm vac}\, (\hat{a} + \hat{a}^{\dagger})$
        on the coupling (middle) dot,   
        essentially as  {\it quantum capacitance couplings}:
        these are (i)
        the dynamical longitudinal coupling,  Eq.~(\ref{dynamic-long-coupling-parallel}),
        $\tilde{g}_{\parallel}\, \sigma_z\, (\hat{a} + \hat{a}^{\dagger}) \sim  \hat{V}_r\, V_m(t)$,
        that arises via additional qubit gate modulation
        %
        and (ii) the always on ``dispersive-like''  (quantum capacitance) static coupling,
        $\delta\omega\, \hat{a}^{\dagger}\hat{a}\,\sigma_z \sim \hat{V}_r^2$, Eq.~(\ref{dispersive-like-coupling}). 
        Both couplings can be expressed via the qubit energy curvature,
        $C_q \equiv e^2 \frac{\partial^2 E_q}{\partial \varepsilon_m^2}
        \propto \frac{t_c^2}{U_{\rm charge}^3}$  that can be significant, $\gtrsim 30\, {\rm aF}$
        (see  Ref.~\onlinecite{RuskovTahan2019PRB99} and Table \ref{tab:rates}),
        for typical dot charging energy $U_{\rm charge} \gtrsim 0.4\, {\rm meV}$  
        and interdot tunnelings $t_c \gtrsim 40\, \mu{\rm eV}$, reachable experimentally.
        For a TQD qubit the relevant voltage detuning is $\varepsilon_m \equiv e [(V_3 + V_1)/2 - V_2]$
        and for a DQD qubit it is $\varepsilon_v \equiv e (V_1 - V_2)$.
        (b)
        Full sweets spot regime of a TQD exchange only qubit (solid, red circle).
        The energy splitting, $E_q(\varepsilon_v,\varepsilon_m)$,  as a function of the
        qubit gate voltage detunnings,
        $\varepsilon_v \equiv e (V_3 - V_1)/2$ and $\varepsilon_m$ 
        (in units of the dots' charging energy, $U_{\rm ch}$). 
        At the full sweet spot\cite{AEON2016} where
        $\frac{\partial E_q}{\partial \varepsilon_v}, \frac{\partial E_q}{\partial \varepsilon_m} = 0$,
        while its transverse dipole moment is also zero\cite{RuskovTahan2019PRB99},
        the TQD qubit is insensitive (in first order) to gate voltage fluctuations, and
        at the same time qubit dephasing through phonon relaxation and coupling to two-level fluctuators
        is also minimized
        as the transition dipole moment is zero\cite{RuskovTahan2019PRB99} as well.
        Similar sweet spot of the DQD qubit (not shown) with respect to the detuning
        $\varepsilon_v \equiv e (V_1 - V_2)$ will be referred as
        a symmetric operating point (SOP)\cite{ReedHunter2016PRL,MartinsKuemethMarcus2016PRL}.
        }
        \label{fig:1}
\end{figure}

In this paper we propose
and fully analyze
(including all loss mechanisms)
an alternative approach
for remote spin-spin entanglement
by establishing
longitudinal {\it curvature} coupling $\sim \tilde{g}_{\parallel} \sigma_z\, (\hat{a} + \hat{a}^{\dagger}$)
of encoded
spin qubits to a superconducting (SC) resonator,
via simultaneous gate voltage modulation of the qubits involved,
even when
the modulation frequency, $\omega_m$, as well as the qubits frequencies, $\omega_q^{(j)}$, are off resonance.
%
The modulated qubits accumulate multi-spin phases
comprising spin-spin unitary gates,
allowing entanglement of various spin clusters, depending on the chosen subset of modulated qubits.
The qubits reside in their {\it full sweet spot to gate voltage fluctuations} (Fig.~\ref{fig:1} a,b),
%
and at the same time ensuring absence of
transverse coupling\cite{RuskovTahan2017preprint1,RuskovTahan2019PRB99}, $g_{\perp} = 0$,
thus minimizing qubits
charge dephasing.
In what follows, we consider both the triple quantum dot (TQD) always on exchange only (AEON) qubit
in the full sweet spot\cite{AEON2016},
and
the double quantum dot (DQD) singlet-triplet (S-T) qubit in its
symmetric operating point\cite{ReedHunter2016PRL,MartinsKuemethMarcus2016PRL} (SOP),
Fig.~\ref{fig:1} a,b,
which are described in the same terms as to their curvature coupling to a SC resonator\cite{RuskovTahan2019PRB99}.

\begin{figure} [t!] 
    \centering
        \includegraphics[width=0.5\textwidth]{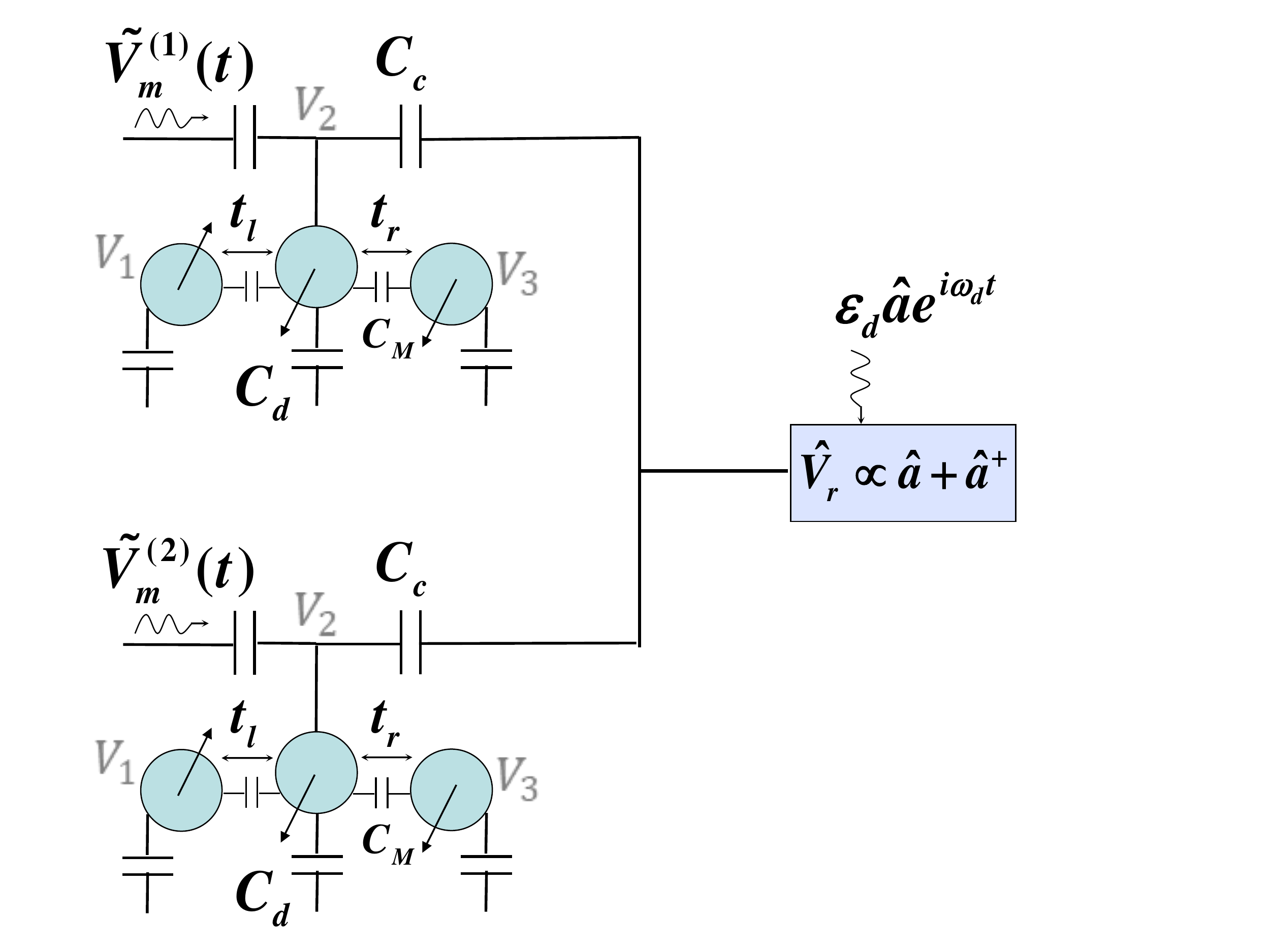}
        \caption{Two TQD spin-qubits
        (or DQD S-T spin-qubits, Fig.~\ref{fig:1}a)
        can be entangled by modulating their respective
        dot gate voltages,
        $V_2^{(j)}$, by a modulation
        $V_m^{(j)}(t) = \tilde{V}_m^{(j)} \cos{(\omega_m t + \varphi_m)}$,
        for a finite gate time, $t_g = 2\pi/\delta \equiv 2\pi/(\omega_r - \omega_m)$,
        where $\delta \equiv \omega_r - \omega_m$ is the frequency detuning.
        For a specially chosen
        detuning, $\delta = \delta_{\pi} \equiv \pi \sqrt{4 N\tilde{g}_{\parallel}^{(1)} \tilde{g}_{\parallel}^{(2)} }$,
        the accumulated geometric phases for each two-qubit spin state amount to a controlled $\pi$-phase gate,
        see Eq.~(\ref{entangling_time}) and Appendix~\ref{app-A: Geometric phases}.
        The defining gate voltages for each qubit as well as the coupling to the (high $Q$-factor)
        resonator can be different in general (see text).
        The qubit frequencies can be different, and they can be strongly detuned
        from the resonator (beyond the usual dispersive limit).
        }
        \label{fig:2}
\end{figure}

The total Hamiltonian of the resonator plus a system of $n$ multi-QD spin qubits
(here, $j=1,\ldots,n$ enumerates the qubits)
at their full sweet spots
with a generic coupling to the environment, ${\cal H}_{\rm env}$,
reads (scf. Ref.~\onlinecite{RuskovTahan2019PRB99}):
\begin{eqnarray}
&& {\cal H}_{\rm tot}/\hbar = \omega_r \hat{a}^{\dagger}\hat{a} + \sum_{j=1}^n \frac{\omega_q^{(j)} }{2}\, \sigma_z^{(j)} + {\cal H}_{\rm env}
\nonumber\\
&&\quad  { } +  \sum_{j=1}^n \delta\omega^{(j)} \sigma_z^{(j)} \left( \hat{a}^{\dagger}\hat{a} + \frac{1}{2} \right)
\nonumber\\
&&\quad  { } + \sum_{j=1}^n  \left[ \tilde{g}_{\parallel}^{(j)}\, \sigma_z^{(j)} + \tilde{g}_0^{(j)} \right]
                  \cos (\omega_m t + \varphi_m) (\hat{a} + \hat{a}^{\dagger} )
\nonumber\\
&&\quad  { } +  2\varepsilon_d \, \cos (\omega_d t + \varphi_d)\ (\hat{a} + \hat{a}^{\dagger} ) .
\label{total_Hamiltonian0}
\end{eqnarray}
Besides the system Hamiltonians, Eq.~(\ref{total_Hamiltonian0}) at the full sweet spot
includes two curvature interactions:
the 
%
``dispersive-like''
static
interaction (second row in Eq.~\ref{total_Hamiltonian0}) 
and the
longitudinal dynamical  
interaction (third row);
while the former is always on,
the latter is on only when the relevant qubits are modulated simultaneously
at a frequency $\omega_m \sim \omega_r$
and phase $\varphi_m$
($|\omega_m - \omega_r| \gg \kappa$, with $\kappa$ the resonator damping),
see Fig.~\ref{fig:1}a and Fig.~\ref{fig:2}.
We also include the driving of the SC resonator with frequency
$\omega_d$ and phase $\varphi_d$.
Other possible interactions, such as the static longitudinal\cite{StaticLongitudinal}
%
%
$(g_{\parallel}^{\rm st}\, \sigma_z + g_0^{\rm st}) (\hat{a} + \hat{a}^{\dagger} )$
and transverse  $g_{\perp}\, \sigma_x  (\hat{a} + \hat{a}^{\dagger} )$ interaction,
were shown to exactly cancel  at the full sweet spot\cite{RuskovTahan2019PRB99}
for each qubit
since, e.g. $g_{\parallel}^{\rm st,(j)} \propto \frac{\partial E_q^{(j)}}{\partial V_m} = 0$,
while the transverse couplings are zeroed due to exact cancelation of contributions
to the transition dipole moment
of the qubit's  
higher excited states \cite{RuskovTahan2019PRB99}.

The main focus in this paper will be on the two types of curvature interactions
that appear through the influence of the resonator quantized voltage
$\hat{V}_r = V_{\rm vac}\, \left( \hat{a} + \hat{a}^{\dagger} \right)$,
Fig.~\ref{fig:1}a,
on the QD qubit levels,
see Refs.~\onlinecite{RuskovTahan2017preprint1,RuskovTahan2019PRB99},
where $V_{\rm vac} = \frac{\hbar\omega_r}{e} \sqrt{\frac{Z_r}{\hbar/e^2}}$
reflects the resonator vacuum voltage fluctuations
(here $Z_r \simeq \omega_r L_r$ is the resonator impedance,
assuming a high quality factor, $Q \equiv \frac{\omega_r L_r}{R_r} \gg 1$).
In what follows,
it is convenient to introduce the dimensionless ratios related to the
qubits-resonator coupling strength:
\begin{equation}
\frac{\eta^{(j)}}{\hbar} \equiv \alpha_c^{(j)} \sqrt{\frac{Z_r}{\hbar/e^2}}
\label{eta-hbar-coupling}
\end{equation}
where
$\alpha_c^{(j)} \simeq \frac{C_c^{(j)}}{C_c^{(j)} + C_d^{(j)}}$ are the  
QDs' lever arms to the SC resonator, Fig.~\ref{fig:1}b.
For a lever arm in the range $\alpha_c \lesssim 0.2$, $\omega_r/2\pi = 10\,{\rm GHz}$,
and reachable resonator impedance of
$L_r \lesssim 50\, {\rm nH}$
one can reach
$\frac{\eta}{\hbar} \lesssim 0.2$.

First, one considers
the dynamical longitudinal (curvature)  Hamiltonian
in Eq.~(\ref{total_Hamiltonian0}),
\begin{equation}
{\cal H}_{\parallel}^{(j)} =
\hbar \left[ \tilde{g}_{\parallel}^{(j)}\, \sigma_z^{(j)} + \tilde{g}_0^{(j)} \right]
\cos (\omega_m t + \varphi_m) (\hat{a} + \hat{a}^{\dagger} )
\label{dynamical-longitud-interaction} ,
\end{equation}
implying the couplings \cite{RuskovTahan2017preprint1,RuskovTahan2019PRB99}
\begin{eqnarray}
&& \tilde{g}_{\parallel}^{(j)}
= \frac{\omega_r}{2} \left(\frac{\eta^{(j)}}{\hbar}\right) \frac{\partial^2 E_q^{(j)}(V^0_m)}{e\, \partial V_m^2} \tilde{V}_m^{(j)}
\label{dynamic-long-coupling-parallel}\\
&& \tilde{g}_{0}^{(j)}
= \omega_r  \left(\frac{\eta^{(j)}}{\hbar}\right) \frac{\partial^2 G_q^{(j)}(V^0_m)}{e\, \partial V_m^2} \tilde{V}_m^{(j)}
\label{dynamic-long-coupling-0} ,
\end{eqnarray}
that appears  under external voltage modulation
of the energy levels of each qubit
with a strength $\tilde{V}_m^{(j)}$
[here, $E_q^{(j)} \equiv E_{+}^{(j)} - E_{-}^{(j)}$ and
$G_q^{(j)} \equiv (E_{+}^{(j)} + E_{-}^{(j)})/2$ are the
$(j)$-qubit energy combinations].

We note that
the spin-independent constants
($\sim \tilde{g}_{0}^{(j)}$)
for the qubits involved in the gate
can be
canceled at once by synchronous resonator driving;
the conditions for this
are
the equal frequencies of driving and modulation,
and special choice of the resonator phase $\varphi_d$ and driving amplitude $\varepsilon_d$:
\begin{equation}
\omega_d = \omega_m, \ \ \varphi_d = \varphi_m + \pi, \ \
\varepsilon_d = \varepsilon_d^0 \equiv \frac{1}{2}\sum_{j=1}^n  \tilde{g}_0^{(j)} .
\label{cancellation-of-g0}
\end{equation}

The always on ``dispersive-like'' (curvature) Hamiltonian in Eq.~(\ref{total_Hamiltonian0})
appears as a second-order effect in $\hat{V}_r$,
\begin{equation}
{\cal H}_{\delta\omega}^{(j)} = \hbar \delta\omega^{(j)} \sigma_z^{(j)} \left( \hat{a}^{+}\hat{a} + \frac{1}{2} \right)
\label{dispersive-like-interaction} ,
\end{equation}
with
\begin{equation}
\delta\omega^{(j)}
= \frac{\hbar \omega_r^2}{2} \left(\frac{\eta^{(j)}}{\hbar}\right)^2  \frac{\partial^2 E_q^{(j)}}{e^2\partial V^2_G}
\label{dispersive-like-coupling} .
\end{equation}
Here, the name ``dispersive-like'' is convenient since this coupling coincides in its form
with the dispersive limit of the J-C transverse coupling, though it has nothing to do with the latter;
in fact the transverse coupling is zeroed  at the full sweet spot \cite{RuskovTahan2019PRB99},
$g_{\perp}^{\rm sweet\ spot} = 0$.
The ``dispersive-like'' interaction ${\cal H}_{\delta\omega}^{(j)}$ causes a
resonator frequency shift $\pm\delta\omega^{(j)}$
depending on the $(j)$-th qubit spin state, $|\uparrow, \downarrow\rangle_{(j)}$,
that can be interpreted as due to the
spin-qubit quantum capacitance \cite{RuskovTahan2019PRB99}.
In fact, it exactly coincides with an analogous expression 
for the quantum capacitance of a Cooper pair box see, e.g.
Refs.~\onlinecite{AverinZorinLikharevZhETF1985,AverinBruderPRL2003,SillanpaaHakonenPRL2005,DutyDelsingPRL2005}.

Since the dynamical longitudinal coupling, Eq.~(\ref{dynamic-long-coupling-parallel}),
is also proportional to the energy curvature,
$\tilde{g}_{\parallel} \propto \frac{\partial^2 E_q}{e^2\partial V^2_G}$, it is
a quantum capacitance related coupling as well.
We also note that using a quantum capacitance approach here is justified
for the resonator/modulation 
frequency range
\begin{equation}
\omega_r,\omega_m \ll U_{\rm charge}
\label{frequency-range} ,
\end{equation}
since at the sweet spot the qubit dipole coupling
is zero, and
one needs to compare
with an energy gap to the qubit higher excited states which is of the order
of the QD's charging energy\cite{RuskovTahan2019PRB99}, $U_{\rm charge}$.
For a typical $U_{\rm charge} \approx 0.4\, {\rm meV} \approx 100\, {\rm GHz}$
this implies  $\omega_r,\omega_m \lesssim 20\, {\rm GHz}$.

With the curvature interactions in Eq.~(\ref{total_Hamiltonian0}),
there are two scenarios to perform
accumulated geometric phase gates on a system of $n$ qubits.
In the first scenario we briefly
consider in the next Sec.~\ref{Sec:no-modulation-accumulated-phase-gate},
the geometric phase gate is based solely on the
always on $\delta\omega^{(j)}$-couplings (we show the entangling rate to be generally small).
The second scenario of a phase gate is
realized via the longitudinal dynamical coupling $\tilde{g}_{\parallel}^{(j)}$,
that is the phase gate of interest
in this paper,
see Secs.~\ref{Sec: modulation-accumulated-phase-gate} and
\ref{ChII-Spin-dependent force: entangling-time}-\ref{ChIV-Discussion-and-Summary}.

\subsection{Multi-spin accumulated phase entangling gates via resonator driving (no qubit gate modulation)}
\label{Sec:no-modulation-accumulated-phase-gate}

In the first scenario, the qubits are not modulated ($\tilde{g}_{\parallel} = 0$),
and the always on  ``dispersive-like''
static
curvature interactions
${\cal H}_{\delta\omega}^{(j)}$ with the couplings $\delta\omega^{(j)}$
can be used to generate spin-dependent geometric phases
via direct driving of the SC resonator\cite{CrossGambetta2015PRA}
at some  detuned frequency $\omega_d \neq \omega_r$,
and using a specially modulated spline microwave pulses, $\varepsilon_d(t)\, e^{i\omega_d t}$,
to suppress
the (remaining) qubit-resonator
entanglement as a possible source of gate infidelity\cite{CrossGambetta2015PRA}.
At the TQD full sweet spot, Fig.~\ref{fig:1}b,
when the dispersive transverse couplings are off, $\chi^{(j)} \equiv (g_{\perp}^{(j)})^2/\Delta^{(j)} = 0$,
one can use the ``dispersive-like'' curvature couplings instead.
Following
the approach as in the superconducting entangling proposal
of Cross and Gambetta\cite{CrossGambetta2015PRA,PaikChow2016PRL},
one is essentially replacing the dispersive couplings, $\chi^{(j)}$
by the ``dispersive-like'' contributions, $\delta\omega^{(j)}$.
Then,  we estimated the entangling rate for two qubits to be:
\begin{equation}
\Gamma_{\rm ent, \delta\omega}^{\rm geom}  \simeq
\frac{|\varepsilon_d|^2 \delta\omega^{(j_1)} \delta\omega^{(j_2)}}{(\omega_d - \omega_r)^3}
\label{Cross-Gambetta-entangling} .
\end{equation}
It is worth comparing of the result of Eq.~(\ref{Cross-Gambetta-entangling})
to a situation with a transverse coupling \cite{CrossGambetta2015PRA,PaikChow2016PRL},
e.g., in a charge degeneracy point\cite{RuskovTahan2019PRB99} (c.d.p.), where
the transition dipole moment is maximal, and correspondingly
$g_{\perp}^{\rm c.d.p.} \approx \frac{\omega_r}{2}\frac{\eta}{\hbar}$.
The entangling gate rate via the ``dispersive-like'' coupling
at zero dipole moment,
will be slower than that based on the dispersive coupling at a c.d.p.,
by a factor of $(\delta\omega /\chi)^2$,
where the coupling ratio
of the two cases is given by \cite{RuskovTahan2019PRB99}
\begin{equation}
\frac{\delta\omega}{\chi_{\rm c.d.p.}}  = \frac{\hbar\Delta}{e^2/2C_q}
\label{entangling-ratio-delta-omega} .
\end{equation}
For a typical
spin-qubit quantum capacitance of $C_q \approx 20\, {\rm aF}$ (see Table \ref{tab:rates} below),
and qubit-resonator detuning, $\Delta \equiv \omega_r - \omega_q \approx 5\,{\rm GHz}$
this amounts to
$\frac{\delta\omega_{\rm sweet\ spot}}{\chi_{\rm c.d.p.}} \approx 1/25$.
The suppression factor can be overcome, however, via increasing the
TQD quantum capacitance, e.g. by using higher interdot tunneling rates, $t_c$,
since $C_q \propto t_c^2$.
Also, the dispersive coupling, $\chi$, is rapidly decreasing
(since one is working out of the c.d.p. to avoid large charge noise),
and thus the
two types of entangling gates can be made comparable in speed.

Overall speed up of
the entangling rate, Eq.~(\ref{Cross-Gambetta-entangling}),
can be achieved by increasing the
ratio $\frac{\eta}{\hbar}$ in QD qubits,
e.g., by using
high kinetic inductance ($L_r \gtrsim 200 \, {\rm nH}$)  resonators \cite{Samkharadze2016,Bylander2018},
higher QDs lever arm $\alpha_c \lesssim 0.5$,
and higher resonator frequencies, $\omega_r \gtrsim 10\, {\rm GHz}$,
where ratio of $\frac{\eta}{\hbar} \sim 1$ may be reached.

In the approach outlined above,
contrary to the transverse J-C coupling case in the
dispersive limit\cite{CrossGambetta2015PRA,PaikChow2016PRL},
where higher photon numbers in the resonator
(higher driving amplitude $\varepsilon_d$)
involves infidelities via spurious photon transitions to
higher system states (see, e.g. Ref.~\onlinecite{SankKorotkov2016PRL}),
here
one is allowed to go to higher photon numbers,
since the higher-curvature Hamiltonian corrections \cite{RuskovTahan2019PRB99}
that arise for $n_{\rm phot} \gtrsim 1$
still commute with the qubit Hamiltonian.
(The role of the higher-curvature Hamiltonians
will be investigated elsewhere).

\subsection{Accumulated phase entangling gates via qubits' longitudinal modulation}
\label{Sec: modulation-accumulated-phase-gate}

In this paper we concentrate on
geometric entangling gates
obtained
when  suitable qubits' gate voltages are modulated
with the {\it same modulation frequency and phase},
$V_m^{(j)}(t) = \tilde{V}_m^{(j)}\, \cos (\omega_m t + \varphi_m)$,
for each qubit $(j)$ participating in the entangling gate, see Fig.~\ref{fig:2}.
The modulation of the qubits' energy levels,
in the presence of a capacitive coupling to the resonator,
leads to the  longitudinal  dynamical
interactions, ${\cal H}_{\parallel}^{(j)}$,
with the (curvature) couplings, $\tilde{g}_{\parallel}^{(j)}$, $\tilde{g}_{0}^{(j)}$,
of Eqs.~(\ref{dynamic-long-coupling-parallel}) and (\ref{dynamic-long-coupling-0}),
where the spin-independent coupling $\tilde{g}_{0}^{(j)}$ plays the role of another channel of resonator driving.
(compare with Eq.~\ref{cancellation-of-g0}).
The longitudinal dynamical coupling ${\cal H}_{\parallel}^{(j)}$
corresponds to a periodic in time spin-dependent ``force''  exerted on the resonator
(see Sec.~\ref{ChII-Spin-dependent force: entangling-time}
and Appendix~\ref{app-A: Geometric phases}).
Similar  approach  was explored in ion traps \cite{Molmer1999PRL,Milburn2000F,Leibried2003N,HaljanMonroe2005PRL},
and recently it was proposed by Kerman \cite{Kerman2013} and
others \cite{Billangeon2015PRB,Didier2015PRL,RicherDiVincenzo2016PRB} 
for superconducting devices.

An ideal multi-qubit entangling gate arises
when simultaneous periodic voltage modulation
is applied to
each qubit $(j)$ of a chosen subset of $n$ qubits,
starting from a product state of the qubits plus resonator:
\begin{equation}
|\psi(0)\rangle = \left( \sum_{s=1}^{2^n} a_s |s \rangle\right) \otimes |0\rangle_{\rm res}
\label{initial-product-state} ,
\end{equation}
where
\begin{equation}
|s\rangle \equiv |i_1\ldots i_j \dots i_n \rangle, \quad  i_j = \pm 1
\label{s-basis-states}
\end{equation}
are the basis $n$-qubit product states of up(down) qubits.

The corresponding dynamical longitudinal couplings, 
\begin{equation}
\tilde{g}^{(j)}_{\parallel} (t) = \tilde{g}^{(j)}_{\parallel} \cos(\omega_m t + \varphi_m)
\label{longitudinal-coupling-modulated}
\end{equation}
will be equivalent to a periodic driving of the resonator with a modulation frequency
$\omega_m$ and with an amplitude 
dependent on the $n$-qubit spin state.
[To get a non-trivial operation, of course, one needs to prepare
the $n$ qubits to a state different from the $n$-qubit ground state,
which can be achieved, e.g., by local qubits' manipulations of their left and right
tunnelings, $t_l$, $t_r$.]
%
Thus, after a time of one full cycle, $T_{\rm cycle} = 2\pi/(\omega_r - \omega_m)$,
when the resonator returns to its initial state (in a rotating frame with $\omega_r$),
the qubits and the resonator become again disentangled,
leading to {\it accumulation of non-trivial geometric phases to the multi-qubit state},
Eq.~(\ref{initial-product-state}).
Various entangling gates can be established by attaching/detaching to the resonator
of some subset of qubits,
by switching on/off particular qubits' modulations,  $V_m^{(j)}(t)$.

A substantial longitudinal (curvature) coupling exists\cite{RuskovTahan2017preprint1,RuskovTahan2019PRB99}
both at the full sweet spot or far from it, e.g.
in the resonant-exchange (RX) qubit regime \cite{Medford2013PRL,TaylorSrinivasa2013PRL}
(near the charge degeneracy point, see Fig.~\ref{fig:1}b),
where the transverse coupling, $g_{\perp}$, is the largest.
In fact, at the charge degeneracy point (c.d.p.) the quantum capacitance $C_q$
increases by a factor of
$\frac{U_{\rm charge}^3}{t_c^3} \approx 10^2 - 10^3$
with respect to the  full sweet spot regime
(here, we have used $U_{\rm charge} = 0.5\, {\rm meV}$ and $t_c = 20 - 40\, \mu{\rm eV}$
for the dot's charging energy and interdot tunneling, respectively).
Since the longitudinal geometric entangling rate is of the order of
$\Gamma_{\rm ent,\tilde{g}_{\parallel}}^{\rm geom} \sim \tilde{g}_{\parallel}$
(Sec.~\ref{ChII-Spin-dependent force: entangling-time}),
a comparison with the standard transverse entangling rate, $\Gamma_{g_{\perp}} \sim g_{\perp}^2/\Delta$,
gives at the c.d.p.:
\begin{equation}
\frac{\Gamma_{\rm ent,\tilde{g}_{\parallel}}^{\rm geom}}{\Gamma_{g_{\perp}}}  \sim
\left. \frac{\tilde{g}_{\parallel}}{g_{\perp}^2/\Delta} \right|_{c.d.p.}
= 2 \left( \frac{\eta}{\hbar} \right)^{-1}\, \frac{C_q^{\rm \, c.d.p.}\,\, \tilde{V}_m}{e}\,
\frac{\Delta}{\omega_r} \gtrsim 10 \, .
\label{entangling-ratio-g_parallel}
\end{equation}
This demonstrates that the longitudinal geometric entangling rate can be substantially larger at the c.d.p.
(note, however, that this particular estimation
implies adiabaticity: $\omega_r \ll \omega_q \sim \frac{t_c}{\hbar}$).
The drawback of the c.d.p. regime is obviously the
large charge noise, see, e.g., Ref.~\onlinecite{Petta2017Science}.

At the full spin-qubit sweet spot,
referred to as  symmetric operating point\cite{ReedHunter2016PRL,MartinsKuemethMarcus2016PRL} (SOP)
of a DQD-ST qubit,
and as the  always-on exchange-only (AEON) regime\cite{AEON2016} of a TQD qubit,
the electric dipole moment goes to zero\cite{RuskovTahan2017preprint1,RuskovTahan2019PRB99},
and the longitudinal (curvature) coupling is
the only remaining\cite{RuskovTahan2017preprint1,RuskovTahan2019PRB99},
while  the charge noise to the qubit is suppressed.
Despite the much smaller quantum capacitance at the sweet spot,
a parameter regime can be provided
where
fast entangling multi-spin gates
can be performed,
{\it while each of the  qubits involved is residing in its  full sweet spot},
with a gate time
of few tens to hundred of ns.

\subsection{Infidelities of the longitudinal entangling geometric phase gate}

In this paper we study various kinds of  imperfections that can deteriorate the
accumulated phase gate performance.
One major imperfection may come from the presence of the always on
``dispersive-like'' couplings
$\delta\omega^{(j)}$, Eq.~(\ref{dispersive-like-interaction}), which leads to
a phase-gate infidelity
since it  leaves small qubit-resonator entanglement at the end of each phase accumulation cycle.
This can be partially canceled applying an
echo-like technique \cite{HaljanMonroe2005PRL,Kerman2013}.
To suppress this infidelity one accumulates the necessary phase
in some {\it even number $N$ of cycles},
where on each next cycle the sign of the dynamical longitudinal coupling,
is changed by proper change of the phase $\varphi_m$ of the qubits' gate modulation:
$\tilde{g}_{\parallel} \to - \tilde{g}_{\parallel}$.
This is, however,
not sufficient to reach very small infidelities ($\lesssim 10^{-3}$)
and one generally requires the ratio of the two curvature couplings
\begin{equation}
\frac{\delta\omega}{\tilde{g}_{\parallel}} = \frac{\eta}{\hbar} \, \frac{\hbar \omega_r}{e \tilde{V}_m}
\label{curvature-couplings-ratio}
\end{equation}
to be small (that is reachable experimentally), implying smaller ratio $\frac{\eta}{\hbar}$
and smaller resonator frequencies.
The small ratio $\frac{\delta\omega}{\tilde{g}_{\parallel}}$
provides  the main restriction on the choice of parameters when one
is aiming to approach
a high-fidelity phase gate.

Another type of phase-gate infidelity  arises from  the resonator
voltage  noise (Johnson noise),
which affects the resonator trajectory in the phase space,
and thus, the accumulated phases.
The gate infidelity shows
two distinct contributions:
(i) due to the
fluctuations of the resonator field that scales
with the number of cycles
as $\sim \sqrt{N}$,
and (ii) due to the
fluctuations of the associated
accumulated phase with a scaling $\sim 1/\sqrt{N}$.
The suppression of the Johnson noise infidelity requires
low enough temperatures and weak
resonator coupling to environment
(i.e., resonator photon leakage, $\kappa$, much smaller than resonator frequency,
$\omega_r$).

The resonator   
{\it photon number fluctuations}  also
cause qubits' dephasing and ac-Stark frequency shifts,
mediated
via the qubit-resonator curvature interactions, ${\cal H}_{\delta\omega}$, ${\cal H}_{\parallel}$,
leading to small phase gate infidelities that
were estimated to be negligible, see Table \ref{tab:rates}.

Finally, we also estimate the phase-gate error due to qubit dephasing via the
qubit gate charge noises,
which seems to be the main obstacle
to obtain high-fidelity two spin-qubit gate.
While charge noise will be minimized at the qubit's  full sweet spot
due to zeroing of the linear QD voltage fluctuations
(e.g., $\sim \frac{\partial E_q}{\partial V_{\varepsilon,m} } \delta V_{\varepsilon,m} = 0$),
there remain two other sources of gate voltage noise:
(i)
{\it quadratic effects},
$\sim \frac{\partial^2 E_q}{\partial V_{\varepsilon,m}^2 } \delta V_{\varepsilon,m}^2$
and
(ii)
{\it tunneling gate voltage fluctuations},
$\sim \frac{\partial E_q}{\partial t_c } \delta t_c$,
both leading to $1/f$-noise\cite{RussBurkard2015PRB,RussGinzelBurkard2016PRB}
that cannot be canceled by the simple echo-like procedure mentioned above.
In the current experiment at the SOP of a DQD qubit\cite{ReedHunter2016PRL,MartinsKuemethMarcus2016PRL}
a dephasing time
of $T_2^* \lesssim 1.5\, \mu{\rm s}$ was measured.
The estimated $1/f$-noise phase gate-infidelity scales as $\sim \left(t_{\rm gate}/T_2^*\right)^2$
and can reach $\lesssim 10^{-2}$ for moderate parameters;
%
this can be improved by making the geometric phase gate faster,
e.g., by increasing the qubit quantum capacitance $C_q$
(respectively, the longitudinal coupling $\tilde{g}_{\|}$).
While this increases the quadratic noise effects, we show, they are still much smaller
than the tunneling gate charge noise, which gives a hope
to make all the qubit phase gate infidelities to reach the level of $10^{-3}$, Table~\ref{tab:rates}.

An additional effect of $1/f$-charge noise on the tunnelings, $t_{l,r}$, Fig.~\ref{fig:1}a,
leads to fluctuation of the longitudinal couplings, $\tilde{g}_{\|}^{(j)}$.
The corresponding infidelity is shown to be negligible, see Table~\ref{tab:rates}.

The remaining of the paper is organized as follows.
In Chapter \ref{ChII-Spin-dependent force: entangling-time}
we consider briefly the ideal case of an $n$-qubit phase gate,
calculating the qubits' phase gate matrix.
Then we specialize on the two-qubit case, calculating
the gate time for a controlled $\pi$-phase gate\cite{Leibried2003N}.
In Chapter \ref{ChIII-Qubit-and-resonator-with-decoherence}
the main results for the various kind of phase gate infidelities
are obtained for the $n$-qubit case.
Consequences for a two-qubit controlled $\pi$-phase gate are considered.
Calculated gate times and infidelities
for experimentally reachable ranges of parameters are summarized in Table~\ref{tab:rates}.
In Chapter \ref{Sec: Relevance to other work} relevance to other works is given.
In Chapter \ref{ChIV-Discussion-and-Summary} we provide discussion and
summary of the obtained results.
Figures 1 through 10 sketch the idea of the longitudinal multiqubit phase gates
and present numerical plots of the leading infidelities vs. chosen range of parameters.
In the Appendices~\ref{app-A: Geometric phases} through
\ref{app E:Infidelity due to linear and quadratic charge noise}
we described important details of the derivations presented in the main text of the paper.

\section{Spin-dependent ``force'': ideal $n$-qubit entangling geometric phase gate}
\label{ChII-Spin-dependent force: entangling-time}

The time modulated longitudinal (curvature) interaction in Eq.~(\ref{total_Hamiltonian0}) for $n$ qubits,
$
{\cal H}_{\parallel} \equiv  \sum_j \hbar \left\{ \tilde{g}^{(j)}_{\parallel}\,  \sigma_z +
              \tilde{g}^{(j)}_0 \right\} \cos (\omega_m t + \varphi_m) (\hat{a} + \hat{a}^{+} )
$,
is generating a spin-dependent ``force'':
$\hat{F}(t) = -\partial {\cal H}_{\parallel}/\partial \hat{x}$, where
$\hat{x} \equiv \Delta x_0 (\hat{a} + \hat{a}^{\dagger})$ is the ``position'' operator,
with $\Delta x_0 = e \sqrt{\frac{\hbar/e^2}{2 \omega_r L_r }}$
being the resonator zero point motion.
In a rotating frame with the resonator frequency $\omega_r$ and in a rotating wave approximation (RWA),
$\hat{F}(t)$  will  drive the resonator at the difference frequency
$\omega_m - \omega_r$
with an amplitude
depending on the $n$-qubit state $|s \rangle$, Eq.~(\ref{s-basis-states}).
The $n$-qubit dependent resonator
driving amplitude $\Omega_{\parallel, s}$
is given by (see Appendix \ref{app-A: Geometric phases}):
\begin{equation}
\Omega_{\parallel, s} \equiv
\langle s| \hat{\Omega}_{\parallel}  | s\rangle
= \langle s|\, \frac{1}{2} \sum_{j=1}^n
\left[ \tilde{g}^{(j)}_{0} + \tilde{g}^{(j)}_{\parallel} \sigma_z^{(j)}  \right] \, | s\rangle .
\label{omega-n-qubit-coupling}
\end{equation}
Starting with an arbitrary resonator coherent state $|\alpha(0)\rangle$,
it evolves for finite time to $|\alpha_s^{\rm id}(t)\rangle$,
with the difference frequency $\delta \equiv \omega_m - \omega_r$:
\begin{equation}
\alpha_s^{\rm id}(t) = \alpha(0) -
\left(\frac{\Omega_{\parallel, s} }{\delta} \right)\, e^{-i\, \varphi_m}\, \left(1 - e^{-i t\delta} \right)
\label{ideal_alpha0}
\end{equation}
(here ``id'' stands for an ideal evolution).
For an initial product state of $n$-qubits plus resonator,
at intermediate times they become entangled:
\begin{equation}
|\Psi_i\rangle = \sum_s a_s |s\rangle |\alpha(0)\rangle  \rightarrow
|\Psi_f (t)\rangle = \sum_s a_s e^{i\Phi_s (t)} |s\rangle |\alpha_s(t)\rangle .
\label{n-qubit-resonator-evolution}
\end{equation}
via the spin-dependent geometric phases $\Phi_s(t)$;
the latter are path-dependent in
the resonator phase space $\{{\rm Re}\alpha, {\rm Im}\alpha\}$
and
for harmonic modulation
read (independent from the initial modulation phase $\varphi_m$)
\begin{equation}
\Phi^{\rm id}_s(t) = {\rm Im}\left[ \int^{\alpha_s(t)} \alpha_s^*(t') d\alpha_s(t') \right] =
\left( \frac{\Omega_{\parallel, s} }{\delta}  \right)^2  \left[ \delta\, t - \sin \delta\, t \right]
\label{ideal_Phi0}
\end{equation}

\begin{figure} [t!] 
    \centering
        \includegraphics[width=0.45\textwidth]{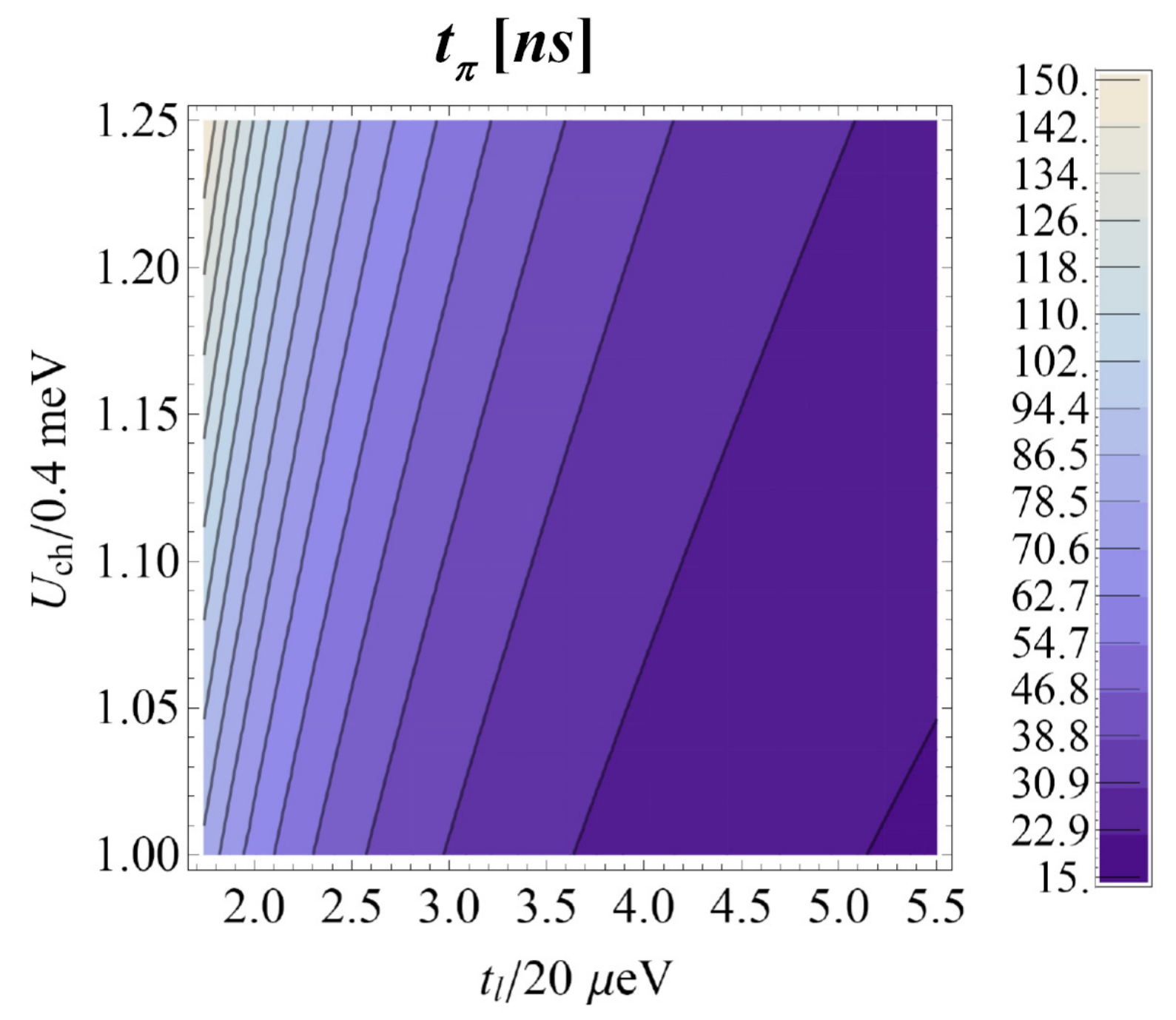}
        \caption{
        Gate time $t_{\pi}$ of a 2-qubit controlled $\pi$-phase gate for TQD exchange only qubits,
        Eq.~(\ref{entangling_time}),
        is plotted vs. interdot tunneling amplitude $t_l=t_r$ and dots' charging energy $U_{\rm ch}$,
        for fixed qubits' modulation, $\tilde{V}_m = 0.1\, {\rm mV}$,
        resonator frequency $\omega_r \simeq 6.3\, {\rm GHz}$,  
        resonator inductance $L_r = 50\, {\rm nH}$
        (impedance $Z_r \simeq \omega_r L_r \simeq 1.85\, {\rm k}\Omega$;
        compare, e.g. with Ref.\onlinecite{Stockklauser-Ihn-Ensslin-Wallraff2017PRX}),
        and qubits'-resonator lever arm
        $\alpha_c \simeq 0.14$,
        that amounts to a reachable coupling ratio of
        $\eta/\hbar = \alpha_c \sqrt{\frac{Z_r}{\hbar/e^2}} \simeq 0.1$.
        The gate time $t_{\pi}$ ranges from 15 ns to 150 ns for $t_l \simeq [35,110]\, \mu{\rm eV}$
        and $U_{\rm ch} \in [0.4,0.5]\, {\rm meV}$, featuring relatively high tunneling amplitudes
        and relatively low dots' charging energies.
        The scaling of the gate time, $t_{\pi} \sim 1/\sqrt{L_r}$,
        leads to a moderate increase for smaller resonator inductances $L_r$.
        By reaching higher experimental values for $\alpha_c$, $\omega_r$, and $\tilde{V}_m$
        one can bring dots' tunneling and charging energy to experimentally more favorable ranges\cite{Petta2017Science}.
        %
        Since for a DQD S-T qubit the scaling of the longitudinal coupling,
        Eq.~(\ref{dynamic-long-coupling-parallel}), with parameters
        is exactly the same\cite{RuskovTahan2019PRB99}, one expects similar ranges for these systems.
        }
        \label{fig:41}
\end{figure}

Using that $\hat{\Omega}_{\parallel,s}$ and $\hat{\Omega}_{\parallel,s}^2$ have the same eigenstates, $| s\rangle$,
one also derives the general {\it accumulated phase matrix for $n$ qubits}
(dropping the common phase, see Appendix  \ref{app-A: Geometric phases}, Eq.~\ref{Phi-matrix}):
%
\begin{eqnarray}
&&\hat{\Phi}(t)   
= \frac{\left( \sin \delta\, t - \delta\, t\right)}{2 \delta^2} \times
\nonumber\\
&& \qquad \times
\left[ \sum_{j<k}^n \tilde{g}^{(j)}_{\parallel} \tilde{g}^{(k)}_{\parallel}\, \sigma_z^{(j)} \otimes \sigma_z^{(k)}
+ \sum_{j=1}^n \tilde{g}^{(j)}_{0}\, \sum_{k=1}^n \tilde{g}^{(k)}_{\parallel}\, \sigma_z^{(k)} \right]
\qquad
\label{Phi-matrix-0} ,
\end{eqnarray}
where the second (double) sum amounts to single qubit operations,
that can be canceled at once by the synchronous resonator driving, Eq.~(\ref{cancellation-of-g0}).
For a gate time of $N$ completed cycles, $t_g = \frac{2\pi N}{\delta}$,
the resonator returns to its initial state,
$|\alpha_{s}(t_g)\rangle = |\alpha(0)\rangle$, independent of the $n$-qubit spin configuration.
Thus, the qubits-resonator state  again becomes disentangled,
while the $n$-qubit state, Eq.~(\ref{n-qubit-resonator-evolution})
acquires the phases $\Phi_s (t_g) = 2\pi (\Omega_{\parallel, s}/\delta)^2$
that comprises an $n$-qubit entangling gate.

Specializing for two qubits (and performing  $N$ cycles),
the spin-dependent accumulated phase
matrix is
\begin{eqnarray}
&& \hat{\Phi}_N = - \frac{2\pi N}{\delta^2}
\left[\tilde{g}_{\parallel}^{(1)} \tilde{g}_{\parallel}^{(2)} \sigma_z^{(1)}\otimes \sigma_z^{(2)} \right.
\nonumber\\
&& \qquad\ \left.
{ } +  \left( \tilde{g}_0^{(1)} + \tilde{g}_0^{(2)} \right)
\left( \tilde{g}_{\parallel}^{(1)}\sigma_z^{(1)} + \tilde{g}_{\parallel}^{(2)}\sigma_z^{(2)} \right) \right]
\label{two-qubit-phase} ,
\end{eqnarray}
which comprises a two-qubit gate up to single-qubit rotations (Appendix  \ref{app-A: Geometric phases}).
For the case of a TQD exchange-only qubit these  rotations can be performed for each qubit $(j)$
while residing in  their full sweet spot,
by simply manipulating the tunneling amplitudes, $t^{(j)}_l$, $t^{(j)}_r$,
see Ref.~\onlinecite{AEON2016,RuskovTahan2019PRB99}.

Using Eq.~(\ref{two-qubit-phase}) for two particular qubits  
for the choice of the frequency detuning $\delta_{\pi}(N)$ such that
$\delta_{\pi}^2 = 4 N \tilde{g}_{\parallel}^{(1)} \tilde{g}_{\parallel}^{(2)}$
one obtains an accumulated two-qubit phase
matrix
$\hat{\Phi}_{\pi} = -\frac{\pi}{4} \sigma_z^{(1)}\otimes \sigma_z^{(2)}$
corresponding to a controlled $\pi$-phase gate\cite{Leibried2003N,Kerman2013}
(Appendix  \ref{app-A: Geometric phases}).
For this ideal situation to happen
one requires that the frequency detuning to be
at least
$\delta_{\pi}(N) \gg \kappa$,
with $\kappa$ being the resonator damping rate, implying a high $Q$-factor SC resonator
($Q \gtrsim 10^3 - 10^6$).
Other restrictions on the resonator and qubit parameters will follow from
minimization of the two-qubit gate infidelities.
%
%
The entangling gate
time reads:
\begin{equation}
t_{\pi} = \frac{2\pi N}{\delta_\pi} = \pi \sqrt{\frac{N}{\tilde{g}_{\parallel}^{(1)} \tilde{g}_{\parallel}^{(2)}} }
\label{entangling_time}
\end{equation}
and reaches $15 - 150\, {\rm ns}$ for the parameters of Fig.~\ref{fig:41},
see also Table~\ref{tab:rates}.

\section{Qubit and resonator with decoherence and the phase gate errors}
\label{ChIII-Qubit-and-resonator-with-decoherence}
\begin{figure} [t!] 
    \centering
        \includegraphics[width=0.45\textwidth]{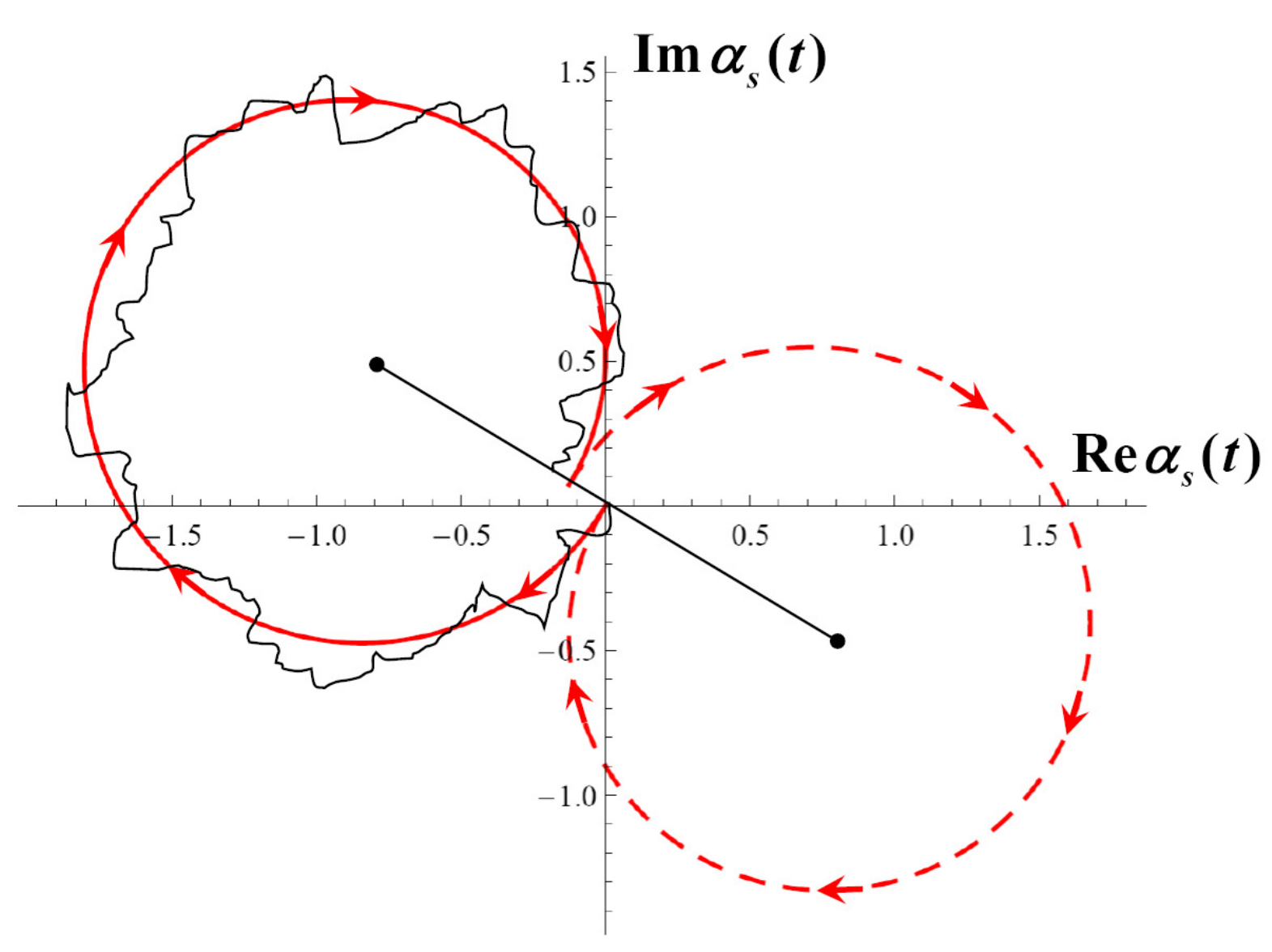}
        \caption{The non-ideal oscillator trajectories in the phase space
        $\{{\rm Re}[\alpha_{s}(t)], {\rm Im}[\alpha_{s}(t)]\}$
        for two subsequent cycles, each of duration $t_g = 2\pi/\delta$.
        The ideal evolution $\alpha^{\rm id}_{s}(t)$ obtains deviations $\delta\alpha_{s}(t)$
        (red solid or dashed decaying circles)
        due to
        (i) resonator damping $\kappa$, Eqs.~(\ref{damping_diffusion-CL}), (\ref{alpha-equation}), and
        (ii) qubit-induced  spin-dependent resonator  frequency shifts, $\delta\omega_s$,
        Eqs.~(\ref{total_Hamiltonian0}), (\ref{damping_diffusion-CL}), (\ref{alpha-equation}),
        via the always on ``dispersive-like'' curvature coupling, Eq.~(\ref{dispersive-like-interaction}).
        By changing the modulation phase for every odd cycle: $\varphi_m' = \varphi_m + \pi$, the spin-dependent
        resonator driving amplitude $\Omega_{\parallel, s}$ flips sign (red, dashed decaying circle; sf. also Fig.~\ref{fig:3circ})
        that allows partial cancelation of the deviations, for small $\kappa$, $\delta\omega_s$.
        The (iii) resonator thermal (Johnson) noise  is shown schematically (black, solid line)
        as a noisy deviation from the ideal trajectory $\alpha^{\rm id}_{s}(t)$.
        The noisy trajectory is generated by a random force Hamiltonian, ${\cal H}_f = -\xi_f(t)\hat{x}$
        [see Eq.(\ref{random-force}) and Appendix \ref{App-D: Johnson nosie infidelity}],
        which is an unraveling of the diffusion term in the ensemble-averaged
        evolution, Eq.~(\ref{damping_diffusion-CL}).
        }
        \label{fig:4circ-non-ideal}
\end{figure}

The qubit entangling gate errors arise from several sources (see Fig.~\ref{fig:4circ-non-ideal}),
including (i) resonator damping $\kappa$,
(ii) qubit-induced spin-dependent resonator  frequency shifts $\delta\omega_s$
via the always on ``dispersive-like'' curvature coupling,
Eqs.~(\ref{dispersive-like-interaction})-(\ref{dispersive-like-coupling}),
(iii) resonator thermal (Johnson) noise,
and
qubit dephasing due to (iv)  photon number fluctuation in the resonator and
due to (v)  charge $1/f$-noise on the qubit defining gates;
(vi) finally,  one also considers the effect charge $1/f$-noise on the qubit's tunneling gates,
causing additional resonator trajectory deviations via a change of the curvature (longitudinal)
coupling, $\tilde{g}_{\parallel}$. 
The effects (iv), (v), and (vi) of qubits' defining gates charge noise 
are considered latter on in Secs.~\ref{Sec: Qubit charge noise phase gate errors}
and \ref{Sec: Infidelity via the charge fluctuations of the longitudinal (curvature) coupling}.

The first three effects, (i)-(iii), as well as (vi),
cause (time dependent) deviations from the ideal {\it resonator trajectory}
in the phase space, Eqs.~(\ref{ideal_alpha0}) and (\ref{ideal_Phi0}),
changing both the (spin-dependent) resonator trajectories, Fig.~\ref{fig:4circ-non-ideal},
\begin{equation}
\alpha_{s}(t) = \alpha^{\rm id}_{s}(t) + \delta\alpha_{s}^{\kappa,\delta\omega} + \delta\alpha_{s}^{\xi}
+ \delta\alpha_{s}^{\tilde{g}_{\parallel},1/f}
\label{delta-alpha}
\end{equation}
and the corresponding accumulated phases,
\begin{equation}
\Phi_{s}(t;[\alpha_{s}(t)]) = \Phi^{\rm id}_{s}(t) + \delta\Phi_{s}^{\kappa,\delta\omega} + \delta\Phi_{s}^{\xi}
+ \delta\Phi_{s}^{\tilde{g}_{\parallel},1/f}
\label{delta-Phi} ,
\end{equation}
the latter causing path-dependent
non-local in time errors.
Correspondingly, as compared to the ideal values,
the change in the final resonator state $|\alpha_s(t_g) \rangle$,
leaves some qubit-resonator entanglement at the end of the phase gate accumulation cycle,
leading to qubits'
phase gate infidelity,
$\delta\varepsilon_{\rm traject}$.
For small deviations of the trajectory:
$\delta\alpha_{s}^{\kappa,\delta\omega}, \delta\alpha_{s}^{\xi}, \delta\alpha_{s}^{\tilde{g}_{\parallel},1/f}  \ll 1$
and
$\delta\Phi_{s}^{\kappa,\delta\omega}, \delta\Phi_{s}^{\xi}, \delta\Phi_{s}^{\tilde{g}_{\parallel},1/f} \ll 1$,
the infidelity splits into  
separate terms
(Appendix \ref{app-C:combined-infidelity-noise-curvature-detuning}):
\begin{equation}
\delta\varepsilon_{\rm traject} \simeq \delta\varepsilon_{\kappa,\delta\omega} + \delta\varepsilon_{\xi}
+ \delta\varepsilon_{\tilde{g}_{\parallel},1/f}
\end{equation}

The superconducting (SC)  resonator-to-environment  interaction at finite temperature   
is described via
the Caldeira-Leggett master equation (ME)\cite{CaldeiraLeggett1983a,CaldeiraCerdeira1989}
and includes damping ($\kappa$) and diffusion ($K_d$) contributions.
The time evolution of the qubit(s)-resonator density matrix reads:
\begin{eqnarray}
&&\frac{d\rho}{dt} = -i \left[{\cal \tilde{H}}_{\rm tot},\rho\right]
-i\frac{\kappa}{2\hbar} \left[\hat{x}, \left\{\hat{p}, \rho \right\}_{+} \right]
-\frac{K_d}{\hbar^2}\, \left[\hat{x}, \left[\hat{x}, \rho \right] \right]   \qquad
\label{damping_diffusion-CL}
\\
&& K_d \equiv \frac{\hbar \omega_r L_r \kappa}{2}\, \coth{\frac{\hbar \omega_r}{2 k_B T_r}} ,
\label{K_d}
\end{eqnarray}
where
$\hat{x}$, $\hat{p}$,
are the ``position'' and ``momentum'' operators \cite{RLC-position-momentum},
Eq.~(\ref{position-momentum}),
and $\{\, ,\, \}_{+}$ is anticommutator.
The last (double commutator) term in Eq.~(\ref{damping_diffusion-CL})
is governed by
a temperature dependent diffusion coefficient,
$K_d$,
and $T_r$ is the resonator temperature\cite{Markovian}.
%
(An analog of the Caldeira-Leggett ME (\ref{damping_diffusion-CL})
is the quantum-optics ME \cite{MilburnWalls-book2008};
the two master equations coincide in the RWA and especially for Gaussian states considered in this paper.
We will use one or another form of the ME for convenience, Appendix \ref{App-B:Equations-of-motion}.)

The qubit-resonator density matrix can be expanded in a complete set of qubit operators
$|s\rangle \langle s'|$  
\cite{WisemanMilburn-book2010}:
\begin{equation}
\rho = \sum_{s,s'} \hat{\rho}_{s,s'} |s\rangle \langle s'|
\label{expanded-density-matrix}
\end{equation}
where
the partial density matrices    
$\hat{\rho}_{s,s'}$  act only on the resonator subspace.
Using Eq.~(\ref{damping_diffusion-CL}), in a general rotating frame with
frequency $\omega_{r'}$
and in RWA, one gets the equation for the spin-diagonal resonator density matrices
(Appendix~\ref{App-B1: Evolution for the partial density matrices}):
\begin{eqnarray}
&& \frac{d\hat{\rho}_{ss}}{dt} = -i \tilde{\omega}_r \left[a^{+} a,\hat{\rho}_{ss}\right]
-i \Omega_{\parallel, s}
\left[\hat{X}_{\varphi_m}(t),\hat{\rho}_{ss}\right]
\nonumber\\
&&\qquad\ \ { } -i \varepsilon_d \left[\hat{X}_{\varphi_d}(t),\hat{\rho}_{ss}\right]
-i \delta \omega_s \left[a^{+} a,\hat{\rho}_{ss}\right]
\nonumber\\
&&\qquad\qquad { } -i\frac{\kappa}{2\hbar} \left[\hat{x} \left\{\hat{p}, \hat{\rho}_{ss} \right\}_{+} \right]
-\frac{K_d}{\hbar^2}\, \left[\hat{x} \left[\hat{x}, \hat{\rho}_{ss} \right] \right] ,
\label{partial-dm-diagonal}
\end{eqnarray}
where the modulating and driving terms, $\hat{X}_{\varphi_m}$, $\hat{X}_{\varphi_d}$, are given by
$\hat{X}_{\varphi}(t) \equiv \left[ \hat{a} e^{i (\tilde{\delta} t + \varphi)} +
\hat{a}^{+} e^{-i (\tilde{\delta} t + \varphi)}\right]$, 
and the detunings are
\begin{equation}
\tilde{\omega}_r = \omega_r - \omega_{r'}, \ \ \tilde{\delta} = \omega_m - \omega_{r'}
\label{detunings} .
\end{equation}
The resonator frequency shift for $n$ qubits in the $|s \rangle$-state is given by:
\begin{equation}
\delta \omega_s =  \langle s| \sum_j \delta\omega^{(j)} \sigma_z^{(j)} | s\rangle
\label{spin-dependent-frequency-shift} ,
\end{equation}
with the individual curvature frequency shifts, $\delta\omega^{(j)}$,
given by Eq.~(\ref{dispersive-like-coupling}).
The spin-diagonal Eq.~(\ref{partial-dm-diagonal})
can be reduced to equations
for the set of moments,
e.g.,
the averages $\bar{x}_s \equiv \langle \hat{x}\rangle_s = {\rm Tr}\left[\hat{x} \hat{\rho}_{s,s}\right]$,
$\bar{p}_s \equiv \langle \hat{p}\rangle_s = {\rm Tr}\left[\hat{p} \hat{\rho}_{s,s}\right]$,
the variances,
$D^{(s)}_x \equiv \langle \hat{x}^2 \rangle_s - \langle \hat{x}\rangle_s^2$,
$D^{(s)}_p \equiv \langle \hat{p}^2 \rangle_s - \langle \hat{p}\rangle_s^2$,
$D^{(s)}_{xp} \equiv
\langle \frac{\hat{x}\hat{p}+ \hat{p}\hat{x}}{2} \rangle_s - \langle \hat{x}\rangle_s \langle \hat{p}\rangle_s$,
etc.,
see Appendix~\ref{App-B2: equations of motion for averages and variances}.
One obtains for the averages:
\begin{eqnarray}
&& \frac{\dot{\bar{x}}_s }{\Delta x_0} =  \frac{\bar{p}_s}{\Delta p_0} \left( \tilde{\omega}_r  - \delta\omega_s \right)
- 2 \Omega_{\parallel, s} \, \sin (\tilde{\delta}  t + \varphi_m)
\label{dx}
\\
&& \frac{\dot{\bar{p}}_s }{\Delta p_0} = -  \frac{\bar{x}_s}{\Delta x_0} \left( \tilde{\omega}_r  - \delta\omega_s \right)
- 2 \Omega_{\parallel, s}\,  \cos (\tilde{\delta}  t + \varphi_m)
\nonumber\\
&& \qquad\qquad { } - 2\varepsilon_d\, e^{-i (\tilde{\delta} t + \varphi_d)}  - \kappa\, \frac{\bar{p}_s }{\Delta p_0} ,
\label{dp}
\end{eqnarray}
We notice several important properties of Eqs.~(\ref{dx}) and (\ref{dp})
that coincide with the single resonator case \cite{RuskovSchwabKorotkov2005}.
First, for zero temperature a coherent resonator state remains coherent, while it is damped to
the ground state $|0\rangle$ at long times, $t \gg 1/\kappa$
(in particular, the state purity is preseved).
%
Secondly, the equations for the averages $\bar{x}$, $\bar{p}$
are not affected by the diffusion term and
decouple from the variances
(Appendix \ref{App-B2: equations of motion for averages and variances}).
In addition, we
show that the variances are not affected by the
longitudinal coupling modulation or by the resonator driving
(Appendix \ref{App-B2: equations of motion for averages and variances}).

The above statements are correct for any resonator state.
In what follows
we consider solutions of Eq.~(\ref{damping_diffusion-CL}) with Gaussian density matrices
since typical initial states are Gaussian (e.g., a coherent or a thermal state) and Gaussian states are
preserved under the evolution of Eq.~(\ref{damping_diffusion-CL}).
They are also preserved by continuous measurements,
(see Ref.~\cite{RuskovSchwabKorotkov2005} and references therein).
Thus, Eq.~(\ref{partial-dm-diagonal})
reduces to equations
for the set of five Gaussian moments, since the higher moments are expressed by the former.

Using the ``field'' variable,
$\alpha_s \equiv \langle \hat{a} \rangle_s =
\frac{1}{2}\left( \frac{\bar{x}_s}{\Delta x_0} + i \frac{\bar{p}_s}{\Delta p_0} \right)$,
Eqs.~(\ref{dx}), (\ref{dp}) are combined to:
\begin{eqnarray}
&& \dot{\alpha}_s = - i \left( \tilde{\omega}_r - \delta\omega_s \right)\, \alpha_s
-i \Omega_{\parallel, s} e^{-i (\tilde{\delta}  t + \varphi_m)} -i \varepsilon_d e^{-i (\tilde{\delta} t + \varphi_d)}
\nonumber\\
&& \qquad  { } - \frac{\kappa}{2} \left( \alpha_s - \alpha^*_s \right) ,
\label{alpha-equation}
\end{eqnarray}
where the difference from the quantum optics equation is the last (contra-rotating) term,
that can be neglected in a RWA.
Note, that disregarding it
will be equivalent
to neglecting the usual resonator frequency shift due to damping.
The Eq.~(\ref{alpha-equation}) for
$\alpha_s$ does not include the thermal diffusion,
the latter is giving a contribution only to the equations for the variances,
$D^{(s)}_x$, $D^{(s)}_p$, $D^{(s)}_{xp}$,
see Appendix \ref{App-B2: equations of motion for averages and variances},
%
that will contribute to the variances of the field, $\langle |\delta\alpha_s(t)|^2\rangle$,
and the accumulated phases, $\langle \delta\Phi_s(t)\,\delta\Phi_{s'}(t)\rangle$,
see Sec.~\ref{Sec: Resonator (Johnson) noise phase gate error}.

Since the gate error is non-local in time
[e.g., via the accumulated phases $\Phi_s(t;[\delta\alpha_s(t)])\propto \int \alpha_s^{*}(t') d\alpha_s(t')$,
Eq.~(\ref{ideal_Phi0})],
it is useful to represent the thermal diffusion term
as originating from
a stochastic time-dependent term (Appendix \ref{App-D: Johnson nosie infidelity}):
\begin{equation}
{\cal H}_f = - \xi_f(t) \hat{x} ,
\label{random-force}
\end{equation}
which represents a random force Hamiltonian;
here
$\xi_f(t)$ is a white noise ``random force'' with spectral density given by
the correlator $\langle \xi_f(t)\xi_f(t')\rangle = \frac{S_f}{2}\delta(t-t')$.
Then, the resulting fluctuations of the (spin-dependent) resonator trajectory, Fig.~\ref{fig:4circ-non-ideal},
will be integrated into the accumulated variances.

Thus, in Eq.~(\ref{damping_diffusion-CL}) one is replacing the last (double commutator) term
by the random force Hamiltonian,
Eq.~(\ref{random-force}).
%
%
As shown in Appendix \ref{App-D: Johnson nosie infidelity}, the two representations are equivalent.
Indeed, by adding the random force Hamiltonian to the equation of motion:
$\frac{d\rho}{dt} \sim -\frac{i}{\hbar}\left[{\cal H}_f, \rho \right]$,
and transforming from Stratonovich form of the equations to its It\^{o} form\cite{RuskovSchwabKorotkov2005},
one reproduces the double commutator term in Eq.~(\ref{damping_diffusion-CL})
if the spectral density is chosen as $S_f = 4 K_d$.
The ``standard'' ensemble averaged evolution, Eq.~(\ref{damping_diffusion-CL}),
is then obtained by
averaging out the noise. 
In what follows, in Sec.~\ref{Sec: Resonator (Johnson) noise phase gate error},
we will use the random force Hamiltonian to calculate the diffusion (Johnson) noise gate error,
by averaging out the noise at the end of the procedure.
In the next Section \ref{Sec:Damping and detuning gate errors}
we first consider the situation, when the ideal evolution in the phase space
is disturbed only by resonator damping and $n$-qubit spin-dependent detuning.

\subsection{Damping and detuning gate errors for $n$ qubits. 2-qubit numerical study}
\label{Sec:Damping and detuning gate errors}
For each $n$-qubit state $|s\rangle$,
damping $\kappa$ and
spin-dependent detuning  $\delta\omega_s$, lead to
shrinking and deviation of the ideal circle in the phase space $\{{\rm Re}\alpha, {\rm Im}\alpha\}$,
leaving the qubits and the resonator entangled
at the time of one cycle, $t_g = 2\pi/\delta$ (see Fig.~\ref{fig:4circ-non-ideal}),
[since $\alpha_s({\scriptscriptstyle \frac{2\pi}{\delta} }) \neq 0$], and disturbing the accumulation phases.
Starting from an initial product state of $n$ qubits and resonator,
$|\psi_i\rangle \equiv \sum_s a_s |s\rangle \, |0\rangle$
[we assume
vacuum resonator initial state, for simplicity],
one ends up in the state
\begin{equation}
|\psi_f\rangle = \sum_s  a_s  e^{i\Phi_{s}(t)} |s \rangle \, |\alpha_{s}(t)\rangle \mid_{t=\frac{2\pi}{\delta}}
\label{final-state}
\end{equation}
where $\alpha_{s}(t) = \alpha^{\rm id}_s(t) + \delta\alpha_s^{\kappa,\delta\omega}(t)$,
and $\Phi_{s}(t) = \Phi^{\rm id}_s(t) + \delta\Phi_s^{\kappa,\delta\omega}(t)$,
differ from the ideal values [scf. Eqs.~(\ref{ideal_alpha}),(\ref{accumulated_phase})]
by the small deviations, $\delta\alpha_s^{\kappa,\delta\omega}(t)$, $\delta\Phi_s^{\kappa,\delta\omega}(t)$,
due to
damping and detuning.
Using the matrix element
\begin{equation}
\langle \psi_f|\psi_f^{\rm id}\rangle =
 \sum_s |a_s|^2 e^{-i\delta\Phi_s (t)} \langle \alpha_s(t) |0 \rangle  ,
\label{fidelity-matrix-element}
\end{equation}
one obtains
for small deviations, $\delta\alpha_s^{\kappa,\delta\omega}$, $\delta\Phi_s^{\kappa,\delta\omega}$,
the gate error (infidelity) for $n$ qubits (Appendix \ref{app-C:combined-infidelity-noise-curvature-detuning}):
\begin{equation}
\delta\epsilon_{\kappa,\delta\omega}^{\rm n\, Qb}  \lesssim  \frac{1}{2^n} \sum_s |\delta\alpha_{s}^{k,\delta\omega}|^2
+ \frac{f_{12}^{(n)}}{2} \sum_{s < s'}
\left( \delta\Phi_{s}^{\kappa,\delta\omega} - \delta\Phi_{s'}^{\kappa,\delta\omega} \right)^2
\label{infidelity-kappa-delta-omega} ,
\end{equation}
where
the averaging over the initial $n$-qubit state,
$f_1^{(n)} = f_s^{(n)} \equiv \overline{|a_s|^2}$,  $f_{12}^{(n)} = f_{s,s'}^{(n)} \equiv \overline{2 |a_s|^2 |a_{s'}|^2}$,
(using uniformity of the averages, see Eq.~(\ref{n-qubit-averages-1}) and
Appendix \ref{app C2: Averaging over the n-qubit initial state})
results in:
\begin{equation}
f_s^{(n)} = f_{s,s'}^{(n)} = \frac{1}{2^n + \frac{2^{2n}-2^n}{2}}
\label{n-qubit-averages-11} .
\end{equation}
To lower the average gate infidelity, $\delta\epsilon_{\kappa,\delta\omega_s}$,
one needs to suppress
$\delta\alpha_s$
and  $\left(\delta\Phi_s - \delta\Phi_{s'}\right)$
at the end of the cycle.

The simplest strategy (Appendix \ref{app-C1:infidelity-damping-detuning})
is to change the sign of the
driving amplitude
$\Omega_{\parallel,s}$ by changing the phase of the qubits gates modulations,
$\varphi_m \to \varphi_m + \pi$, see Eq.~(\ref{alpha-equation}).
By performing a second cycle, with flipped sign of
$\Omega_{\|,s}$
an opposite shrinking   of the second cycle,
Fig.~\ref{fig:4circ-non-ideal},
will (partially)
compensate the first one (scf. Refs.~\cite{HayesMonroe2012PRL,Kerman2013}).
In the simple case, when detunings are neglected, $\delta\omega_s = 0$,
at the end of the second cycle
one gets, starting at $\alpha_s(0)=0$:
\begin{equation}
\delta\alpha_s({\scriptscriptstyle \frac{4\pi}{\delta}}) \equiv
\tilde{\alpha}_s({\scriptscriptstyle \frac{4\pi}{\delta}})
\simeq
-\alpha_s({\scriptscriptstyle \frac{2\pi}{\delta}})\, \left(\frac{\pi \kappa}{\delta}\right)
\label{alpha-two-cycles}
\end{equation}
i.e.,
the $\alpha$-deviation for two cycles
is suppressed by extra power of $\kappa/\delta$, for $\kappa \ll \delta$.
Here $\tilde{\alpha}_s(t)$ denotes a time evolution via Eq.~(\ref{alpha-equation})
with flipping sign of the modulation strength, $\Omega_{\parallel,s}$, after each cycle.
For the realistic case, when $\delta\omega_s \gg \kappa$,
one solves Eq.~(\ref{alpha-equation})
iteratively, for $N$ cycles ($N=1,2,3,\dots$)
with every even cycle with the sign of $\Omega_{\parallel,s}$ flipped,
and obtains for the deviation of the resonator variable,
$\delta\alpha_s$ in Eq.~(\ref{infidelity-kappa-delta-omega}):
\begin{equation}
\delta\alpha_s^{\kappa,\delta\omega} \equiv
\tilde{\alpha}_s\left( N\frac{2\pi}{\delta} \right) = \alpha_s\left( \frac{2\pi}{\delta} \right) \,
\frac{ e^{b_s^* \, N \frac{2\pi}{\delta}} - (-1)^N}{e^{b_s^* \frac{2\pi}{\delta}} + 1}
\label{alpha-s-deviation-0} ,
\end{equation}
where $b_s \equiv -i [\delta - \delta\omega_s] - \frac{\kappa}{2}$.

\begin{figure} [t!] 
    \centering
        \includegraphics[width=0.41\textwidth]{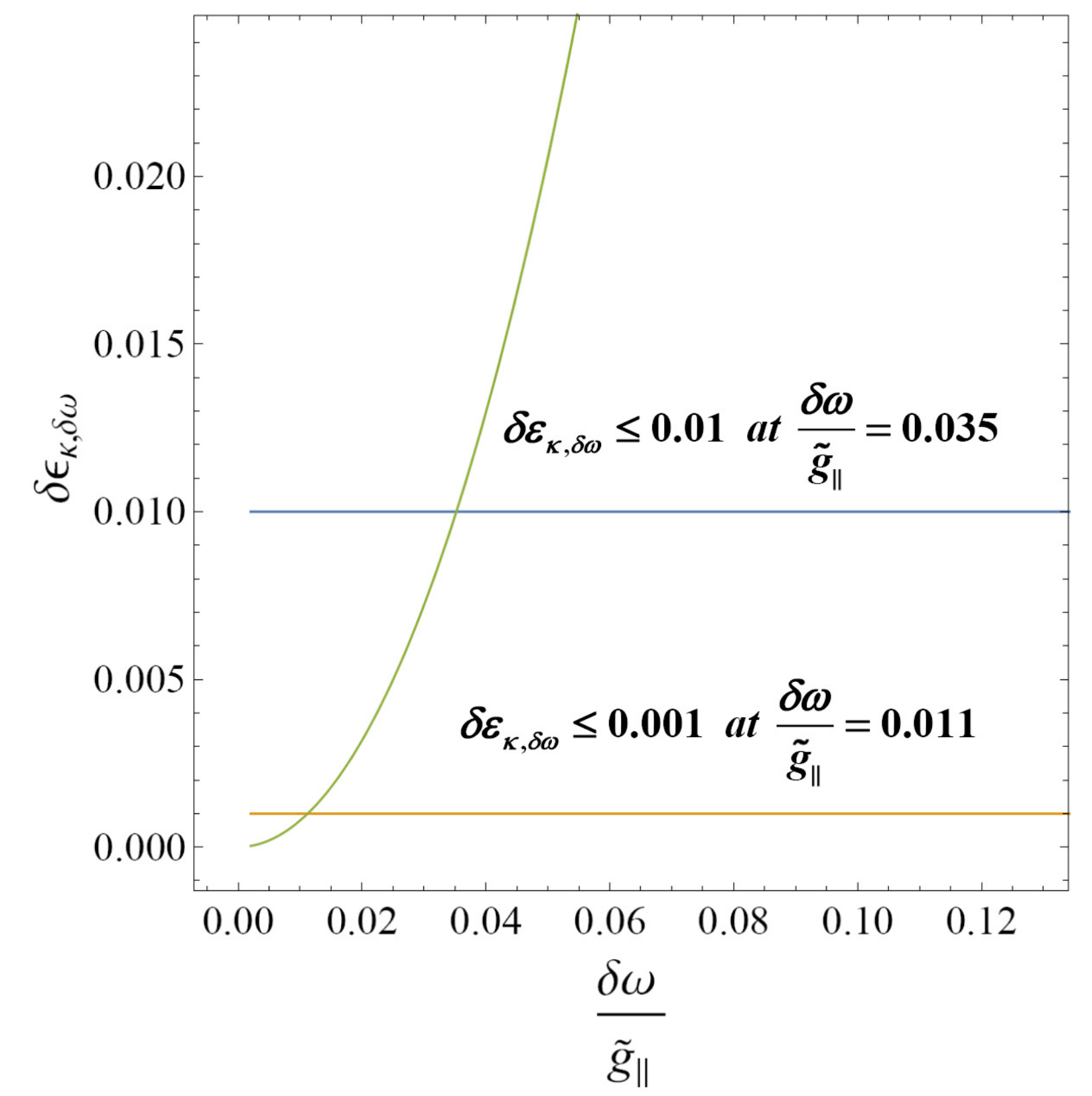}
        \caption{The infidelity
        $\delta\epsilon_{\kappa,\delta\omega}^{\rm 2\, Qb}\left(\frac{\delta\omega}{\tilde{g}_{\parallel}}, 0 \right)$,
        Eqs.~(\ref{infidelity-kappa-delta-omega}) and (\ref{scalings-of-infidelity-kappa-domega}),
        is a growing function of the ratio
        $\frac{\delta\omega}{\tilde{g}_{\parallel}} = \frac{\eta}{\hbar}\, \frac{\hbar\omega_r}{e\tilde{V}_m}$.
        The values of $\frac{\delta\omega}{\tilde{g}_{\parallel}} = 0.035,\ 0.011$
        at which $\delta\epsilon_{\kappa,\delta\omega} = 0.01,\ 0.001$,
        respectively, are calculated for resonator driving and phase, Eq.~(\ref{cancellation-of-g0})
        at which the qubits' spin-independent curvature couplings,
        $\tilde{g}_0^{(j)}$ are canceled, see Eq.~(\ref{omega-n-qubit-coupling-with driving-0}).
        }
        \label{fig:42a}
\end{figure}

The expressions for the calculation of $|\delta\alpha_{s}^{k,\delta\omega}|^2$ and
$\left( \delta\Phi_{s}^{\kappa,\delta\omega} - \delta\Phi_{s'}^{\kappa,\delta\omega} \right)^2$
are cumbersome and are presented in Appendix \ref{app-C1:infidelity-damping-detuning},
by Eqs.~(\ref{alpha-s-for-the-gates})-(\ref{delta-Phi-kappa-delta-omega}).
The $n$-qubit entangling gate is essentially driven only by the
spin-dependent longitudinal couplings, $\tilde{g}_{\parallel}^{(j)}$,
see Eq.~(\ref{Phi-matrix-0}).
The modulation of the qubits gates creates, however, {\it the spin-independent couplings}, $\tilde{g}_{0}^{(j)}$,
which affect the infidelity considerably.
Aiming to cancel the $\tilde{g}_{0}^{(j)}$ couplings of the qubits involved into the entangling gate,
one is driving the resonator with the same frequency as the modulation, $\omega_d = \omega_m$,
and with a phase $\varphi_d = \varphi_m + \pi$, see Eq.~(\ref{cancellation-of-g0}).
Then the quantities of interest are shown to depend on a
modified spin-dependent driving strength,
\begin{equation}
\Omega_{\parallel,s}^{\varepsilon} \equiv \Omega_{\parallel,s} - \varepsilon_d
\label{omega-n-qubit-coupling-with driving-0} ,
\end{equation}
and benefit from the cancelation of the $\tilde{g}_{0}^{(j)}$,
see Eqs.~(\ref{contribution-1-int})-(\ref{contribution-3-int}).
Numerically, for the choice of parameters 
of Figs.~\ref{fig:41} and \ref{fig:42a}, and Table~\ref{tab:rates},
we have shown that
the infidelity $\delta\epsilon_{\kappa,\delta\omega}$ improves more than 30 times
for the conditions of Eq.~(\ref{cancellation-of-g0}).

\begin{figure} [t!] 
    \centering
        \includegraphics[width=0.45\textwidth]{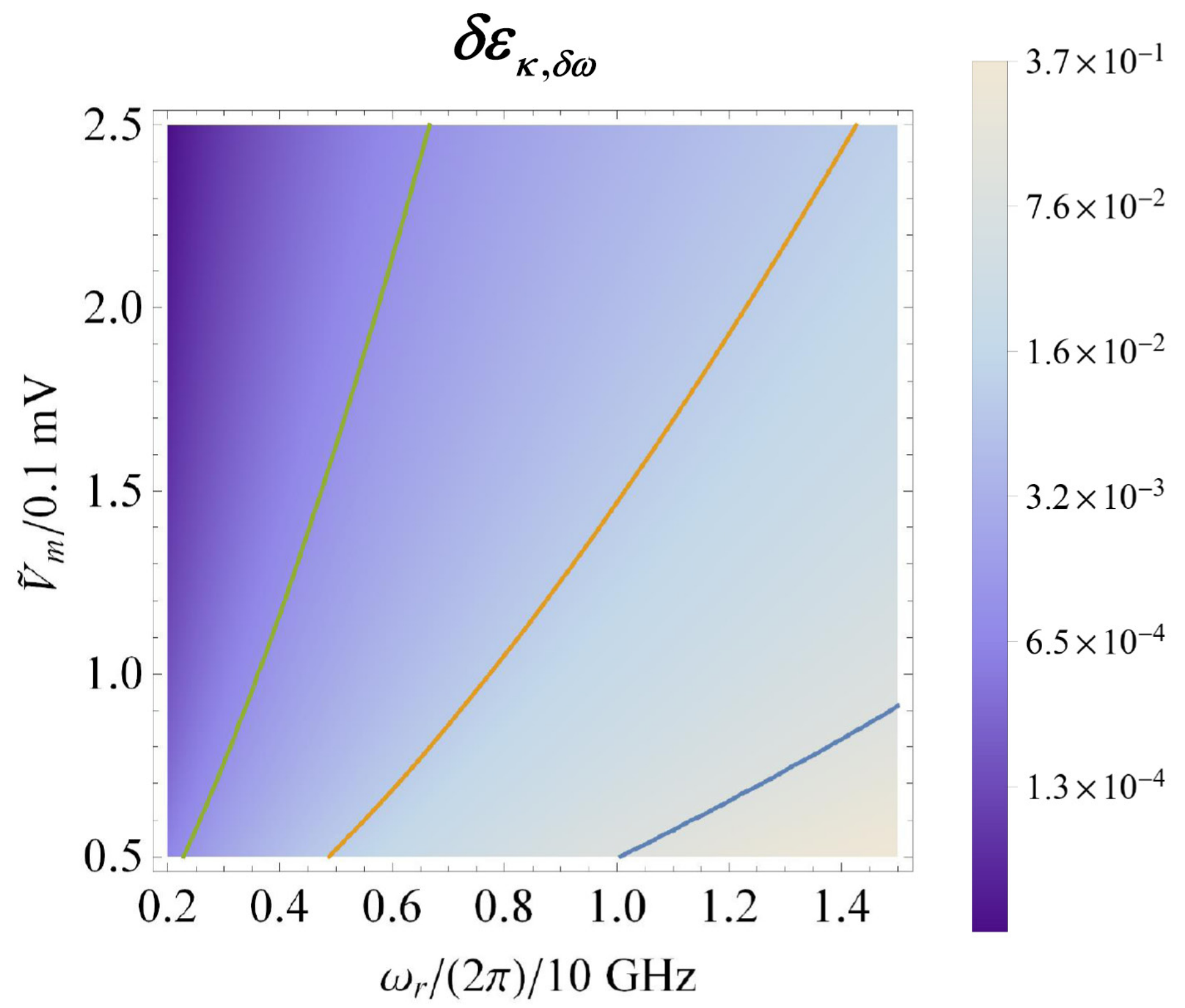}
        \caption{The infidelity $\delta\epsilon_{\kappa,\delta\omega}^{\rm 2\, Qb}(\omega_r, \tilde{V}_m)$
        for the range of resonator frequency $\omega_r \in [2, 10]\, {\rm GHz}$
        and qubits gate modulation voltage (amplitude), $\tilde{V}_m \in [0.05, 0.25]\, {\rm mV}$.
        Contours where the infidelity reaches the values of $0.1$, $0.01$, $0.001$ are also shown.
        To keep $\delta\epsilon_{\kappa,\delta\omega}$ constant requires keeping the ratio, $\frac{\omega_r^{3/2}}{\tilde{V}_m}$, constant
        (see text).
        }
        \label{fig:42b}
\end{figure}

While $\delta\alpha_s$  is  effectively
suppressed by increasing the number of (pair of) cycles,
the accumulated phase deviations
$\delta\Phi_s$  may remain large for finite detunings\cite{Walsh},
$\delta\omega_s$,
and can be suppressed only by suitable choice of the parameters.
In what follows, we consider a two-qubit gate infidelity,
taking identical curvature couplings, $\delta\omega^{(1)} = \delta\omega^{(2)}$,
and
$\tilde{g}_{\parallel}^{(1)} = \tilde{g}_{\parallel}^{(2)}$.
From dimensional considerations
the infidelity
$\delta\epsilon_{\kappa,\delta\omega}^{\rm 2\, Qb}$,
see Eq.~(\ref{infidelity-kappa-delta-omega}),
is a function of two dimensionless ratios:
\begin{equation}
\delta\epsilon_{\kappa,\delta\omega} =
\delta\epsilon_{\kappa,\delta\omega}
\left(\frac{\delta\omega}{\tilde{g}_{\parallel}}, \frac{\kappa}{\tilde{g}_{\parallel}} \right)
\simeq  \delta\epsilon_{\kappa,\delta\omega}\left(\frac{\delta\omega}{\tilde{g}_{\parallel}}, 0 \right) ,
\label{scalings-of-infidelity-kappa-domega}
\end{equation}
where, for high $Q \simeq 10^5 - 10^6$
the ratio $\frac{\kappa}{\tilde{g}_{\parallel}}$ can be set to zero.
Since $\frac{\delta\omega}{\tilde{g}_{\parallel}} = \frac{\eta}{\hbar}\, \frac{\hbar\omega_r}{e\tilde{V}_m}$,
Eq.~(\ref{curvature-couplings-ratio}), the infidelity is independent of
the qubits' energy curvature (quantum capacitance).
On Fig.~\ref{fig:42a} is shown numerically the dependence
of $\delta\epsilon_{\kappa,\delta\omega}^{\rm 2\, Qb}\left(\frac{\delta\omega}{\tilde{g}_{\parallel}}, 0 \right)$,
which grows rapidly with $\frac{\delta\omega}{\tilde{g}_{\parallel}}$.
For the values of $\frac{\delta\omega}{\tilde{g}_{\parallel}} = 0.035,\ 0.011$,
the infidelity $\delta\epsilon_{\kappa,\delta\omega}$
reaches the levels $0.01$ or $0.001$, respectively.

On Fig.~\ref{fig:42b} is shown numerically the dependence
of $\delta\epsilon_{\kappa,\delta\omega}^{\rm 2\, Qb}(\omega_r, \tilde{V}_m)$ for the range of
resonator frequency $\omega_r \in [2, 10]\, {\rm GHz}$
and qubits gate modulation voltage (amplitude), $\tilde{V}_m \in [0.05, 0.25]\, {\rm mV}$.
Contours where the infidelity reaches the values of $0.1$, $0.01$, $0.001$ are also shown.
To reach a level of $10^{-3}$, this infidelity requires higher $\tilde{V}_m \gtrsim 2\, {\rm mV}$,
and relatively low $\omega_r \lesssim 6 - 10\, {\rm GHz}$.

In what follows, for the calculation of the other infidelities, Table~\ref{tab:rates},
our strategy is to fix the ratio  
$\frac{\delta\omega}{\tilde{g}_{\parallel}} \simeq 0.026$
for which the infidelity $\delta\epsilon_{\kappa,\delta\omega}^{\rm 2\, Qb}$ reaches  
$\simeq 5\times 10^{-3}$.
%
For the chosen ratio, $\frac{\delta\omega}{\tilde{g}_{\parallel}}$,
an experimentally reachable lever arm of  
$\alpha_c \simeq 0.14$, resonator inductance,
$L_r = 50\, {\rm nH}$, and a voltage modulation amplitude
of $\tilde{V}_m = 0.1\, {\rm mV}$, one requires a resonator frequency
$\omega_r/2\pi \simeq 6.3\, {\rm GHz}$, see Table~\ref{tab:rates}.

It will be beneficial to increase the voltage modulation amplitude $\tilde{V}_m$, in order to increase the
longitudinal coupling $\tilde{g}_{\parallel}$ (e.g., to compensate the smallness of the lever arm, $\alpha_c$,
respectively of $\frac{\eta}{\hbar}$),
reaching smaller gate times.
Then, in order to keep the infidelity $\delta\epsilon_{\kappa,\delta\omega}$ constant, Fig.~\ref{fig:42b},
one will also need
a higher resonator frequency, so that to keep the ratio $\frac{\omega_r^{3/2}}{\tilde{V}_m}$ constant,
Eq.~(\ref{curvature-couplings-ratio}).
Thus, for $\tilde{V}_m = 0.1,\, 0.2,\, 0.3\, {\rm mV}$ one would obtain
$\omega_r/2\pi \simeq  6.3,\, 10.,\, 13.1 {\rm GHz}$,
which is beneficial
for suppressing the charge noise gate error,
see Eq.~(\ref{infidelity-phi-2qb-one-over-f-scaling-0}) and Table~\ref{tab:rates}.

By increasing the modulation amplitude $\tilde{V}_m$, however, one is increasing
the relative contribution of the higher-curvature corrections \cite{RuskovTahan2019PRB99}
(for the sweet spot they are proportional to the higher energy curvature, $\frac{\partial^4 E_q}{\partial V_m^4}$).
Some of these corrections \cite{RuskovTahan2019PRB99}  
just change $\tilde{g}_{\parallel}$ and $\delta\omega_s$
(relative corrections are of the order of $\propto \left[\frac{e \tilde{V}_m}{U_{\rm ch}}\right]^2$),
keeping their ratio nearly the same.
In addition, the higher-curvature corrections generate
the non-linear Hamiltonians \cite{RuskovTahan2019PRB99},
${\cal H}_{\hat{n}^2} = \hbar(\zeta_0 + \zeta_{\parallel}\, \sigma_z) \hat{n}^2$,
${\cal H}_{\hat{n}a} = \hbar(\xi_0 + \xi_{\parallel}\, \sigma_z) (\hat{n}\hat{a} + \hat{a}^{\dagger}\hat{n})\, \cos(\omega_m t + \varphi_m)$,
with a relative strength suppressed by the factors,
$\langle \hat{n} \rangle \, \left[\frac{\hbar\omega_r}{U_{\rm ch}}\right]^2 \left[\frac{\eta}{\hbar}\right]^2$,
which makes them of the order of $\lesssim 10^{-4}$ since the average photon number in the resonator
is of order of $\langle \hat{n} \rangle \lesssim 1$,
see Table~\ref{tab:rates} and Eq.~(\ref{photon-number}), Appendix \ref{Coherent-state-ansatz-for-Pss'}.

\subsection{Resonator (Johnson) noise phase gate error for $n$-qubits. 2-qubit numerics}
\label{Sec: Resonator (Johnson) noise phase gate error}
Consider now the situation, when one performs ideal evolution in the phase space
disturbed only via the random force Hamiltonian ${\cal H}_f$, Eq.~(\ref{random-force}) and Fig.~\ref{fig:4circ-non-ideal}
(assuming small noise deviations, Appendix \ref{app-C:combined-infidelity-noise-curvature-detuning}).
The random force makes the trajectory in the phase space noisy,
by adding a new term in the equation of motion, see Eq.~(\ref{alpha-equation-with-noise}):
$\frac{d\alpha_s(t)}{dt} = \cdots + i\, \frac{\xi_f(t)}{2\Delta p_0} e^{i \omega_r t}$.
Correspondingly, as
$\alpha_{s}(t) = \alpha^{\rm id}_s(t) + \delta\alpha_s^{\xi}(t)$,
and $\Phi_{s}(t) = \Phi^{\rm id}_s(t) + \delta\Phi_s^{\xi}(t)$,
the noise
affects the final resonator state (for each $n$-qubit state, $|s\rangle$),
as well as the accumulated phase,
so that some
qubit-resonator entanglement still remains at the time of one cycle, $t_g = \frac{2\pi}{\delta}$.
Starting from an initial product state of $n$ qubits and resonator,
one obtains the fidelity $F \equiv |\langle \Psi_f | \Psi_f^{\rm id} \rangle|^2$,
similar to Eq.~(\ref{fidelity-matrix-element}),
averaging on the $n$-qubit initial states and on the noise.
The average infidelity $\delta\epsilon_{\xi}$ reads (Appendix \ref{app-C:combined-infidelity-noise-curvature-detuning}):
\begin{eqnarray}
&&  \delta\epsilon_{\xi}^{n\, Qb}  =  1 - \overline{ |\langle \Psi_f | \Psi_f^{\rm id} \rangle|^2 } \rangle_\xi
\nonumber
\\
&&  { } = \langle |\delta\alpha_{s}^{\xi}|^2 \rangle_\xi + \frac{f_{12}^{(n)}}{2}
\sum_{s < s'} \left\langle \left(\delta\Phi_{s}^{\xi} - \delta\Phi_{s'}^{\xi} \right)^2 \right\rangle_\xi
\label{infidelity-xi-0} ,
\end{eqnarray}
where we have used that the variance of $\delta\alpha_{s}^{\xi}(t)$,
and initial state averages
$f_s^{(n)} \equiv \overline{|a_s|^2}$,  $f_{s,s'}^{(n)} \equiv \overline{2 |a_s|^2 |a_{s'}|^2}$,
are spin-independent,
see Eqs.~(\ref{n-qubit-averages-0}) and (\ref{n-qubit-averages-1}),
Appendix \ref{app C2: Averaging over the n-qubit initial state}.

To calculate the variances, Appendix \ref{App-D: Johnson nosie infidelity},
one uses
the noise contribution, $\delta\alpha_s(t)$, in rotating frame with $\omega_r$,
\begin{equation}
\delta\alpha_s(t) = i \frac{\Delta x_0}{\hbar} \int_0^t dt' \xi_f(t')  e^{i \omega_r t'}
\label{alpha-noise}
\end{equation}
and obtains a spin-independent variance of
(otherwise spin-dependent evolution of) $\alpha_s(t)$:
\begin{eqnarray}
&& \langle |\delta\alpha_s(t)|^2\rangle_{\xi} =
\left( \frac{\Delta x_0}{\hbar}\right)^2 \,
\int_0^t \int_0^t dt' dt'' e^{i \omega_r (t'-t'')} \langle \xi_f(t') \xi_f(t'')\rangle_{\xi}
\nonumber\\
&& \qquad  { } =
\frac{K_d}{2 (\Delta p_0)^2} \ t \equiv C_0\, t
\label{alpha-variance}
\end{eqnarray}
that is linear in time $t$, as expected for a diffusion process.

To obtain the phase variance,
$\langle \delta\Phi_s(t)^2\rangle_{\xi} = \langle \left[ {\rm Im}\, \delta I_{\Phi,s}\right]^2\rangle_{\xi}$
one considers the variation
of the accumulated phase integral [scf. Eq.~(\ref{accumulated_phase})],
assuming small variations:
\begin{equation}
\delta I_{\Phi,s}(t) \simeq \int_0^t dt'
\left[ \delta \alpha_{s}^*(t') \frac{d\alpha^{\rm id}_{s}(t')}{dt'}
+ \alpha^{\rm id\, *}_{s}(t') \frac{d\delta\alpha_{s}(t')}{dt'} \right] ,
\label{delta-I}
\end{equation}
and averages over the noise, similar to Eq.~(\ref{alpha-variance}).

Averaging over the noise,
the variance of the accumulated phase differences
for $N$ cycles, at $t=\frac{2\pi N}{\delta}$ reads:
\begin{eqnarray}
&& \langle \left[  \delta\Phi_{s}^{\xi}(t) - \delta\Phi_{s'}^{\xi}(t)  \right]^2 \rangle_{\xi}
= 4 C_0\,
\frac{[\Omega_{\parallel,s}^{\varepsilon} - \Omega_{\parallel,s'}^{\varepsilon}]^2}{\delta^2} \,
\frac{2\pi N}{\delta}, \qquad
\label{delta-Phi-variance} .
\end{eqnarray}
At arbitrary time $t$ the corrections to Eq.~(\ref{delta-Phi-variance}) are of the
order of ${\cal O}\left( \frac{\delta}{\omega_d} \right)$
(some of them are zeroed at completed cycles),
and oscillate with frequencies,
$\delta$, and $2\delta$,
see Appendix \ref{app D2: Johnson-noise-variances}.
Notice, that the variance $\langle \left[  \delta\Phi_{s}(t) - \delta\Phi_{s'}(t)  \right]^2 \rangle_{\xi}$
is linear in time units of $\frac{2\pi}{\delta}$ for integer number of cycles $N$.

The $n$-qubit infidelity due to resonator (Johnson) noise for $N$ cycles
is then obtained from Eq.~(\ref{infidelity-xi-0}):
\begin{eqnarray}
&& \delta \epsilon^{\rm n\,Qb}_{\xi} \lesssim \frac{2\pi N}{\delta}\, \frac{\kappa}{2}
\coth{\left(\frac{\hbar \omega_r}{2 k_B T_r}\right)}
\nonumber\\
&& \qquad\quad  \times \left[ 1 +
2 f_{12}^{(n)} \sum_{s<s'}
\frac{[\Omega_{\parallel,s}^{\varepsilon} - \Omega_{\parallel,s'}^{\varepsilon}]^2}{\delta^2} \right]
\label{infidelity-xi-n-qubits} ,
\end{eqnarray}
where $f_{12}^{(n)}$ is given by Eq.~(\ref{n-qubit-averages-11}) and
in the modulation strength difference
the spin-independent terms $\sim \tilde{g}_{0}^{(j)}$, and the driving $\varepsilon_d$,
are separately canceled:
\begin{equation}
\Omega_{\parallel,s}^{\varepsilon} - \Omega_{\parallel,s'}^{\varepsilon}
= \frac{1}{2} \sum_j \tilde{g}_{\parallel}^{(j)}
\left( \langle s| \sigma_z^{(j)} |s\rangle - \langle s'| \sigma_z^{(j)} |s'\rangle \right)
\label{modulation-strength-difference} .
\end{equation}

\begin{figure} [t!] 
    \centering
        \includegraphics[width=0.45\textwidth]{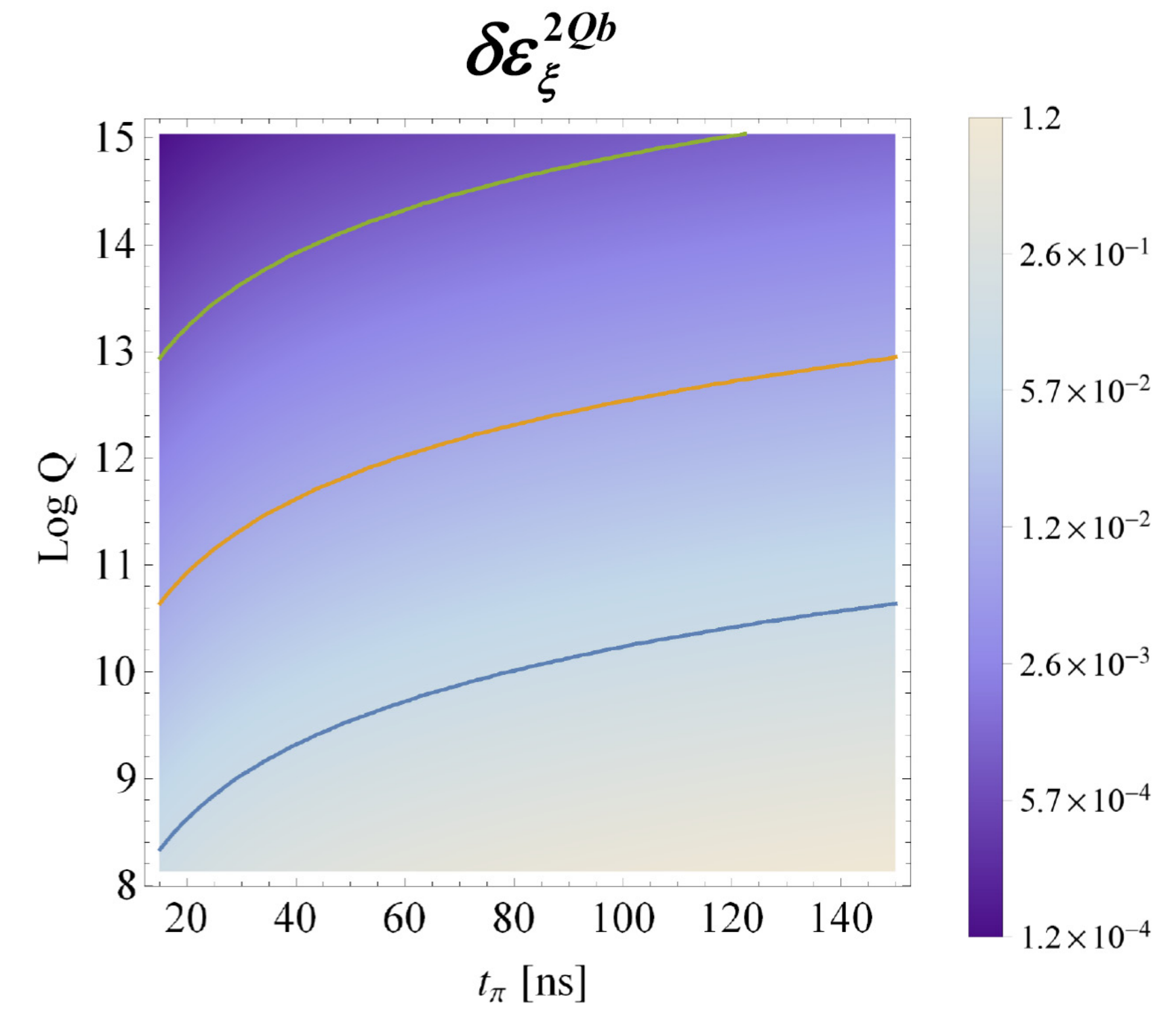}
        \caption{Density plot of the resonator (Johnson) noise
        infidelity $\delta\epsilon_{\xi}(t_\pi, \log Q)$ for a resonator frequency 
        $\omega_r/2 \pi \simeq 6.3\, {\rm GHz}$,
        range of the two-qubit gate time $t_{\pi} \in [15,150]\, {\rm ns}$
        and resonator $Q$-factor of $Q \in [3.4\, 10^3, 3.4\, 10^6]$.
        Contours where the infidelity reaches the values of $0.1$, $0.01$, $0.001$ are also shown.
        }
        \label{fig:43}
\end{figure}

For the two-qubit controlled $\pi$-phase gate considered in this paper, the
modulation-to-resonator detunnig
is chosen such that
$\delta_\pi = 2\sqrt{ N \tilde{g}^{(1)}_{\parallel} \tilde{g}^{(2)}_{\parallel} }$
so that the variances of interest scale differently with $N$:
$\langle |\delta\alpha_s|^2\rangle \propto \sqrt{N}$, while
$\langle \left[  \delta\Phi_{s}(t) - \delta\Phi_{s'}(t)  \right]^2 \rangle_{\xi} \propto \frac{1}{\sqrt{N}}$.
The two-qubit gate error due to resonator noise then reads:
%
\begin{eqnarray}
&& \delta \epsilon^{\rm 2Qb}_{\xi} \lesssim
\frac{\pi \sqrt{N}}{\sqrt{ \tilde{g}_{\parallel}^{(1)} \tilde{g}_{\parallel}^{(2)} } }
\frac{\kappa}{2} \coth{\left(\frac{\hbar \omega_r}{2 k_B T_r}\right)}
\nonumber\\
&& \qquad\quad
\times \left[ 1 + \frac{4 f_{12}^{(2)} }{N}\,
\frac{ \left(\tilde{g}_{\parallel}^{(1)}\right)^2 +  \left(\tilde{g}_{\parallel}^{(2)}\right)^2  }
{ \tilde{g}_{\parallel}^{(1)} \tilde{g}_{\parallel}^{(2)} }
\right]
\label{noise-infidelity-2Qb}
\end{eqnarray}
%
For two qubits one is using the average over the initial state
$f_{12}^{(2)} = \overline{2 |a_s|^2 |a_{s'}|^2} = 1/10$
(Eq.~\ref{n-qubit-averages-11} and Appendix \ref{app C2: Averaging over the n-qubit initial state}).
We notice, that both terms in Eq.~(\ref{noise-infidelity-2Qb}) are important, especially for small $N$.
For equal couplings, $\tilde{g}_{\parallel}^{(1)} = \tilde{g}_{\parallel}^{(2)}$
the second term was derived in Ref.~\onlinecite{Kerman2013}.
It gives 40\% of the first term for $N=2$.
In what follows, higher $N>2$ are not welcomed since both the gate time and the error scales up
as $\sim \sqrt{N}$.

Taking, e.g. a temperature of $T_r = 40\,{\rm mK}$ and $\omega_r/2\pi \in [5,10]\, {\rm GHz}$
notice, that the two-qubit gate noise error (infidelity) up to a factor of order 1
is given by:
$\delta \epsilon^{\rm 2Qb}_{\xi} \approx  t_{\pi} \frac{\omega_r}{2 Q}$.
On Fig.~\ref{fig:43} is shown a density plot of $\delta \epsilon^{\rm 2Qb}_{\xi}(t_{\pi},\log Q)$,
for the range of parameters that fixes the range of gate times,
$t_{\pi} \in [15,150]\, {\rm ns}$
(that implies $N=2$, $\tilde{V}_m = 0.1\, {\rm mV}$ and
$\omega_r/2\pi \simeq 6.3\, {\rm GHz}$,
see also Fig.~\ref{fig:41}).
One obtains ranges of the resonator $Q$-factor that provide corresponding level of
the resonator noise two-qubit phase gate error:
$\delta \epsilon^{\rm 2Qb}_{\xi} \leq 0.1$ for $Q \in [3.4\, 10^3, 3.4\, 10^4]$,
$\delta \epsilon^{\rm 2Qb}_{\xi} \leq 0.01$ for $Q \in [3.4\, 10^4, 3.4\, 10^5]$,
and
$\delta \epsilon^{\rm 2Qb}_{\xi} \leq 0.001$ for $Q \in [3.4\, 10^5, 3.4\, 10^6]$.

\subsection{Qubit charge noise phase gate errors}
\label{Sec: Qubit charge noise phase gate errors}

Due to qubits' gate charge noise,
the actual final state of $n$ qubits acquires the random phases $\Delta\phi_s(t)$:
\begin{equation}
|\psi_f \rangle = \sum_s a_s e^{i\Phi_s(t)}\, e^{-i \omega_s t} \, e^{-i\, \Delta \phi_s(t)} |s\rangle |0\rangle
\label{actual-final-state-charge-noise}
\end{equation}
with the spin-dependent geometric phase, $\Phi_s(t)$,  and the regular qubits phase, $\omega_s\, t$.
The $n$-qubit random phase, $\Delta \phi_s(t)$ is:
\begin{equation}
\Delta \phi_s(t) = \sum_j \langle s | \frac{\Delta \phi^{(j)}(t) }{2} \sigma_z^{(j)} | s \rangle,
\ \ \ \Delta \phi^{(j)}(t) \equiv \int_0^t dt' \, \delta \omega_q^{(j)}(t') ,
\label{random-phase-s-state}
\end{equation}
where the random phase $\Delta \phi^{(j)}(t)$ is accumulated via
the $(j)$-th qubit frequency fluctuation $\delta \omega_q^{(j)}(t)$.
Averaging over the initial qubits' state and over the random phases $\Delta\phi_s(t)$
(assuming Gaussian distributed $\Delta\phi_s(t)$,
see Appendix \ref{app E:Infidelity due to linear and quadratic charge noise})
one obtains the $n$-qubit charge noise infidelity in general form:
\begin{equation}
\delta\varepsilon_\phi^{\rm n\, Qb}
= f_{12}^{(n)} \sum_{s<s'} \frac{1}{2}  \langle \left(\Delta\phi_{s}(t) - \Delta\phi_{s'}(t) \right)^2 \rangle_{\xi_q}
\label{infidelity-phi-0} .
\end{equation}
Essentially, the noise average,
$\langle \left(\Delta\phi_{s}(t) - \Delta\phi_{s'}(t) \right)^2 \rangle_{\xi_q}$, will lead to
$n$-qubit dephasing, and below we consider several charge noise dephasing mechanisms.

\subsubsection{Curvature couplings induced qubit dephasing via the resonator shot noise}
One mechanism is via the qubits {\it curvature coupling} to the resonator.
Weak leakage of photons from the resonator (shot noise)
will lead to an $n$-qubit dephasing with rates $\Gamma_{ss'}^{\rm shot}$ and
ac-Stark frequency shifts $\delta\omega_{ss'}^{\rm shot}$.
These are derived from the Caldeira-Leggett master equation for the $n$-qubits plus resonator density matrix,
Eq.~(\ref{damping_diffusion-CL}), at zero resonator temperature, via tracing out the resonator.
In a long-time limit, $t \gg 1/\kappa$,
the $n$-qubit density matrix acquires a dephasing term
(Appendix \ref{app: dephasing rates from resonator photon shot noise})
\begin{equation}
\rho^q_{ss'}(t) = \rho^q_{ss'}(0)\, e^{- \Gamma^{\rm shot}_{\phi,ss'}\,\, t}\, e^{i\, \delta\omega^{\rm shot}_{ss'}\, t}
\label{white-noise-depahsing} .
\end{equation}
For the dephasing rates one obtains
\begin{eqnarray}
&&  \Gamma^{\rm shot}_{\phi,ss'} \equiv \delta\gamma_{\rm disp,ss'} + \delta\gamma_{\rm long,ss'}
\label{delta-gamma-0}
\\
&&  \delta\gamma_{\rm disp,ss'} =
\frac{(2 A_{-,ss'})^2(\kappa/2)}{(\delta_{s}^2 + \frac{\kappa^2}{4})(\delta_{s'}^2 + \frac{\kappa^2}{4})}
\Omega_{\parallel,s}^{\varepsilon} \,\Omega_{\parallel,s'}^{\varepsilon}
\label{delta-gamma-disp-0}
\\
&&  \delta\gamma_{\rm long,ss'} =
B_{-,ss'} (\kappa/2)
\left[ \frac{\Omega_{\parallel,s}^{\varepsilon} }{\delta_{s}^2 + \frac{\kappa^2}{4}} -
       \frac{\Omega_{\parallel,s'}^{\varepsilon} }{\delta_{s'}^2 + \frac{\kappa^2}{4}}  \right]
\label{delta-gamma-long-0}
\end{eqnarray}
and similarly, for the ac-Stark frequency shifts, $\delta\omega_{ss'}^{\rm shot}$,
\begin{eqnarray}
&&  \delta\omega^{\rm shot}_{ss'} \equiv \delta\omega_{\rm disp,ss'} + \delta\omega_{\rm long,ss'}
\label{delta-omega-0}
\\
&&  \delta\omega_{\rm disp,ss'} =
-2 A_{-,ss'} [\delta_{s} \delta_{s'} + \frac{\kappa^2}{4}]
\nonumber\\
&& \qquad\qquad \times
\frac{\Omega_{\parallel,s}^{\varepsilon}\, \Omega_{\parallel,s'}^{\varepsilon}}
{(\delta_{s}^2 + \frac{\kappa^2}{4}) (\delta_{s'}^2 + \frac{\kappa^2}{4})}
\label{delta-omega-disp-0}
\\
&&  \delta\omega_{\rm long,ss'} =
B_{-,ss'}
\left[ \delta_{s}\, \frac{\Omega_{\parallel,s}^{\varepsilon} }{\delta_{s}^2 + \frac{\kappa^2}{4}} +
       \delta_{s'}\,
       \frac{\Omega_{\parallel,s'}^{\varepsilon} }{\delta_{s'}^2 + \frac{\kappa^2}{4}}  \right] ,
       \quad
\label{delta-omega-long-0}
\end{eqnarray}
where we have introduced the shortcomings,
\begin{eqnarray}
&&\delta_{s} \equiv \delta - \delta\omega_s \equiv \omega_m - \omega_r - \delta\omega_s
\\
&& A_{-,ss'} \equiv \frac{\delta\omega_s - \delta\omega_{s'} }{2},\ \
B_{-,ss'} \equiv \Omega_{\parallel,s}^{\varepsilon}  - \Omega_{\parallel,s'}^{\varepsilon}
\label{shortcomings}
\end{eqnarray}
see Appendix \ref{App-B1: Evolution for the partial density matrices}.
The expressions, Eqs.~(\ref{delta-gamma-0})-(\ref{delta-omega-long-0}),
are derived with the effective spin-dependent driving strength,
$\Omega_{\parallel,s}^{\varepsilon}$,
Eq.~(\ref{omega-n-qubit-coupling-with driving-0}) and
Appendix \ref{app: dephasing rates from resonator photon shot noise},
and a proper choice of the resonator driving, Eq.~(\ref{cancellation-of-g0}),
can significantly decrease the gate errors.

In the short-time limit of the geometric phase gates one has typical time scales:
$t^* \sim \frac{1}{\delta} \ll \frac{1}{\delta\omega_s} \ll \frac{1}{\kappa}$,
and the dephasing factors differ from that of Eq.~(\ref{white-noise-depahsing}).
In particular, for intermediate times within the cycle, $0 < t < t_g$,
the exponents are non-linear in time and oscillate with a period, $\frac{2\pi}{\delta}$,
changing considerably
(Appendix \ref{app: dephasing rates from resonator photon shot noise}).
At time moments $t_g = \frac{2\pi N}{\delta}$ one is left, however, with
linear in time
(but modified) exponents of Eq.~(\ref{white-noise-depahsing})
with dephasing rates and frequency shifts given by:
\begin{eqnarray}
&&\tilde{\Gamma}^{\rm shot}_{\phi,ss'} \equiv \delta\gamma_{\rm disp,ss'} + 2\, \delta\gamma_{\rm long,ss'}
\label{delta-gamma-1}\\
&&\delta{\tilde\omega}^{\rm shot}_{ss'} \equiv \delta\omega_{\rm disp,ss'} + 2\, \delta\omega_{\rm long,ss'}
\label{delta-omega-1} ,
\end{eqnarray}
instead of Eqs.~(\ref{delta-gamma-0}) and (\ref{delta-omega-0}).
In its turn, the frequency shifts, $\delta\tilde{\omega}^{\rm shot}_{ss'}$
can be represented as
\begin{equation}
\delta\tilde{\omega}^{\rm shot}_{ss'} \equiv
\frac{(\Omega_{\parallel,s}^{\varepsilon})^2  - (\Omega_{\parallel,s'}^{\varepsilon})^2}{\delta}
+ \delta{\tilde{\tilde\omega}}^{\rm shot}_{ss'}
\label{delta-omega-12} ,
\end{equation}
where the first term of the order of $\tilde{g}_{\parallel}$ can be absorbed in the
redefinition of the $n$-qubit energy levels, $\omega_s$,
see Eqs.~(\ref{Hamiltonian-qubit-evolution}) and (\ref{ss'-transition-n-qubit-frequency})
(essentially, because  
$|s\rangle$ are eigenstates of $\hat{\Omega}_{\|}^2$).
The remaining frequency shifts, $\delta{\tilde{\tilde\omega}}^{\rm shot}_{ss'} \sim \delta\omega_s$,
cause a quadratic in time infidelity
(Appendix \ref{app E:Infidelity due to linear and quadratic charge noise}):
\begin{equation}
\delta\varepsilon_{\delta\omega,\rm shot}^{\rm n\, Qb}
=  \frac{f_{12}^{(n)} }{2}\,  \sum_{s<s'} \left( \delta{\tilde{\tilde\omega}}^{\rm shot}_{ss'}\, t_g \right)^2
\label{infidelity-delta-omega-uncorrelated-white-noise-0} .
\end{equation}

The explicit form of the $n$-qubit shot noise rates, $\tilde{\Gamma}^{\rm shot}_{\phi,ss'}$,
cannot be expressed as a sum of individual qubit dephasings.
This is equivalent to say that the phase average in Eq.~(\ref{infidelity-phi-0}) is represented
via {\it correlated qubits white noises}, see Eq.~(\ref{correlated-white-qubit-noises}).
For small times, $t_g \ll 1/\Gamma_{\phi,ss'}$,
the $n$-qubit infidelity is linear in the gate time $t_g$ and reads:
\begin{equation}
\delta\varepsilon_{\phi,\rm shot}^{\rm n\, Qb}  =  f_{12}^{(n)} \sum_{s<s'} \tilde{\Gamma}_{\phi,ss'}^{\rm shot}\,\, t_g
\label{infidelity-phi-uncorrelated-white-noise-0} ,
\end{equation}
substituting in it the shot noise dephasings, $\tilde{\Gamma}^{\rm shot}_{\phi,ss'}$, Eq.~(\ref{delta-gamma-1}).
Using the expressions of Eqs.~(\ref{infidelity-delta-omega-uncorrelated-white-noise-0}) and
(\ref{infidelity-phi-uncorrelated-white-noise-0}),
one can show that
these infidelities are negligible for a reasonable set of parameters,
see Table~\ref{tab:rates}.

\subsubsection{Uncorrelated white noise gate infidelity}
A second charge noise dephasing mechanism is via the voltage fluctuations of the qubit's defining gates.
For a TQD qubit these are the gates defining the dots and the gates defining the interdot tunneling,
see Fig.~\ref{fig:1}a.

It is worth to mention, that for a model with {\it uncorrelated white noises}
the $n$-qubit dephasings can be represented as a sum of appropriate individual qubit dephasings,
see Appendix \ref{app: Uncorrelated white noises} and Fig.~\ref{fig:431}.
Then, for equal qubits dephasings,
$\Gamma^{(1)}_{\phi} = \Gamma^{(2)}_{\phi} \equiv \Gamma^{\rm white}_{\phi}$,
the two-qubit infidelity
is obtained from Eq.~(\ref{infidelity-phi-uncorrelated-white-noise-0})
using $f_{12}^{(2)} = 1/10$ and a gate time $t_g$ 
given by Eq.~(\ref{entangling_time}):
\begin{eqnarray}
&& \delta\varepsilon_\phi^{\rm 2Qb} = \frac{8}{10}\, \frac{\pi\sqrt{N}\,\, \Gamma^{\rm white}_{\phi} }{\tilde{g}_{\|}}
\label{2-qubit-dephasing-infidelity-white-0} ,
\\
&& \Gamma^{\rm white}_{\phi} =
\sum_{x = \varepsilon_v,\varepsilon_m,t_l,t_r} \frac{S_x}{4 \hbar^2} \, \left( \frac{\partial E_q}{\partial x} \right)^2
\label{single-qubit-dephasing-white-0} ,
\end{eqnarray}
that is minimized at a full sweet spot,
where $\frac{\partial E_q}{\partial \varepsilon_v} = \frac{\partial E_q}{\partial \varepsilon_m} = 0$.

\subsubsection{Uncorrelated $1/f$-noise phase gate infidelity}
\label{sec: Uncorrelated $1/f$-noise phase gate infidelity}

The dephasing measured in actual experiments,
performed for a DQD Singlet-Triplet qubit in its symmetric
operating point\cite{ReedHunter2016PRL,MartinsKuemethMarcus2016PRL},
shows that this model is unrealistic:
the experiment is featuring an $1/f$ charge noise spectrum of the qubits' gate noise,
leading to a Gaussian (quadratic) dephasing exponent:    
\begin{equation}
e^{-\frac{1}{2}  \langle \left(\Delta\phi_{s'}(t) - \Delta\phi_{s'}(t) \right)^2 \rangle_{\xi_q} }
= e^{-(\tilde{\Gamma}_{ss'}\,\, t)^2}
\label{dephasing-factor-one-over-f-0} ,
\end{equation}
compare with Eq.~(\ref{white-noise-depahsing}).
In the case of uncorrelated charge noise to each of the qubits
one obtains the relation
\begin{eqnarray}
&& \left( \tilde{\Gamma}_{ss'} \right)^2
= \frac{1}{2} \, \sum_{j=1}^n \left(\tilde{\Gamma}^{(j)}\right)^2 \, (1 - i_j\, i'_j)
\label{dephasing-rate-ss'-one-over-f-0} ,
\end{eqnarray}
where $\tilde{\Gamma}^{(j)}$ is the individual $j$-th qubit $1/f$-noise dephasing rate,
Appendix \ref{app: one-over-f-carge-noise-infidelity}.
For small fluctuations,
$\langle \left(\Delta\phi_{s'}(t) - \Delta\phi_{s'}(t) \right)^2 \rangle_{\xi_q} \ll 1$,
one obtains the $n$-qubit infidelity:
\begin{equation}
\delta\varepsilon_{\phi, 1/f}^{\rm n\, Qb}
= f_{12}^{(n)} \sum_{s<s'} (\tilde{\Gamma}_{ss'}\,\, t)^2
\label{infidelity-phi-one-over-f-0} ,
\end{equation}
that is growing quadratically with time.
Since one is assuming
$t \ll 1/\tilde{\Gamma}_{ss'}$,
the quadratic dependence is
beneficial for suppressing the infidelity.

For the two-qubit case one gets from Eq.~(\ref{dephasing-rate-ss'-one-over-f-0}) the relations
\begin{eqnarray}
&& \tilde{\Gamma}_{12} =  \tilde{\Gamma}_\phi^{(2)}, \ \ \tilde{\Gamma}_{13} =  \tilde{\Gamma}_\phi^{(1)}
\\
&& \tilde{\Gamma}_{14} =
\sqrt{ \left(\tilde{\Gamma}_\phi^{(1)}\right)^2 + \left(\tilde{\Gamma}_\phi^{(2)}\right)^2  }, \ {\rm etc.}
\label{2-qubit-rates-one-over-f-noise-0}
\end{eqnarray}
Note that as compared to the uncorrelated white noise case, Eq.~(\ref{2-qubit-rates-white-noise}),
here $\tilde{\Gamma}_{14} \neq \tilde{\Gamma}_{24} + \tilde{\Gamma}_{34}
\equiv \tilde{\Gamma}_\phi^{(1)} + \tilde{\Gamma}_\phi^{(2)}$,
see Fig.~\ref{fig:431}.
One then obtains the two-qubit gate infidelity for a gate time $t_g$
\begin{eqnarray}
&& \delta\varepsilon_{\phi, 1/f}^{\rm 2Qb}
= \frac{4}{10} \, \left[\left(\tilde{\Gamma}_\phi^{(1)}\right)^2 +
\left(\tilde{\Gamma}_\phi^{(2)}\right)^2 \right]\, t_g^2
\label{2qb-infidelity-Gamma1-Gamma2}
\\
&& \qquad\quad { } = \frac{8}{10} \, (\tilde{\Gamma}_{\phi}\,\, t_g)^2
\label{2qb-infidelity-Gamma1=Gamma2} ,
\end{eqnarray}
where the second equality is for equal qubits' dephasing,
$\tilde{\Gamma}_\phi^{(1)} = \tilde{\Gamma}_\phi^{(2)} \equiv \tilde{\Gamma}_\phi$.
%

\begin{figure} [t!] 
    \centering
        \includegraphics[width=0.45\textwidth]{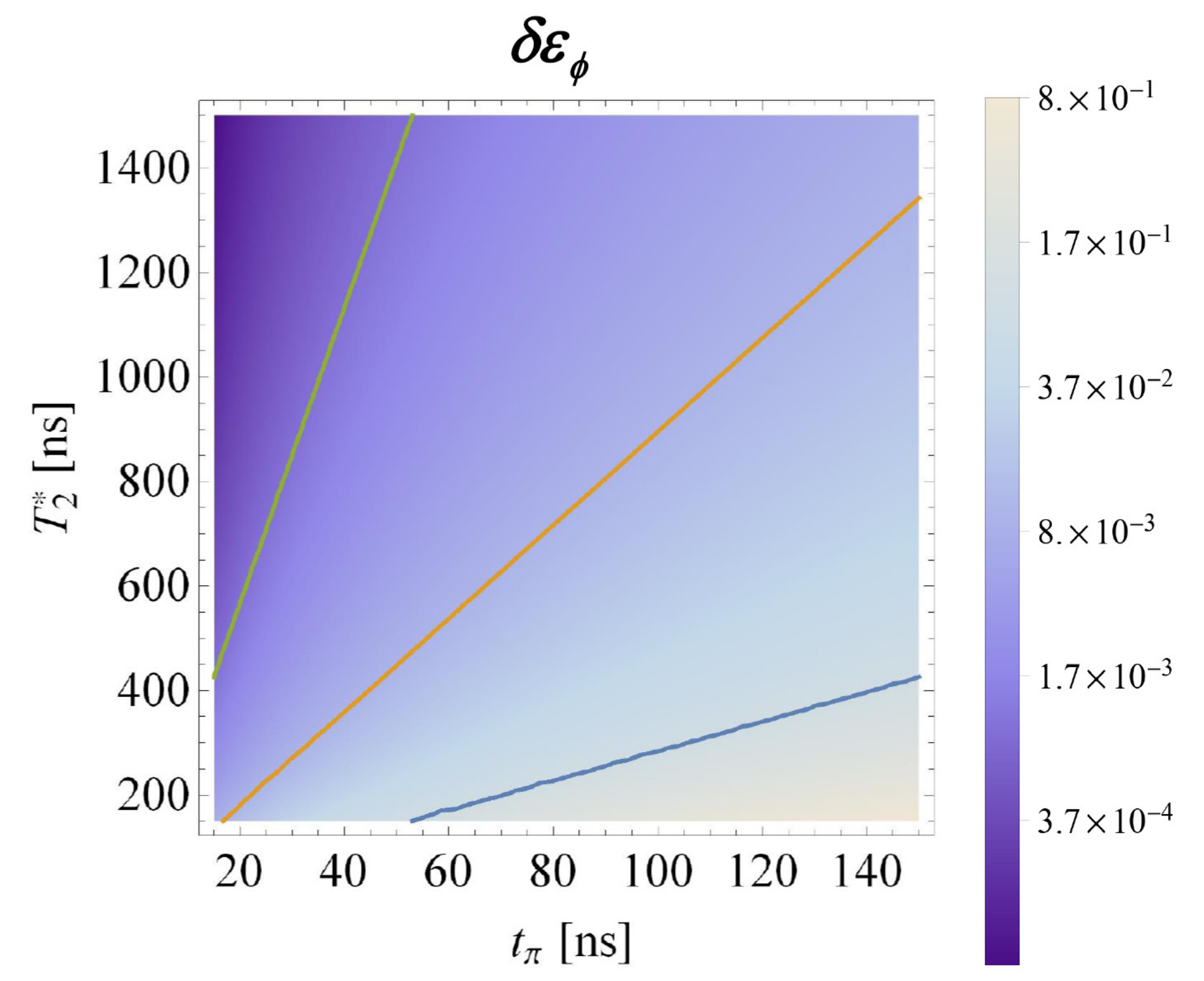}
        \caption{Density plot of the two-qubit controlled $\pi$-phase gate $1/f$ charge noise
        infidelity $\delta\varepsilon_{\phi,1/f}^{\rm 2Qb}(t_\pi, T^*_2)$, Eq.~(\ref{2qb-infidelity-Gamma1=Gamma2}),
        for a resonator frequency $\omega_r/2 \pi \simeq 6.3\, {\rm GHz}$, 
        range of the two-qubit ($\pi$-phase) gate time $t_{\pi} \in [15,150]\, {\rm ns}$ and
        qubit's dephasing times $T_2^* \in [200,1500]\, {\rm ns}$, where $T_2^* \equiv 1/\tilde{\Gamma}_\phi$,
        Eq.~(\ref{dephasing-one-over-f-leading-symm-0}).
        Contours where the infidelity reaches the values of $0.1$, $0.01$, $0.001$ are shown,
        that corresponds to the relations, $t_{\pi} = 0.354 T_2^*, \, 0.112 T_2^*, \, 0.0354 T_2^*$, respectively.
        }
        \label{fig:44}
\end{figure}

\begin{figure} [t!] 
    \centering
        \includegraphics[width=0.35\textwidth]{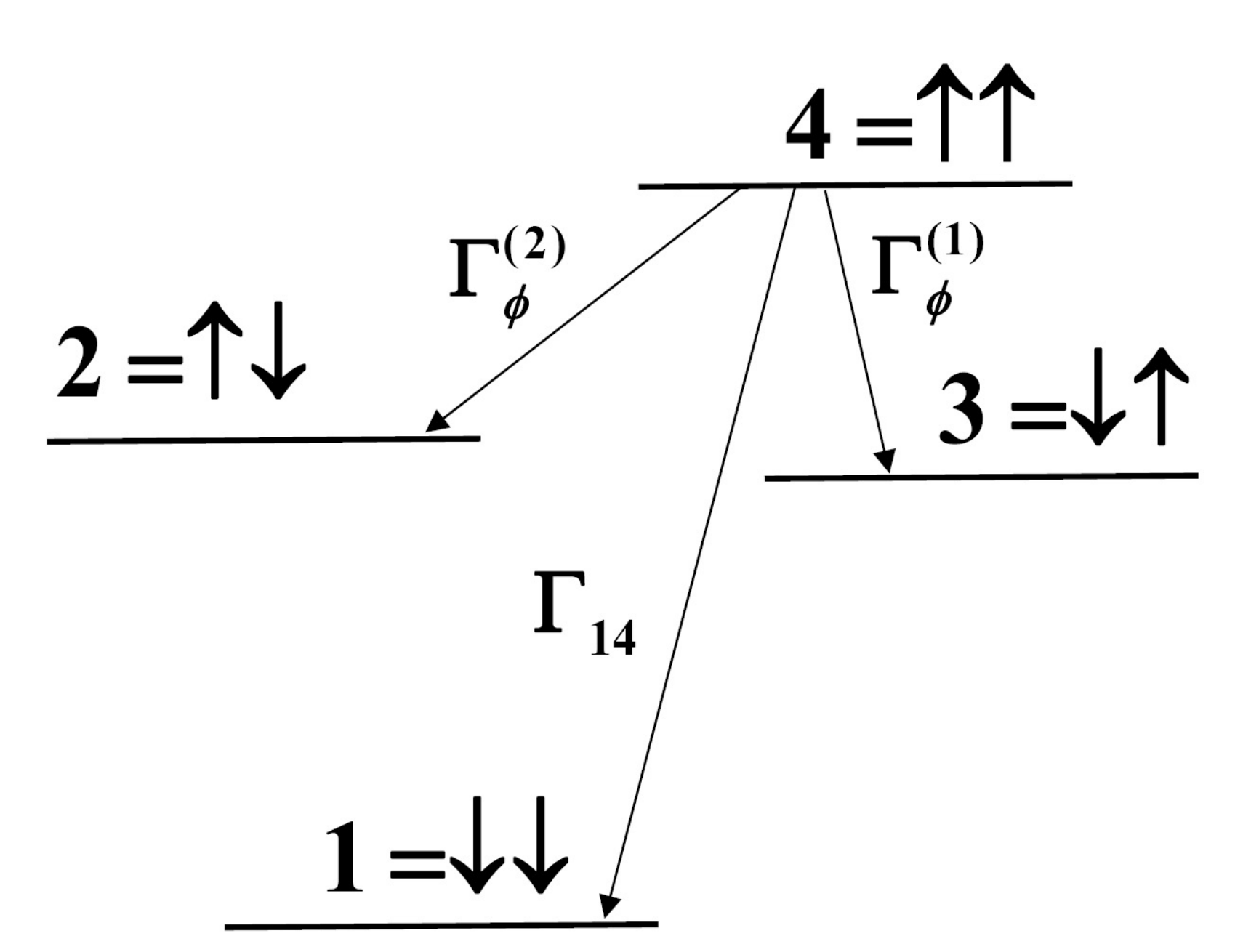}
        \caption{Schematic of the two-qubit energy levels and the associated
        charge noise dephasing rates. The individual qubit dephasings are
        $\tilde{\Gamma}_\phi^{(1)} = \tilde{\Gamma}_{34}$ and
        $\tilde{\Gamma}_\phi^{(2)} = \tilde{\Gamma}_{24}$.
        For uncorrelated white noise,
        $\tilde{\Gamma}_{14} = \tilde{\Gamma}_\phi^{(1)} + \tilde{\Gamma}_\phi^{(2)}$,
        see Eq.~(\ref{2-qubit-dephasing-infidelity-white-0}) and Appendix \ref{app: Uncorrelated white noises}.
        For correlated white noise and for uncorrelated $1/f$-noise,
        $\tilde{\Gamma}_{14} \neq \tilde{\Gamma}_\phi^{(1)} + \tilde{\Gamma}_\phi^{(2)}$,
        see Eqs.~(\ref{delta-gamma-0}) and (\ref{2-qubit-rates-one-over-f-noise-0}), respectively.
        }
        \label{fig:431}
\end{figure}

On Fig.~\ref{fig:44} is shown a density plot of the $1/f$-noise infidelity
$\delta\varepsilon_{\phi,1/f}^{\rm 2Qb}(t_\pi, T^*_2)$,
Eq.~(\ref{2qb-infidelity-Gamma1=Gamma2}),
for a range of gate times $t_\pi \in [15,150]\, {\rm ns}$,
and qubit's dephasing times $T_2^* \in [200,1500]\, {\rm ns}$.
The infidelity levels of $\delta\varepsilon_{\phi,1/f}^{\rm 2Qb} \lesssim 0.1,\, 0.01,\, 0.001$
can be reached for gate times, $t_{\pi} \lesssim 0.354 T_2^*, \, 0.112 T_2^*, \, 0.0354 T_2^*$.

In the symmetric case of a TQD
(Appendix \ref{app: $1/f$-dephasing: Scaling of parameters}),
one obtains for the
single qubit $1/f$-noise dephasing rate
\begin{equation}
 \tilde{\Gamma}_\phi \simeq \frac{1}{\hbar} \sqrt{ \log r_{c} \,\, S_{t_l} }  \, \frac{4 t_l}{U_{\rm ch}}
\label{dephasing-one-over-f-leading-symm-0} ,
\end{equation}
where $r_{c} \equiv \omega_{UV}/\omega_{IR}$ is the ratio of the noise ultraviolet-to-infrared
frequency cutoffs \cite{RussGinzelBurkard2016PRB}, and $S_{t_l}$ is the spectral density constant
of the $1/f$-noise associated with the gates forming the interdot tunnelings $t_l,t_r$.

The curvature contributions to $\tilde{\Gamma}_\phi$ are shown to give a negligible
effect with respect to the noise from the tunneling gates,
Appendix \ref{app: $1/f$-dephasing: Scaling of parameters},
that allows to increase the quantum capacitance,
respectively to decrease the gate time $t_\pi$ without additional $1/f$-noise, see
Table~\ref{tab:rates} and Appendix \ref{app: $1/f$-dephasing: Scaling of parameters}.

The scaling of the two-qubit gate infidelity for $1/f$-noise with
all relevant parameters is
obtained as
(Appendix \ref{app: $1/f$-dephasing: Scaling of parameters}):
\begin{equation}
\delta\varepsilon_{\phi, 1/f}^{\rm 2Qb}
= \frac{4 \pi^2 N}{5 \hbar^2}\,
\frac{\log (\omega_{UV}/\omega_{IR})\, S_{t_l}}{\omega_r^2\, (\eta/\hbar)^2 \, (e \tilde{V}_m)^2} \,
\frac{U_{\rm ch}^4}{t_l^2}
\label{infidelity-phi-2qb-one-over-f-scaling-0} .
\end{equation}

It is worth to comment the important features of this expression:\\
(1) The smaller is the spectral density constant $S_{t_l}$ of the interdot tunneling $t_l$ the better.
Here the spectral density of the $1/f$-noise is defined as $S(\omega) = \frac{S_{t_l}}{|\omega|}$.
One can estimate the spectral density constant from current experiments
with DQD Singlet-triplet qubit at the symmetric operating point \cite{ReedHunter2016PRL,MartinsKuemethMarcus2016PRL}
as $S_{t_l} \simeq 10^{-5} (\mu{\rm eV})^2$,
see also Appendix \ref{app: Spectral density constant of tunnelings from experiment}.
\\
(2) The dot-to-resonator coupling ratio should be small, $\eta/\hbar \sim 0.1$,
in order the make the infidelity  $\delta\varepsilon_{\kappa,\delta\omega}$ small,
see Appendix \ref{app-C1:infidelity-damping-detuning}.
In order to compensate for this smallness one needs to increase the gate voltage modulation amplitude $\tilde{V}_m$,
and simultaneously to increase the resonator frequency $\omega_r$,
so that to keep $\delta\varepsilon_{\kappa,\delta\omega}$ fixed.
The simultaneous increase of these parameters is beneficial for the
suppression of the charge noise infidelity $\delta\varepsilon_{\phi,1/f}$.
\\
(3)  The charge noise infidelity is critically sensitive to the dot's charging energy
$U_{\rm ch}$ and the tunneling amplitude $t_l$,
featuring relatively small $U_{\rm ch}$ and relatively large $t_l$,
see Table~\ref{tab:rates}.

\begin{table*}[t]
\begin{ruledtabular}
\begin{tabular}{c|c|c|c|c|c|c|c||c|c|c|c|c|c||c}
$t_l$          & $E_q$       & $C_q$      & $\eta/\hbar$ & $\delta\omega$  & $\tilde{g}_{\parallel}$  &   
$\tilde{V}_m$   &  $\omega_r$  &  $\delta\varepsilon_{\kappa,\delta\omega}$ & $\delta\varepsilon_{\xi}$  &  
$\delta\varepsilon_{\phi,1/f}$ & $\delta\varepsilon_{\phi,\rm shot}$  &  $\delta\varepsilon_{\delta\omega,\rm shot}$  & 
$\delta\epsilon_{\tilde{g}_{\parallel},1/f}$ &  $t_\pi$  \\ 
$\rm (\mu eV)$ & $\rm (GHz)$ & $\rm (aF)$ &              & $\rm (MHz)$     &    $\rm (MHz)$           &    
$\rm (mV)$      & $\rm (GHz)$  &                                            &                            &   
                           &                                      &                                               & 
                                             &\rm (ns)   \\   
 \hline
40 &3.9 &32.  &      &0.16 &6.2  &    &    &               &$3.1\,10^{-3}$ &$0.53$         &$2.2\,10^{-4}$ &              & $< 10^{-9} - 10^{-10}$ &113.6  \\
60 &8.7 &72.1 &0.1   &0.36 &14.  &0.1 & 6.3&$5.3\, 10^{-3}$&$1.4\,10^{-3}$ &$0.23$         &$1.\,10^{-4}$  &$1.6\,10^{-4}$&  "              &50.5   \\
80 &15.5&128.2&      &0.64 &24.9 &    &    &               &$7.8\,10^{-4}$ &$0.13$         &$5.6\,10^{-5}$ &              &  "              &28.4   \\
 \hline
40 &3.9 &32.  &      &0.64 &24.9 &    &    &               &$1.2\,10^{-3}$ &$3.3\,10^{-2}$ &$8.9\,10^{-5}$ &              &  "              &28.4   \\
60 &8.7 &72.1 &0.125 &1.44 &56.  &0.2 & 10 &$5.3\, 10^{-3}$&$5.5\,10^{-4}$ &$1.5\,10^{-2}$ &$4.\,10^{-5}$  &$1.6\,10^{-4}$&  "              &12.6   \\
80 &15.5&128.2&      &2.57 &99.6 &    &    &               &$3.1\,10^{-4}$ &$8.2\,10^{-3}$ &$2.2\,10^{-5}$ &              &  "             &7.1    \\
 \hline
40 &3.9 &32.  &      &1.44 &56.  &    &    &               &$7.2\,10^{-4}$ &$6.5\,10^{-3}$ &$5.2\,10^{-5}$  &             &  "             &12.6   \\
60 &8.7 &72.1 &0.143 &3.25 &126.1&0.3 &13.1&$5.3\, 10^{-3}$&$3.2\,10^{-4}$ &$2.9\,10^{-3}$ &$2.3\,10^{-5}$ &$1.6\,10^{-4}$&  "             &5.6    \\
80 &15.5&128.2&      &5.77 &224.2&    &    &               &$1.8\,10^{-4}$ &$1.6\,10^{-3}$ &$1.3\,10^{-5}$ &              &  "             &3.2    \\
 \hline
40 &3.9 &32.  &      &0.08 &8.8  &    &    &               &$1.8\,10^{-3}$ &$0.26$         &$1.3\,10^{-4}$ &              &  "             &80.    \\
60 &8.7 &72.1 &0.088 &0.18 &19.9 &0.2 & 5  &$6.6\, 10^{-4}$&$7.9\,10^{-4}$ &$0.12$         &$5.5\,10^{-5}$ &$2.1\,10^{-5}$&  "             &35.6   \\
80 &15.5&128.2&      &0.32 &35.4 &    &    &               &$4.4\,10^{-4}$ &$6.5\,10^{-2}$ &$3.1\,10^{-5}$ &              &  "             &20.
\end{tabular}
\end{ruledtabular}
\caption{\label{tab:rates}
The leading two-qubit controlled $\pi$-phase gate infidelities,
$\delta\varepsilon_{\kappa,\delta\omega}$, $\delta\varepsilon_{\xi}$,
$\delta\varepsilon_{\phi,1/f}$, $\delta\varepsilon_{\phi,\rm shot}$, $\delta\varepsilon_{\delta\omega,\rm shot}$,
and $\delta\epsilon_{\tilde{g}_{\parallel},1/f}$
are calculated
for a range of parameters, see
Eqs.~(\ref{infidelity-kappa-delta-omega}), (\ref{noise-infidelity-2Qb}),
(\ref{infidelity-phi-2qb-one-over-f-scaling-0}), (\ref{infidelity-phi-uncorrelated-white-noise-0}),
(\ref{infidelity-delta-omega-uncorrelated-white-noise-0}),
and (\ref{infidelity-gparallel-charge-0}),
respectively.
For an
experimentally reachable\cite{Stockklauser-Ihn-Ensslin-Wallraff2017PRX,Petta2017Science}
dot-resonator lever arm, $\alpha_c \equiv \frac{C_c}{C_c + C_d} \simeq 0.14$,
resonator inductance,
 $L_r = 50\, {\rm nH}$,
and a $Q$-factor, $Q=10^6$, one chooses a dot charging energy, $U_{\rm ch} \simeq 0.4\, {\rm meV}$.
By setting the error $\delta\varepsilon_{\kappa,\delta\omega} \simeq 5\, 10^{-3}$, one sets the
ratio $\frac{\delta\omega}{\tilde{g}_{\parallel}} \simeq 2.6\, 10^{-2}$, that is independent of the
QD system quantum capacitance, $C_q \propto \frac{t_{l,r}^2}{U_{\rm ch}^3}$.
Since the scaling of $\frac{\delta\omega}{\tilde{g}_{\parallel}} \propto \omega_r^{3/2}/\tilde{V}_r$,
Eq.~(\ref{curvature-couplings-ratio}),
the increase of $\tilde{V}_r$ requires the moderate increase of $\omega_r$ to keep
the infidelity $\delta\varepsilon_{\kappa,\delta\omega}$ constant in Table \ref{tab:rates},
see also Fig.~\ref{fig:42a}.
The scaling, $\eta/\hbar \propto \sqrt{L_r} \omega_r^{3/2}$, leads to
the important scalings,
$\delta\varepsilon_{\kappa,\delta\omega} \propto L_r \omega_r^3$,
$\delta\varepsilon_{\xi} \propto 1/(\sqrt{L_r} \omega_r^{3/2})$,
$\delta\varepsilon_{\phi,1/f} \propto 1/(L_r \omega_r^3)$, so that a decrease of $L_r$ can be
compensated by a moderate increase of $\omega_r$.
The Johnson noise error,  $\delta\varepsilon_{\xi}$, easily reaches a level $\lesssim 10^{-3}$
for higher $C_q$ (respectively, higher tunneling, $t_l$) and/or higher modulation voltage, $\tilde{V}_m$.
The main obstacle is the charge noise error, $\delta\varepsilon_{\phi,1/f}$, due to qubit gate $1/f$ charge noise
which scales with the gate time as $\propto \tilde{\Gamma}_\phi^2\, t_{\pi}^2$,
Eq.~(\ref{2qb-infidelity-Gamma1=Gamma2}).
Generally, to reach an error level of $\delta\varepsilon_{\phi,1/f} \lesssim 10^{-3}$
pushes the dots'-resonator parameters towards
relatively low charging energy,
$U_{\rm ch}$, relatively high interdot tunnelings, $t_{l,r}$,
a smaller $1/f$-noise spectral density constant, $S_{t_l}$
and higher resonator frequency, $\omega_r$, and higher gate voltage modulation amplitude, $\tilde{V}_r$,
see Eq.~(\ref{infidelity-phi-2qb-one-over-f-scaling-0})
[We used $S_{t_l} \simeq 10^{-5}\, (\mu{\rm eV})^2$ that is taken from the experiment\cite{ReedHunter2016PRL},
see Sec.~\ref{sec: Uncorrelated $1/f$-noise phase gate infidelity} and
Appendix \ref{app: Spectral density constant of tunnelings from experiment}].
As an illustration,
on the forth group of rows of the Table we take $\omega_r = 5\, {\rm GHz}$ while the other parameters
are as on the second group of rows.
The above scaling of $\delta\varepsilon_{\phi,1/f}$ with $\omega_r$ leads to an order of magnitude error increase,
which can be compensated by an increase of the lever arm to $0.4$ if lower $\omega_r$ is needed.
}
\end{table*}

\subsection{Infidelity via the charge fluctuations of the longitudinal (curvature) coupling $\tilde{g}_{\parallel}$}
\label{Sec: Infidelity via the charge fluctuations of the longitudinal (curvature) coupling}

Charge noise fluctuations of the (TQD) qubit tunnelings, $\delta t_{l,r}^{(j)}$, causes fluctuations
of the longitudinal (curvature) couplings, $\delta\tilde{g}_{\parallel}^{(j)}$,
and, respectively, of the $n$-qubit spin-dependent resonator driving strength,
$\delta \Omega_{\parallel,s}^{\varepsilon}$,
scf. Eqs.~(\ref{omega-n-qubit-coupling}) and (\ref{omega-n-qubit-coupling-with driving-0}).
Averaging over the charge noise and assuming
uncorrelated fluctuations for the different qubits, one obtains for the
correlation function
\begin{eqnarray}
&& \langle\delta \Omega_{\parallel,s}^{\varepsilon}(t')\, \delta \Omega_{\parallel,s}^{\varepsilon}(t'')\rangle_{\xi_q}
\equiv K_{ss'}(t' - t'')
\nonumber\\
&& K_{ss'}(t' - t'') = \sum_{j=1}^{n} \left[ \frac{\tilde{g}_{\parallel}^{(j)} }{t_l^{(j)}} \right]^2
i_j \, i'_j \ \langle \delta t_l^{(j)}(t')\, \delta t_l^{(j)}(t'')\rangle_{\xi_q}
\label{one-over-f-correlation}
\\
&& \langle \delta t_l^{(j)}(t')\, \delta t_l^{(j)}(t'')\rangle_{\xi_q}
= \int_{-\infty}^{+\infty} d\omega \, e^{i \omega (t' - t'')}\, S_{t_l}^{(j)}(\omega) ,
\label{tunnelings-correlation}
\end{eqnarray}
where $|s\rangle \equiv |i_1,\ldots,i_j,\ldots,i_n \rangle$,
and
$S_{t_l}^{(j)}(\omega)$ is the
spectral density of the tunneling fluctuations.
While some of the results below are correct for a general spectral density,
for further numerical estimations an $1/f$-noise is assumed: $S_{t_l}^{(j)}(\omega) = \frac{S_{t_l}^{(j)}}{|\omega|}$,
with the spectral density constants, $S_{t_l}^{(j)}$ extracted from the experiment \cite{ReedHunter2016PRL},
see Appendix \ref{app: Spectral density constant of tunnelings from experiment}.

The corresponding fluctuation in the   
resonator trajectory,
$\delta\alpha_{s}^{\xi_q} \equiv \delta\alpha_{s}^{\tilde{g}_{\parallel},1/f}$,
scf. Eq.~(\ref{delta-alpha}),
then leads to an infidelity
similar to the Johnson noise, Eq.~(\ref{infidelity-xi-0}),
but with an averaging over the charge noise, $\xi_q$
(see Appendix \ref{app-C:combined-infidelity-noise-curvature-detuning}):
\begin{eqnarray}
&&  \delta\epsilon_{\tilde{g}_{\parallel},1/f}^{n\, Qb}
= \frac{1}{2^n} \sum_s \langle |\delta\alpha_{s}^{\xi_q}|^2 \rangle_{\xi_q}
\nonumber
\\
&&  \qquad\qquad { }
+ \frac{f_{12}^{(n)}}{2}
\sum_{s < s'} \left\langle \left(\delta\Phi_{s}^{\xi_q} - \delta\Phi_{s'}^{\xi_q} \right)^2 \right\rangle_{\xi_q}
\label{infidelity-gparallel-charge-0} .
\end{eqnarray}
The average of the trajectory fluctuation is spin-independent
and is obtained as:
\begin{eqnarray}
&&  \langle |\delta\alpha_{s}^{\xi_q}(t)|^2 \rangle_{\xi_q}
= \int_0^t\int_0^t dt' dt'' \, K_{ss}(t' - t'')\, e^{-i \delta (t' - t'')}
\nonumber
\\
&& \qquad\quad { }
= t^2\, \sum_{j=1}^{n} \left[ \frac{\tilde{g}_{\parallel}^{(j)} }{t_l^{(j)}} \right]^2
\int_{-\infty}^{\infty} d\omega \ S_{t_l}^{(j)}(\omega) \ f_{\alpha}\left(\frac{\omega t}{2},\frac{\delta t}{2}\right)
\qquad
\label{delta-alpha-s-xiq-0}
\\
&& f_{\alpha}(x,y) \equiv {\rm sinc}^2\left(x - y\right) ,
\end{eqnarray}
where ${\rm sinc}(x) \equiv \frac{\sin(x)}{x}$.

For the contribution of the accumulated phase fluctuations one proceeds
similar to the Johnson noise case
(see Appendices \ref{app-C:combined-infidelity-noise-curvature-detuning}
and \ref{app D2: Johnson-noise-variances}).
The accumulated phase fluctuation average is then obtained as:
\begin{eqnarray}
&& \langle \delta\Phi_s(t)\, \delta\Phi_{s'}(t) \rangle_{\xi_q}
= \Omega_{\parallel,s}^{\varepsilon}\, \Omega_{\parallel,s'}^{\varepsilon}\,
\frac{t^4}{4}
\nonumber\\
&& \qquad\quad  \times \sum_{j=1}^{n} \left[ \frac{\tilde{g}_{\parallel}^{(j)} }{t_l^{(j)}} \right]^2\, i_j \, i'_j
\int_{-\infty}^{\infty} d\omega \ S_{t_l}^{(j)}(\omega) \ f_{\Phi}\left(\frac{\omega t}{2},\frac{\delta t}{2}\right)
\qquad
\label{delta-Phi-s-xiq-0}
\\
&& f_{\Phi}(x,y) \equiv
\frac{\left[ xy\cos x \sin 2y + 2(x^2 \sin^2 y - y^2)\sin x\right]^2}{x^2y^2 (x^2 - y^2)^2}
\quad
\end{eqnarray}
(note that $f_{\Phi}(x,y)$ is smooth at $x=y$).
Integrating over frequencies $\omega$
(using infrared and ultraviolet cutoffs, $\omega_{IR} < |\omega| < \omega_{UV}$,
scf. Appendix \ref{app: $1/f$-dephasing: Scaling of parameters}),
one can show numerically that the resulting functions
in Eqs.~(\ref{delta-alpha-s-xiq-0}) and (\ref{delta-Phi-s-xiq-0}) are decreasing
(and oscillating) functions
of $\frac{\delta t}{2}$
and of the order of $\lesssim 1$
(also obtaining additional suppression at complete cycles, $t_g = \frac{2\pi N}{\delta}$).
%
Since $\{\Omega_{\parallel,s}, \delta \} \sim \tilde{g}_{\parallel}$, the order of magnitude of the contributions,
Eqs.~(\ref{delta-alpha-s-xiq-0}) and (\ref{delta-Phi-s-xiq-0}) is given by the
ratio ${S_{t_l}^{(j)}}/{t_l^2} \sim 10^{-5} - 10^{-9}$,
for $S_{t_l} \simeq 10^{-2} - 10^{-5} (\mu{\rm eV})^2$
and $t_l = 40 - 80\, \mu{\rm eV}$
that is,
the infidelity, $\delta\epsilon_{\tilde{g}_{\parallel},1/f}^{2\, Qb}$,
is strongly suppressed,
Table ~\ref{tab:rates}.

\subsection{Switching off the modulation}

After completing the $n$-qubit phase gate,
when the (distant) qubits become entangled,
the modulation is switched off.
Assuming no driving and modulation off,
$\varepsilon_d = 0$, $\tilde{g}_{\|}=0$, $\tilde{g}_0 = 0$,
the dephasing is via the  ``dispersive-like''
coupling ${\cal H}_{\delta\omega}$  only,
due to leakage of resonator thermal photons. 
Estimating the single-qubit dephasing rate\cite{ClerkUtami2012PRA},
$\Gamma_{\phi,{\rm th}} \simeq \kappa \bar{n}_{\rm th}$,
for $T = 40\, {\rm mK}$ one obtains
$\Gamma_{\phi,th} \simeq 8\pi\, 10^{-2}\, {\rm s}^{-1}$,
so pure thermal dephasing is negligible.

At first glance
the always on ``dispersive-like'' coupling ${\cal H}_{\delta\omega}$ may change the $n$-qubit entangled state
via free evolution with the qubits' frequency shifts, $\delta\omega^{(j)}$, see Eqs.~(\ref{total_Hamiltonian0})
and (\ref{Hamiltonian-qubit-evolution}).
For the two-qubit system in an arbitrary state (pure or mixed)
we show that
the state change
can be
corrected by a local rotation
of one of the qubits.
%
Indeed, the   two-spin resonator frequency shifts
$\delta\omega_s$
are opposite in sign
for the relevant $|s \rangle$-states
(here $s = 1,2,3,4 \equiv \downarrow\downarrow, \downarrow\uparrow, \uparrow\downarrow, \uparrow\uparrow$),
namely:
$\delta\omega_{1} = -(\delta\omega^{(1)} + \delta\omega^{(2)}) $,
$\delta\omega_{2} = (-\delta\omega^{(1)} + \delta\omega^{(2)}) $,
$\delta\omega_{3} = - \delta\omega_{2}$,
$\delta\omega_{4} = - \delta\omega_{1}$.
Then, the phases acquired by the two-qubit state
amplitudes $a_s$,
\begin{equation}
|\psi \rangle = \sum_1^4 a_s e^{-i\, \delta\omega_s\, t} |s \rangle
\label{phases-2-qubit-state}
\end{equation}
(assuming the qubits are disentangled from the resonator), 
can be partially compensated via a
$\sigma_z$-rotation of the first qubit:
\begin{equation}
U_{z1} =
\left(
\begin{array}{cc}
e^{-i\Delta\omega_z\, t} & 0 \\
0     &  e^{i\Delta\omega_z\, t}
\end{array}
\right)
\label{elimination_domega_s} ,
\end{equation}
with
${\scriptsize \Delta\omega_z \equiv -\frac{\delta\omega_1 + \delta\omega_2}{2} }$.
After this transformation,
the state amplitudes become transformed to:
\begin{equation}
a_s' = a_s\  e^{(-1)^s i \frac{\left( \delta\omega_1 - \delta\omega_2  \right)}{2} \, t},\ \ \
s=1,\ldots,4
\label{non-abelian_gauge}
\end{equation}
which is a pure gauge phase factor,
i.e.
the two-qubit density matrix elements, see Eqs.~(\ref{x0-2qb})-(\ref{5D-Bloch-sphere}), remain intact
\cite{MatrixPhase}.

For the general case of $n$-qubits ($n > 2$), one can
preserve the entangled $n$-qubit state
by decreasing the tunnelings $t_{l,r}^{(j)}$
for each qubit,
to decrease couplings to the resonator
($\delta\omega^{(j)} \propto \left( t_{l,r}^{(j)} \right)^2 \to 0$).
Another possibility to correct the
acquired phases in Eq.~(\ref{phases-2-qubit-state})
is via a  spin-echo technique:
for the phases acquired after a time interval $\Delta t \ll 1/\kappa$,
one performs simultaneous
$\pi$-pulse on all qubits involved into the phase gate
(by manipulating qubits' interdot tunnelings, $t_{l,r}^{(j)}$,
while still remaining in the sweet-spot for each qubit,
see Ref.~\onlinecite{AEON2016,RuskovTahan2019PRB99}).
Then one waits for a second time interval $\Delta t$,
and performs a $\pi$-pulse again
so that
the effect of the frequency shifts
is canceled out.

\section{Relevance to other work}
\label{Sec: Relevance to other work}
%
%
Similar entangling proposals via a modulated longitudinal coupling,
based on a specially designed superconducting qubits\cite{Royer2017Quantum}
or a double quantum dot singlet-triplet qubits\cite{HarveyYacoby2018PRB}
have been proposed.
While these works utilize
essentially the same dynamical longitudinal coupling
$\tilde{g}_{\parallel}$, Eq.~(\ref{dynamic-long-coupling-parallel}),
as we discussed
previously\cite{Delft2016-Ruskov-poster,QCPR2016-Alexandria-Tahan,RuskovTahan2017preprint1,RuskovTahan2019PRB99},
they have ignored
the other (curvature) ``dispersive-like'' coupling\cite{RuskovTahan2017preprint1,RuskovTahan2019PRB99},
$\delta\omega_s$, which is essential in the estimation of accumulated phase gate infidelities,
as shown in the present paper.
Another missing ingredient in their analysis is the spin-independent modulation coupling,
$\tilde{g}_0$, Eq.~(\ref{dynamic-long-coupling-0}).
Both these ingredients are important for the
associated infidelity, $\delta\varepsilon_{\kappa,\delta\omega}$,
Eq.~(\ref{infidelity-kappa-delta-omega}).
As we have shown, taking into account the infidelity,
$\delta\varepsilon_{\kappa,\delta\omega}$,
essentially restricts the field of available parameters.

It is also worthwhile to compare
the (Johnson noise) infidelity, Eq.~(\ref{noise-infidelity-2Qb}),
with an analogous infidelity of Eq.~(8) of Ref.~\onlinecite{HarveyYacoby2018PRB}.
The latter consists of two terms that scale with the number of cycles, $N$,
as $\sim \sqrt{N}$ and $\sim 1/\sqrt{N}$, respectively.
%
The first term, $\sim \sqrt{N}$, exactly corresponds to
an uncorrelated white gate noise infidelity\cite{Delft2016-Ruskov-poster,QCPR2016-Alexandria-Tahan}
described by Eq.~(\ref{2-qubit-dephasing-infidelity-white-0});
this term should be zero in a sweet spot.
The second term, $\sim 1/\sqrt{N}$, exactly corresponds to our term
$\sim \left\langle \left(\delta\Phi_{s}^{\xi} - \delta\Phi_{s'}^{\xi} \right)^2 \right\rangle_\xi$,
scf. Eq.~(\ref{infidelity-xi-0}).
However, the other (leading) term of Eq.~(\ref{infidelity-xi-0}),
that is proportional to the field variance
$\langle |\delta\alpha_{s}^{\xi}|^2 \rangle_\xi$
and also scales as $\sim \sqrt{N}$, is missing
in the analysis of Ref.~\onlinecite{HarveyYacoby2018PRB}.

As to the $1/f$ charge noise dephasing infidelity, Eq.~(\ref{infidelity-phi-2qb-one-over-f-scaling-0}),
we have analyzed the situation of an encoded spin-qubit residing in its sweet spot,
while
Ref.~\onlinecite{HarveyYacoby2018PRB} deals with
a working bias point that is generally not a sweet spot (not an SOP),
scf. Eq.~(9) of Ref.~\onlinecite{HarveyYacoby2018PRB}.
While such a working point is eligible to consider,
it would imply in addition
a non-zero (transverse) dipole coupling
that would be essential for the analysis,
see, e.g. Ref.\cite{RuskovTahan2019PRB99}.
It is also obvious
(see Appendix \ref{app: $1/f$-dephasing: Scaling of parameters})
that
it would be beneficial to work in an SOP\cite{MartinsKuemethMarcus2016PRL,ReedHunter2016PRL},
where the charge noise is minimized
and the $1/f$ noise will originate
only from the fluctuations of the tunneling gate voltages
that are generally much weaker\cite{MartinsKuemethMarcus2016PRL,ReedHunter2016PRL},
as is considered in this paper.

\section{Discussion and Summary}
\label{ChIV-Discussion-and-Summary}

In this paper we have presented a careful study
of the geometric $n$-qubit (2-qubit) phase gates based on the
modulated longitudinal
coupling $\sim \tilde{g}_{\parallel} \sigma_z (\hat{a} + \hat{a}^{\dag})$,
including the phase gate error mechanisms.
The results for the various kinds of infidelities presented in Table \ref{tab:rates}
imply that infidelities of the order of $\sim 10^{-3}$ are reachable for a range of parameters,
including relatively small charging energy, $U_{\rm charge} \lesssim 0.4\, {\rm meV}$
(see, e.g. Ref.\onlinecite{Xiao2014NC})
and relatively high qubit interdot tunnelings, $t_{l,r} \sim 40 - 80\, \mu{\rm eV}$,
e.g. giving a larger
quantum capacitance, $C_q \propto \frac{\partial^2 E_q}{\partial V_m^2}$.

Increasing
the dynamical longitudinal coupling $\tilde{g}_{\parallel}$,
Eq.~(\ref{dynamic-long-coupling-parallel}),
also suppresses
the resonator (Johnson) noise infidelity $\delta\varepsilon_{\xi}$,
Eq.~(\ref{noise-infidelity-2Qb}),
while decreasing the gate time $t_{\pi}$, Eq.~(\ref{entangling_time}).
The smallness of $\delta\varepsilon_{\xi}$
implies a
resonator $Q$-factor
of
$Q \gtrsim 10^4 - 10^6$, see Fig.~\ref{fig:43}.
Increasing the charging energy from $0.4\, {\rm meV}$ to
currently available values  $\approx 1\, {\rm meV}$
(see, e.g. Refs.~\onlinecite{Maune2012N,FreyWallraff2012PRL,ReedHunter2016PRL,
MiCadyPetta2017SiCQED,Petta2017Science,
Stockklauser-Ihn-Ensslin-Wallraff2017PRX} )
will require  
a 4 times increase of the tunneling rate to keep $C_q$ constant.
Such high tunneling rates of $\sim 160\,\mu{\rm eV}$ were demonstrated recently
for QDs filled with three electrons\cite{WestDzurak2019NNano}.

At a full sweet spot one
avoids
dealing with the qubit's electric dipole moment,
however,
a static
curvature interaction appear even without qubit gate modulation,
that is the
always on ``dispersive-like''
(or quantum capacitance)
interaction ${\cal H}_{\delta\omega}$, Eq.~(\ref{dispersive-like-interaction}).
%
While this interaction
could be interesting for entangling gates on its own,
see Sec.~\ref{Sec:no-modulation-accumulated-phase-gate},
${\cal H}_{\delta\omega}$ is an obstacle for the accumulated phase gates discussed in this paper
and need
to
be suppressed, as no simple cancelation scheme exists
for the case of interest, $\delta\omega_s \gg \kappa$,
see Appendix~\ref{app-C1:infidelity-damping-detuning}.

In order to suppress the infidelity $\delta\varepsilon_{\kappa,\delta\omega}$,
Eqs.~(\ref{infidelity-kappa-delta-omega}) and (\ref{scalings-of-infidelity-kappa-domega}),
one requires a small ratio of
$\frac{\delta\omega_s}{\tilde{g}_{\parallel}}
\equiv \frac{\eta}{\hbar} \, \frac{\hbar \omega_r}{e \tilde{V}_m}$
(effect of $\kappa$ is negligible for high $Q$-factor resonator),
see Eq.~(\ref{curvature-couplings-ratio}),
see Figs.~\ref{fig:42a} and \ref{fig:42b},
and Table \ref{tab:rates}.
This implies a smaller coupling ratio $\frac{\eta}{\hbar}$, Eq.~(\ref{eta-hbar-coupling}),
smaller resonator frequency, $\omega_r$, and larger qubit gate modulation amplitude, $\tilde{V}_m$,
while this infidelity is independent of $C_q$.
The smaller coupling ratio of $\frac{\eta}{\hbar}$, however, will generally make the
entangling gate slower,
which can be compensated only by
larger modulation, $\tilde{V}_m$.

The curvature interactions (``dispersive-like'' and longitudinal) also induce
qubit dephasings and ac-Stark frequency shifts via the resonator shot noise,
implying the infidelities
$\delta\varepsilon_{\phi,\rm shot}$,  $\delta\varepsilon_{\delta\omega,\rm shot}$,
Eqs.~(\ref{infidelity-phi-uncorrelated-white-noise-0}) and (\ref{infidelity-delta-omega-uncorrelated-white-noise-0}).
In addition, an infidelity  $\delta\epsilon_{\tilde{g}_{\parallel},1/f}$, Eq.~(\ref{infidelity-gparallel-charge-0}),
induced via charge noise fluctuation of the
longitudinal coupling, $\tilde{g}_{\parallel}$, is considered.
All these infidelities are shown to be of the order of $10^{-4} - 10^{-5}$ for a range of parameters,
Table \ref{tab:rates}.

The largest infidelity $\delta\varepsilon_{\phi,1/f}$, see Table \ref{tab:rates},
is due to qubit gate $1/f$ charge noise
which scales with the qubit charge dephasing rate and gate time as
$\delta\varepsilon_{\phi,1/f} \propto (\tilde{\Gamma}_\phi t_{\pi})^2$.
With the scalings of the charge noise infidelity     
with $t_{l}$ and $U_{\rm charge}$,
see Eq.~(\ref{infidelity-phi-2qb-one-over-f-scaling-0}),
the increase of $U_{\rm charge}$ by two times (to $\approx 1\, {\rm meV}$)
will require an increase of tunneling by $\approx 6$ times,
which could be experimentally challenging.
While the charge noise infidelity is also quadratically suppressed by $\tilde{V}_m$, Table \ref{tab:rates},
too high modulation amplitude will require including of higher-curvature corrections \cite{RuskovTahan2019PRB99}.
As mentioned at the end of Sec.~\ref{Sec:Damping and detuning gate errors}, these corrections are not harmful:
on one hand corrections to $\delta\omega_s$ and $\tilde{g}_{\parallel}$ could be significant \cite{RuskovTahan2019PRB99},
however the ratio $\frac{\delta\omega_s}{\tilde{g}_{\parallel}}$ will remain approximately the same;
on the other hand, the generated higher-order non-linear Hamiltonians\cite{RuskovTahan2019PRB99},
in addition to the lowest-curvature one, Eqs.~(\ref{dynamical-longitud-interaction}) and (\ref{dispersive-like-interaction}),
brings only small correction of the order of $\lesssim 10^{-4}$.

In the estimation of the charge noise infidelity in Table \ref{tab:rates}, we have used for the
$1/f$-noise spectral density constant a value, $S_{t_l} \simeq 10^{-5} (\mu{\rm eV})^2$,
extracted from the experiment, see Appendix \ref{app: Spectral density constant of tunnelings from experiment}.
Here, the qubit defining gate voltages (as $V_{1,2,3}$ and $V_{t_l}$, $V_{t_r}$ on Fig.~\ref{fig:1}a)
have a typical
spectral density constant $S_{V_{\rm gate}} \lesssim 1\,(\mu{\rm eV})^2$
for Si heterostructures\cite{FreemanJiang2016APL,ConnorsNichol2019PRB}.
From a simple biquadratic model of a DQD (see, e.g., Ref.~\cite{Culcer2010PRB,Xiao2014NC})
one can relate $S_{V_{\rm gate}}$ to the spectral density constants of interest, $S_{t_l}$:
\begin{equation}
S_{t_l} = \left(\frac{t_l}{U_{\rm charge} } \right)^2 S_{V_{\rm gate}} .
\label{spectral-density-const-relations}
\end{equation}
Thus, $S_{t_l}$ can be decreased either by decreasing the ratio $t_l/U_{\rm charge}$,
or by decreasing $S_{V_{\rm gate}}$.
Recent experiment with holes in a SiGe/Ge/SiGe  heterostructure 
shows
a 2-4 times improvement for $S_{V_{\rm gate}}$    
with respect to a Si quantum well 2DEG system\cite{ScapucciVeldhorst2020-preprint}.
This
would make the realization
of our proposal for
{\it remote geometric phase gate entanglement of encoded spin qubits
via  longitudinal couplings to a SC resonator}
possible in
the
near future, with a proposed target infidelity of $10^{-3}$.

Longitudinal coupling of a TQD
or DQD
spin-qubit to a SC resonator
is a viable route to medium distance range ($l \sim 1\, {\rm mm}$)
quantum gates across/off chip.
It offers a way to couple always-on exchange-only TQD qubits
while staying
at their charge dephasing sweet spot.
All the above analysis is applicable to DQD Singlet-Triplet qubits
at the symmetric operating point (see Ref.~\cite{RuskovTahan2019PRB99}).
The modulation scheme allows selectivity
via   a potentially large on/off coupling ratio
(by setting off the gate modulation of relevant qubits, $\tilde{V}^{(j)}_r \to 0$).
In addition, by setting relevant dots' tunnelings to zero, $t_{l,r}^{(j)} \to 0$,
one can switch off the curvature couplings
and
the tunneling gate charge noise,
since
$C_q \equiv \frac{\partial^2 E_q}{\partial \varepsilon_G^2} \propto t_{l,r}^2$, and
$S_{t_l} \propto t_{l,r}^2$, respectively.

The longitudinal coupling rates of tens to hundred MHz can be larger
than the best transverse couplings for a similar TQD system,
where the latter needs a large electric dipole moment,
and therefore are subject to
charge noise dephasing.
In addition,
here
the qubits can be of low frequency
(e.g., highly detuned from the resonator).
Although
entangling
via geometric
phase,
as studied in this paper,
requires some overhead in cycles and correction strategy, it may be useful
over a more simple circuit-QED like coupling
scheme
due to potentially much lower values of the qubit gates' charge noise.

In previous publications\cite{RuskovTahan2017preprint1,RuskovTahan2019PRB99},
we have shown how both curvature couplings, $\tilde{g}_{\parallel}$, $\delta\omega_s$,
can be used for a potentially quantum-limited
QND
measurement of an encoded spin-qubit,
while at the full qubit sweet spot.
In a forthcoming work we will consider $n$-qubit entanglement preparation via
joint qubits measurement
(as discussed preliminary here\cite{RuskovTahan2017preprint1}),
that is based on an extension of previous
``entanglement-by-measurement'' proposals
\cite{RuskovKorotkov2003PRB67,RuskovKorotkovMizel2006PRB72,Lalumiere2010}.

{\bf Acknowledgments:}
The authors are thankful for useful conversations with A. Hunter, J. Kerckhoff, T. Ladd,
M. House, P. Scarlino, and K. Ensslin
for discussion of currently available qubit and resonator parameters.
The authors also appreciate useful discussions with A. Kerman, J. Petta,
G. Burkard, M. Russ, Sue Coppersmith and M. Friesen
on theoretical issues of the presented approach.

\appendix

\section{Geometric phases}
\label{app-A: Geometric phases}

\subsection{Driving a resonator. Single resonator phase}
\label{app-A: driving-a-resonator}
The dynamical longitudinal Hamiltonian, Eq.~(\ref{dynamical-longitud-interaction}),
is  providing a spin dependent ``force'':
$\hat{F} = -\partial {\cal H}_{\parallel}/\partial \hat{x}$, where
$\hat{x} \equiv \Delta x_0 (\hat{a} + \hat{a}^{\dagger})$ is the ``position'' operator,
see Eqs.~(\ref{position-momentum}) and (\ref{zero-point-fluctuations}) below.

To get an intuition, we  first consider  a constant force $F_0$  applied to the resonator:
${\cal H} = \hbar \omega_r \hat{a}^{\dagger}\hat{a} - \Delta x_0 (\hat{a} + \hat{a}^{\dagger}) F_0$.
Diagonalizing  ${\cal H}$  by the transformation, $\hat{b} = \hat{a} - \frac{\Delta x_0 F_0}{\hbar \omega_r}$,
gives the Hamiltonian, ${\cal H} \rightarrow \tilde{{\cal H}} = \hbar \omega_r \hat{b}^{\dagger}\hat{b}$.
The new vacuum state $\hat{b} |0'\rangle = 0$ is a coherent state:
$\hat{a} |0'\rangle = \frac{\Delta x_0 F_0}{\hbar \omega_r} |0'\rangle = \alpha_0 |\alpha_0\rangle$,
displaced from the ground state of an unbiased resonator by
$\alpha_0 \equiv \frac{\Delta x_0 F_0}{\hbar \omega_r}$.
This is represented via a displacement operator $D(\alpha)$:
$|\alpha_0\rangle \equiv D(\alpha_0) |0\rangle \equiv \exp(\alpha_0 \hat{a}^{\dagger} - \alpha_0^{*} \hat{a})|0\rangle$.

\begin{figure} [t!] 
    \centering
        \includegraphics[width=0.45\textwidth]{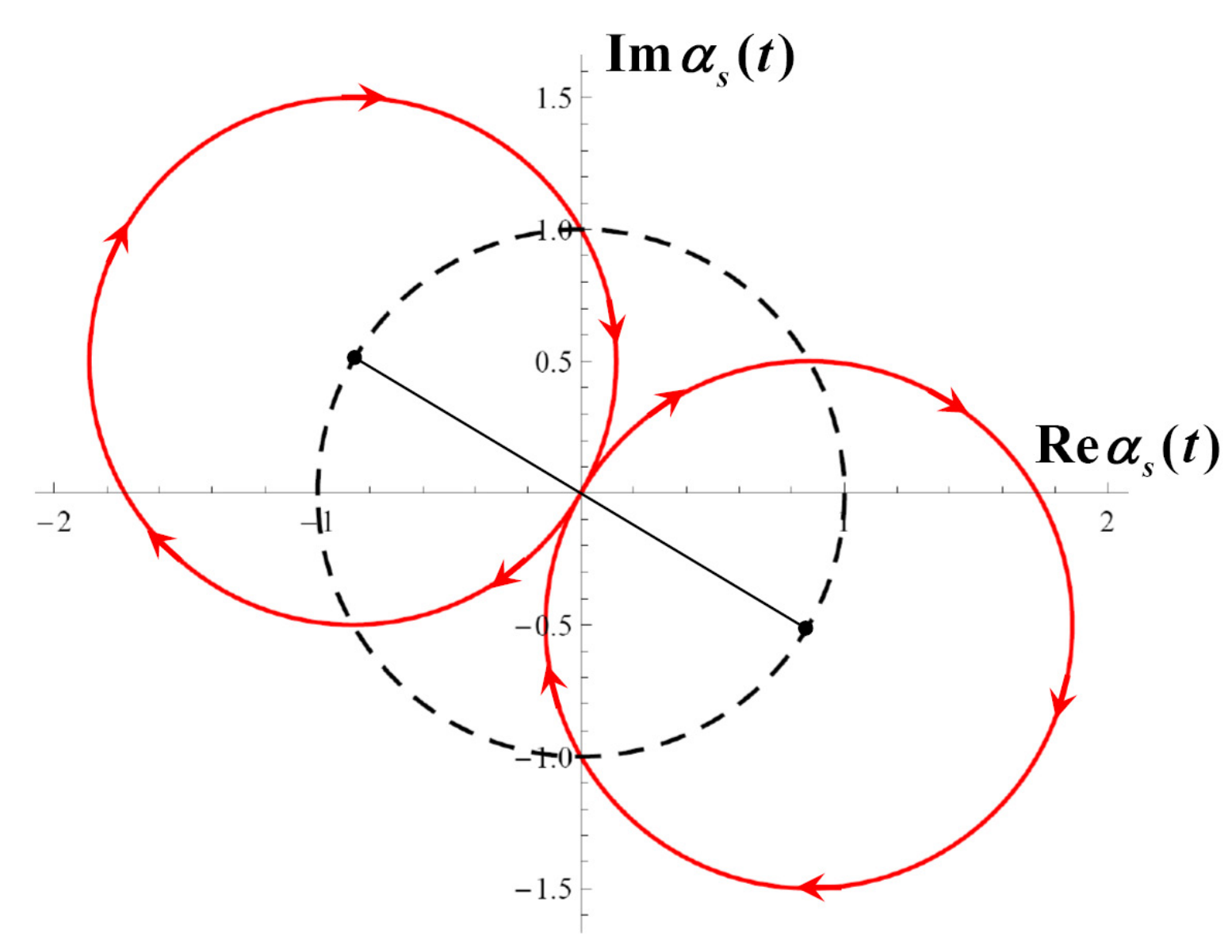}
        \caption{The ideal circular oscillator trajectories for
        two different initial phases: $\varphi_m$ and $\varphi_m' = \varphi_m + \pi$ (red, solid circles).
        The radius of the circles is $\frac{\Omega}{\delta}$ and their centers
        lie on a circle with the same radius put at the origin (black, dashed circle).
        Interchanging the phase $\varphi_m$ at each odd circle allows cancelation
        of small imperfections of the ideal evolution (see Fig.~\ref{fig:3circ} and Appendix B).
        }
        \label{fig:3circ}
\end{figure}

For a force modulated in time, $F(t) = F_0 \cos (\omega_m t + \varphi_m)$,
the above Hamiltonian can be rewritten \cite{Leibried2003N,Roos2008NJP}
in a rotating frame (with $\omega_r$)
and in a rotating wave approximation (RWA) as
\begin{equation}
\tilde{\cal H}_m(t) = \hbar \Omega \left(\hat{a} e^{i (\delta t + \varphi_m)} + \hat{a}^{\dagger} e^{-i (\delta t + \varphi_m)}\right)
\label{driving_resonator} ,
\end{equation}
where we defined $\hbar \Omega \equiv -\frac{\Delta x_0 F_0}{2}$, and
$\delta = \omega_m - \omega_r$ is the detuning of the modulation frequency from the resonance.
The  evolution due to $\tilde{\cal H}_m(t)$  for small time step $dt$ is an infinitesimal displacement
\begin{eqnarray}
&&
e^{-i \frac{\tilde{\cal H}_m(t) dt}{\hbar}} = D[d\alpha(t)] \equiv \exp\left[ d\alpha(t) \hat{a}^{\dagger} - d\alpha^{*}(t) \hat{a} \right]
\qquad
\label{infinitesimal_displacement}
\end{eqnarray}
with $d\alpha(t) = -i\,\Omega e^{-i (\delta t + \varphi_m)} dt$.
Integrating for finite times one gets
the (ideal) evolution of a resonator under a driving periodic force:
\begin{equation}
\alpha^{\rm id}(t) = \alpha(0) - \left(\frac{\Omega}{\delta} \right) e^{-i \varphi_m} \left(1 - e^{-i t\delta} \right)
\label{ideal_alpha} .
\end{equation}
In the phase space of $\{{\rm Re}\alpha(t), {\rm Im}\alpha(t)\}$
this describes a clock-wise rotating circle path starting at the origin (for $\alpha(0)=0$),
with radius $R = \frac{\Omega}{\delta}$ and center
$O_m = \frac{\Omega}{\delta}(-\cos \varphi_m, \sin \varphi_m)$:
\begin{eqnarray}
&& {\rm Re}\alpha(t) = \left(\frac{\Omega}{\delta} \right) \left[-\cos \varphi_m + \cos(\delta t + \varphi_m)  \right]
\label{ut}\\
&& {\rm Im}\alpha(t) = \left(\frac{\Omega}{\delta} \right) \left[\sin \varphi_m - \sin(\delta t + \varphi_m)  \right]
\label{vt}
\end{eqnarray}
For further reference, on Fig.~\ref{fig:3circ} we show a full circle (for a gate time $t_g = \frac{2\pi}{\delta}$)
with an initial phase $\varphi_m$ and a second circle with $\varphi_m' = \varphi_m + \pi$ .

Using the standard relation for displacement operators
\begin{equation}
D(\alpha) D(\beta) = D(\alpha+\beta) e^{i\, {\rm Im}(\alpha \beta^*)}
\end{equation}
one obtains
the total displacement for a finite time in the form \cite{Leibried2003N,Roos2008NJP}:
\begin{equation}
D_{\rm tot} = \underset{n \to \infty}{\rm lim}\, D[d\alpha(t_n)] \ldots D[d\alpha(0)] = D[\alpha(t)]\ e^{i \Phi(t)}
\end{equation}
substituting for an ideal  accumulated (geometric) phase $\Phi^{\rm id}(t)$:
%
\begin{eqnarray}
&& \Phi^{\rm id}(t) = {\rm Im}\left[ \int^t_0 \alpha^*(t') d\alpha(t') \right] =
\left( \frac{\Omega}{\delta} \right)^2 \left[ \sin \delta\, t - \delta\, t\right] .
\qquad
\label{accumulated_phase}
\end{eqnarray}
For a gate time $t_g = 2\pi/\delta$, when $\alpha(t)$  makes a full circle  in the phase space
$\{ {\rm Re}\alpha, {\rm Im}\alpha\}$, Fig.~\ref{fig:3circ}, 
the accumulated phase is
$\Phi_g = 2\pi \left( \Omega/\delta \right)^2$
which is twice the encircled area
of radius $\Omega/\delta$ and is independent of the initial phase $\varphi_m$.

\subsection{Accumulated phases for resonator with $n$ qubits}

One considers the Hamiltonian of a resonator with $n$ modulated qubits,
Eq.~(\ref{dynamical-longitud-interaction}),
\begin{equation}
{\cal H}/\hbar =
\omega_r \hat{a}^{\dagger}\hat{a} +
\sum_{j=1}^n  \left[ \tilde{g}_{\parallel}^{(j)}\, \sigma_z^{(j)} + \tilde{g}_0^{(j)} \right]
                  \cos (\omega_m t + \varphi_m) (\hat{a} + \hat{a}^{\dagger} )
\end{equation}
for a chosen $n$-qubit spin state $|s\rangle \equiv |i_1, i_2, \ldots, i_n \rangle$
with $|i_k \rangle = |\uparrow \rangle {\rm or} |\downarrow \rangle$ being the eigenstates
of the k-th qubit, $\sigma_z^{(k)} |i_k \rangle = i_k  |i_k \rangle$.
Since $|s\rangle$ are eigenstates of ${\cal H}_{\parallel}$,
for each particular spin state the driving strength of the resonator $\Omega$ in Eq.~(\ref{driving_resonator})
is replaced by
\begin{eqnarray}
&&\Omega_{\parallel, s} \equiv
\langle s| \hat{\Omega}_{\parallel}  | s\rangle
\equiv \langle s|\, \frac{1}{2} \sum_{k=1}^n
\left[ \tilde{g}^{(k)}_{0} + \tilde{g}^{(k)}_{\parallel} \sigma_z^{(k)}  \right] \, | s\rangle
\nonumber\\
&& { } = \frac{1}{2} \sum_{k=1}^n \left[ \tilde{g}^{(k)}_{0} + \tilde{g}^{(k)}_{\parallel} i_{k}  \right]
\label{Omega_s} .
\end{eqnarray}
See also Eqs.~(\ref{omega-n-qubit-coupling})-(\ref{ideal_Phi0}) of the main text.

One can also derive the {\it accumulated phase matrix for $n$ qubits}, since
$\hat{\Omega}_{\parallel}$ and $\hat{\Omega}_{\parallel}^2$ have the same eigenstates $| s\rangle$.
After little algebra and dropping the common phase,
the accumulated phase matrix for time $t$ reads:
%
\begin{eqnarray}
&&\hat{\Phi}(t)   
= \frac{\left( \sin \delta\, t - \delta\, t\right)}{2 \delta^2} \times
\nonumber\\
&& \qquad \times
\left[ \sum_{j<k}^n \tilde{g}^{(j)}_{\parallel} \tilde{g}^{(k)}_{\parallel}\, \sigma_z^{(j)} \otimes \sigma_z^{(k)}
+ \sum_{j=1}^n \tilde{g}^{(j)}_{0}\, \sum_{k=1}^n \tilde{g}^{(k)}_{\parallel}\, \sigma_z^{(k)} \right]
\qquad
\label{Phi-matrix} .
\end{eqnarray}

For further applications one considers a gate time with $N$ cycles, $t_g = \frac{2\pi N}{\delta}$.
Up to single-qubit operations, the accumulated phase for $n$ qubits and for $N$ cycles 
becomes
\begin{equation}
\hat{\Phi}_N
= -\frac{\pi N}{\delta^2}
\sum_{j<k}^n \tilde{g}^{(j)}_{\parallel} \tilde{g}^{(k)}_{\parallel}\, \sigma_z^{(j)} \otimes \sigma_z^{(k)}
\label{Phi_N} .
\end{equation}

\subsection{Controlled $\pi$-phase gate for two qubits}
For two qubits one requires the relation
\begin{equation}
\frac{\pi N \tilde{g}^{(1)}_{\parallel} \tilde{g}^{(2)}_{\parallel} }{\delta^2} = \frac{\pi}{4}
\label{pi-4} ,
\end{equation}
which sets the required frequency difference
$\delta = 2\sqrt{N \tilde{g}^{(1)}_{\parallel} \tilde{g}^{(2)}_{\parallel} }$
to obtain the phase matrix
$\hat{\Phi}_N = -\frac{\pi}{4}\, \sigma_z^{(1)} \otimes \sigma_z^{(2)}$.
One can show that this is equivalent to a controlled $\pi$-phase gate
(up to single-qubit operations)
\begin{eqnarray}
&& | \uparrow \uparrow \rangle \to  | \uparrow \uparrow \rangle
\nonumber\\
&& | \uparrow \downarrow \rangle \to  e^{i\, \frac{\pi}{2}} \, | \uparrow \downarrow \rangle
\nonumber\\
&& | \downarrow \uparrow \rangle \to   e^{i\, \frac{\pi}{2}} \, | \downarrow \uparrow \rangle
\nonumber\\
&& | \downarrow \downarrow \rangle \to   | \downarrow \downarrow \rangle
= e^{-i\, \pi} \, \left( e^{i\ \frac{\pi}{2}} | \downarrow_1 \rangle \right) \, \left( e^{i\ \frac{\pi}{2}} | \downarrow_2 \rangle \right)
\label{controlled-pi-phase} .
\end{eqnarray}

\section{Equations of motion for the partial density matrices $\hat{\rho}{ss}(t)$, $\hat{\rho}{ss'}(t)$,
and respective quantum averages, such as $\alpha_s(t)$, etc.}
\label{App-B:Equations-of-motion}

One starts from the Caldeira-Leggett master equation
that was derived \cite{CaldeiraLeggett1983a,CaldeiraLeggett1983b}
for superconducting Josephson circuits in the context of searching for the macroscopic quantum coherence.
While the first derivation was only for the high temperature limit,
the result was later extended to zero temperatures as well \cite{CaldeiraCerdeira1989}.
The time evolution of the $n$ qubits plus a SC resonator density matrix reads
(${\cal \tilde{H}}_{\rm tot}$ is in the rotating frame with $\omega_d$):
\begin{equation}
\frac{d\rho}{dt} = -i \left[{\cal \tilde{H}}_{\rm tot},\rho\right]
-i\frac{\kappa}{2\hbar} \left[\hat{x}, \left\{\hat{p}, \rho \right\}_{+} \right]  
-\frac{K_d}{\hbar^2}\, \left[\hat{x}, \left[\hat{x}, \rho \right] \right] ,
\label{damping_diffusion-CL1}
\end{equation}
where
$\left\{\hat{p}, \rho \right\}_{+} \equiv \hat{p} \rho +  \rho \hat{p}$ is an anticommutator,
\begin{equation}
\hat{x} \equiv \Delta x_0 (\hat{a} + \hat{a}^{\dagger} ), \ \
\hat{p} \equiv -i \Delta p_0 (\hat{a} - \hat{a}^{\dagger} )
\label{position-momentum}
\end{equation}
are the ``position'' and ``momentum'' operators,
and the zero-point fluctuations
are given by
\begin{equation}
\Delta x_0 \equiv \sqrt{\frac{\hbar}{2\omega_r L_r}},\ \ \
\Delta p_0 \equiv \sqrt{\frac{\hbar\omega_r L_r}{2}}
\label{zero-point-fluctuations}
\end{equation}
($\Delta x_0 \Delta p_0 = \frac{\hbar}{2}$).
Note, that as
inductance plays the role of {\it a mass},
$\Delta x_0$ has dimension of charge
\cite{RLC-position-momentum}.
Using Eq.~(\ref{position-momentum})
one can show that  Eq.~(\ref{damping_diffusion-CL1}) coincides with the
analogouse equation of the quantum optics \cite{MilburnWalls-book2008} in the RWA.
Namely, the damping and diffusion term can be reduced in the RWA to the familiar quantum optics decoherence terms
expressed via $\hat{a}$, $\hat{a}^{\dagger}$:
\begin{eqnarray}
&& -i\frac{\kappa}{2\hbar} \left[\hat{x}, \left\{\hat{p}, \rho \right\}_{+} \right]
-\frac{K_d}{\hbar^2}\, \left[\hat{x}, \left[\hat{x}, \rho \right] \right] \simeq
\nonumber\\
&&\qquad\qquad\qquad\qquad { } \simeq
\frac{\kappa}{2} \left( n_{\rm th} + 1 \right)\,
\left(2 \hat{a}\rho \hat{a}^{\dagger} -  \hat{a}^{\dagger}\hat{a}\rho  -  \rho \hat{a}^{\dagger}\hat{a}\right)
\nonumber\\
&&\qquad\qquad\qquad\qquad\ \  { }
+
\frac{\kappa}{2} n_{\rm th} \,
\left(2 \hat{a}^{\dagger}\rho \hat{a} -  \hat{a}\hat{a}^{\dagger}\rho  -  \rho \hat{a}\hat{a}^{\dagger}\right)
\label{Caldeira-Leggett-to-quantum-optics} ,
\end{eqnarray}
where $n_{\rm th} \equiv \frac{1}{2}( \coth\frac{\hbar\omega_r}{2T_r} - 1)$.
In what follows, we will use both the Caldeira-Leggett and the quantum optics forms
depending on the case of study.

\subsection{Evolution for the partial density matrices $\hat{\rho}{ss}(t)$, $\hat{\rho}{ss'}(t)$}
\label{App-B1: Evolution for the partial density matrices}
By expanding the qubit-resonator density matrix in the complete set of qubit operators
$|s\rangle \langle s'|$
\cite{WisemanMilburn-book2010}:
\begin{equation}
\rho = \sum_{s,s'} \hat{\rho}_{s,s'} |s\rangle \langle s'|
\label{expanded-density-matrix1}
\end{equation}
one is to obtain equations for the
partial density matrices
$\hat{\rho}_{s,s'}$
by substituting into Eq.~(\ref{damping_diffusion-CL1})
and finding the respective (anti)commutators.

Starting with the unitary evolution, $\propto \left[{\cal \tilde{H}}_{\rm tot},\rho\right]$,
the Hamiltonian ${\cal \tilde{H}}_{\rm tot}$,
Eq.~(\ref{total_Hamiltonian0}),
contains the linear form
$\hat{A}(\hat{a},\hat{a}^{\dagger})
\equiv \hat{a} e^{i(\tilde{\delta}t + \varphi_d)} + \hat{a}^{\dagger} e^{-i(\tilde{\delta}t + \varphi_d)}$
and the higher operators
\begin{eqnarray}
&& \hat{n} \equiv \hat{a}^{\dagger}\hat{a}    
\\
&& \hat{X}_{\varphi} \, \hat{\Omega}_{\parallel} \equiv
\left[ \hat{a} e^{i\, (\tilde{\delta} t + \varphi)} +
\hat{a}^{\dagger} e^{-i\, (\tilde{\delta} t + \varphi)} \right]\, \hat{\Omega}_{\parallel}
\\
&& \delta \hat{\omega}\, \hat{a}^{\dagger}\hat{a}
\label{operators-of-Hamiltonian} .
\end{eqnarray}
Here and below,
the frequency differences are in general rotating frame with $\omega_{r'}$
and we assume $\omega_d = \omega_m$, Eq.~(\ref{cancellation-of-g0}),
\begin{equation}
\tilde{\omega}_r = \omega_r - \omega_{r'}, \ \ \   \tilde{\delta} = \omega_m - \omega_{r'} .
\label{frequency-differences}
\end{equation}
%
The $n$-qubit operators and their eigenvalues are denoted as:
\begin{eqnarray}
&& \hat{\Omega}_{\parallel} \equiv \frac{1}{2} \sum_j  
\left[ \tilde{g}^{(j)}_{0} + \tilde{g}^{(j)}_{\parallel} \sigma_z^{(j)}  \right],\ \
\hat{\Omega}_{\parallel} |s \rangle = \Omega_{\parallel,s} |s \rangle
\label{n-qubit-longitud-operators-of-Hamiltonian}\\
&& \delta \hat{\omega} \equiv \sum_j \delta\omega^{(j)} \sigma_z^{(j)},\ \
\delta \hat{\omega} |s \rangle = \delta\omega_s |s \rangle
\label{n-qubit-freq-shift-operator-of-Hamiltonian} .
\end{eqnarray}
The essential commutators are calculated as:
\begin{eqnarray}
&& \left[\hat{A}(\hat{a},\hat{a}^{\dagger}),\, \hat{\rho}_{s,s'}\, |s\rangle \langle s'| \right] =
\left[\hat{A}(\hat{a},\hat{a}^{\dagger}),\, \hat{\rho}_{s,s'} \right]\, |s\rangle \langle s'|
\label{commutators-of-Hamiltonian-1}\\
&& \left[\delta \hat{\omega}\, \hat{a}^{\dagger}\hat{a},\, \hat{\rho}_{s,s'}\, |s\rangle \langle s'| \right] =
(\hat{n}\, \hat{\rho}_{s,s'}\, \delta\omega_s - \hat{\rho}_{s,s'}\, \hat{n}\, \delta\omega_{s'} )\, |s\rangle \langle s'|
\label{commutators-of-Hamiltonian-2}\\
&& \left[\hat{X}_{\varphi_m} \, \hat{\Omega}_{\parallel},\,  \hat{\rho}_{s,s'}\, |s\rangle \langle s'| \right] =
( \hat{a}\,  \hat{\rho}_{s,s'}\, \Omega_{\parallel,s}
- \hat{\rho}_{s,s'}\, \hat{a}\, \Omega_{\parallel,s'} )\, |s\rangle \langle s'|
\nonumber\\
&& \qquad\qquad\qquad\qquad\qquad { } \times e^{i\, (\tilde{\delta} t + \varphi_m)}
\label{commutators-of-Hamiltonian-3}\\
&& \qquad\qquad\qquad\qquad\ { }
+(\tilde{\delta} \to -\tilde{\delta}, \varphi_m \to -\varphi_m, \hat{a} \to \hat{a}^{\dagger})
\qquad
\label{commutators-of-Hamiltonian-4} .
\end{eqnarray}

By introducing the shortcomings
\begin{equation}
B_{\pm,ss'} \equiv \Omega_{\parallel,s} \pm \Omega_{\parallel,s'}, \ \
A_{\pm,ss'} \equiv \frac{\delta\omega_s \pm \delta\omega_{s'} }{2}
\label{A-B-combinations}
\end{equation}
one obtains the following equation of motion for $\hat{\rho}_{ss'}(t)$
in a rotating frame with $\omega_{r'}$:
\begin{eqnarray}
&& \frac{d\hat{\rho}_{ss'} }{dt} = -i\, \tilde{\omega}_r \left[\hat{n}, \hat{\rho}_{s,s'} \right]
\nonumber\\
&&\qquad\quad  { } -\frac{i}{2}\, B_{+,ss'}\, \left[\hat{X}_{\varphi_m},  \hat{\rho}_{s,s'} \right]
-\frac{i}{2}\, B_{-,ss'}\, \left\{ \hat{X}_{\varphi_m},  \hat{\rho}_{s,s'} \right\}_{+}
\nonumber\\
&&\qquad\quad  { } -i\, A_{+,ss'}\, \left[\hat{n},  \hat{\rho}_{s,s'} \right]
+i\, A_{-,ss'}\, \left\{ \hat{n},  \hat{\rho}_{s,s'} \right\}_{+}
\nonumber\\
&&\qquad\quad  { }
-i\, \varepsilon_d\, \left[\hat{X}_{\varphi_d},  \hat{\rho}_{s,s'} \right]
+ {\rm \{ qubits\ evolution\} }
\nonumber\\
&&\qquad\quad  { } -i\frac{\kappa}{2\hbar} \left[\hat{x}, \left\{\hat{p}, \hat{\rho}_{ss'} \right\}_{+} \right]
-\frac{K_d}{\hbar^2}\, \left[\hat{x}, \left[\hat{x}, \hat{\rho}_{ss'} \right] \right]
\label{partial-dm-non-diagonal} ,
\end{eqnarray}
For $s=s'$  one recovers the evolution of the diagonal  
partial density matrix, $\hat{\rho}_{ss}(t)$, Eq.~(\ref{partial-dm-diagonal}) of the main text.

\subsection{Uncorrelated $T_1$, $T_2$ processes}

The term ${\rm \{ qubits\ evolution \}}$ contains the collection of qubits Hamiltonians,
$\frac{{\cal H}_q}{\hbar} = \sum_{j=1}^n \frac{\omega_q^{(j)} }{2}\, \sigma_z^{(j)}$,
and qubits relaxation and dephasing.
Using the notation for the $n$-qubits spin states,
\begin{eqnarray}
\lefteqn{
|s\rangle \equiv |i_1\ldots i_j \dots i_n \rangle, \ \ \ |s'\rangle \equiv |i'_1 \ldots i'_j \dots i'_n \rangle,}
\label{ss'-n-qubit-product-states} \\
&& (i_j, i'_j = \pm 1)
\qquad\qquad\qquad\qquad\qquad\qquad\qquad
\nonumber
\end{eqnarray}
one obtains for the $s,s'$-term in the expansion
of Eq.~(\ref{expanded-density-matrix1}):
\begin{eqnarray}
&& \langle s| \frac{(-i)}{\hbar} \left[ {\cal H}_q, \rho\right] |s' \rangle 
= \hat{\rho}_{s,s'} \, \sum_j  (-i)\frac{\omega_q^{(j)} }{2}\, (i_j - i'_j)
\label{Hamiltonian-qubit-evolution}\\
&& \langle s|  \sum_j \gamma^{(j)}_{1}  {\cal D}[\sigma^{(j)}_{-}] \rho |s' \rangle
\label{T1-qubit-evolution}\\
&& \langle s|  \sum_j \frac{\gamma^{(j)}_{\phi} }{2} {\cal D}[\sigma^{(j)}_z] \rho  |s' \rangle
= -\hat{\rho}_{s,s'}\, \sum_{j\subset \{i_j \neq i'_j \} }\, \gamma^{(j)}_{\phi}
\label{T2-qubit-evolution} .
\end{eqnarray}
Eq.~(\ref{T1-qubit-evolution}) is the contribution generated by the qubits relaxation ($T_1$-process), which we
will not show explicitly here. Nevertheless, it is worth mentioning that
the dephasing of the $s,s'$ subspace generated by the relaxations
can be calculated via the following mnemonic rule:
``For each population that leaves the state $s$ or the state $s'$ with rate $\gamma_1^{(j)} \equiv 1/T_1^{(j)}$
one obtains
the dephasing contribution $- \frac{\gamma_1^{(j)}}{2} \rho_{q,ss'}$''
summed up over all such cases:
\begin{equation}
\frac{d \rho_{q,ss'} }{dt} \propto  -\sum_{j \subset \{{\rm subset}\} } \frac{\gamma_1^{(j)}}{2} \rho_{q,ss'}
\label{contribution-of-relaxation-to-dephasing} .
\end{equation}
The above relaxation and dephasing contributions are given here for further reference.
In a full sweet spot for each of the qubits involved,
both relaxation $\gamma_1^{(j)}$ and charge dephasing $\gamma^{(j)}_{\phi}$
arising via the transverse and longitudinal dipole moments
{\it vanish in this regime}.

\subsection{Degression on the equations of motion for averages and variances. Gaussian resonator states}
\label{App-B2: equations of motion for averages and variances}

For each partial density matrix $\hat{\rho}_{s,s'}$
one can derive equations for the averages and variances of the
position and momentum operators \cite{RuskovSchwabKorotkov2005}.
One defines the averages and the variances
(here $\sigma \equiv \{s,s'\}$ is a compound index):
\begin{eqnarray}
&& \bar{x}_{\sigma} \equiv \langle \hat{x}\rangle_{\sigma} = {\rm Tr}\left[\hat{x} \hat{\rho}_{\sigma}\right],\ \
\bar{p}_{\sigma} \equiv \langle \hat{p}\rangle_{\sigma} = {\rm Tr}\left[\hat{p} \hat{\rho}_{\sigma}\right]
\label{x-av-p-av}\\
&& D^{({\sigma})}_x \equiv \langle \hat{x}^2 \rangle_{\sigma} - \langle \hat{x}\rangle_{\sigma}^2
\label{Dx}\\
&& D^{({\sigma})}_p \equiv \langle \hat{p}^2 \rangle_{\sigma} - \langle \hat{p}\rangle_{\sigma}^2
\label{Dp}\\
&& D^{({\sigma})}_{xp} \equiv
\frac{ \left\langle \hat{x}\hat{p} + \hat{p}\hat{x} \right\rangle_{\sigma} }{2}
- \langle \hat{x}\rangle_{\sigma} \langle \hat{p}\rangle_{\sigma}
\label{Dxp}
\end{eqnarray}

Below we consider only the diagonal density matrices and replace
$\{s,s'\} \to s$
(the non-diagonal case can be calculated similarly).
In a general rotating frame with
the frequency $\omega_{r'}$ one gets for the evolution equations
for the averages and variances,
the latter are made dimensionless
via the
zero-point fluctuations, $\Delta x_0$, $\Delta p_0$:
$d^{(s)}_x \equiv D^{(s)}_x/(\Delta x_0)^2$,
$d^{(s)}_p \equiv D^{(s)}_p/(\Delta p_0)^2$,
$d^{(s)}_{xp} \equiv D^{(s)}_{xp}/(\Delta x_0 \Delta p_0)$.
%
%
\begin{eqnarray}
&& \frac{\dot{\bar{x}}_s }{\Delta x_0} =  \frac{\bar{p}_s}{\Delta p_0} \left( \tilde{\omega}_r  - \delta\omega_s \right)
- 2 \Omega_{\parallel, s} \, \sin (\tilde{\delta} t + \varphi_m)
\label{dx1}
\\
&& \frac{\dot{\bar{p}}_s }{\Delta p_0} = -  \frac{\bar{x}_s}{\Delta x_0} \left( \tilde{\omega}_r  - \delta\omega_s \right)
- 2 \Omega_{\parallel, s}\,  \cos (\tilde{\delta} t + \varphi_m)
\nonumber\\
&& \qquad\qquad { } - 2\varepsilon_d e^{-i (\tilde{\delta}t + \varphi_d})  - \kappa\, \frac{\bar{p}_s }{\Delta p_0} ,
\label{dp1}
\\
&& \dot{d^{(s)}}_x = \left( \tilde{\omega}_r  - \delta\omega_s \right)\, 2d^{(s)}_{xp}
\label{dDx}
\\
&& \dot{d^{(s)}}_p = - \left( \tilde{\omega}_r  - \delta\omega_s \right)\, 2d^{(s)}_{xp}
- 2\kappa\, d^{(s)}_p + 2 \frac{K_d}{(\Delta p_0)^2}
\label{dDp}
\\
&& \dot{d^{(s)}}_{xp} = \left( \tilde{\omega}_r  - \delta\omega_s \right)
\left\{ d^{(s)}_p -  d^{(s)}_x\right\}  - \kappa\, d^{(s)}_{xp}
\label{dDxp} .
\end{eqnarray}
Eqs.~(\ref{dx1})-(\ref{dDxp}) are correct {\it for any density matrix}
and the derivation
is straightforward from Eq.~(\ref{partial-dm-non-diagonal}):
one is just using the commutation relations (e.g. $[\hat{x},\hat{p}]=i\hbar$, etc.)
and the cyclic property of the trace.

For a Gaussian density  matrix 
the higher moments are expressed via
the five moments, Eqs.~(\ref{x-av-p-av})-(\ref{Dxp}),
and the evolution of the state is completely described by them.
%
%
It is important to mention that Eq.~(\ref{partial-dm-non-diagonal}) as well as Eq.~(\ref{damping_diffusion-CL1})
preserves the Gaussianity of the state.
Moreover, under continuous quantum measurement of a resonator
a non-Gaussian state rapidly goes to a Gaussian state
(see, e.g. Ref.~\onlinecite{RuskovSchwabKorotkov2005}).

It is worth to mention that
Eqs.~(\ref{dx1})-(\ref{dp1}) are exactly the classical Hamilton equations of a damped driven oscillator.
By combining them,
we reproduce
the more familiar
equation of motion for the
``field'' variable
$\alpha_s \equiv \langle \hat{a} \rangle_s =
\frac{1}{2}\left( \frac{\bar{x}_s}{\Delta x_0} + i \frac{\bar{p}_s}{\Delta p_0} \right)$:
%
\begin{eqnarray}
&& \dot{\alpha}_s = - i \left( \tilde{\omega}_r - \delta\omega_s \right)\, \alpha_s
-i \Omega_{\parallel, s} e^{-i (\tilde{\delta} t + \varphi_m)} -i \varepsilon_d e^{-i(\tilde{\delta}t + \varphi_d )}
\nonumber\\
&&\qquad\ { } - \frac{\kappa}{2} \left( \alpha_s - \alpha^*_s \right) ,
\label{alpha-equation1}
\end{eqnarray}
where
the difference from the quantum optics equation is the last (contra-rotating) term,
that can be neglected in a RWA.
For a coherent state $|\alpha\rangle$ (defined via $\hat{a}|\alpha\rangle = \alpha |\alpha\rangle$)
its density matrix has minimal variances $d_x = d_p =1$, $d_{xp}=0$.
This property is preserved
by Eqs.~(\ref{dDx})-(\ref{dDxp}) at zero resonator temperature $T_r=0$.
The variances are not affected by the resonator driving ($\sim \varepsilon_d$)
and qubit modulation $\sim \Omega_{\parallel, s}$.

Also one mentions that the diffusion term,
$\sim K_d\, \left[\hat{x}, \left[\hat{x}, \hat{\rho}_{ss'} \right] \right]$
enters only in the equations for the variances.
This last property will be used below to calculate the resonator noise
induced $n$-qubit phase gate error (see Appendix \ref{App-D: Johnson nosie infidelity}).

\subsection{Positive $P^{(+)}$-representation and the derivation of the partial dephasing rates $\Gamma_{s,s'}$
for an $n$-qubit system coupled to a resonator at $T_r=0$}

For the derivation of the dephasing rates we will assume  zero resonator temperature, $T_r=0$,
as a good approximation
(since $\hbar\omega_r \gg k_B T_r$ for a $5-10\, {\rm GHz}$ resonator and
typical SC resonator temperatures $T_r$ of $20-50\, {\rm mK}$).


The partial density  matrices\cite{WisemanMilburn-book2010} $\hat{\rho}_{s,s'}$
of the expansion Eq.~(\ref{expanded-density-matrix1})
can be represented using the positive $P^{(+)}$-representation \cite{GardinerZoller-book2000}
\begin{equation}
\hat{\rho}_{ss'}(t)
= \int d^2\alpha\, d^2\beta \,
\frac{|\alpha \rangle \langle \beta^* |}{\langle \beta^* |\alpha \rangle} \, P_{ss'}(\alpha,\beta,t)
\label{positive-P}
\end{equation}
where $|\alpha \rangle$  is a coherent state. 

From the definition of coherent states one gets the relations
\begin{eqnarray}
&& \hat{a}\,|\alpha \rangle = \alpha\,|\alpha \rangle
\\
&&  \hat{a}^{\dagger}\,|\alpha \rangle =
\left(\frac{\partial }{\partial \alpha} + \frac{\alpha^*}{2} \right)\,|\alpha \rangle
\label{Gardiner-relations} .
\end{eqnarray}
One then derives useful correspondences between various terms of the
equation of motion, Eq.~(\ref{partial-dm-non-diagonal}), and the corresponding
$P^{(+)}$-representation kernels:
\begin{eqnarray}
&& \frac{d \hat{\rho}_{ss'} }{dt} \to \frac{d P_{ss'}(\alpha,\beta,t)}{dt}
\label{drho-dt}\\
&&  \left[\hat{a}^{\dagger}\hat{a}, \hat{\rho}_{ss'} \right] \to \left( \beta \frac{\partial }{\partial \beta} - \alpha \frac{\partial }{\partial \alpha}\right)  P_{ss'}(\alpha,\beta)
\\
&&  \left\{\hat{a}^{\dagger}\hat{a}, \hat{\rho}_{ss'} \right\}_{+} \to  \left[ 2\alpha\beta
-  \frac{\partial }{\partial \alpha} \left( \alpha\, \cdot \right)
-  \frac{\partial }{\partial \beta} \left( \beta\, \cdot \right) \right]  P_{ss'}(\alpha,\beta) \qquad\ \
\\
&&  \left[(\hat{a}^{\dagger} + \hat{a}), \hat{\rho}_{ss'} \right] \to \left(\frac{\partial }{\partial \beta} - \frac{\partial }{\partial \alpha}\right) P_{ss'}(\alpha,\beta)
\\
&& \left[\hat{X}_{\varphi}, \hat{\rho}_{ss'} \right] \to \left( e^{i\, \varphi} \frac{\partial }{\partial \beta} - e^{-i\, \varphi}  \frac{\partial }{\partial \alpha}\right)  P_{ss'}(\alpha,\beta)
\\
&& \left\{\hat{X}_{\varphi}, \hat{\rho}_{ss'} \right\}_{+} \to \left( 2\alpha e^{i\, \varphi} + 2\beta e^{-i\, \varphi} \right.
\\
&& \left. \qquad\qquad\qquad { }- e^{i\, \varphi} \frac{\partial }{\partial \beta} - e^{-i\, \varphi}  \frac{\partial }{\partial \alpha}\right)  P_{ss'}(\alpha,\beta)
\\
&& {\cal D}[\hat{a}] \hat{\rho}_{ss'} \to \frac{1}{2}\left[ \frac{\partial }{\partial \alpha} \left( \alpha\, \cdot \right)
+  \frac{\partial }{\partial \beta} \left( \beta\, \cdot \right) \right]  P_{ss'}(\alpha,\beta)
\label{Ruskov-Gardiner-relations} .
\end{eqnarray}

Substituting Eqs.~(\ref{drho-dt})-(\ref{Ruskov-Gardiner-relations}) into Eq.~(\ref{partial-dm-non-diagonal})
one derives equation for the positive kernel $P_{ss'}(\alpha,\beta)$:
\begin{eqnarray}
&&  \frac{d P_{ss'}(\alpha,\beta)}{dt} =
\nonumber\\
&&
\frac{i\, \partial }{\partial \alpha}
\left[ \left( (\tilde{\omega}_r + \delta\omega_s) \alpha + \tilde{\varepsilon}_d
+ \Omega_{\parallel, s}\, e^{-i\, (\tilde{\delta} t + \varphi_m)} - i\, \frac{\kappa\alpha}{2}
\right) P_{ss'} \right]
\nonumber\\
&&
{ } -
\frac{i\, \partial }{\partial \beta}
\left[ \left( (\tilde{\omega}_r + \delta\omega_{s'}) \beta + \tilde{\varepsilon}_d^*
+ \Omega_{\parallel, s'}\, e^{ i\, (\tilde{\delta} t + \varphi_m)} + i\, \frac{\kappa\beta}{2}
\right) P_{ss'} \right]
\nonumber\\
&& { } - i\, B_{-,ss'}\, \left(\alpha\, e^{ i\, (\tilde{\delta} t + \varphi_m)}
+ \beta\, e^{ -i\, (\tilde{\delta} t + \varphi_m)} \right)\, P_{ss'}
\nonumber\\
&& { } - i\, A_{-,ss'}\, 2\alpha\beta\, P_{ss'}
\label{Positive-kernel-Pss'} ;
\end{eqnarray}
here
$\tilde{\varepsilon}_d \equiv \varepsilon_d\, e^{-i(\tilde{\delta}t + \varphi_d )}$
and
$A_{-,ss'}$, $B_{-,ss'}$ are
from Eq.~(\ref{A-B-combinations}).

For the diagonal kernel $P_{ss}(\alpha,\beta)$ one then obtains:
\begin{eqnarray}
&&  \frac{d P_{ss}(\alpha,\beta)}{dt} =
\nonumber\\
&&
\frac{i\, \partial }{\partial \alpha}
\left[ \left( (\tilde{\omega}_r + \delta\omega_s) \alpha + \tilde{\varepsilon}_d
+ \Omega_{\parallel, s}\, e^{-i\, (\tilde{\delta} t + \varphi_m)} - i\, \frac{\kappa\alpha}{2}
\right) P_{ss} \right]
\nonumber\\
&& { } -
\frac{i\, \partial }{\partial \beta}
\left[ \left( (\tilde{\omega}_r + \delta\omega_s) \beta + \tilde{\varepsilon}_d^*
+ \Omega_{\parallel, s}\, e^{ i\, (\tilde{\delta} t + \varphi_m)} + i\, \frac{\kappa\beta}{2}
\right) P_{ss} \right] .
\nonumber\\
{ } \label{Positive-kernel-Pss}
\end{eqnarray}
%

\subsection{Coherent state ansatz  for  $P_{ss'}(\alpha,\beta)$  and solutions of the equations of motion}
\label{Coherent-state-ansatz-for-Pss'}

For $T_r=0$,
we consider coherent (pure Gaussian) states
and Eqs.~(\ref{Positive-kernel-Pss'}), (\ref{Positive-kernel-Pss}) can be solved
via coherent state ansatz
\cite{Gambetta2008PRA}    
\begin{eqnarray}
&& P_{ss'}(\alpha,\beta) =
\rho^q_{ss'}(t) \delta^{(2)}\left(\alpha - \alpha_{s}(t) \right)\, \delta^{(2)}\left(\beta - \alpha_{s'}^*(t) \right) .
\quad
\label{P_ss'}
\end{eqnarray}
In this case, by substituting into Eq.~(\ref{positive-P})
one obtains
\begin{eqnarray}
&& \hat{\rho}_{s,s'} =
\int d^2\alpha\, d^2\beta \, \frac{|\alpha \rangle \langle \beta^* |}{\langle \beta^* |\alpha \rangle} \,
\rho^q_{ss'}(t) \,
\nonumber\\
&& \qquad  \times \delta^{(2)}\left(\alpha - \alpha_{s}(t) \right)\, \delta^{(2)}\left(\beta - \alpha_{s'}^*(t) \right)
\nonumber\\
&&  \qquad { } = \rho^q_{ss'}(t) \,
\frac { |\alpha_{s}(t) \rangle \langle \alpha_{s'}(t) | }{\langle \alpha_{s'}(t) |\alpha_{s}(t) \rangle } ,
\label{coherent-rho_ss'}
\end{eqnarray}
and the total qubits-resonator density matrix becomes
\begin{equation}
\rho = \sum_{s,s'} \hat{\rho}_{s,s'} |s\rangle \langle s'| =
\sum_{s,s'}  \rho^q_{ss'}(t) \,
\frac { |\alpha_{s}(t) \rangle \langle \alpha_{s'}(t) | }{\langle \alpha_{s'}(t) |\alpha_{s}(t) \rangle }\,
|s\rangle \langle s'|
\label{coherent-qubits-resonator-dm}
\end{equation}
It is straightforward to obtain from this representation the reduced density matrix of the $n$ qubits
or of the resonator by tracing out the other degrees of freedom:
\begin{eqnarray}
&& \rho_{\rm qb} = {\rm Tr}_{\rm res} [\rho] = \sum_{s,s'}  \rho^q_{ss'}(t) \, |s\rangle \langle s'|
\label{reduced-rho-q}\\
&& \rho_{\rm res} = {\rm Tr}_{\rm qb} [\rho] = \sum_{s}  \rho^q_{ss}(t) \, |\alpha_{s}(t) \rangle \langle \alpha_{s}(t) |
\label{reduced-rho-res} ,
\end{eqnarray}
where for the trace over the resonator we have used
${\rm Tr}_{\rm res} [\ldots] = \frac{1}{\pi} \int d^2 \alpha_1 \langle \alpha_1| \ldots | \alpha_1 \rangle$.
It is interesting to note that the reduced resonator density matrix $\rho_{\rm res}$ is a mixed state
with weights being the diagonal qubits density matrix elements, $\rho^q_{ss}(t)$.
Note, however, that $|\alpha_{s}(t) \rangle$-states are not orthogonal in general:
$\langle \alpha_{s'}(t) |\alpha_{s}(t) \rangle \neq 0$.

From Eq.~(\ref{Positive-kernel-Pss}) for  $\dot{P}_{ss}(\alpha,\beta)$ one obtains for
the quantum average of the field, $\alpha_{s}(t) \equiv \langle \alpha_{s}(t) |\hat{a} | \alpha_{s}(t) \rangle$,
the equations in RWA (in a rotating frame with $\omega_{r'}$):
\begin{equation}
\dot{\alpha}_s = - i \left( \tilde{\omega}_r + \delta\omega_s -i\, \frac{\kappa}{2} \right)\, \alpha_s
- i\, \varepsilon_d \, e^{-i (\tilde{\delta}t + \varphi_d )}
-i \Omega_{\parallel, s} e^{-i (\tilde{\delta} t + \varphi_m)}
\label{alpha-equation2-quant-optics}  ,
\end{equation}
where $\tilde{\omega}_r \equiv \omega_r - \omega_{r'}$, $\tilde{\delta} \equiv \omega_m - \omega_{r'}$.
As expected, Eq.~(\ref{alpha-equation2-quant-optics}) that is derived for coherent states ($T_r = 0$),
coincides with the general Eq.~(\ref{alpha-equation1}), the latter being true for general Gaussian or non-Gaussian states at any $T_r \neq 0$.
For the sake of further reference we write here the general solution of Eq.~(\ref{alpha-equation2-quant-optics}).
\begin{eqnarray}
%
&& \alpha_s(t) =
\left( \alpha_s(0) -
\frac{ \Omega_{\parallel,s}\, e^{-i\,\varphi_m} + \varepsilon_d\, e^{-i\,\varphi_d}}
     {\tilde{\delta} - o_s} \right)\, e^{-i o_s t}
\nonumber\\
&&\qquad\  { } +
\left( \frac{\Omega_{\parallel,s}\, e^{-i\,\varphi_m} + \varepsilon_d\, e^{-i\,\varphi_d} }{\tilde{\delta} - o_s} \right)
\, e^{-i \tilde{\delta} t}
\label{alpha-solution-quant-optics}  ,
\end{eqnarray}
where $o_s \equiv \tilde{\omega}_r - \delta\omega_s -i \frac{\kappa}{2}$.

With this solution the average photon number in the resonator can be calculated
for times $t\lesssim \frac{2\pi}{\delta}$ (assuming $\alpha_s(0)=0$):
\begin{equation}
\langle \hat{n} \rangle = |\alpha_s(t)|^2 \simeq
2 \left( \frac{\Omega^{\varepsilon}_{\parallel,s}}{\delta} \right)^2 (1 - \cos \delta t)
\leq 4 \left( \frac{\Omega^{\varepsilon}_{\parallel,s}}{\delta} \right)^2 ,
\label{photon-number}
\end{equation}
where we have used that in the parameter regime of interest, see Table~\ref{tab:rates},
$\delta \sim \tilde{g}_{\parallel} \gg \delta\omega_s \gg \kappa$.
For a two-qubit controlled $\pi$-phase gate one gets $\langle \hat{n} \rangle \leq \frac{1}{2}$.

From Eq.~(\ref{Positive-kernel-Pss'}) for  $\dot{P}_{ss'}(\alpha,\beta)$, and using the solution
of Eq.~(\ref{alpha-equation2-quant-optics}) for $\alpha_s(t)$
one obtains equation for the reduced density matrix of the $n$ qubits:
\begin{eqnarray}
&& \frac{d\rho^q_{ss'}(t)}{dt} =  -i [\omega_{q,ss'} - i \gamma_{2,ss'} ]\, \rho^q_{ss'}
-2 i A_{-,ss'}  \alpha_{s}(t)\, \alpha_{s'}^*(t)\, \rho^q_{ss'}
\nonumber\\
&&\quad  { }
- i B_{-,ss'} \left( \alpha_{s}(t)\, e^{i (\tilde{\delta} t + \varphi_m)}
+  \alpha_{s'}^*(t)\, e^{-i (\tilde{\delta} t + \varphi_m)} \right)
 \, \rho^q_{ss'}.  \qquad\qquad
\label{dot-rho-q_ss'}
\end{eqnarray}
%
In the derivation of the equation of motion  we have used a relation for the Dirac delta-function:
\begin{equation}
x \frac{\partial \delta(x - x_0)}{\partial x} = x_0 \frac{\partial \delta(x - x_0)}{\partial x} - \delta(x - x_0)
\label{derivative-delta-function-relation} .
\end{equation}
In Eq.~(\ref{dot-rho-q_ss'}),
the first term includes the $s,s'$-transition frequency, Eq.~(\ref{Hamiltonian-qubit-evolution}),
\begin{equation}
\omega_{q,ss'} = \sum_j \frac{\omega_q^{(j)} }{2}\, (i_j - i'_j)
\label{ss'-transition-n-qubit-frequency} ,
\end{equation}
and the linear $s,s'$-dephasing is arising as a particular sum of all qubits internal dephasings,
Eqs.~(\ref{T2-qubit-evolution}),(\ref{contribution-of-relaxation-to-dephasing}):
\begin{equation}
\gamma_{2,ss'} = \sum_{j\subset \{i_j \neq i'_j \} }\, \gamma^{(j)}_{\phi}
+ \sum_{j \subset \{{\rm subset}\} } \frac{\gamma_1^{(j)}}{2}
\label{linear-dephasing-n-qubits} .
\end{equation}

\subsection{$n$-qubit dephasing rates and frequency shifts mediated by the resonator photon shot noise}
\label{app: dephasing rates from resonator photon shot noise}

The second and third term of Eq.~(\ref{dot-rho-q_ss'}) will provide
the qubits dephasing due to resonator leakage (photon shot noise)
{\it mediated by the  curvature (quantum capacitance) interactions with the resonator}.
Indeed, by integrating Eq.~(\ref{dot-rho-q_ss'}) one obtains the solution:
%
\begin{eqnarray}
&& \rho^q_{ss'}(t) =  \rho^q_{ss'}(0) \,\,
\scalebox{1.1}{e}^{\textstyle -i [\omega_{q,ss'} - i \gamma_{2,ss'} ]\, t}
\nonumber\\
&&\quad\ \ { }\times
\scalebox{1.1}{e}^{\textstyle -2 i A_{-,ss'}  \int_0^t dt' \alpha_{s}(t')\, \alpha_{s'}^*(t') }
\nonumber\\
&&\quad\ \ { }
\times \scalebox{1.1}{e}^{\textstyle - i B_{-,ss'}
\int_0^t dt' \left( \alpha_{s}(t')\, \scalebox{1.1}{e}^{\scriptstyle i (\tilde{\delta} t' + \varphi_m)}  +
\alpha_{s'}^*(t')\, \scalebox{1.1}{e}^{\scriptstyle -i (\tilde{\delta} t' + \varphi_m)} \right) }
\nonumber\\
{ }
\label{rho-q_ss'}
\end{eqnarray}
For the accumulated phase gates of this paper we consider
the equal modulation and driving frequencies,
$\omega_m = \omega_d$, in order to fulfil the cancelation of the spin-independent curvature couplings
$\tilde{g}_0^{(j)}$, see Eq.~(\ref{cancellation-of-g0}).
For a rotating frame with $\omega_r$,
$\tilde{\delta} \equiv \omega_m - \omega_r \equiv \delta$ and
$\tilde{\omega}_r = 0$.
Using then the solution of Eq.~(\ref{alpha-equation2-quant-optics})
for $t \gg 1/\kappa$,
\begin{equation}
\alpha_s^{t\gg} =
\frac{\Omega_{\parallel,s}\, e^{-i\,\varphi_m} + \varepsilon_d\, e^{-i\,\varphi_d} }
{\delta - \delta\omega_s -i\frac{\kappa}{2} }
\, e^{-i \delta t}
\equiv \alpha_s^{\rm st}\, e^{-i \delta t}
\label{alpha-stationary-solution} ,
\end{equation}
one obtains the qubit non-diagonal
density matrix evolution in the long-time limit of Eq.~(\ref{rho-q_ss'}):
%
\begin{eqnarray}
&& \rho^q_{ss'}(t) =  \rho^q_{ss'}(0) \,\,
%
\scalebox{1.1}{e}^{\textstyle -i \left[\omega_{q,ss'} + \delta\omega^{\rm shot}_{ss'} \right] t}
\nonumber\\
&&\qquad\quad\ { }\times
\scalebox{1.1}{e}^{\textstyle  - \left[\gamma_{2,ss'} + \Gamma^{\rm shot}_{\phi,ss'} \right] t}
%
%
\label{shot-noise-dephasing} ,
\end{eqnarray}
where the resonator induced {\it shot noise}
qubits dephasings and qubits frequency shifts (via curvature coupling)
are given by:
%
\begin{eqnarray}
&&  \Gamma^{\rm shot}_{\phi,ss'} \equiv - 2 A_{-,ss'} {\rm Im} [\alpha_s^{\rm st} \alpha_{s'}^{\rm st *}]
\nonumber\\
&&\qquad\quad\  { } - B_{-,ss'} {\rm Im}  [ \alpha_{s}^{\rm st}\, e^{i \varphi_m}  +
                                             \alpha_{s'}^{\rm st *}\, e^{-i \varphi_m} ]
\label{shot-noise-curvature-induced-qubit-dephasing}
\end{eqnarray}
and
\begin{eqnarray}
&&  \delta\omega^{\rm shot}_{ss'} \equiv - 2 A_{-,ss'} {\rm Re} [\alpha_s^{\rm st} \alpha_{s'}^{\rm st *}]
\nonumber\\
&&\qquad\quad\ \ { }
- B_{-,ss'} {\rm Re}  [ \alpha_{s}^{\rm st}\, e^{i \varphi_m}  +  \alpha_{s'}^{\rm st *}\, e^{-i \varphi_m} ]
{ } \label{shot-noise-curvature-induced-qubit-freq-shift} ,
\end{eqnarray}
respectively.
By choosing in the above equations $\varphi_m=0$ and $\varphi_d =\varphi_m + \pi$,
one arrives at the long-time dephasing rates and frequency shifts of
Eqs.~(\ref{delta-gamma-0})-(\ref{delta-omega-long-0}) of the main text.

In the short-time limit, where
$t \lesssim t_g = \frac{2\pi}{\delta} \ll \frac{1}{\delta\omega_s} \ll \frac{1}{\kappa}$,
one explicitly integrates Eq.~(\ref{rho-q_ss'})
to obtain for the first time integral in the exponent:
\begin{eqnarray}
&&
- i \int_0^t dt' \alpha_{s}(t')\, \alpha_{s'}^*(t')
\simeq - 2 \left( t - \frac{\sin \delta t}{\delta} \right)\, \, f_s\, f_{s'}\,
\nonumber\\
&& \times
\left\{ \frac{\kappa}{2} (\delta\omega_s - \delta\omega_{s'})
 + i\left[ (\delta - \delta\omega_s)(\delta - \delta\omega_{s'}) + \frac{\kappa^2}{4} \right]\right\} ,
 \qquad
\label{1st-exponent-finite-t}
\end{eqnarray}
where we denote
$f_s \equiv \frac{\Omega_{\parallel,s}^{\varepsilon}}{(\delta - \delta\omega_s)^2 + \frac{\kappa^2}{4}}$.
For the second time integral in the exponent in Eq.~(\ref{rho-q_ss'})
one obtains ($\varphi_m = 0$):
\begin{eqnarray}
&&
- i \int_0^t dt' \left(\alpha_{s}(t')\, \scalebox{1.1}{e}^{\scriptstyle i \delta t' }  +
\alpha_{s'}^*(t')\, \scalebox{1.1}{e}^{\scriptstyle -i \delta t' } \right)
\simeq \left\{ \left( t - \frac{\sin \delta t}{\delta} \right) \right.
\nonumber\\
&&
\times \left. \left[ -\frac{\kappa}{2} (f_s - f_{s'})
- i [f_s (\delta - \delta\omega_s) + f_{s'} (\delta - \delta\omega_{s'})] \right] \right.
\nonumber\\
&& \left.
+ \frac{\cos {\delta t} - 1}{\delta} \right.
\nonumber\\
&& \left.
\times \left[ [f_s (\delta - \delta\omega_s) - f_{s'} (\delta - \delta\omega_{s'})]
- i \frac{\kappa}{2} (f_s + f_{s'}) \right] \right\} .
\label{2nd-exponent-finite-t}
\end{eqnarray}
The linear in time 
expressions in Eqs.~(\ref{1st-exponent-finite-t}) and (\ref{2nd-exponent-finite-t}),
that survive for $t=t_g = \frac{2\pi N}{\delta}$, contribute to
the effective dephasing rates, $\tilde{\Gamma}^{\rm shot}_{\phi,ss'}$ of Eq.~(\ref{delta-gamma-1}),
and ac Stark shifts, $\delta{\tilde{\tilde\omega}}^{\rm shot}_{ss'}$ of Eq.~(\ref{delta-omega-12}),
of the main text.


\section{Gate infidelity due to resonator damping, always on curvature ``dispersive-like'' coupling,
 and due to resonator noise}
%

\subsection{Combined gate infidelity}
\label{app-C:combined-infidelity-noise-curvature-detuning}
For $n$-qubits plus resonator one starts with an initial product state
$|\psi_i\rangle \equiv \sum_s a_s |s\rangle \, |0\rangle$
[we assume
vacuum resonator initial state, $|\alpha_{s}(0)\rangle =|0\rangle$, for simplicity].
For an ideal evolution, performing
complete $N$ cycles at a gate time $t_g=\frac{2\pi N}{\delta}$, Fig.~\ref{fig:3circ}
one ends up in a product state:
\begin{equation}
|\psi_f^{\rm id}\rangle
= \sum_s a_s e^{i\Phi_{s}^{\rm id}(t)} |s \rangle \, |\alpha_{s}^{\rm id}(t)\rangle \mid_{t=\frac{2\pi N}{\delta}}
\label{final-product-state-n-qubits}
\end{equation}
since $\alpha_{s}^{\rm id}(\frac{2\pi N}{\delta}) = 0$ and
$\Phi_{s}^{\rm id}(\frac{2\pi N}{\delta}) = 2\pi N \left( \frac{\Omega_{\parallel,s}^{\varepsilon}}{\delta} \right)^2$
is the ideal phase given by Eq.~(\ref{ideal_Phi0}).

For a non-ideal evolution, at the end of the gate cycle the resonator trajectory
in the phase space obtains non-zero contributions from the resonator noise $\xi_f(t)$,
the resonator damping $\kappa$, and from the spin-dependent frequency shift, $\delta\omega_s$,
Eq.~(\ref{spin-dependent-frequency-shift}),
denoted as
\begin{eqnarray}
&&\delta\alpha_{s}(t) = \delta\alpha_{s}^{\xi}(t) + \delta\alpha_{s}^{k,\delta\omega}(t)
\label{delta-alpha1}
\\
&& \delta\Phi_{s}(t;[\alpha_{s}(t)]) = \delta\Phi_{s}^{\xi}(t) + \delta\Phi_{s}^{k,\delta\omega}(t)
\label{delta-Phi1} ,
\end{eqnarray}
the latter being path-dependent functionals of $\alpha_{s}(t)$.
These contributions at gate time $t_g$ lead to $|\alpha_s(t_g) \rangle \neq |0 \rangle$,
and thus leave some qubit-resonator entanglement 
leading to qubits' gate infidelity.
Also, there appear non-local in time errors via the accumulated phases.

The fidelity of the actual final state with respect to the ideal final state
is then expressed for a particular trajectory $\alpha_{s}(t)$,
\begin{eqnarray}
&& |\langle \Psi_f | \Psi_f^{\rm id} \rangle|^2
= \sum_s  |a_s|^4 \, e^{ -|\delta\alpha_{s}(t)|^2 }
\nonumber\\
&& \quad { } + \sum_{s < s'} 2 |a_s|^2 |a_{s'}|^2  \,
e^{ -\frac{1}{2} (|\delta\alpha_{s}|^2 + |\delta\alpha_{s'}|^2 ) } \,
\cos (\delta\Phi_{s} - \delta\Phi_{s'} ) .  \qquad\ \ \
\label{fidelity-noise-kappa-delta-omega}
\end{eqnarray}
Averaging over all initial $n$-qubit states
leads to the replacements
$|a_s|^4  \to \overline{|a_s|^4 } \equiv f_1$,
$2 |a_s|^2 |a_{s'}|^2   \to \overline{2 |a_s|^2 |a_{s'}|^2} \equiv f_{12}$,
with $f_1 = f_{12}$ given by Eq.~(\ref{n-qubit-averages-1}),
Appendix \ref{app C2: Averaging over the n-qubit initial state}

Averaging over the noise is using the concavity of the exponent,
$\langle e^A \rangle \geq e^{\langle A \rangle}$,
the relation (for small fluctuations)
%
$\langle e^A \cos B \rangle_\xi \gtrsim
\langle e^A  e^{-\frac{B^2}{2} } \rangle_\xi >
e^{\langle A \rangle_\xi} e^{-\langle B^2 \rangle_\xi/2 } $,
and the zero noise averages $\langle \delta\alpha_s^{\xi} \rangle_\xi = 0$,
$\langle \delta\Phi_{s}^{\xi} \rangle_\xi = 0$.
Thus, for the combined average fidelity
one obtains
\begin{eqnarray}
&& F_{\xi,\kappa,\delta\omega} \equiv
\langle \overline{ |\langle \Psi_f | \Psi_f^{\rm id} \rangle|^2 } \rangle_\xi \gtrsim
f_1 \, \sum_s \, \scalebox{1.1}{e}^{\scriptstyle -\left( \langle |\delta\alpha_{s}^{\xi}|^2 \rangle_\xi + |\delta\alpha_{s}^{k,\delta\omega}|^2 \right)} \qquad\ \
\nonumber\\
&& \qquad\quad { } + f_{12} \, \sum_{s < s'} \,
\scalebox{1.1}{e}^{\scriptstyle  -\frac{1}{2} \left( \langle|\delta\alpha_{s}^{\xi}|^2 \rangle_\xi + |\delta\alpha_{s}^{k,\delta\omega}|^2 +
\langle|\delta\alpha_{s'}^{\xi}|^2 \rangle_\xi + |\delta\alpha_{s'}^{k,\delta\omega}|^2 \right)}
\nonumber\\
&& \qquad\qquad  { } \times
\scalebox{1.1}{e}^{\textstyle -\frac{ \left\langle \left(\delta\Phi_{s}^{\xi} - \delta\Phi_{s'}^{\xi} \right)^2 \right\rangle_\xi}{2} } \
\scalebox{1.1}{e}^{\textstyle -\frac{ \left( \delta\Phi_{s}^{\kappa,\delta\omega} - \delta\Phi_{s'}^{\kappa,\delta\omega} \right)^2 }{2} }
\label{combined-average-fidelity}  .
\end{eqnarray}

For small fluctuations,
$\langle |\delta\alpha_{s}^{\xi}|^2 \rangle_\xi,  |\delta\alpha_{s}^{k,\delta\omega}|^2 \ll 1$
and
$\langle |\delta\Phi_{s}^{\xi}|^2 \rangle_\xi,  |\delta\Phi_{s}^{k,\delta\omega}|^2 \ll 1$
the infidelity $\delta\epsilon \equiv 1 - F$ splits into
two independent contributions:
\begin{eqnarray}
&&  \delta\epsilon_{\xi,\kappa,\delta\omega}  =
1 - F_{\xi,\kappa,\delta\omega} \cong  \delta\epsilon_{\kappa,\delta\omega} + \delta\epsilon_{\xi}
\label{infidelity-splits}
\\
&&  \delta\epsilon_{\kappa,\delta\omega} = \frac{1}{2^n} \sum_s |\delta\alpha_{s}^{k,\delta\omega}|^2
+ \frac{f_{12}^{(n)}}{2} \sum_{s < s'}
\left( \delta\Phi_{s}^{\kappa,\delta\omega} - \delta\Phi_{s'}^{\kappa,\delta\omega} \right)^2
\qquad\ \
\label{infidelity-kappa-delta-omega-1}
\\
&&  \delta\epsilon_{\xi} =
\frac{1}{2^n} \sum_s
\langle |\delta\alpha_{s}^{\xi}|^2 \rangle_\xi +
\frac{f_{12}^{(n)}}{2} \sum_{s < s'}
\left\langle \left(\delta\Phi_{s}^{\xi} - \delta\Phi_{s'}^{\xi} \right)^2 \right\rangle_\xi
\label{infidelity-xi}  .
\end{eqnarray}
Here, Eq.~(\ref{infidelity-kappa-delta-omega-1}) for $\delta\epsilon_{\kappa,\delta\omega}$
and Eq.~(\ref{infidelity-xi}) for $\delta\epsilon_{\xi}$,
reproduce
Eqs.~(\ref{infidelity-kappa-delta-omega})
and (\ref{infidelity-xi-0})
of the main text, respectively.
In Eq.~(\ref{infidelity-kappa-delta-omega-1}) we have used that
$f_1^{(n)} + f_{12}^{(n)} \frac{2^n-1}{2} = \frac{1}{2^n}$,
that follows from Eq.~(\ref{n-qubit-averages-0}).

Similarly, one can consider the deviations of the resonator trajectory
due to charge ($\xi_q$) noise fluctuations of the qubit tunnelings, $\delta t_{l,r}(t)$.
The corresponding infidelity is calculated in
Sec.~\ref{Sec: Infidelity via the charge fluctuations of the longitudinal (curvature) coupling},
similar
to the Johnson noise, Eq.~(\ref{infidelity-xi}),
see Appendix \ref{app D2: Johnson-noise-variances}.

\subsection{Averaging over the $n$-qubit initial state}
\label{app C2: Averaging over the n-qubit initial state}
With the initial state
\begin{equation}
|\Psi_i \rangle = \sum_s a_s |s \rangle
\label{initial_state}
\end{equation}
one needs to find the averages
\begin{equation}
f_s \equiv \overline{|a_s|^4}, \ \ \ f_{s,s'} \equiv \overline{2 |a_s|^2 |a_s'|^2}
\label{n-qubit-averages}
\end{equation}
where $\overline{(\ldots)}$ denotes averaging on the initial $n$-qubit state.
Since the averages are invariant under any unitary transformation of the basis states
$|s\rangle$, one can argue that the averages are $s,s'$-independent
and we denote: $f_1^{(n)} = f_s^{(n)}$, $f_{s,s'}^{(n)} = f_{12}^{(n)}$, for any $s,s'$.
Moreover, we will conjecture the {\it uniformity condition}
\begin{equation}
f_1^{(n)} = f_{12}^{(n)}
\label{uniformity-condition}
\end{equation}
(for the 1-qubit and 2-qubit case see derivation below).
Starting with the normalization condition
$\sum_s |a_s|^2 = 1$ one obtains
\begin{eqnarray}
&& 1 = \overline{\left(\sum_s |a_s|^2\right)^2}
= \sum_{s=1}^m \overline{|a_s|^4} + \sum_{s<s'}^m \overline{2 |a_s|^2 |a_s'|^2}
\nonumber\\
&& \qquad { } = m f_1^{(n)} + \frac{m^2-m}{2} f_{12}^{(n)}
\label{n-qubit-averages-0} ,
\end{eqnarray}
where $m \equiv 2^n$ is the $n$-qubit space dimension.
Assuming the uniformity condition one obtains
\begin{equation}
f_1^{(n)} = f_s^{(n)} = f_{12}^{(n)} = f_{s,s'}^{(n)} = \frac{1}{2^n + \frac{2^{2n}-2^n}{2}}
\label{n-qubit-averages-1} .
\end{equation}

\subsubsection{One-qubit and two-qubit cases}
Eq.~(\ref{n-qubit-averages-1}) can be  confirmed for the one-qubit and two-qubit cases by
explicit averaging over the corresponding Bloch sphere.

Indeed, for the one-qubit (pure state) density matrix one is
using $S^2$ Bloch sphere representation (in 3D)
with the amplitudes of
$|\Psi_i \rangle = \sum_s a_s |s \rangle = a_1 |\uparrow\rangle + a_2 |\downarrow\rangle$,
obtaining:
\begin{eqnarray}
&& x_0 = \cos \phi_0 = |a_1|^2 - |a_2|^2   
\\
&& x_1 = \sin \phi_0 \cos \phi_1 = 2 {\rm Re} (a_1 a_2^*)   
\\
&& x_2 = \sin \phi_0 \sin \phi_1 = 2 {\rm Im} (a_1 a_2^*)
\label{3D-Bloch-sphere}
\\
&& \phi_0 \in [0,\pi], \ \phi_1 \in [0,2\pi)
\nonumber ,
\end{eqnarray}
with the $S^2$ area element, $dS_2 = d\phi_0 d\phi_1 \sin \phi_0$,  and the total area of $S_2 = 4\pi$.
Noting that $|a_1|^4 = \frac{1 + x_0^2 + 2x_0}{4}$, $|a_2|^4 = \frac{1 + x_0^2 - 2x_0}{4}$,
and $2 |a_1|^2 |a_2|^2 = \frac{x_1^2 + x_2^2}{2}$,
one obtains the averages
\begin{eqnarray}
&& \overline{|a_1|^4} = \overline{|a_2|^4} = \frac{1 + \overline{x_0^2}}{4} = \frac{1}{3}
\label{3D-sphere-averaging-a1}\\
&& \overline{2|a_1|^2 |a_2|^2} = \frac{\overline{x_1^2 + x_2^2}}{2} = \frac{1}{3}
\label{3D-sphere-averaging-a1-a2} ,
\end{eqnarray}
where the averaging over the $S^2$-sphere is represented by
\begin{equation}
\overline{(\ldots)} = \frac{1}{S_2} \int dS_2 \, (\ldots)
\label{S2-averaging}
\end{equation}
The result of Eqs.~(\ref{3D-sphere-averaging-a1}) and (\ref{3D-sphere-averaging-a1-a2})
is in agreement with Eq.~(\ref{n-qubit-averages-1})
for $n=1$.
(More comprehensive one-qubit averages can be found in Ref.~\onlinecite{KeaneKorotkov2012PRA}).

For the two-qubit (pure state)
$|\Psi_i \rangle = a_1 |\uparrow\uparrow\rangle + a_2 |\uparrow\downarrow\rangle
                 + a_3 |\downarrow\uparrow\rangle + a_4 |\downarrow\downarrow\rangle$
one is using the $S^4$-Bloch sphere
coordinates (in 5D), related to the density matrix elements:
(see, e.g. Ref.~\onlinecite{Mosseri2001}):
\begin{eqnarray}
&& x_0 = \cos \phi_0 = |a_1|^2 + |a_2|^2 - |a_3|^2 - |a_4|^2
\label{x0-2qb}
\\
&& x_1 = \sin \phi_0 \cos \phi_1 = 2 {\rm Re} (a_1^* a_3 + a_2^* a_4)
\label{x1-2qb}
\\
&& x_2 = \sin \phi_0 \sin \phi_1 \cos \phi_2 = 2 {\rm Im} (a_1^* a_3 + a_2^* a_4)
\label{x2-2qb}
\\
&& x_3 = \sin \phi_0 \sin \phi_1 \sin \phi_2 \cos \phi_3 = 2 {\rm Re} (a_1 a_4 - a_2 a_3)
\label{x3-2qb}
\\
&& x_4 = \sin \phi_0 \sin \phi_1 \sin \phi_2 \sin \phi_3 = 2 {\rm Im} (a_1 a_4 - a_2 a_3) \qquad\quad
\label{5D-Bloch-sphere}
\\
&& \phi_0, \phi_1, \phi_2, \in [0,\pi], \  \phi_3 \in [0,2\pi)
\nonumber .
\end{eqnarray}
The area element is $dS_4 = d\phi_0 d\phi_1 d\phi_2 d\phi_3 \sin^3 \phi_0 \sin^2 \phi_1 \sin \phi_2$
and the total area of the $S^4$-Bloch sphere is $S_4 = \frac{8\pi^2}{3}$.
Using the 5D Bloch sphere representation,  
it is straightforward to show that
\begin{equation}
f_s \equiv \overline{|a_s|^4} = f_{s,s'} \equiv \overline{2 |a_s|^2 |a_s'|^2}
= \frac{1}{10}, \ \ \forall\, s,s' = 1,2,3,4
\label{2-qubit-averages} ,
\end{equation}
in agreement with Eq.~(\ref{n-qubit-averages-1})
for $n=2$.

\subsubsection{The $n$-qubit case}
In the general $n$-qubit case an explicit averaging may be cumbersome.
Instead, one can use symmetry arguments.
Indeed, the $n$-qubit density matrix,
$\rho = \sum_{s,s'} a_s a_{s'}^* |s \rangle \langle s'|$, can be expanded in the
$2^{2n} - 1$  basis operators (Kronecker products of Pauli matrices:
here, $\sigma_0 \equiv I_2$, $\sigma_{1,2,3} \equiv \sigma_{x,y,z}$),
$\hat{f}_l = \sigma_{\mu_1}\otimes \ldots \otimes \sigma_{\mu_n}$, $\mu_i = 0,1,2,3, \ \forall i$.
\begin{equation}
\rho = \frac{1}{2^n} \left(1 + \sum_{l=1}^{2^{2n} - 1} w_l \hat{f}_l  \right)
\label{n-qubit-density-matrix expansion} .
\end{equation}
This expansion can be performed by writing the operators $|s \rangle \langle s'|$
as Kronecker products:
\begin{equation}
|s \rangle \langle s'| \equiv |i_1,\ldots,i_n \rangle \langle i'_1,\ldots,i'_n |
= |i_1 \rangle \langle i'_1| \otimes \ldots \otimes |i_n \rangle \langle i'_n | , \qquad
\end{equation}
and mentioning that (${k}$ enumerates the qubits)
\begin{eqnarray}
&& |i_k \rangle \langle i'_k| = \frac{1}{2}\,(\sigma_0^{(k)} \pm \sigma_3^{(k)})\ {\rm for}\ i_k = i'_k
\\
&& |i_k \rangle \langle i'_k| = \frac{1}{2}\,(\sigma_1^{(k)} \pm i\,\sigma_2^{(k)})\ {\rm for}\ i_k \neq i'_k
\label{reexpansion-denstiy-matrix-operators} .
\end{eqnarray}
Using symmetry arguments along the line of Ref.~\onlinecite{CabreraBaylis2007PL},
one can show that averaging over the $n$-qubit initial state for the expansion coefficients
$w_k$ leads to
\begin{equation}
\overline{w_l} = 0, \ \ \ \overline{w_l^2} = \frac{1}{1 + 2^n}
\label{average-w_k^2}
\end{equation}
On the other hand $w_l^2$ can be re-expanded as linear combinations
of $|a_s|^4$ and $2 |a_s|^2 |a_{s'}|^2$, $s,s' \in \{1,\dots, 2^n\}$.
Solving these equations one can re-establish
the uniformity condition, Eq.~(\ref{n-qubit-averages-1}), for the $n$-qubit case.

\subsection{The infidelity $\delta\epsilon_{\kappa,\delta\omega}$ in a simple strategy}
\label{app-C1:infidelity-damping-detuning}
In this section we perform exact calculations of the infidelity
$\delta\epsilon_{\kappa,\delta\omega}$ in a simple strategy when
the modulation amplitude
$\Omega_{\parallel,s}$ changes sign on each subsequent cycle, see Fig.~\ref{fig:4circ-non-ideal}.
This can be achieved by changing the phase of the qubits gate modulations,
$\varphi_m \to \varphi_m + \pi$, and simultaneously changing the phase
of the resonator driving, $\varphi_d \to \varphi_d + \pi$, while keeping the
relation $\varphi_d = \varphi_m + \pi$, see Eq.~(\ref{cancellation-of-g0}).
In Sec. \ref{Sec:Damping and detuning gate errors} of the main text we have shown that this simple strategy
works well when the energy curvature resonator shift,
$\delta\omega_s$, Eq.~(\ref{spin-dependent-frequency-shift}) is small
or can be neglected with respect to resonator damping $\kappa$.
In general, one deals with the opposite case of $\delta\omega_s \gg \kappa$.

For the $n$-qubit state $|s\rangle$ one uses
the general solution for the $\alpha_s(t)$, Eq.~(\ref{alpha-solution-quant-optics}),
with $\omega_d=\omega_m$ and $\varphi_d = \varphi_m + \pi$
in a rotating frame with $\omega_r$ (see Appendix \ref{app-A: Geometric phases})
to obtain
\begin{eqnarray}
&&\alpha_s(t) = \alpha_s(0) \, e^{-i (\delta\omega_s - i \frac{\kappa}{2})\, t}
\nonumber\\
&& \qquad\quad { } -\frac{\Omega_{\parallel,s}^{\varepsilon}\,
e^{-i\,\varphi_m}}{[\delta - \delta\omega_s] + i\frac{\kappa}{2}}
\left[ e^{ -i\,( \delta\omega_s - i\frac{\kappa}{2} )\, t } - e^{-i\delta\, t} \right]
\nonumber\\
&& \qquad { } \equiv  A_s(t) + B_s(t)
\label{alpha-s-for-the-gates}
\end{eqnarray}
where
$\delta = \omega_m - \omega_r$,
and $\Omega_{\parallel,s}^{\varepsilon}$ is defined in Eq.~(\ref{omega-n-qubit-coupling-with driving-0}).
We note that by changing the modulation phase $\varphi_m \to \varphi_m + \pi$  after each cycle
the sign of $\Omega_{\parallel,s}^{\varepsilon}$ flips,
allowing for essential cancelation of the effects of damping $\kappa$ and energy curvature detuning $\delta\omega_s$,
see Fig.~\ref{fig:4circ-non-ideal}.

Starting with $\alpha_s(0)=0$, the deviation from zero after one cycle is:
\begin{equation}
\alpha_s\left( \frac{2\pi}{\delta} \right)
= \frac{-i \Omega_{\parallel,s}^{\varepsilon}}{b_s^*} \left[ e^{b_s^* \frac{2\pi}{\delta}} - 1 \right]
\label{alpha-s-2pid} ,
\end{equation}
where we denoted $b_s \equiv -i [\delta - \delta\omega_s] - \frac{\kappa}{2}$,
For the deviation of $\alpha_s(t)$ accumulated after the $N$-th cycle ($N=1,2,3,\ldots$)
one then obtains,
using recurrences
(the deviation at the end of each cycle is an initial condition for the next cycle):
\begin{equation}
\delta\tilde{\alpha}_s\left( N\frac{2\pi}{\delta} \right) = \alpha_s\left( \frac{2\pi}{\delta} \right) \,
\frac{ e^{b_s^* N \frac{2\pi}{\delta}} - (-1)^N}{e^{b_s^* \frac{2\pi}{\delta}} + 1}
\label{alpha-s-deviation} .
\end{equation}
Here $\tilde{\alpha}_s(t)$ denotes a time evolution with flipping sign of the modulation strength,
$\Omega_{\parallel,s}^{\varepsilon}$.
The quantity of interest that enters the gate error $\delta\varepsilon_{\kappa,\delta\omega}$
is then given by
\begin{eqnarray}
&& |\delta\alpha_s^{\kappa,\delta\omega}|^2 \equiv |\delta\tilde{\alpha}_s\left( N \frac{2\pi}{\delta} \right)|^2
\nonumber\\
&& \qquad\quad { } = \frac{(\Omega_{\parallel,s}^{\varepsilon})^2}{|b_s|^2}\,
\frac{\left[\cosh\left( \frac{\pi\kappa}{\delta} \right) - \cos\left( \frac{2\pi\delta\omega_s}{\delta} \right) \right]}
{\left[\cosh\left( \frac{\pi\kappa}{\delta} \right) + \cos\left( \frac{2\pi\delta\omega_s}{\delta} \right) \right]}
\, e^{-\frac{N\pi\kappa}{\delta}} \,
\nonumber\\
&& \qquad\quad { } \times
\left[\cosh\left( \frac{N \pi\kappa}{\delta} \right) - (-1)^N \cos\left( \frac{N 2\pi\delta\omega_s}{\delta} \right) \right] .
\qquad\
\label{delta-alpha-squared}
\end{eqnarray}

For the accumulated phases one uses the equation
$\dot{\alpha}_s = -i (\delta\omega_s - i \frac{\kappa}{2}) \, \alpha_s(t)
- i \Omega_{\parallel,s}^{\varepsilon}\, e^{-i\,\varphi_m}\, e^{-i\delta\, t}$
and its solution, Eq.~(\ref{alpha-s-for-the-gates}).
Thus, for the accumulated phase integral one obtains three contributions
for the time of the $n_1$-cycle:
\begin{eqnarray}
&&  I_{\Phi,s} = \int_{t_1}^{t_2} dt' \alpha_s^*(t') \dot{\alpha}_s(t')
\nonumber
\\
&& \ \quad { } = \int_{t_1}^{t_2} dt' \, \left\{-i (\delta\omega_s - i \frac{\kappa}{2})\, | A_s(t') + B_s(t')|^2  \right.
\label{contribution-1}
\\
&& \qquad\qquad\qquad
\left. { } -i A_s^*(t')\, \Omega_{\parallel,s}^{\varepsilon}\, e^{-i\,\varphi_m}\, e^{-i\delta\, t'} \right.
\label{contribution-2}
\\
&& \qquad\qquad\qquad
\left. { } -i B_s^*(t')\, \Omega_{\parallel,s}^{\varepsilon}\, e^{-i\,\varphi_m}\, e^{-i\delta\, t'}  \right\}
\label{contribution-3}
\\
&& \ \quad { } \equiv  I_{\Phi,s}^{(1)} + I_{\Phi,s}^{(2)} + I_{\Phi,s}^{(3)}
\end{eqnarray}
where $t_1 = (n_1 -1) \frac{2\pi}{\delta}$ and $t_2 = n_1 \frac{2\pi}{\delta}$,
and $\Delta t \equiv t_2 - t_1 = \frac{2\pi}{\delta}$.
For the sake of further use we write down the result of the integration:
\begin{eqnarray}
&& I_{\Phi,s}^{(1)} =
-i \left(\delta\omega_s - i \frac{\kappa}{2} \right)\, \int_{t_1}^{t_2} dt' \,  | A_s(t') + B_s(t')|^2
\nonumber\\
&& \quad\  { } = -i (\delta\omega_s - i \frac{\kappa}{2}) \left\{ |\alpha_s(0)|^2 \frac{2\pi}{\delta} \right.
\nonumber\\
&& \qquad \left. { } + \frac{\left( \Omega_{\parallel,s}^{\varepsilon} \right)^2}{|b_s|^2} \,
\left[ \Delta t  + \frac{1}{b_s} e^{b_s t_1} \left( 1 - e^{b_s \Delta t} \right)
  \right.\right.
\nonumber\\
&& \left.\left.
{ } + \frac{1}{b_s^*} e^{b_s^* t_1} \left( 1 - e^{b_s^* \Delta t} \right)
+ \frac{1}{b_s + b_s^*} e^{(b_s + b_s^*) t_1} \left( e^{(b_s + b_s^*) \Delta t} - 1 \right)
\right] \right.
\nonumber\\
&& \qquad  \left.
{ } + \frac{i \alpha_s(0) \Omega_{\parallel,s}^{\varepsilon}}{b_s} \,
\left[ \frac{1}{b_s + b_s^*} e^{(b_s + b_s^*) t_1} \left( e^{(b_s + b_s^*) \Delta t} - 1 \right) \right.\right.
\nonumber\\
&& \left.\left.
\qquad\qquad\qquad  { } - \frac{1}{b_s^*} e^{b_s^* t_1} \left( e^{b_s^* \Delta t} - 1 \right) \right]
+ {\rm c.c.}  \right\}
\label{contribution-1-int}
\end{eqnarray}
\begin{eqnarray}
&& I_{\Phi,s}^{(2)} = -i \Omega_{\parallel,s}^{\varepsilon}\, e^{-i\,\varphi_m}\, \int_{t_1}^{t_2} dt' \,  A_s^*(t')\,  e^{-i\delta\, t'}
\nonumber\\
&& \qquad\qquad { } = \frac{-i \alpha_s(0)^* \Omega_{\parallel,s}^{\varepsilon}}{b_s} \, e^{b_s t_1} \left( e^{b_s \Delta t} -1 \right)
\qquad\qquad\qquad
\label{contribution-2-int}
\end{eqnarray}

\begin{eqnarray}
&& I_{\Phi,s}^{(3)} = -i \Omega_{\parallel,s}^{\varepsilon}\, e^{-i\,\varphi_m}\, \int_{t_1}^{t_2} dt' \,  B_s^*(t')\, e^{-i\delta\, t'}
\nonumber\\
&& \qquad\qquad { } = \frac{ \left( \Omega_{\parallel,s}^{\varepsilon} \right)^2}{b_s} \,
\left[ \frac{1}{b_s}  e^{b_s t_1} \left( e^{b_s \Delta t} -1 \right) - \Delta t \right]
\qquad\qquad
\label{contribution-3-int} .
\end{eqnarray}
The accumulated phase error $\delta \Phi_s^{\kappa\delta\omega}$
is then given by:
\begin{equation}
\delta \Phi_s^{\kappa,\delta\omega} = {\rm Im} I_{\Phi,s} -  {\rm Im} I_{\Phi,s} \Big\vert_{\kappa=0,\delta\omega_s=0}
\label{delta-Phi-kappa-delta-omega} .
\end{equation}


\section{Multi-qubit phase gate infidelity due to resonator (Johnson) noise}
\label{App-D: Johnson nosie infidelity}

\subsection{Diffusion term in the Caldeira-Leggett master equation as generated via a random force Hamiltonian}

Here we show that the random force Hamiltonian
${\cal H}_f = - \xi_f(t) \hat{x}$, Eq.(\ref{random-force}), is an unraveling of the diffusion term
in the ensemble-averaged Caldeira-Leggett  Eq.~(\ref{damping_diffusion-CL}).
By definition, the (single-sided) spectral density $S_f$ is defined by the correlator:
\begin{equation}
\langle \xi_f(t)\xi_f(t')\rangle_\xi = \frac{S_f}{2}\delta(t-t')
\label{white-noise-correlator} ,
\end{equation}
where $\langle \ldots \rangle_\xi$ denotes averaging over realizations of the noise process.
It will be useful to work in the position representation, so that the generic density matrix  element
is $\rho(x,x') \equiv \rho_{xx'}$, with the position being a continuous index.
By adding the random force Hamiltonian in the Stratonovich form of the equations of motion
(as for any physical interaction, see e.g. Refs.~\onlinecite{WisemanMilburn1994PRA,WisemanMilburn-book2010})
one obtains:
\begin{eqnarray}
&&\frac{d\rho_{xx'}}{dt} = -\frac{i}{\hbar} [{\cal H}_{\rm tot},\rho]_{xx'} - \frac{i}{\hbar} [{\cal H}_{f},\rho]_{xx'}
\nonumber\\
&& \qquad\ { } \equiv G_{\rm Strat}(\rho_{xx'}) + F(\rho_{xx'})\, \xi_f(t)
\label{Stratonovich} ,
\end{eqnarray}
where $G(\rho_{xx'})$ and $F(\rho_{xx'})\, \xi_f(t)$ are the regular and the noise part, respectively.
The noise part is calculated from the commutator $[\hat{x},\rho]_{xx'} = (x - x')\,\rho_{xx'}$:
\begin{equation}
F(\rho_{xx'}) = \frac{i}{\hbar}\, (x - x')\rho_{xx'} .
\end{equation}
The transition to the It\^{o} form of the equation of motion follows the prescription
of Refs.~\onlinecite{Oksendal-book1998,KorotkovPRB2001}
for a system of differential equations,
however with the replacing of a discrete index ``i'' (that enumerates the number of equations)
with the continuous index $(x,x')$, and by replacing the partial derivatives $\frac{\partial F_i}{\partial \rho_k}$
with a {\it functional derivative} (let $x < x'$):
\begin{equation}
\frac{\delta F(\rho(x,x'))}{\delta \rho(x_1,x_2)} = \frac{i}{\hbar}\, (x - x')\, \delta(x - x_1)\, \delta(x' - x_2)
\label{functional_derivative} ,
\end{equation}
with $\delta(x)$ being the Dirac delta-function.
Thus, the regular part in the It\^{o} form is given by (see also Ref.~\cite{RuskovSchwabKorotkov2005}):
\begin{eqnarray}
&& G_{\rm Ito}(\rho_{xx'}) = G_{\rm Strat}(\rho_{xx'}) +
\frac{S_f}{4} \int_{-\infty}^{\infty}\, \int_{-\infty}^{x_2} dx_1 dx_2
\nonumber\\
&& \qquad\qquad\quad\ { } \times
\frac{\delta F(\rho(x,x'))}{\delta \rho(x_1,x_2)} F(\rho(x_1,x_2)) \qquad
\label{G-Ito-continuous-general}\\
&& \qquad\quad\ \  { } = G_{\rm Strat}(\rho_{xx'}) - \frac{S_f}{4\hbar^2}\, (x - x')^2\, \rho_{xx'}
\label{G-Ito-regular}
\end{eqnarray}
and
the equation of motion for the density matrix in It\^{o} form is given by:
\begin{eqnarray}
&& d\rho_{xx'} = -\frac{i}{\hbar} [{\cal H}_{\rm tot},\rho]_{xx'}
- \frac{S_f}{4\hbar^2}\, (x - x')^2\, \rho_{xx'}\, dt
\nonumber\\
&& \qquad\qquad { } + \frac{i}{\hbar}\, (x - x')\rho_{xx'}\, \xi_f(t)\, dt
\label{rho-Ito} .
\end{eqnarray}
Averaging over the noise in Eq.~(\ref{rho-Ito}) (by simply dropping the noise term)
one can identify the second term in Eq.~(\ref{rho-Ito}) with the diffusion term in Eq.~(\ref{damping_diffusion-CL})
by choosing the noise spectral density as
\begin{equation}
S_f = 4 K_d \equiv 2\hbar\omega_r L_r\, \kappa\, \coth \frac{\hbar\omega_r}{2k_B T}
\label{S_f-K_d-correspondence}
\end{equation}
and
by noting that the double commutator in position representation is
given by
\begin{equation}
[\hat{x},[\hat{x},\rho]]_{xx'} = (x - x')^2\, \rho_{xx'} .
\end{equation}

By using the random force Hamiltonian one calculates its contribution
to the equations of motion for the average position and momentum via Eq.~(\ref{rho-Ito}).
(Below, we have dropped the index ``s'' enumerating different $n$-qubit states):
\begin{eqnarray}
&& d\bar{x} = {\rm Tr} [ \hat{x}\, d\rho ] = \int dx\, x\, d\rho_{xx} =
\int dx\, x (-\frac{i}{\hbar}) [{\cal H}_{\rm tot},\rho]_{xx}\, dt
\nonumber\\
&& { } = -\frac{i}{\hbar}\, {\rm Tr} \left( \hat{x}\, [{\cal H}_{\rm tot},\rho] \right) dt
\label{x-average-from-rho-Ito}\\
&& d\bar{p} = -\frac{i}{\hbar}\, {\rm Tr} [ \hat{p}\, d\rho ] =
{\rm Tr} \left( \hat{p} \, [{\cal H}_{\rm tot},\rho] \right) dt + \xi_f(t)\, dt
\label{p-average-from-rho-Ito} ,
\end{eqnarray}
where we have used the momentum operator in position representation:
\begin{equation}
(\hat{p})_{x\tilde{x}} = [-i\hbar \delta(x-\tilde{x})] \frac{\partial}{\partial \tilde{x}} .
\end{equation}

Using Eqs.~(\ref{x-average-from-rho-Ito}) and (\ref{p-average-from-rho-Ito})
one obtains an additional noise term in the equation of motion of the ``field'' variable
$\alpha_s \equiv \langle \hat{a} \rangle_s =
\frac{1}{2}\left( \frac{\bar{x}_s}{\Delta x_0} + i \frac{\bar{p}_s}{\Delta p_0} \right)$.
%
Notice, that for the averages (first moments), It\^{o} and Stratonovich forms of the equations coincide,
%
and by going to the rotating frame with $\omega_{r'}$ one obtains:
\begin{eqnarray}
&&\dot{\alpha}_s = - i \left( \tilde{\omega}_r - \delta\omega_s \right)\, \alpha_s
-i \Omega_{\parallel, s} e^{-i (\tilde{\delta} t + \varphi_m)}
-i \varepsilon_d e^{-i (\tilde{\delta} t + \varphi_d)}
\nonumber\\
&& \qquad { } - \frac{\kappa}{2} \left( \alpha_s - \alpha^*_s \right)
+ i\, \frac{\xi_f(t)}{2\Delta p_0}\, e^{i \omega_{r'} t}
\label{alpha-equation-with-noise}  ,
\end{eqnarray}
to be compared with Eq.~(\ref{alpha-equation}).
The noisy evolution of $\alpha_s(t)$ due to the last stochastic term  is shown schematically
on Fig.~\ref{fig:4circ-non-ideal}.
(It cannot be neglected in a rotating wave approximation since the white noise $\xi_f(t)$ contains
a frequency component that eliminates the fast rotating factor).

\subsection{The variances $\langle | \delta\alpha_s |^2 \rangle_{\xi}$
and $\langle \delta\Phi_s \,\delta\Phi_{s'} \rangle_{\xi}$}
\label{app D2: Johnson-noise-variances}

Calculation of infidelity caused by the resonator (Johnson) noise
requires the knowledge of the variances $\langle | \delta\alpha_s |^2 \rangle_{\xi}$,
and $\langle \delta\Phi_s \,\delta\Phi_{s'} \rangle_{\xi}$, see Eqs.~(\ref{infidelity-xi-0})
or (\ref{infidelity-xi}).
The resonator trajectory in the phase space, $\alpha_s(t)$ obtains a fluctuating term
$\delta\alpha_s(t)$, see Fig.~\ref{fig:4circ-non-ideal} and Eq.~(\ref{alpha-equation-with-noise}),
with a zero average over the realizations of the noise process, $\langle \delta\alpha_s(t) \rangle_{\xi} = 0$.
Similarly, the average of the accumulated phase fluctuation is zero
over the realizations, $\langle \delta\Phi_s(t) \rangle_{\xi} = 0$, see below.
The fluctuation at time $t$ is obtained by integration of the last term in Eq.~(\ref{alpha-equation-with-noise})
thus obtaining Eq.~(\ref{alpha-noise}) of the main text
(we have used that the vacuum fluctuations satisfy $\Delta x_0\, \Delta p_0 = \hbar/2$).
%
From  Eq.~(\ref{alpha-equation-with-noise}) and in rotation frame with
$\omega_r$ the variance of $\alpha_s(t)$ is:
\begin{eqnarray}
&& \langle |\delta\alpha_s(t)|^2\rangle_{\xi} =    
\nonumber\\
&& \qquad  \left( \frac{\Delta x_0}{\hbar}\right)^2 \,
\int_0^t \int_0^t dt' dt'' e^{i \omega_r (t'-t'')} \langle \xi_f(t') \xi_f(t'')\rangle_{\xi}  \qquad
\label{alpha-variance1}
\end{eqnarray}
that is spin-independent.
%
By introducing the shortcoming for the noise:
$r(t) \equiv - \xi_f(t) \Delta x_0 /\hbar$
one writes the white noise average:
\begin{eqnarray}
&& \langle r(t') \,r(t'')  \rangle_{\xi} =
\frac{S_f (\Delta x_0)^2}{2 \hbar^2}\, \delta(t' - t'') \equiv C_0 \, \delta(t' - t'')   \qquad
\label{white-noise-correlator}
\\
&& C_0 \equiv \frac{\kappa}{2} \coth\left( \frac{\hbar\omega_r}{2 k_B T} \right)
\label{C0-coefficient} ,
\end{eqnarray}
also using the relations, Eqs.~(\ref{position-momentum}) and (\ref{S_f-K_d-correspondence}).

The accumulated phase variance
$\langle \delta\Phi_s(t)^2\rangle_{\xi} = \langle \left[ {\rm Im}\, \delta I_{\Phi,s}\right]^2\rangle_{\xi}$
is obtained via
the fluctuations
of the accumulated phase integral [scf. Eq.~(\ref{accumulated_phase})],
assuming small variations $\delta \alpha_{s}(t)$.
Expanding $I_{\Phi,s}(t)$ to first order one obtains:
\begin{eqnarray}
&& \delta I_{\Phi,s}(t) \simeq \int_0^t dt'
\left[ \delta \alpha_{s}^*(t') \frac{d\alpha^{\rm id}_{s}(t')}{dt'}
+ \alpha^{\rm id\, *}_{s}(t') \frac{d\delta\alpha_{s}(t')}{dt'} \right]  \qquad
\nonumber\\
&& \qquad\quad\  \equiv   \delta A_{1,s}  + \delta A_{2,s}   \qquad
\label{delta-I}
\\
&&   A_{1s} \equiv \int_0^t dt'
\delta \alpha_{s}^*(t') \frac{d\alpha^{\rm id}_{s}(t')}{dt'}
\label{A1s}
\\
&&   A_{2s} \equiv \int_0^t dt' \alpha^{\rm id\, *}_{s}(t') \frac{d\delta\alpha_{s}(t')}{dt'}
\label{A2s}
\end{eqnarray}
%
where the ideal resonator evolution is given by Eq.~(\ref{ideal_alpha0}) or Eq.~(\ref{ideal_alpha}),
see Fig.~\ref{fig:3circ},
i.e. here we do not take into account higher corrections due to the resonator damping
$\kappa$ and the always on curvature ``dispersive-like'' resonator frequency shifts, $\delta\omega$,
given by Eq.~(\ref{dispersive-like-interaction}).
Indeed, the error of such approximation is of second order in the
small resonator trajectory deviations, e.g., $\sim \delta \alpha_{s}^{\xi}\, \delta \alpha_{s}^{\kappa,\delta\omega}$.
Thus, the $s$-dependence of the accumulated phase variation
will come only through the
{\it modulation driving strength} $\Omega_{\parallel,s}$, see Eq.~(\ref{omega-n-qubit-coupling})
or (\ref{n-qubit-longitud-operators-of-Hamiltonian}).

For the resonator driving conditions of Eq.~(\ref{cancellation-of-g0}),
one is replacing the modulation strength by $\Omega_{\parallel, s}^{\varepsilon}$,
Eq.~(\ref{omega-n-qubit-coupling-with driving-0}),
and
the phase of the modulation
is chosen as $\varphi_m = 0$.
The ideal evolution
(neglecting the $\kappa$ and $\delta\omega_s$ terms)
is recast to  
$\dot{\alpha}^{\rm id}_s = - i\Omega_{\parallel, s}^{\varepsilon} e^{-i \delta t }$.

The fluctuation of the accumulated phase is then
\begin{eqnarray}
&& \delta \Phi_s =
{\rm Im} \left( \delta I_{\Phi_s} \right) \equiv {\rm Im} \left(  A_{1s}\right) + {\rm Im} \left(  A_{2s}\right)
\\
&& {\rm Im} \left(  A_{1s} \right) = - \Omega_{\parallel,s}^{\varepsilon} \, \int_0^t dt' \int_0^{t'} dt{''}
r(t'') \sin(\delta t' + \omega_r t'' )    \qquad\
\label{Im-A1s}
\\
&& {\rm Im} \left(  A_{2s} \right) = \frac{\Omega_{\parallel,s}^{\varepsilon}}{\delta}
\int_0^t dt' r(t') \left[ \cos(\omega_r t') - \cos(\omega_m t') \right] .  \qquad\
\label{Im-A2s}
\end{eqnarray}
The variance of the accumulated phase, $\langle \delta\Phi_s \,\delta\Phi_{s'} \rangle_{\xi}$,
is obtained via explicit time integration with Eqs.~(\ref{Im-A1s}) and (\ref{Im-A2s}),
using Eq.~(\ref{white-noise-correlator}) for the noise average:
%
\begin{eqnarray}
&& \langle {\rm Im} \left(  A_{1s}\right) \, {\rm Im} \left(  A_{1s'}\right) \rangle_{\xi}
\simeq  \left( \Omega_{\parallel,s}^{\varepsilon} \Omega_{\parallel,s'}^{\varepsilon} \right)
C_0  \qquad\qquad
\nonumber\\
&& \qquad\qquad\qquad\qquad\quad
\times \left[ \frac{t}{\delta^2} - \frac{\sin(\delta t)}{\delta^3}
  \right]  \qquad
\label{A1s-A1s'}
\\
&& \langle {\rm Im} \left(  A_{1s}\right) \, {\rm Im} \left(  A_{2s'}\right) \rangle_{\xi}
\simeq  - \left( \Omega_{\parallel,s}^{\varepsilon} \frac{\Omega_{\parallel,s'}^{\varepsilon}}{\delta} \right)
C_0  \qquad\qquad
\nonumber\\
&& \qquad\qquad\qquad\qquad\quad
\times \left\{ \frac{t}{2\delta} \, \left[ 1 + \cos(\delta t) \right]
\right\} \quad
\label{A1s-A2s'}
\\
&& \langle {\rm Im} \left(  A_{2s}\right) \, {\rm Im} \left(  A_{2s'}\right) \rangle_{\xi}
\simeq  \left( \frac{\Omega_{\parallel,s}^{\varepsilon} \Omega_{\parallel,s'}^{\varepsilon}}{\delta^2} \right)
C_0  \qquad\qquad
\nonumber\\
&& \qquad\qquad\qquad\qquad\quad
\times \left[ t - \frac{4\sin(\delta t)}{\delta} \right] ,  \qquad
\label{A2s-A2s'}
\end{eqnarray}
where only leading contributions are shown,
with corrections of the order of ${\cal O}\left( \frac{\delta}{\omega_r} \right)$.
%
One can see that these averages, Eqs.~(\ref{A1s-A1s'})-(\ref{A2s-A2s'}),
oscillate in time, as expected for variances
of a modulating resonator.
For a complete number of cycles,
$\delta\, t = 2 \pi N,\ N=2,4,\ldots$  some of these are zeroed or minimized,
and the average of interest  is obtained
\begin{eqnarray}
&& \langle \left[  \delta\Phi_{s}(t) - \delta\Phi_{s'}(t)  \right]^2 \rangle_{\xi}
= 4 C_0\,
\frac{[\Omega_{\parallel,s}^{\varepsilon} - \Omega_{\parallel,s'}^{\varepsilon}]^2}{\delta^2} \,
\frac{2\pi N}{\delta}, \qquad
\label{delta-Phi-variance-1} ,
\end{eqnarray}
as is Eq.~(\ref{delta-Phi-variance}) of the main text.
The variance of the accumulated phase fluctuations is linear in time units of $t = \frac{2 \pi N}{\delta}$.

\section{Infidelity of $n$-qubit phase gate  due to qubits' charge noise}
\label{app E:Infidelity due to linear and quadratic charge noise}

For an initial product state of $n$-qubits plus resonator
the ideal final state after one cycle, $t=\frac{2\pi}{\delta}$,
is defined by the accumulated (geometric) phases and the frequency $\omega_s$ of the $|s\rangle$-state:
$|\psi_f^{\rm id} \rangle = \sum_s a_s e^{i\Phi_s(t)}\, e^{-i \omega_s t} |s\rangle |0\rangle$,
where
\begin{equation}
\omega_s = \sum_j  \langle s | \frac{\omega_q^{(j)} }{2} \sigma_z^{(j)} | s \rangle
\label{omega-s}
\end{equation}
is the frequency of the $|s \rangle$-state.
Since the qubit energy defining parameters (voltage gates) fluctuate,
an additional fluctuating phase is accumulated.
Thus, for the $(j)$-th  qubit  with a fluctuation,
$\omega_q^{(j)} + \delta \omega_q^{(j)}(t)$
one obtains the phase factor
\begin{equation}
e^{-i \omega_q^{(j)}\, t} \, e^{-i \Delta \phi^{(j)}(t)}, \ \ \
\Delta \phi^{(j)}(t) \equiv \int_0^t dt' \, \delta \omega_q^{(j)}(t')
\label{noise-single-phase}
\end{equation}
(we consider only longitudinal noise, see, e.g. Ref.\onlinecite{RussBurkard2015PRB}).
The accumulated phase noise of the $|s\rangle$-state is then expressed
via individual qubit  noises
\begin{equation}
\Delta \phi_s(t) = \sum_j \langle s | \frac{\Delta \phi^{(j)}(t) }{2} \sigma_z^{(j)} | s \rangle
\label{noise-n-qubit-phase} ,
\end{equation}
and the actual final state acquires noisy phases:
$|\psi_f \rangle = \sum_s a_s e^{i\Phi_s(t)}\, e^{-i \omega_s t} \, e^{-i \Delta \phi_s(t)} |s\rangle |0\rangle$.
%
The fidelity of a single realization of the noise process is obtained as:
\begin{eqnarray}
&& F = |\langle \psi_f^{\rm id} | \psi_f \rangle|^2 =
\sum_s |a_s|^4
\nonumber\\
&&
\qquad { } + \sum_{s<s'} 2 |a_s|^2 |a_{s'}|^2 e^{-i \left(\Delta\phi_{s}(t) - \Delta\phi_{s'}(t) \right)}
\, \cos{\delta\tilde{\tilde{\omega}}_{ss'}\, t}
\label{fidelity-phi-single-realization} ,
\end{eqnarray}
where the deterministic qubit evolutions is due to possibly induced ac Stark shifts,
see Eq.~(\ref{delta-omega-12}),
and Appendix \ref{app: dephasing rates from resonator photon shot noise}, Eq.~(\ref{2nd-exponent-finite-t}).
One then averages over the initial $n$-qubit state and over the noise realizations
to get
\begin{eqnarray}
&& \overline{\langle F \rangle}_{\xi_q} = 2^n f_1^{(n)}
+ f_{12}^{(n)} \sum_{s<s'} \langle e^{-i \left(\Delta\phi_{s}(t) - \Delta\phi_{s'}(t) \right)} \rangle_{\xi_q}
\, \cos{\delta\tilde{\tilde{\omega}}_{ss'}\, t}  \qquad
\label{fidelity-phi-averaged} .
\end{eqnarray}
Here
$f_1 \equiv \overline{|a_s|^4} = f_{12} \equiv \overline{2 |a_s|^2 |a_{s'}|^2}$
are the averages over the initial qubit states,
given by Eq.~(\ref{n-qubit-averages-1}).

One calculates the dephasing factor assuming that the random variable
$X_{ss'} \equiv \left(\Delta\phi_{s}(t) - \Delta\phi_{s'}(t) \right)$ has zero mean
and is Gaussian distributed and obtains
\begin{equation}
\langle e^{-i \left(\Delta\phi_{s}(t) - \Delta\phi_{s'}(t) \right)} \rangle_{\xi_q}
= e^{-\frac{1}{2}  \langle \left(\Delta\phi_{s}(t) - \Delta\phi_{s'}(t) \right)^2 \rangle_{\xi_q} }
\label{dephasing-factor} .
\end{equation}
Assuming small variances and using Eq.~(\ref{n-qubit-averages-0})
one obtains the $n$-qubit infidelity due to charge noise in a general form:
\begin{equation}
\delta\varepsilon_\phi^{\rm n\, Qb} \equiv 1 - \overline{\langle F \rangle}_{\xi_q}
= f_{12}^{(n)} \sum_{s<s'} \frac{1}{2}  \langle \left(\Delta\phi_{s}(t) - \Delta\phi_{s'}(t) \right)^2 \rangle_{\xi_q}
\label{infidelity-phi} .
\end{equation}

\subsection{Correlated white noises}
\label{Correlated white noises}
If the noises impinged on the qubits are white noise correlated 
(with correlation matrix $A^{jk}$)
\begin{equation}
\langle \delta\omega_q^{(j)}(t') \, \delta\omega_q^{(j)}(t'') \rangle = A^{jk} \delta(t'-t'')
\label{correlated-white-qubit-noises} ,
\end{equation}
the correlation of interest in Eq.~(\ref{infidelity-phi}) can be represented as
\begin{equation}
\frac{1}{2}  \langle \left[\Delta\phi_{s}(t) - \Delta\phi_{s'}(t) \right]^2 \rangle_{\xi_q} = \Gamma_{\phi,ss'}\,\, t
\label{dephasing-rates-s-sprime-ingeneral} .
\end{equation}
To see this, one rewrites the random variable $\Delta\phi_{s}(t) - \Delta\phi_{s'}(t)$ as 
\begin{eqnarray}
&& \Delta\phi_{s}(t) - \Delta\phi_{s'}(t)
= \int_0^t dt' \sum_j \frac{\delta\omega_q^{(j)}(t')}{2} (i_j - i'_j)  \qquad\
\\
&& |s \rangle \equiv |i_1, \ldots, i_n \rangle,\ \  |s' \rangle \equiv |i'_1, \ldots, i'_n \rangle,
\label{n-qubit-states}
\\
&& \langle s| \sigma_z^{(j)} |s\rangle = i_j,\ \ \langle s'| \sigma_z^{(j)} |s'\rangle = i'_j
\label{random-phase-difference-X}
\end{eqnarray}
and obtain the average via the $\delta$-correlation, Eq.~(\ref{correlated-white-qubit-noises}):
\begin{eqnarray}
&& \frac{1}{2}  \langle \left[ \Delta\phi_{s}(t) - \Delta\phi_{s'}(t) \right]^2 \rangle =
\sum_{j,k} (i_j - i'_j)\,(i_k - i'_k) \frac{A^{jk}}{8}\,\, t
\label{averaged-noisy-phase-difference-s-sprime} . \qquad
\end{eqnarray}
Thus, one obtains the $n$-qubit dephasing rates for correlated white noise:
\begin{eqnarray}
&&  \Gamma_{\phi,ss'}^{\rm n\, Qb} = \sum_{j,k} (i_j - i'_j)\,(i_k - i'_k) \frac{A^{jk}}{8}
\label{dephasing-rates-s-sprime-via-corraltion-matrix} .
\end{eqnarray}
The dephasing rates cannot be represented as a sum of individual qubit rates
(for uncorrelated white noises see next Section).
This is, e.g., the case of collective qubits dephasing due to resonator shot noise,
considered in Appendix \ref{app: dephasing rates from resonator photon shot noise}.

The infidelity, Eq.~(\ref{infidelity-phi}), is recast to
\begin{equation}
\delta\varepsilon_{\phi,\rm shot}^{\rm n\, Qb}  =  f_{12}^{(n)} \sum_{s<s'} \tilde{\Gamma}_{\phi,ss'}^{\rm shot}\,\, t
\label{infidelity-phi-uncorrelated-white-noise} .
\end{equation}
From Eq.~(\ref{fidelity-phi-averaged}) one gets additional infidelity at small $t$
due to the ac Stark shifts:
\begin{equation}
\delta\varepsilon_{\delta\omega,\rm shot}^{\rm n\, Qb}
=  \frac{f_{12}^{(n)} }{2}\,  \sum_{s<s'} \left( \delta{\tilde{\tilde\omega}}^{\rm shot}_{ss'}\, t \right)^2
\label{infidelity-delta-omega-uncorrelated-white-noise-1} .
\end{equation}
The dephasing rates, $\tilde{\Gamma}^{\rm shot}_{\phi,ss'}$,
and frequency shifts, $\delta{\tilde{\tilde\omega}}^{\rm shot}_{ss'}$,
for whole time periods, $t = t_g = \frac{2\pi N}{\delta}$,
are given by Eq.~(\ref{delta-gamma-1}) and Eq.~(\ref{delta-omega-12}), respectively;
see also Appendix \ref{app: dephasing rates from resonator photon shot noise}.
%

\subsection{Uncorrelated white noises}
\label{app: Uncorrelated white noises}
For uncorrelated white noise impinged on the qubits frequencies
one gets the relation
\begin{equation}
\langle \delta\omega_q^{(j)}(t') \, \delta\omega_q^{(j)}(t'') \rangle_{\xi_q} = A_j \, \delta^{jk} \, \delta(t'-t'')
\label{uncorrelated-white-qubit-noises} ,
\end{equation}
where the constants $A_j$ can be represented via the spectral densities $S_{x^{(j)}}$ of the gate voltages
of the $(j)$-th qubit:
\begin{equation}
A_j =  \sum_x  \frac{S_{x^{(j)}} }{2 \hbar^2} \, \left( \frac{\partial E_q^{(j)}}{\partial x^{(j)}} \right)^2
\label{uncorrelated-white-qubit-noises-spectral-constant} .
\end{equation}
Here we have used that the frequency fluctuations are expressed via the qubit's gate voltage fluctuations
$\delta\omega_q^{(j)}(t') = \frac{1}{\hbar} \sum_x \frac{\partial E_q^{(j)}}{\partial x^{(j)}} \, \delta x^{(j)}$,
with  $x^{(j)} = \{\varepsilon_v^{(j)},\varepsilon_m^{(j)}, t_l^{(j)}, t_r^{(j)}\}$
being the qubit's definig gate voltage differences and interdot tunneling amplitudes (for a TQD qubit).
We assumed for simplicity that these variables are mutually uncorrelated,
while each is white noise correlated
\begin{equation}
\langle \delta x^{(j)}(t') \delta x^{(j)}(t'') \rangle_{\xi_q}  = \frac{S_{x^{(j)}} }{2} \, \delta(t' - t'')
\end{equation}
where $S_{x^{(j)}}$ are the (single-sided) white noise spectral densities for each of the variables of the $(j)$-th qubit.

The resulting $n$-qubit dephasing rates are then given by
\begin{equation}
\Gamma_{\phi,ss'} = \sum_j \frac{A_j}{4} (1 - i_j i'_j)
\label{dephasing-rates-s-sprime-uncorrelated noises} ,
\end{equation}
which allows to express the $n$-qubit dephasing rates via the single-qubit one.
The $n$-qubit infidelity, $\delta\varepsilon_\phi^{\rm n\, Qb}$,
is expressed by the same Eq.~(\ref{infidelity-phi-uncorrelated-white-noise}).

The single-qubit dephasing rate is then a sum of contributions,
\begin{equation}
\Gamma_{\phi}^{\rm white}
= \frac{A_1}{2}
= \sum_{x = \varepsilon_v,\varepsilon_m,t_l,t_r}
\frac{S_x}{4 \hbar^2} \, \left( \frac{\partial E_q}{\partial x} \right)^2
\label{white-noise-dephasing-rate} ,
\end{equation}
and is minimized,
as $\frac{\partial E_q}{\partial \varepsilon_v} = \frac{\partial E_q}{\partial \varepsilon_m} = 0$
at a full sweet spot.

For two-qubit states, $|1\rangle = |\downarrow\downarrow\rangle$, $|2\rangle = |\downarrow\uparrow\rangle$
$|3\rangle = |\uparrow\downarrow\rangle$, $|4\rangle = |\uparrow\uparrow\rangle$,
one renders via Eq.(\ref{dephasing-rates-s-sprime-uncorrelated noises})
the two-qubit rates expressed via the single-qubit one:
\begin{eqnarray}
&& \Gamma_{\phi,12} = \frac{A_2}{2} = \Gamma_\phi^{(2)}, \ \ \
\Gamma_{\phi,13}  = \frac{A_1}{2} = \Gamma_\phi^{(1)}  \qquad
\\
&& \Gamma_{\phi,14} = \frac{A_1}{2} +  \frac{A_2}{2}
=  \Gamma_\phi^{(1)} + \Gamma_\phi^{(2)} , \ {\rm etc.}
\label{2-qubit-rates-white-noise}
\end{eqnarray}
which is also illustrated on Fig.~\ref{fig:431}.
The two-qubit infidelity in this case is expressed via the sum
$\sum_{s<s'} \Gamma_{ss'}\,\, t = 4 \left[\Gamma_\phi^{(1)} + \Gamma_\phi^{(2)} \right] \, t$.
For equal qubit dephasings, $\Gamma_\phi^{(1)} = \Gamma_\phi^{(2)} = \Gamma_{\phi}^{\rm white}$ one obtains
\begin{equation}
\delta\varepsilon_\phi^{\rm 2Qb} = \frac{8}{10}\, \Gamma_{\phi}^{\rm white}\,\, t
\label{2-qubit-dephasing-infidelity-white} ,
\end{equation}
where $t = t_g = \frac{2\pi N}{\delta}$ is the gate time.
The result of Eq.~(\ref{2-qubit-dephasing-infidelity-white}) was first presented
at the QCPR in August 2016, and also at the 2017 APS March meeting.

\subsection{Phase gate infidelity in the case of uncorrelated $1/f$ charge noise}
\label{app: one-over-f-carge-noise-infidelity}
One is assuming that the accumulated random phases of individual qubits are uncorrelated
\begin{equation}
\langle \Delta\phi^{(j)}(t) \, \Delta\phi^{(k)}(t) \rangle_{\xi_q} \propto \delta^{jk}
\label{uncorrelated}
\end{equation}
(that is more general than Eq.~\ref{uncorrelated-white-qubit-noises}).
Using the representation of the $n$-qubit states,
$|s \rangle = |i_1,\ldots,i_n \rangle$, Eq.~(\ref{n-qubit-states}),
one obtains from Eq.~(\ref{uncorrelated})
%
\begin{equation}
\left\langle \left(\Delta\phi_{s}(t) - \Delta\phi_{s'}(t) \right)^2 \right\rangle_{\xi_q}
= \frac{1}{2} \sum_{j=1}^n \langle \left(\Delta\phi^{(j)}(t)\right)^2\rangle_{\xi_q} \, (1 - i_j\, i'_j)
\label{uncorrelated-average} ,
\end{equation}
which allows to express the $n$-qubit dephasing rates via the single-qubit one.

For noise fluctuations subject to $1/f$-noise spectrum
the dephasing factor in Eq.~(\ref{dephasing-factor}) has a Gaussian time dependence
\begin{equation}
e^{-\frac{1}{2}  \langle \left(\Delta\phi_{s'}(t) - \Delta\phi_{s'}(t) \right)^2 \rangle_{\xi_q} }
= e^{-(\tilde{\Gamma}_{ss'}\,\, t)^2}
\label{dephasing-factor-one-over-f} .
\end{equation}
(Here $\tilde{\Gamma}$ denotes an $1/f$-noise dephasing rate).
In the case of uncorrelated charge noise to each of the qubits, applying Eq.~(\ref{uncorrelated-average}),
one obtains the relation
\begin{eqnarray}
&& (\tilde{\Gamma}_{ss'}\,\, t)^2 =
\frac{1}{4} \, \sum_{j=1}^n \langle \left(\Delta\phi^{(j)}(t)\right)^2\rangle \, (1 - i_j\, i'_j)
\nonumber\\
&&\qquad { } = \frac{1}{2} \, \sum_{j=1}^n (\tilde{\Gamma}^{(j)}\,\, t)^2 \, (1 - i_j\, i'_j)
\label{dephasing-rate-ss'-one-over-f} ,
\end{eqnarray}
where $\tilde{\Gamma}^{(j)}$ is the single-qubit $1/f$-noise dephasing rate (for the $j$-th qubit).
%
The $n$-qubit infidelity is then obtained:
\begin{equation}
\delta\varepsilon_{\phi, 1/f}^{\rm n\, Qb}
= f_{12}^{(n)} \sum_{s<s'} (\tilde{\Gamma}_{ss'}\,\, t)^2
\label{infidelity-phi-one-over-f} .
\end{equation}

For the two-qubit case, analogous to Eq.~(\ref{2-qubit-rates-white-noise}), one gets the relations
\begin{eqnarray}
&& \tilde{\Gamma}_{12} =  \tilde{\Gamma}_\phi^{(2)}, \ \ \tilde{\Gamma}_{13} =  \tilde{\Gamma}_\phi^{(1)}
\\
&& \tilde{\Gamma}_{14} = \sqrt{ \left(\tilde{\Gamma}_\phi^{(1)}\right)^2 + \left(\tilde{\Gamma}_\phi^{(2)}\right)^2  }, \ {\rm etc.}
\label{2-qubit-rates-one-over-f-noise}
\end{eqnarray}
Assuming equal qubits' dephasing, $\tilde{\Gamma}_\phi^{(1)} = \tilde{\Gamma}_\phi^{(2)} \equiv \tilde{\Gamma}_\phi$
one obtains the two-qubit infidelity
%
\begin{equation}
\delta\varepsilon_{\phi, 1/f}^{\rm 2Qb}
= \frac{8}{10} \, (\tilde{\Gamma}_{\phi}\,\, t)^2
\label{infidelity-phi-2qb-one-over-f} .
\end{equation}

\subsection{$1/f$-dephasing: Scaling of parameters}
\label{app: $1/f$-dephasing: Scaling of parameters}
For the discussion here we use the expression for the single-qubit dephasing,
$\tilde{\Gamma}_\phi$, calculated
in Ref.~\onlinecite{RussBurkard2015PRB,RussGinzelBurkard2016PRB}
(see also Refs.~\onlinecite{MakhlinShnirman2004PRL,Ithier2005PRB}):
\begin{eqnarray}
&& \tilde{\Gamma}_\phi = \frac{1}{\hbar}
\left[ I + II + III + IV \right]^{1/2}
\\
&& I \equiv  \frac{1}{2}
\sum_k  \left( \frac{\partial E_q}{\partial \varepsilon_k} \right)^2 \,
S_{\varepsilon_k}\, \log r_{c}
\label{I-expression}
\\
&& II \equiv  \frac{1}{4}  \sum_k  \left( \frac{\partial^2 E_q}{\partial \varepsilon_k^2} \right)^2 \,
S_{\varepsilon_k}^2\, \log^2 r_{c}
\\
&& III \equiv \frac{1}{2}  \sum_{k\neq l}
\left( \frac{\partial^2 E_q}{\partial \varepsilon_k \partial \varepsilon_l} \right)^2 \,
S_{\varepsilon_k}\, S_{\varepsilon_l}\,  \log^2 r_{c}
\\
&& IV \equiv \frac{1}{8}  \sum_{k\neq l}  \left( \frac{\partial^2 E_q}{\partial \varepsilon_k^2} \right)\,
\left( \frac{\partial^2 E_q}{\partial \varepsilon_l^2} \right) \, S_{\varepsilon_k}\, S_{\varepsilon_l}
\label{single-qubit-one-over-f-rate} ,
\end{eqnarray}
where $\varepsilon_k = \varepsilon_v, \varepsilon_m$ are the TQD qubit energy detunings
$\varepsilon_v \equiv e(V_1 - V_3)$ and $\varepsilon_m \equiv e[(V_1 + V_3)/2 - V_2]$,
and $\varepsilon_k = t_l, t_r$ are the left and right tunneling amplitudes.
The ratio $r_{\rm c} \equiv \frac{\omega_{UV}}{\omega_{IR}}$
includes the ultraviolet and infrared frequency cutoffs
needed to deal with $1/f$-noise spectral density.
The spectral density of the $1/f$-noise is defined as $S(\omega) = \frac{S_{\varepsilon_k}}{|\omega|}$,
where the spectral density constants $S_{\varepsilon_k}$
%
%
%
%
are subject to experimental determination
\cite{MartinsKuemethMarcus2016PRL,ReedHunter2016PRL}.
For illustration purposes we will assume $S_{\varepsilon_v} \approx S_{\varepsilon_m}$
and $S_{\varepsilon_v}, S_{\varepsilon_m} \gg S_{t_l}, S{t_r}$ (see below).
For $r_{c} = 10^6$ one can safely neglect the term $IV$.
Below we argue that $II$ and $III$ contributes only small corrections
of the order of $10^{-3}$ to the leading term $I$.

At the full sweet spot the term $I$ is minimized since
$\frac{\partial E_q}{\partial \varepsilon_v}, \frac{\partial E_q}{\partial \varepsilon_m} = 0$.
Despite that $\log r_{c} \ll \log^2 r_{c}$  and the smallness of
the spectral density constants $S_{t_l}, S{t_r} \ll S_{\varepsilon_v}\, S_{\varepsilon_m}$ (see below)
it turns out that for typical parameters $I \gg II \sim III$.
%
To see this it is useful to write down the corresponding first derivatives and second derivatives (energy curvatures)
for the TQD qubit at the full sweet spot \cite{RuskovTahan2019PRB99}.
By using the expressions for a TQD qubit energy \cite{RuskovTahan2019PRB99}, $E_q(\varepsilon_v,\varepsilon_m,t_l,t_r)$
one obtains at the full sweet spot:
\begin{equation}
E_q = \frac{8 t_l^2}{a_l} \sqrt{1 - r + r^2}, \qquad \ \ r \equiv \frac{t_r^2 a_l}{t_l^2 a_r}
\label{Eq-TQD-DQD}
\end{equation}
\begin{equation}
\frac{\partial E_q}{\partial t_l} = \frac{16 t_l}{a_l} \frac{1-r/2}{\sqrt{1 - r + r^2}} 
\end{equation}
\begin{equation}
\frac{\partial^2 E_q}{\partial \varepsilon_v^2} = \frac{\partial^2 E_q}{\partial \varepsilon_m^2}
= \frac{64 t_l^2}{a_l^3}\, \frac{\left\{1 - \frac{r}{2}\left(1 + \frac{a_l^2}{a_r^2} \right) + r^2 \frac{a_l^2}{a_r^2}  \right\}}{\sqrt{1 - r + r^2}}
\end{equation}
\begin{equation}
\frac{\partial^2 E_q}{\partial \varepsilon_v \partial \varepsilon_m}
= \frac{64 t_l^2}{a_l^3}\, \frac{\left\{-1 + \frac{r}{2}\left(1 - \frac{a_l^2}{a_r^2} \right) + r^2 \frac{a_l^2}{a_r^2}  \right\}}{\sqrt{1 - r + r^2}}
\qquad
\end{equation}
\begin{equation}
\frac{\partial^2 E_q}{\partial t_l^2} = \frac{8}{a_l} \, \frac{(2 - 3 r + 6 r^2 - r^3)}{(1 - r + r^2)^{3/2}}
\end{equation}
\begin{equation}
\frac{\partial^2 E_q}{\partial t_l \partial t_r} = -\frac{24}{a_l}\, \frac{t_l}{t_r}\, \frac{r^2}{(1 - r + r^2)^{3/2}}
\label{curvatures} .
\end{equation}
Here and in the following, $a_l \equiv \tilde{U}_1 + \tilde{U}_2^{'}$ and $a_r \equiv \tilde{U}_2 + \tilde{U}_3$
are the combinations of the charging energy costs $\tilde{U}_i$ to fill the i-th dot with 2 electrons starting from the
$(1,1,1)$-configuration\cite{RuskovTahan2019PRB99}.
In this TQD model some of the curvatures are zero always, e.g.,
$\frac{\partial^2 E_q}{\partial \varepsilon_v \partial t_r} =
\frac{\partial^2 E_q}{\partial \varepsilon_m \partial t_r} =0$,
while $\frac{\partial^2 E_q}{\partial \varepsilon_v \partial \varepsilon_m} = 0$ in
the symmetric case $a_l = a_r$ and $r \equiv \frac{t_r^2 a_l}{t_l^2 a_r} = 1$.
By taking the ratios of the surviving terms at the full sweet spot one obtains
the important scalings:
\begin{equation}
\frac{{II}'}{I}
= \frac{\frac{1}{2} \left( \frac{\partial^2 E_q}{\partial \varepsilon_v^2} \right)^2 S_{\varepsilon_v}^2 \log r_{c} }
                     {\left( \frac{\partial E_q}{\partial t_l} \right)^2 S_{t_l}} = \frac{\log r_{c}}{2}\,
                     \frac{ S_{\varepsilon_v}^2 t_l^2}{S_{t_l} a_l^4} \approx  10^{-3}  
\end{equation}
\begin{equation}
\frac{{II}''}{I}
= \frac{\frac{1}{2} \left( \frac{\partial^2 E_q}{\partial t_l^2} \right)^2 S_{t_l}^2 \log r_{c} }
                     {\left( \frac{\partial E_q}{\partial t_l} \right)^2 S_{t_l}} =
                     8\log r_{c}\, \frac{ S_{t_l} }{ t_l^2}\approx  10^{-3}  
\end{equation}
\begin{equation}
\frac{{III}}{I}
= \frac{\left( \frac{\partial^2 E_q}{\partial t_l \partial t_r} \right)^2 S_{t_l} S_{t_r} \log r_{c} }
                     {\left( \frac{\partial E_q}{\partial t_l} \right)^2 S_{t_l}} =
                     9\log r_{c}\, \frac{ S_{t_r} }{ t_r^2}\approx  10^{-3}  
\label{scalings} .
\end{equation}
In these estimations we have used tunnelings $t_{l,r} = 40\, \mu{\rm eV}$,
a dot charging energy, $U_{\rm ch} \approx \tilde{U}_i \approx 0.4\, {\rm meV}$,
$S_{\varepsilon_v} = S_{\varepsilon_m} = (1\, \mu{\rm eV})^2$ and $S_{t_l} = S_{t_r} = 10^{-2} S_{\varepsilon_v}$.
By increasing tunneling\cite{WestDzurak2019NNano,KratochwilEnslin2020-preprint}
to $t_{l,r} = 160\, \mu{\rm eV}$ and $U_{\rm ch} = 0.8\, {\rm meV}$
which is beneficial for larger quantum capacitance (see below) the smallness of these ratios remains a fact.
Thus, to a very good approximation one can write for the $1/f$-dephasing rate
\begin{eqnarray}
&& \tilde{\Gamma}_\phi \simeq \frac{1}{\hbar} \sqrt{ I }
\\
&& \quad { } = \frac{1}{\hbar} \left\{ \frac{\log r_{c}}{2}  \,
\left[ \left( \frac{\partial E_q}{\partial t_l} \right)^2 \, S_{t_l} +
\left( \frac{\partial E_q}{\partial t_r} \right)^2 \, S_{t_r} \right] \right\}^{1/2} \qquad
\\
&& \quad { } \simeq \frac{16}{\hbar} \left\{ \frac{\log r_{c} \,\, S_{t_l}}{2 (1 - r + r^2)}  \right.
\nonumber\\
&& \quad \left. {  } \times \left[ \left( \frac{t_l}{a_l} \right)^2 \left( 1 - \frac{r}{2} \right)^2  +
\left( \frac{t_r}{a_r} \right)^2 \left( r - \frac{1}{2} \right)^2 \right] \right\}^{1/2}
\label{dephasing-one-over-f-leading} .
\end{eqnarray}
In the symmetric case, $a_l = a_r$ and $r \equiv \frac{t_r^2 a_l}{t_l^2 a_r} = 1$,
one obtains
for the single-qubit $1/f$-dephasing rate, Eq.~(\ref{dephasing-one-over-f-leading-symm-0})
of the main text.
Using Eqs.~(\ref{infidelity-phi-2qb-one-over-f}) and (\ref{dephasing-one-over-f-leading-symm-0})
one obtains the 2-qubit infidelity for $1/f$-noise expressed through all relevant parameters,
see Eq.~(\ref{infidelity-phi-2qb-one-over-f-scaling-0}) of the main text.

\subsection{Spectral density constant of tunnelings from the experiment\cite{ReedHunter2016PRL}}
\label{app: Spectral density constant of tunnelings from experiment}
The spectral density constant $S_{t_l}$ can be extracted from the
experiment.
To this end one uses typical tunneling $t_l = 10 - 20\, \mu{\rm eV}$,
dot charging energy $U_{\rm ch} = 0.6 - 2.4\, {\rm meV}$,
and ratio $r_{\rm c} \equiv \frac{\omega_{UV}}{\omega_{IR}} = 10^6$.
These parameters
fit Eqs.~(\ref{Eq-TQD-DQD}) and (\ref{dephasing-one-over-f-leading-symm-0})
with the experimental values for qubit splitting, $E_q \simeq 160\, {\rm MHz}$,
and measured dephasing time (Rabi oscillations)
of the current experiment\cite{ReedHunter2016PRL} (DQD Singlet-triplet qubit at the symmetric operating point),
which assumes $S_{t_l} \simeq 10^{-4} - 10^{-5} (\mu{\rm eV})^2$.


\begin{thebibliography}{115}%
\makeatletter
\providecommand \@ifxundefined [1]{%
 \@ifx{#1\undefined}
}%
\providecommand \@ifnum [1]{%
 \ifnum #1\expandafter \@firstoftwo
 \else \expandafter \@secondoftwo
 \fi
}%
\providecommand \@ifx [1]{%
 \ifx #1\expandafter \@firstoftwo
 \else \expandafter \@secondoftwo
 \fi
}%
\providecommand \natexlab [1]{#1}%
\providecommand \enquote  [1]{``#1''}%
\providecommand \bibnamefont  [1]{#1}%
\providecommand \bibfnamefont [1]{#1}%
\providecommand \citenamefont [1]{#1}%
\providecommand \href@noop [0]{\@secondoftwo}%
\providecommand \href [0]{\begingroup \@sanitize@url \@href}%
\providecommand \@href[1]{\@@startlink{#1}\@@href}%
\providecommand \@@href[1]{\endgroup#1\@@endlink}%
\providecommand \@sanitize@url [0]{\catcode `\\12\catcode `\$12\catcode
  `\&12\catcode `\#12\catcode `\^12\catcode `\_12\catcode `\%12\relax}%
\providecommand \@@startlink[1]{}%
\providecommand \@@endlink[0]{}%
\providecommand \url  [0]{\begingroup\@sanitize@url \@url }%
\providecommand \@url [1]{\endgroup\@href {#1}{\urlprefix }}%
\providecommand \urlprefix  [0]{URL }%
\providecommand \Eprint [0]{\href }%
\providecommand \doibase [0]{http://dx.doi.org/}%
\providecommand \selectlanguage [0]{\@gobble}%
\providecommand \bibinfo  [0]{\@secondoftwo}%
\providecommand \bibfield  [0]{\@secondoftwo}%
\providecommand \translation [1]{[#1]}%
\providecommand \BibitemOpen [0]{}%
\providecommand \bibitemStop [0]{}%
\providecommand \bibitemNoStop [0]{.\EOS\space}%
\providecommand \EOS [0]{\spacefactor3000\relax}%
\providecommand \BibitemShut  [1]{\csname bibitem#1\endcsname}%
\let\auto@bib@innerbib\@empty
\bibitem [{\citenamefont {Petta}\ \emph {et~al.}(2005)\citenamefont {Petta},
  \citenamefont {Johnson}, \citenamefont {Taylor}, \citenamefont {Laird},
  \citenamefont {Yacoby}, \citenamefont {Lukin}, \citenamefont {Marcus},
  \citenamefont {Hanson},\ and\ \citenamefont {Gossard}}]{Petta2005S}%
  \BibitemOpen
  \bibfield  {author} {\bibinfo {author} {\bibfnamefont {J.~R.}\ \bibnamefont
  {Petta}}, \bibinfo {author} {\bibfnamefont {A.~C.}\ \bibnamefont {Johnson}},
  \bibinfo {author} {\bibfnamefont {J.~M.}\ \bibnamefont {Taylor}}, \bibinfo
  {author} {\bibfnamefont {E.~A.}\ \bibnamefont {Laird}}, \bibinfo {author}
  {\bibfnamefont {A.}~\bibnamefont {Yacoby}}, \bibinfo {author} {\bibfnamefont
  {M.~D.}\ \bibnamefont {Lukin}}, \bibinfo {author} {\bibfnamefont {C.~M.}\
  \bibnamefont {Marcus}}, \bibinfo {author} {\bibfnamefont {M.~P.}\
  \bibnamefont {Hanson}}, \ and\ \bibinfo {author} {\bibfnamefont {A.~C.}\
  \bibnamefont {Gossard}},\ }\href@noop {} {\bibfield  {journal} {\bibinfo
  {journal} {Science}\ }\textbf {\bibinfo {volume} {309}},\ \bibinfo {pages}
  {2180} (\bibinfo {year} {2005})}\BibitemShut {NoStop}%
\bibitem [{\citenamefont {Maune}\ \emph {et~al.}(2012)\citenamefont {Maune},
  \citenamefont {Borselli}, \citenamefont {Huang}, \citenamefont {Ladd},
  \citenamefont {Deelman}, \citenamefont {Holabird}, \citenamefont {Kiselev},
  \citenamefont {Alvarado-Rodriguez}, \citenamefont {Ross}, \citenamefont
  {Schmitz}, \citenamefont {Sokolich}, \citenamefont {Watson}, \citenamefont
  {Gyure},\ and\ \citenamefont {Hunter}}]{Maune2012N}%
  \BibitemOpen
  \bibfield  {author} {\bibinfo {author} {\bibfnamefont {B.~M.}\ \bibnamefont
  {Maune}}, \bibinfo {author} {\bibfnamefont {M.~G.}\ \bibnamefont {Borselli}},
  \bibinfo {author} {\bibfnamefont {B.}~\bibnamefont {Huang}}, \bibinfo
  {author} {\bibfnamefont {T.~D.}\ \bibnamefont {Ladd}}, \bibinfo {author}
  {\bibfnamefont {P.~W.}\ \bibnamefont {Deelman}}, \bibinfo {author}
  {\bibfnamefont {K.~S.}\ \bibnamefont {Holabird}}, \bibinfo {author}
  {\bibfnamefont {A.~A.}\ \bibnamefont {Kiselev}}, \bibinfo {author}
  {\bibfnamefont {I.}~\bibnamefont {Alvarado-Rodriguez}}, \bibinfo {author}
  {\bibfnamefont {R.~S.}\ \bibnamefont {Ross}}, \bibinfo {author}
  {\bibfnamefont {A.~E.}\ \bibnamefont {Schmitz}}, \bibinfo {author}
  {\bibfnamefont {M.}~\bibnamefont {Sokolich}}, \bibinfo {author}
  {\bibfnamefont {C.~A.}\ \bibnamefont {Watson}}, \bibinfo {author}
  {\bibfnamefont {M.~F.}\ \bibnamefont {Gyure}}, \ and\ \bibinfo {author}
  {\bibfnamefont {A.~T.}\ \bibnamefont {Hunter}},\ }\href@noop {} {\bibfield
  {journal} {\bibinfo  {journal} {Nature}\ }\textbf {\bibinfo {volume} {481}},\
  \bibinfo {pages} {344} (\bibinfo {year} {2012})}\BibitemShut {NoStop}%
\bibitem [{\citenamefont {Medford}\ \emph
  {et~al.}(2013{\natexlab{a}})\citenamefont {Medford}, \citenamefont {Beil},
  \citenamefont {Taylor}, \citenamefont {Rashba}, \citenamefont {Lu},
  \citenamefont {Gossard},\ and\ \citenamefont {Marcus}}]{Medford2013PRL}%
  \BibitemOpen
  \bibfield  {author} {\bibinfo {author} {\bibfnamefont {J.}~\bibnamefont
  {Medford}}, \bibinfo {author} {\bibfnamefont {J.}~\bibnamefont {Beil}},
  \bibinfo {author} {\bibfnamefont {J.~M.}\ \bibnamefont {Taylor}}, \bibinfo
  {author} {\bibfnamefont {E.~I.}\ \bibnamefont {Rashba}}, \bibinfo {author}
  {\bibfnamefont {H.}~\bibnamefont {Lu}}, \bibinfo {author} {\bibfnamefont
  {A.~C.}\ \bibnamefont {Gossard}}, \ and\ \bibinfo {author} {\bibfnamefont
  {C.~M.}\ \bibnamefont {Marcus}},\ }\href@noop {} {\bibfield  {journal}
  {\bibinfo  {journal} {Phys. Rev. Lett.}\ }\textbf {\bibinfo {volume} {111}},\
  \bibinfo {pages} {050501} (\bibinfo {year} {2013}{\natexlab{a}})}\BibitemShut
  {NoStop}%
\bibitem [{\citenamefont {Veldhorst}\ \emph
  {et~al.}(2015{\natexlab{a}})\citenamefont {Veldhorst}, \citenamefont {Yang},
  \citenamefont {Hwang}, \citenamefont {Huang}, \citenamefont {Dehollain},
  \citenamefont {Muhonen}, \citenamefont {Simmons}, \citenamefont {Laucht},
  \citenamefont {Hudson}, \citenamefont {Itoh}, \citenamefont {Morello},\ and\
  \citenamefont {Dzurak}}]{Veldhorst2015N}%
  \BibitemOpen
  \bibfield  {author} {\bibinfo {author} {\bibfnamefont {M.}~\bibnamefont
  {Veldhorst}}, \bibinfo {author} {\bibfnamefont {C.~H.}\ \bibnamefont {Yang}},
  \bibinfo {author} {\bibfnamefont {J.}~\bibnamefont {Hwang}}, \bibinfo
  {author} {\bibfnamefont {W.}~\bibnamefont {Huang}}, \bibinfo {author}
  {\bibfnamefont {J.}~\bibnamefont {Dehollain}}, \bibinfo {author}
  {\bibfnamefont {J.}~\bibnamefont {Muhonen}}, \bibinfo {author} {\bibfnamefont
  {S.}~\bibnamefont {Simmons}}, \bibinfo {author} {\bibfnamefont
  {A.}~\bibnamefont {Laucht}}, \bibinfo {author} {\bibfnamefont
  {F.}~\bibnamefont {Hudson}}, \bibinfo {author} {\bibfnamefont
  {K.}~\bibnamefont {Itoh}}, \bibinfo {author} {\bibfnamefont {A.}~\bibnamefont
  {Morello}}, \ and\ \bibinfo {author} {\bibfnamefont {A.}~\bibnamefont
  {Dzurak}},\ }\href@noop {} {\bibfield  {journal} {\bibinfo  {journal}
  {Nature}\ }\textbf {\bibinfo {volume} {526}},\ \bibinfo {pages} {410}
  (\bibinfo {year} {2015}{\natexlab{a}})}\BibitemShut {NoStop}%
\bibitem [{\citenamefont {E.~Kawakami}\ \emph {et~al.}(2016)\citenamefont
  {E.~Kawakami}, \citenamefont {Jullien}, \citenamefont {Scarlino},
  \citenamefont {Ward}, \citenamefont {Savage}, \citenamefont {Lagally},
  \citenamefont {Dobrovitski}, \citenamefont {Friesen}, \citenamefont
  {Coppersmith}, \citenamefont {Eriksson},\ and\ \citenamefont
  {Vandersypen}}]{Kawakami2016PNAS}%
  \BibitemOpen
  \bibfield  {author} {\bibinfo {author} {\bibfnamefont {E.}~\bibnamefont
  {E.~Kawakami}}, \bibinfo {author} {\bibfnamefont {T.}~\bibnamefont
  {Jullien}}, \bibinfo {author} {\bibfnamefont {P.}~\bibnamefont {Scarlino}},
  \bibinfo {author} {\bibfnamefont {D.~R.}\ \bibnamefont {Ward}}, \bibinfo
  {author} {\bibfnamefont {D.~E.}\ \bibnamefont {Savage}}, \bibinfo {author}
  {\bibfnamefont {M.~G.}\ \bibnamefont {Lagally}}, \bibinfo {author}
  {\bibfnamefont {V.~V.}\ \bibnamefont {Dobrovitski}}, \bibinfo {author}
  {\bibfnamefont {M.}~\bibnamefont {Friesen}}, \bibinfo {author} {\bibfnamefont
  {S.~N.}\ \bibnamefont {Coppersmith}}, \bibinfo {author} {\bibfnamefont
  {M.~A.}\ \bibnamefont {Eriksson}}, \ and\ \bibinfo {author} {\bibfnamefont
  {L.~M.~K.}\ \bibnamefont {Vandersypen}},\ }\href@noop {} {\bibfield
  {journal} {\bibinfo  {journal} {PNAS}\ }\textbf {\bibinfo {volume} {113}},\
  \bibinfo {pages} {11738} (\bibinfo {year} {2016})}\BibitemShut {NoStop}%
\bibitem [{\citenamefont {Ward}\ \emph {et~al.}(2016)\citenamefont {Ward},
  \citenamefont {Kim}, \citenamefont {Savage}, \citenamefont {Lagally},
  \citenamefont {Foote}, \citenamefont {Friesen}, \citenamefont {Coppersmith},\
  and\ \citenamefont {Eriksson}}]{WardQuantInf2016}%
  \BibitemOpen
  \bibfield  {author} {\bibinfo {author} {\bibfnamefont {D.~R.}\ \bibnamefont
  {Ward}}, \bibinfo {author} {\bibfnamefont {D.}~\bibnamefont {Kim}}, \bibinfo
  {author} {\bibfnamefont {D.~E.}\ \bibnamefont {Savage}}, \bibinfo {author}
  {\bibfnamefont {M.~G.}\ \bibnamefont {Lagally}}, \bibinfo {author}
  {\bibfnamefont {R.~H.}\ \bibnamefont {Foote}}, \bibinfo {author}
  {\bibfnamefont {M.}~\bibnamefont {Friesen}}, \bibinfo {author} {\bibfnamefont
  {S.~N.}\ \bibnamefont {Coppersmith}}, \ and\ \bibinfo {author} {\bibfnamefont
  {M.~A.}\ \bibnamefont {Eriksson}},\ }\href@noop {} {\bibfield  {journal}
  {\bibinfo  {journal} {Quant. Inf.}\ }\textbf {\bibinfo {volume} {2}},\
  \bibinfo {pages} {16032} (\bibinfo {year} {2016})}\BibitemShut {NoStop}%
\bibitem [{\citenamefont {Pla}\ \emph {et~al.}(2012)\citenamefont {Pla},
  \citenamefont {Tan}, \citenamefont {Dehollain}, \citenamefont {Lim},
  \citenamefont {Morton}, \citenamefont {Jamieson}, \citenamefont {Dzurak},\
  and\ \citenamefont {Morello}}]{Pla2012N}%
  \BibitemOpen
  \bibfield  {author} {\bibinfo {author} {\bibfnamefont {J.~J.}\ \bibnamefont
  {Pla}}, \bibinfo {author} {\bibfnamefont {K.~Y.}\ \bibnamefont {Tan}},
  \bibinfo {author} {\bibfnamefont {J.~P.}\ \bibnamefont {Dehollain}}, \bibinfo
  {author} {\bibfnamefont {W.~H.}\ \bibnamefont {Lim}}, \bibinfo {author}
  {\bibfnamefont {J.~J.~L.}\ \bibnamefont {Morton}}, \bibinfo {author}
  {\bibfnamefont {D.~N.}\ \bibnamefont {Jamieson}}, \bibinfo {author}
  {\bibfnamefont {A.~S.}\ \bibnamefont {Dzurak}}, \ and\ \bibinfo {author}
  {\bibfnamefont {A.}~\bibnamefont {Morello}},\ }\href@noop {} {\bibfield
  {journal} {\bibinfo  {journal} {Nature}\ }\textbf {\bibinfo {volume} {489}},\
  \bibinfo {pages} {541} (\bibinfo {year} {2012})}\BibitemShut {NoStop}%
\bibitem [{\citenamefont {Muhonen}\ \emph {et~al.}(2014)\citenamefont
  {Muhonen}, \citenamefont {Dehollain}, \citenamefont {Laucht}, \citenamefont
  {Hudson}, \citenamefont {Kalra}, \citenamefont {Sekiguchi}, \citenamefont
  {Itoh}, \citenamefont {Jamieson}, \citenamefont {McCallum}, \citenamefont
  {Dzurak},\ and\ \citenamefont {Morello}}]{Muhonen2014NN}%
  \BibitemOpen
  \bibfield  {author} {\bibinfo {author} {\bibfnamefont {J.~T.}\ \bibnamefont
  {Muhonen}}, \bibinfo {author} {\bibfnamefont {J.~P.}\ \bibnamefont
  {Dehollain}}, \bibinfo {author} {\bibfnamefont {A.}~\bibnamefont {Laucht}},
  \bibinfo {author} {\bibfnamefont {F.~E.}\ \bibnamefont {Hudson}}, \bibinfo
  {author} {\bibfnamefont {R.}~\bibnamefont {Kalra}}, \bibinfo {author}
  {\bibfnamefont {T.}~\bibnamefont {Sekiguchi}}, \bibinfo {author}
  {\bibfnamefont {K.~M.}\ \bibnamefont {Itoh}}, \bibinfo {author}
  {\bibfnamefont {D.~N.}\ \bibnamefont {Jamieson}}, \bibinfo {author}
  {\bibfnamefont {J.~C.}\ \bibnamefont {McCallum}}, \bibinfo {author}
  {\bibfnamefont {A.~S.}\ \bibnamefont {Dzurak}}, \ and\ \bibinfo {author}
  {\bibfnamefont {A.}~\bibnamefont {Morello}},\ }\href@noop {} {\bibfield
  {journal} {\bibinfo  {journal} {Nature Nanotech.}\ }\textbf {\bibinfo
  {volume} {9}},\ \bibinfo {pages} {986} (\bibinfo {year} {2014})}\BibitemShut
  {NoStop}%
\bibitem [{\citenamefont {Mi}\ \emph {et~al.}(2017{\natexlab{a}})\citenamefont
  {Mi}, \citenamefont {Cady}, \citenamefont {Zajac}, \citenamefont {Deelman},\
  and\ \citenamefont {Petta}}]{Petta2017Science}%
  \BibitemOpen
  \bibfield  {author} {\bibinfo {author} {\bibfnamefont {X.}~\bibnamefont
  {Mi}}, \bibinfo {author} {\bibfnamefont {J.~V.}\ \bibnamefont {Cady}},
  \bibinfo {author} {\bibfnamefont {D.~M.}\ \bibnamefont {Zajac}}, \bibinfo
  {author} {\bibfnamefont {P.~W.}\ \bibnamefont {Deelman}}, \ and\ \bibinfo
  {author} {\bibfnamefont {J.~R.}\ \bibnamefont {Petta}},\ }\href@noop {}
  {\bibfield  {journal} {\bibinfo  {journal} {Science}\ }\textbf {\bibinfo
  {volume} {355}},\ \bibinfo {pages} {156} (\bibinfo {year}
  {2017}{\natexlab{a}})}\BibitemShut {NoStop}%
\bibitem [{\citenamefont {Mi}\ \emph {et~al.}(2018)\citenamefont {Mi},
  \citenamefont {Benito}, \citenamefont {Putz}, \citenamefont {Zajac},
  \citenamefont {Taylor}, \citenamefont {Burkard},\ and\ \citenamefont
  {Petta}}]{MiPetta2018N}%
  \BibitemOpen
  \bibfield  {author} {\bibinfo {author} {\bibfnamefont {X.}~\bibnamefont
  {Mi}}, \bibinfo {author} {\bibfnamefont {M.}~\bibnamefont {Benito}}, \bibinfo
  {author} {\bibfnamefont {S.}~\bibnamefont {Putz}}, \bibinfo {author}
  {\bibfnamefont {D.~M.}\ \bibnamefont {Zajac}}, \bibinfo {author}
  {\bibfnamefont {J.~M.}\ \bibnamefont {Taylor}}, \bibinfo {author}
  {\bibfnamefont {G.}~\bibnamefont {Burkard}}, \ and\ \bibinfo {author}
  {\bibfnamefont {J.~R.}\ \bibnamefont {Petta}},\ }\href@noop {} {\bibfield
  {journal} {\bibinfo  {journal} {Nature}\ }\textbf {\bibinfo {volume} {555}},\
  \bibinfo {pages} {599} (\bibinfo {year} {2018})}\BibitemShut {NoStop}%
\bibitem [{\citenamefont {Landig}\ \emph {et~al.}(2018)\citenamefont {Landig},
  \citenamefont {Koski}, \citenamefont {Scarlino}, \citenamefont {Mendes},
  \citenamefont {Blais}, \citenamefont {Reichl}, \citenamefont {Wegscheider},
  \citenamefont {Wallraff}, \citenamefont {Ensslin},\ and\ \citenamefont
  {Ihn}}]{Landig-Wallraff-Ensslin-Ihn2018N}%
  \BibitemOpen
  \bibfield  {author} {\bibinfo {author} {\bibfnamefont {A.~J.}\ \bibnamefont
  {Landig}}, \bibinfo {author} {\bibfnamefont {J.~V.}\ \bibnamefont {Koski}},
  \bibinfo {author} {\bibfnamefont {P.}~\bibnamefont {Scarlino}}, \bibinfo
  {author} {\bibfnamefont {U.~C.}\ \bibnamefont {Mendes}}, \bibinfo {author}
  {\bibfnamefont {A.}~\bibnamefont {Blais}}, \bibinfo {author} {\bibfnamefont
  {C.}~\bibnamefont {Reichl}}, \bibinfo {author} {\bibfnamefont
  {W.}~\bibnamefont {Wegscheider}}, \bibinfo {author} {\bibfnamefont
  {A.}~\bibnamefont {Wallraff}}, \bibinfo {author} {\bibfnamefont
  {K.}~\bibnamefont {Ensslin}}, \ and\ \bibinfo {author} {\bibfnamefont
  {T.}~\bibnamefont {Ihn}},\ }\href@noop {} {\bibfield  {journal} {\bibinfo
  {journal} {Nature}\ }\textbf {\bibinfo {volume} {560}},\ \bibinfo {pages}
  {179} (\bibinfo {year} {2018})}\BibitemShut {NoStop}%
\bibitem [{\citenamefont {Nowack}\ \emph {et~al.}(2007)\citenamefont {Nowack},
  \citenamefont {Koppens}, \citenamefont {Nazarov},\ and\ \citenamefont
  {Vandersypen}}]{Nowack2007S}%
  \BibitemOpen
  \bibfield  {author} {\bibinfo {author} {\bibfnamefont {K.~C.}\ \bibnamefont
  {Nowack}}, \bibinfo {author} {\bibfnamefont {F.~H.~L.}\ \bibnamefont
  {Koppens}}, \bibinfo {author} {\bibfnamefont {Y.~V.}\ \bibnamefont
  {Nazarov}}, \ and\ \bibinfo {author} {\bibfnamefont {L.~M.~K.}\ \bibnamefont
  {Vandersypen}},\ }\href@noop {} {\bibfield  {journal} {\bibinfo  {journal}
  {Science}\ }\textbf {\bibinfo {volume} {318}},\ \bibinfo {pages} {1430}
  (\bibinfo {year} {2007})}\BibitemShut {NoStop}%
\bibitem [{\citenamefont {Yang}\ \emph {et~al.}(2013)\citenamefont {Yang},
  \citenamefont {Rossi}, \citenamefont {Ruskov}, \citenamefont {Lai},
  \citenamefont {Mohiyaddin}, \citenamefont {Lee}, \citenamefont {Tahan},
  \citenamefont {Klimeck}, \citenamefont {Morello},\ and\ \citenamefont
  {Dzurak}}]{Yang2013NC}%
  \BibitemOpen
  \bibfield  {author} {\bibinfo {author} {\bibfnamefont {C.~H.}\ \bibnamefont
  {Yang}}, \bibinfo {author} {\bibfnamefont {A.}~\bibnamefont {Rossi}},
  \bibinfo {author} {\bibfnamefont {R.}~\bibnamefont {Ruskov}}, \bibinfo
  {author} {\bibfnamefont {N.~S.}\ \bibnamefont {Lai}}, \bibinfo {author}
  {\bibfnamefont {F.~A.}\ \bibnamefont {Mohiyaddin}}, \bibinfo {author}
  {\bibfnamefont {S.}~\bibnamefont {Lee}}, \bibinfo {author} {\bibfnamefont
  {C.}~\bibnamefont {Tahan}}, \bibinfo {author} {\bibfnamefont
  {G.}~\bibnamefont {Klimeck}}, \bibinfo {author} {\bibfnamefont
  {A.}~\bibnamefont {Morello}}, \ and\ \bibinfo {author} {\bibfnamefont
  {A.~S.}\ \bibnamefont {Dzurak}},\ }\href@noop {} {\bibfield  {journal}
  {\bibinfo  {journal} {Nature Communcations}\ }\textbf {\bibinfo {volume}
  {4}},\ \bibinfo {pages} {2069} (\bibinfo {year} {2013})}\BibitemShut
  {NoStop}%
\bibitem [{\citenamefont {Veldhorst}\ \emph
  {et~al.}(2015{\natexlab{b}})\citenamefont {Veldhorst}, \citenamefont
  {Ruskov}, \citenamefont {Yang}, \citenamefont {Hwang}, \citenamefont
  {Hudson}, \citenamefont {Flatt\'{e}}, \citenamefont {Tahan}, \citenamefont
  {Itoh}, \citenamefont {Morello},\ and\ \citenamefont
  {Dzurak}}]{VeldhorstRuskov2015PRB}%
  \BibitemOpen
  \bibfield  {author} {\bibinfo {author} {\bibfnamefont {M.}~\bibnamefont
  {Veldhorst}}, \bibinfo {author} {\bibfnamefont {R.}~\bibnamefont {Ruskov}},
  \bibinfo {author} {\bibfnamefont {C.~H.}\ \bibnamefont {Yang}}, \bibinfo
  {author} {\bibfnamefont {J.~C.~C.}\ \bibnamefont {Hwang}}, \bibinfo {author}
  {\bibfnamefont {F.~E.}\ \bibnamefont {Hudson}}, \bibinfo {author}
  {\bibfnamefont {M.~E.}\ \bibnamefont {Flatt\'{e}}}, \bibinfo {author}
  {\bibfnamefont {C.}~\bibnamefont {Tahan}}, \bibinfo {author} {\bibfnamefont
  {K.~M.}\ \bibnamefont {Itoh}}, \bibinfo {author} {\bibfnamefont
  {A.}~\bibnamefont {Morello}}, \ and\ \bibinfo {author} {\bibfnamefont
  {A.~S.}\ \bibnamefont {Dzurak}},\ }\href@noop {} {\bibfield  {journal}
  {\bibinfo  {journal} {Phys. Rev. B}\ }\textbf {\bibinfo {volume} {92}},\
  \bibinfo {pages} {201401(R)} (\bibinfo {year}
  {2015}{\natexlab{b}})}\BibitemShut {NoStop}%
\bibitem [{\citenamefont {DiVincenzo}\ \emph {et~al.}(2000)\citenamefont
  {DiVincenzo}, \citenamefont {Bacon}, \citenamefont {Kempe}, \citenamefont
  {Burkard},\ and\ \citenamefont {Whaley}}]{DiVincenzo2000N}%
  \BibitemOpen
  \bibfield  {author} {\bibinfo {author} {\bibfnamefont {D.~P.}\ \bibnamefont
  {DiVincenzo}}, \bibinfo {author} {\bibfnamefont {D.}~\bibnamefont {Bacon}},
  \bibinfo {author} {\bibfnamefont {J.}~\bibnamefont {Kempe}}, \bibinfo
  {author} {\bibfnamefont {G.}~\bibnamefont {Burkard}}, \ and\ \bibinfo
  {author} {\bibfnamefont {K.~B.}\ \bibnamefont {Whaley}},\ }\href@noop {}
  {\bibfield  {journal} {\bibinfo  {journal} {Nature}\ }\textbf {\bibinfo
  {volume} {408}},\ \bibinfo {pages} {339} (\bibinfo {year}
  {2000})}\BibitemShut {NoStop}%
\bibitem [{\citenamefont {Fong}\ and\ \citenamefont
  {Wandzura}(2011)}]{FongQuInfComp2011}%
  \BibitemOpen
  \bibfield  {author} {\bibinfo {author} {\bibfnamefont {B.~H.}\ \bibnamefont
  {Fong}}\ and\ \bibinfo {author} {\bibfnamefont {S.~M.}\ \bibnamefont
  {Wandzura}},\ }\href@noop {} {\bibfield  {journal} {\bibinfo  {journal}
  {Quantum Inf. Comput.}\ }\textbf {\bibinfo {volume} {11}},\ \bibinfo {pages}
  {1003} (\bibinfo {year} {2011})}\BibitemShut {NoStop}%
\bibitem [{\citenamefont {Gaudreau}\ \emph {et~al.}(2006)\citenamefont
  {Gaudreau}, \citenamefont {Studenikin}, \citenamefont {Sachrajda},
  \citenamefont {Zawadzki}, \citenamefont {Kam}, \citenamefont {Lapointe},
  \citenamefont {Korkusinski},\ and\ \citenamefont
  {Hawrylak}}]{Gaudreau2006PRL}%
  \BibitemOpen
  \bibfield  {author} {\bibinfo {author} {\bibfnamefont {L.}~\bibnamefont
  {Gaudreau}}, \bibinfo {author} {\bibfnamefont {S.~A.}\ \bibnamefont
  {Studenikin}}, \bibinfo {author} {\bibfnamefont {A.~S.}\ \bibnamefont
  {Sachrajda}}, \bibinfo {author} {\bibfnamefont {P.}~\bibnamefont {Zawadzki}},
  \bibinfo {author} {\bibfnamefont {A.}~\bibnamefont {Kam}}, \bibinfo {author}
  {\bibfnamefont {J.}~\bibnamefont {Lapointe}}, \bibinfo {author}
  {\bibfnamefont {M.}~\bibnamefont {Korkusinski}}, \ and\ \bibinfo {author}
  {\bibfnamefont {P.}~\bibnamefont {Hawrylak}},\ }\href@noop {} {\bibfield
  {journal} {\bibinfo  {journal} {Phys. Rev. Lett.}\ }\textbf {\bibinfo
  {volume} {97}},\ \bibinfo {pages} {036807} (\bibinfo {year}
  {2006})}\BibitemShut {NoStop}%
\bibitem [{\citenamefont {Gaudreau}\ \emph {et~al.}(2009)\citenamefont
  {Gaudreau}, \citenamefont {Kam}, \citenamefont {Granger}, \citenamefont
  {Studenikin}, \citenamefont {Zawadzki},\ and\ \citenamefont
  {Sachrajda}}]{Gaudreau2009APL}%
  \BibitemOpen
  \bibfield  {author} {\bibinfo {author} {\bibfnamefont {L.}~\bibnamefont
  {Gaudreau}}, \bibinfo {author} {\bibfnamefont {A.}~\bibnamefont {Kam}},
  \bibinfo {author} {\bibfnamefont {G.}~\bibnamefont {Granger}}, \bibinfo
  {author} {\bibfnamefont {S.~A.}\ \bibnamefont {Studenikin}}, \bibinfo
  {author} {\bibfnamefont {P.}~\bibnamefont {Zawadzki}}, \ and\ \bibinfo
  {author} {\bibfnamefont {A.~S.}\ \bibnamefont {Sachrajda}},\ }\href@noop {}
  {\bibfield  {journal} {\bibinfo  {journal} {Appl. Phys. Lett.}\ }\textbf
  {\bibinfo {volume} {95}},\ \bibinfo {pages} {193101} (\bibinfo {year}
  {2009})}\BibitemShut {NoStop}%
\bibitem [{\citenamefont {Laird}\ \emph {et~al.}(2010)\citenamefont {Laird},
  \citenamefont {Taylor}, \citenamefont {DiVincenzo}, \citenamefont {Marcus},
  \citenamefont {Hanson},\ and\ \citenamefont {Gossard}}]{LairdMarcus2010PRB}%
  \BibitemOpen
  \bibfield  {author} {\bibinfo {author} {\bibfnamefont {E.~A.}\ \bibnamefont
  {Laird}}, \bibinfo {author} {\bibfnamefont {J.~M.}\ \bibnamefont {Taylor}},
  \bibinfo {author} {\bibfnamefont {D.~P.}\ \bibnamefont {DiVincenzo}},
  \bibinfo {author} {\bibfnamefont {C.~M.}\ \bibnamefont {Marcus}}, \bibinfo
  {author} {\bibfnamefont {M.~P.}\ \bibnamefont {Hanson}}, \ and\ \bibinfo
  {author} {\bibfnamefont {A.~C.}\ \bibnamefont {Gossard}},\ }\href@noop {}
  {\bibfield  {journal} {\bibinfo  {journal} {Phys. Rev. B}\ }\textbf {\bibinfo
  {volume} {82}},\ \bibinfo {pages} {075403} (\bibinfo {year}
  {2010})}\BibitemShut {NoStop}%
\bibitem [{\citenamefont {Gaudreau}\ \emph {et~al.}(2011)\citenamefont
  {Gaudreau}, \citenamefont {Granger}, \citenamefont {Kam}, \citenamefont
  {Aers}, \citenamefont {Studenikin}, \citenamefont {Zawadzki}, \citenamefont
  {Pioro-Ladrie`re}, \citenamefont {Wasilewski},\ and\ \citenamefont
  {Sachrajda}}]{Gaudreau2011NP}%
  \BibitemOpen
  \bibfield  {author} {\bibinfo {author} {\bibfnamefont {L.}~\bibnamefont
  {Gaudreau}}, \bibinfo {author} {\bibfnamefont {G.}~\bibnamefont {Granger}},
  \bibinfo {author} {\bibfnamefont {A.}~\bibnamefont {Kam}}, \bibinfo {author}
  {\bibfnamefont {G.~C.}\ \bibnamefont {Aers}}, \bibinfo {author}
  {\bibfnamefont {S.~A.}\ \bibnamefont {Studenikin}}, \bibinfo {author}
  {\bibfnamefont {P.}~\bibnamefont {Zawadzki}}, \bibinfo {author}
  {\bibfnamefont {M.}~\bibnamefont {Pioro-Ladrie`re}}, \bibinfo {author}
  {\bibfnamefont {Z.~R.}\ \bibnamefont {Wasilewski}}, \ and\ \bibinfo {author}
  {\bibfnamefont {A.~S.}\ \bibnamefont {Sachrajda}},\ }\href@noop {} {\bibfield
   {journal} {\bibinfo  {journal} {Nat. Phys.}\ }\textbf {\bibinfo {volume}
  {8}},\ \bibinfo {pages} {54} (\bibinfo {year} {2011})}\BibitemShut {NoStop}%
\bibitem [{\citenamefont {Medford}\ \emph
  {et~al.}(2013{\natexlab{b}})\citenamefont {Medford}, \citenamefont {Beil},
  \citenamefont {Taylor}, \citenamefont {Bartlett}, \citenamefont {Doherty},
  \citenamefont {Rashba}, \citenamefont {DiVincenzo}, \citenamefont {Lu},
  \citenamefont {Gossard},\ and\ \citenamefont {Marcus}}]{Medford2013NatNano}%
  \BibitemOpen
  \bibfield  {author} {\bibinfo {author} {\bibfnamefont {J.}~\bibnamefont
  {Medford}}, \bibinfo {author} {\bibfnamefont {J.}~\bibnamefont {Beil}},
  \bibinfo {author} {\bibfnamefont {J.~M.}\ \bibnamefont {Taylor}}, \bibinfo
  {author} {\bibfnamefont {S.~D.}\ \bibnamefont {Bartlett}}, \bibinfo {author}
  {\bibfnamefont {A.~C.}\ \bibnamefont {Doherty}}, \bibinfo {author}
  {\bibfnamefont {E.~I.}\ \bibnamefont {Rashba}}, \bibinfo {author}
  {\bibfnamefont {D.~P.}\ \bibnamefont {DiVincenzo}}, \bibinfo {author}
  {\bibfnamefont {H.}~\bibnamefont {Lu}}, \bibinfo {author} {\bibfnamefont
  {A.~C.}\ \bibnamefont {Gossard}}, \ and\ \bibinfo {author} {\bibfnamefont
  {C.~M.}\ \bibnamefont {Marcus}},\ }\href@noop {} {\bibfield  {journal}
  {\bibinfo  {journal} {Nature Nanotechnology}\ }\textbf {\bibinfo {volume}
  {8}},\ \bibinfo {pages} {654} (\bibinfo {year}
  {2013}{\natexlab{b}})}\BibitemShut {NoStop}%
\bibitem [{\citenamefont {Eng}\ \emph {et~al.}(2015)\citenamefont {Eng},
  \citenamefont {Ladd}, \citenamefont {Smith}, \citenamefont {Borselli},
  \citenamefont {Kiselev}, \citenamefont {Fong}, \citenamefont {Holabird},
  \citenamefont {Hazard}, \citenamefont {Huang}, \citenamefont {Deelman},
  \citenamefont {Milosavljevic}, \citenamefont {Schmitz}, \citenamefont {Ross},
  \citenamefont {Gyure},\ and\ \citenamefont {Hunter}}]{EngHRLs2015SAdv}%
  \BibitemOpen
  \bibfield  {author} {\bibinfo {author} {\bibfnamefont {K.}~\bibnamefont
  {Eng}}, \bibinfo {author} {\bibfnamefont {T.~D.}\ \bibnamefont {Ladd}},
  \bibinfo {author} {\bibfnamefont {A.}~\bibnamefont {Smith}}, \bibinfo
  {author} {\bibfnamefont {M.~G.}\ \bibnamefont {Borselli}}, \bibinfo {author}
  {\bibfnamefont {A.~A.}\ \bibnamefont {Kiselev}}, \bibinfo {author}
  {\bibfnamefont {B.~H.}\ \bibnamefont {Fong}}, \bibinfo {author}
  {\bibfnamefont {K.~S.}\ \bibnamefont {Holabird}}, \bibinfo {author}
  {\bibfnamefont {T.}~\bibnamefont {Hazard}}, \bibinfo {author} {\bibfnamefont
  {B.}~\bibnamefont {Huang}}, \bibinfo {author} {\bibfnamefont {P.~W.}\
  \bibnamefont {Deelman}}, \bibinfo {author} {\bibfnamefont {I.}~\bibnamefont
  {Milosavljevic}}, \bibinfo {author} {\bibfnamefont {A.~E.}\ \bibnamefont
  {Schmitz}}, \bibinfo {author} {\bibfnamefont {R.~S.}\ \bibnamefont {Ross}},
  \bibinfo {author} {\bibfnamefont {M.~F.}\ \bibnamefont {Gyure}}, \ and\
  \bibinfo {author} {\bibfnamefont {A.~T.}\ \bibnamefont {Hunter}},\
  }\href@noop {} {\bibfield  {journal} {\bibinfo  {journal} {Sci. Adv.}\
  }\textbf {\bibinfo {volume} {1}},\ \bibinfo {pages} {e1500214} (\bibinfo
  {year} {2015})}\BibitemShut {NoStop}%
\bibitem [{\citenamefont {Taylor}\ \emph {et~al.}(2013)\citenamefont {Taylor},
  \citenamefont {Srinivasa},\ and\ \citenamefont
  {Medford}}]{TaylorSrinivasa2013PRL}%
  \BibitemOpen
  \bibfield  {author} {\bibinfo {author} {\bibfnamefont {J.~M.}\ \bibnamefont
  {Taylor}}, \bibinfo {author} {\bibfnamefont {V.}~\bibnamefont {Srinivasa}}, \
  and\ \bibinfo {author} {\bibfnamefont {J.}~\bibnamefont {Medford}},\
  }\href@noop {} {\bibfield  {journal} {\bibinfo  {journal} {Phys. Rev. Lett.}\
  }\textbf {\bibinfo {volume} {111}},\ \bibinfo {pages} {050502} (\bibinfo
  {year} {2013})}\BibitemShut {NoStop}%
\bibitem [{\citenamefont {Russ}\ and\ \citenamefont
  {Burkard}(2015)}]{RussBurkard2015PRB}%
  \BibitemOpen
  \bibfield  {author} {\bibinfo {author} {\bibfnamefont {M.}~\bibnamefont
  {Russ}}\ and\ \bibinfo {author} {\bibfnamefont {G.}~\bibnamefont {Burkard}},\
  }\href@noop {} {\bibfield  {journal} {\bibinfo  {journal} {Phys. Rev. B}\
  }\textbf {\bibinfo {volume} {92}},\ \bibinfo {pages} {205412} (\bibinfo
  {year} {2015})}\BibitemShut {NoStop}%
\bibitem [{\citenamefont {Russ}\ \emph {et~al.}(2016)\citenamefont {Russ},
  \citenamefont {Ginzel},\ and\ \citenamefont
  {Burkard}}]{RussGinzelBurkard2016PRB}%
  \BibitemOpen
  \bibfield  {author} {\bibinfo {author} {\bibfnamefont {M.}~\bibnamefont
  {Russ}}, \bibinfo {author} {\bibfnamefont {F.}~\bibnamefont {Ginzel}}, \ and\
  \bibinfo {author} {\bibfnamefont {G.}~\bibnamefont {Burkard}},\ }\href@noop
  {} {\bibfield  {journal} {\bibinfo  {journal} {Phys. Rev. B}\ }\textbf
  {\bibinfo {volume} {94}},\ \bibinfo {pages} {165411} (\bibinfo {year}
  {2016})}\BibitemShut {NoStop}%
\bibitem [{\citenamefont {Shim}\ and\ \citenamefont {Tahan}(2016)}]{AEON2016}%
  \BibitemOpen
  \bibfield  {author} {\bibinfo {author} {\bibfnamefont {Y.-P.}\ \bibnamefont
  {Shim}}\ and\ \bibinfo {author} {\bibfnamefont {C.}~\bibnamefont {Tahan}},\
  }\href@noop {} {\bibfield  {journal} {\bibinfo  {journal} {Phys. Rev. B}\
  }\textbf {\bibinfo {volume} {93}},\ \bibinfo {pages} {121410(R)} (\bibinfo
  {year} {2016})}\BibitemShut {NoStop}%
\bibitem [{\citenamefont {Ruskov}\ \emph {et~al.}(2018)\citenamefont {Ruskov},
  \citenamefont {Veldhorst}, \citenamefont {Dzurak},\ and\ \citenamefont
  {Tahan}}]{RuskovVeldhorst2018PRB}%
  \BibitemOpen
  \bibfield  {author} {\bibinfo {author} {\bibfnamefont {R.}~\bibnamefont
  {Ruskov}}, \bibinfo {author} {\bibfnamefont {M.}~\bibnamefont {Veldhorst}},
  \bibinfo {author} {\bibfnamefont {A.~S.}\ \bibnamefont {Dzurak}}, \ and\
  \bibinfo {author} {\bibfnamefont {C.}~\bibnamefont {Tahan}},\ }\href@noop {}
  {\bibfield  {journal} {\bibinfo  {journal} {Phys. Rev. B}\ }\textbf {\bibinfo
  {volume} {98}},\ \bibinfo {pages} {245424} (\bibinfo {year}
  {2018})}\BibitemShut {NoStop}%
\bibitem [{\citenamefont {Reed}\ \emph {et~al.}(2016)\citenamefont {Reed},
  \citenamefont {Maune}, \citenamefont {Andrews}, \citenamefont {Borselli},
  \citenamefont {Eng}, \citenamefont {Jura}, \citenamefont {Kiselev},
  \citenamefont {Ladd}, \citenamefont {Merkel}, \citenamefont {Milosavljevic},
  \citenamefont {Pritchett}, \citenamefont {Rakher}, \citenamefont {Ross},
  \citenamefont {Schmitz}, \citenamefont {Smith}, \citenamefont {Wright},
  \citenamefont {Gyure},\ and\ \citenamefont {Hunter}}]{ReedHunter2016PRL}%
  \BibitemOpen
  \bibfield  {author} {\bibinfo {author} {\bibfnamefont {M.~D.}\ \bibnamefont
  {Reed}}, \bibinfo {author} {\bibfnamefont {B.~M.}\ \bibnamefont {Maune}},
  \bibinfo {author} {\bibfnamefont {R.~W.}\ \bibnamefont {Andrews}}, \bibinfo
  {author} {\bibfnamefont {M.~G.}\ \bibnamefont {Borselli}}, \bibinfo {author}
  {\bibfnamefont {K.}~\bibnamefont {Eng}}, \bibinfo {author} {\bibfnamefont
  {M.~P.}\ \bibnamefont {Jura}}, \bibinfo {author} {\bibfnamefont {A.~A.}\
  \bibnamefont {Kiselev}}, \bibinfo {author} {\bibfnamefont {T.~D.}\
  \bibnamefont {Ladd}}, \bibinfo {author} {\bibfnamefont {S.~T.}\ \bibnamefont
  {Merkel}}, \bibinfo {author} {\bibfnamefont {I.}~\bibnamefont
  {Milosavljevic}}, \bibinfo {author} {\bibfnamefont {E.~J.}\ \bibnamefont
  {Pritchett}}, \bibinfo {author} {\bibfnamefont {M.~T.}\ \bibnamefont
  {Rakher}}, \bibinfo {author} {\bibfnamefont {R.~S.}\ \bibnamefont {Ross}},
  \bibinfo {author} {\bibfnamefont {A.~E.}\ \bibnamefont {Schmitz}}, \bibinfo
  {author} {\bibfnamefont {A.}~\bibnamefont {Smith}}, \bibinfo {author}
  {\bibfnamefont {J.~A.}\ \bibnamefont {Wright}}, \bibinfo {author}
  {\bibfnamefont {M.~F.}\ \bibnamefont {Gyure}}, \ and\ \bibinfo {author}
  {\bibfnamefont {A.~T.}\ \bibnamefont {Hunter}},\ }\href@noop {} {\bibfield
  {journal} {\bibinfo  {journal} {Phys. Rev. Lett.}\ }\textbf {\bibinfo
  {volume} {116}},\ \bibinfo {pages} {110402} (\bibinfo {year}
  {2016})}\BibitemShut {NoStop}%
\bibitem [{\citenamefont {Martins}\ \emph {et~al.}(2016)\citenamefont
  {Martins}, \citenamefont {Malinowski}, \citenamefont {Nissen}, \citenamefont
  {Barnes}, \citenamefont {Fallahi}, \citenamefont {Gardner}, \citenamefont
  {Manfra}, \citenamefont {Marcus},\ and\ \citenamefont
  {Kuemmeth}}]{MartinsKuemethMarcus2016PRL}%
  \BibitemOpen
  \bibfield  {author} {\bibinfo {author} {\bibfnamefont {F.}~\bibnamefont
  {Martins}}, \bibinfo {author} {\bibfnamefont {F.~K.}\ \bibnamefont
  {Malinowski}}, \bibinfo {author} {\bibfnamefont {P.~D.}\ \bibnamefont
  {Nissen}}, \bibinfo {author} {\bibfnamefont {E.}~\bibnamefont {Barnes}},
  \bibinfo {author} {\bibfnamefont {S.}~\bibnamefont {Fallahi}}, \bibinfo
  {author} {\bibfnamefont {G.~C.}\ \bibnamefont {Gardner}}, \bibinfo {author}
  {\bibfnamefont {M.~J.}\ \bibnamefont {Manfra}}, \bibinfo {author}
  {\bibfnamefont {C.~M.}\ \bibnamefont {Marcus}}, \ and\ \bibinfo {author}
  {\bibfnamefont {F.}~\bibnamefont {Kuemmeth}},\ }\href@noop {} {\bibfield
  {journal} {\bibinfo  {journal} {Phys. Rev. Lett.}\ }\textbf {\bibinfo
  {volume} {116}},\ \bibinfo {pages} {116801} (\bibinfo {year}
  {2016})}\BibitemShut {NoStop}%
\bibitem [{\citenamefont {Pioro-Ladri{\`e}re}\ \emph
  {et~al.}(2008)\citenamefont {Pioro-Ladri{\`e}re}, \citenamefont {Obata},
  \citenamefont {Tokura}, \citenamefont {Shin}, \citenamefont {Kubo},
  \citenamefont {Yoshida}, \citenamefont {Taniyama},\ and\ \citenamefont
  {Tarucha}}]{Pioro-Ladriere2008NP}%
  \BibitemOpen
  \bibfield  {author} {\bibinfo {author} {\bibfnamefont {M.}~\bibnamefont
  {Pioro-Ladri{\`e}re}}, \bibinfo {author} {\bibfnamefont {T.}~\bibnamefont
  {Obata}}, \bibinfo {author} {\bibfnamefont {Y.}~\bibnamefont {Tokura}},
  \bibinfo {author} {\bibfnamefont {Y.-S.}\ \bibnamefont {Shin}}, \bibinfo
  {author} {\bibfnamefont {T.}~\bibnamefont {Kubo}}, \bibinfo {author}
  {\bibfnamefont {K.}~\bibnamefont {Yoshida}}, \bibinfo {author} {\bibfnamefont
  {T.}~\bibnamefont {Taniyama}}, \ and\ \bibinfo {author} {\bibfnamefont
  {S.}~\bibnamefont {Tarucha}},\ }\href@noop {} {\bibfield  {journal} {\bibinfo
   {journal} {Nat. Phys.}\ }\textbf {\bibinfo {volume} {4}},\ \bibinfo {pages}
  {776} (\bibinfo {year} {2008})}\BibitemShut {NoStop}%
\bibitem [{\citenamefont {Kawakami}\ \emph {et~al.}(2014)\citenamefont
  {Kawakami}, \citenamefont {Scarlino}, \citenamefont {Ward}, \citenamefont
  {Braakman}, \citenamefont {Savage}, \citenamefont {Lagally}, \citenamefont
  {Friesen}, \citenamefont {Coppersmith}, \citenamefont {Eriksson},\ and\
  \citenamefont {Vandersypen}}]{Kawakami2014Nn}%
  \BibitemOpen
  \bibfield  {author} {\bibinfo {author} {\bibfnamefont {E.~E.}\ \bibnamefont
  {Kawakami}}, \bibinfo {author} {\bibfnamefont {P.}~\bibnamefont {Scarlino}},
  \bibinfo {author} {\bibfnamefont {D.~R.}\ \bibnamefont {Ward}}, \bibinfo
  {author} {\bibfnamefont {F.~R.}\ \bibnamefont {Braakman}}, \bibinfo {author}
  {\bibfnamefont {D.~E.}\ \bibnamefont {Savage}}, \bibinfo {author}
  {\bibfnamefont {M.~G.}\ \bibnamefont {Lagally}}, \bibinfo {author}
  {\bibfnamefont {M.}~\bibnamefont {Friesen}}, \bibinfo {author} {\bibfnamefont
  {S.~N.}\ \bibnamefont {Coppersmith}}, \bibinfo {author} {\bibfnamefont
  {M.~A.}\ \bibnamefont {Eriksson}}, \ and\ \bibinfo {author} {\bibfnamefont
  {L.~M.~K.}\ \bibnamefont {Vandersypen}},\ }\href@noop {} {\bibfield
  {journal} {\bibinfo  {journal} {Nature Nanotechnology}\ }\textbf {\bibinfo
  {volume} {9}},\ \bibinfo {pages} {666} (\bibinfo {year} {2014})}\BibitemShut
  {NoStop}%
\bibitem [{\citenamefont {Dial}\ \emph {et~al.}(2013)\citenamefont {Dial},
  \citenamefont {Shulman}, \citenamefont {Harvey}, \citenamefont {Bluhm},
  \citenamefont {Umansky},\ and\ \citenamefont {Yacoby}}]{DialYacoby2013PRL}%
  \BibitemOpen
  \bibfield  {author} {\bibinfo {author} {\bibfnamefont {O.~E.}\ \bibnamefont
  {Dial}}, \bibinfo {author} {\bibfnamefont {M.~D.}\ \bibnamefont {Shulman}},
  \bibinfo {author} {\bibfnamefont {S.~P.}\ \bibnamefont {Harvey}}, \bibinfo
  {author} {\bibfnamefont {H.}~\bibnamefont {Bluhm}}, \bibinfo {author}
  {\bibfnamefont {V.}~\bibnamefont {Umansky}}, \ and\ \bibinfo {author}
  {\bibfnamefont {A.}~\bibnamefont {Yacoby}},\ }\href@noop {} {\bibfield
  {journal} {\bibinfo  {journal} {Phys. Rev. Lett.}\ }\textbf {\bibinfo
  {volume} {110}},\ \bibinfo {pages} {146804} (\bibinfo {year}
  {2013})}\BibitemShut {NoStop}%
\bibitem [{\citenamefont {Doherty}\ and\ \citenamefont
  {Wardrop}(2013)}]{Doherty2013PRL}%
  \BibitemOpen
  \bibfield  {author} {\bibinfo {author} {\bibfnamefont {A.~C.}\ \bibnamefont
  {Doherty}}\ and\ \bibinfo {author} {\bibfnamefont {M.~P.}\ \bibnamefont
  {Wardrop}},\ }\href@noop {} {\bibfield  {journal} {\bibinfo  {journal} {Phys.
  Rev. Lett.}\ }\textbf {\bibinfo {volume} {111}},\ \bibinfo {pages} {050503}
  (\bibinfo {year} {2013})}\BibitemShut {NoStop}%
\bibitem [{\citenamefont {Burkard}\ and\ \citenamefont
  {Imamoglu}(2006)}]{BurkardIamoglu2006PRB}%
  \BibitemOpen
  \bibfield  {author} {\bibinfo {author} {\bibfnamefont {G.}~\bibnamefont
  {Burkard}}\ and\ \bibinfo {author} {\bibfnamefont {A.}~\bibnamefont
  {Imamoglu}},\ }\href@noop {} {\bibfield  {journal} {\bibinfo  {journal}
  {Phys. Rev. B}\ }\textbf {\bibinfo {volume} {74}},\ \bibinfo {pages}
  {041307(R)} (\bibinfo {year} {2006})}\BibitemShut {NoStop}%
\bibitem [{\citenamefont {Frey}\ \emph {et~al.}(2012)\citenamefont {Frey},
  \citenamefont {Leek}, \citenamefont {Beck}, \citenamefont {Blais},
  \citenamefont {Ihn}, \citenamefont {Ensslin},\ and\ \citenamefont
  {Wallraff}}]{FreyWallraff2012PRL}%
  \BibitemOpen
  \bibfield  {author} {\bibinfo {author} {\bibfnamefont {T.}~\bibnamefont
  {Frey}}, \bibinfo {author} {\bibfnamefont {P.~J.}\ \bibnamefont {Leek}},
  \bibinfo {author} {\bibfnamefont {M.}~\bibnamefont {Beck}}, \bibinfo {author}
  {\bibfnamefont {A.}~\bibnamefont {Blais}}, \bibinfo {author} {\bibfnamefont
  {T.}~\bibnamefont {Ihn}}, \bibinfo {author} {\bibfnamefont {K.}~\bibnamefont
  {Ensslin}}, \ and\ \bibinfo {author} {\bibfnamefont {A.}~\bibnamefont
  {Wallraff}},\ }\href@noop {} {\bibfield  {journal} {\bibinfo  {journal}
  {Phys. Rev. Lett.}\ }\textbf {\bibinfo {volume} {108}},\ \bibinfo {pages}
  {046807} (\bibinfo {year} {2012})}\BibitemShut {NoStop}%
\bibitem [{\citenamefont {Petersson}\ \emph {et~al.}(2012)\citenamefont
  {Petersson}, \citenamefont {McFaul}, \citenamefont {Schroer}, \citenamefont
  {Jung}, \citenamefont {Taylor}, \citenamefont {Houck},\ and\ \citenamefont
  {Petta}}]{Petersson2012N}%
  \BibitemOpen
  \bibfield  {author} {\bibinfo {author} {\bibfnamefont {K.~D.}\ \bibnamefont
  {Petersson}}, \bibinfo {author} {\bibfnamefont {L.~W.}\ \bibnamefont
  {McFaul}}, \bibinfo {author} {\bibfnamefont {M.~D.}\ \bibnamefont {Schroer}},
  \bibinfo {author} {\bibfnamefont {M.}~\bibnamefont {Jung}}, \bibinfo {author}
  {\bibfnamefont {J.~M.}\ \bibnamefont {Taylor}}, \bibinfo {author}
  {\bibfnamefont {A.~A.}\ \bibnamefont {Houck}}, \ and\ \bibinfo {author}
  {\bibfnamefont {J.~R.}\ \bibnamefont {Petta}},\ }\href@noop {} {\bibfield
  {journal} {\bibinfo  {journal} {Nature}\ }\textbf {\bibinfo {volume} {490}},\
  \bibinfo {pages} {380} (\bibinfo {year} {2012})}\BibitemShut {NoStop}%
\bibitem [{\citenamefont {Tosi}\ \emph {et~al.}(2014)\citenamefont {Tosi},
  \citenamefont {Mohiyaddin}, \citenamefont {Huebl},\ and\ \citenamefont
  {Morello}}]{TosiMorello2014AIP}%
  \BibitemOpen
  \bibfield  {author} {\bibinfo {author} {\bibfnamefont {G.}~\bibnamefont
  {Tosi}}, \bibinfo {author} {\bibfnamefont {F.~A.}\ \bibnamefont
  {Mohiyaddin}}, \bibinfo {author} {\bibfnamefont {H.}~\bibnamefont {Huebl}}, \
  and\ \bibinfo {author} {\bibfnamefont {A.}~\bibnamefont {Morello}},\
  }\href@noop {} {\bibfield  {journal} {\bibinfo  {journal} {AIP Advances}\
  }\textbf {\bibinfo {volume} {4}},\ \bibinfo {pages} {087122} (\bibinfo {year}
  {2014})}\BibitemShut {NoStop}%
\bibitem [{\citenamefont {Viennot}\ \emph {et~al.}(2015)\citenamefont
  {Viennot}, \citenamefont {Dartiailh}, \citenamefont {Cottet},\ and\
  \citenamefont {Kontos}}]{Viennot2015Sci}%
  \BibitemOpen
  \bibfield  {author} {\bibinfo {author} {\bibfnamefont {J.~J.}\ \bibnamefont
  {Viennot}}, \bibinfo {author} {\bibfnamefont {M.~C.}\ \bibnamefont
  {Dartiailh}}, \bibinfo {author} {\bibfnamefont {A.}~\bibnamefont {Cottet}}, \
  and\ \bibinfo {author} {\bibfnamefont {T.}~\bibnamefont {Kontos}},\
  }\href@noop {} {\bibfield  {journal} {\bibinfo  {journal} {Science}\ }\textbf
  {\bibinfo {volume} {349}},\ \bibinfo {pages} {408} (\bibinfo {year}
  {2015})}\BibitemShut {NoStop}%
\bibitem [{\citenamefont {Blais}\ \emph {et~al.}(2004)\citenamefont {Blais},
  \citenamefont {Huang}, \citenamefont {Wallraff}, \citenamefont {Girvin},\
  and\ \citenamefont {Schoelkopf}}]{Blais2004PRA}%
  \BibitemOpen
  \bibfield  {author} {\bibinfo {author} {\bibfnamefont {A.}~\bibnamefont
  {Blais}}, \bibinfo {author} {\bibfnamefont {R.-S.}\ \bibnamefont {Huang}},
  \bibinfo {author} {\bibfnamefont {A.}~\bibnamefont {Wallraff}}, \bibinfo
  {author} {\bibfnamefont {S.~M.}\ \bibnamefont {Girvin}}, \ and\ \bibinfo
  {author} {\bibfnamefont {R.~J.}\ \bibnamefont {Schoelkopf}},\ }\href@noop {}
  {\bibfield  {journal} {\bibinfo  {journal} {Phys. Rev. A}\ }\textbf {\bibinfo
  {volume} {69}},\ \bibinfo {pages} {062320} (\bibinfo {year}
  {2004})}\BibitemShut {NoStop}%
\bibitem [{\citenamefont {Majer}\ \emph {et~al.}(2007)\citenamefont {Majer},
  \citenamefont {Chow}, \citenamefont {Gambetta}, \citenamefont {Koch},
  \citenamefont {Johnson}, \citenamefont {Schreier}, \citenamefont {Frunzio},
  \citenamefont {Schuster}, \citenamefont {Houck}, \citenamefont {Wallraff},
  \citenamefont {Blais}, \citenamefont {Devoret}, \citenamefont {Girvin},\ and\
  \citenamefont {Schoelkopf}}]{Majer2007N}%
  \BibitemOpen
  \bibfield  {author} {\bibinfo {author} {\bibfnamefont {J.}~\bibnamefont
  {Majer}}, \bibinfo {author} {\bibfnamefont {J.~M.}\ \bibnamefont {Chow}},
  \bibinfo {author} {\bibfnamefont {J.~M.}\ \bibnamefont {Gambetta}}, \bibinfo
  {author} {\bibfnamefont {J.}~\bibnamefont {Koch}}, \bibinfo {author}
  {\bibfnamefont {B.~R.}\ \bibnamefont {Johnson}}, \bibinfo {author}
  {\bibfnamefont {J.~A.}\ \bibnamefont {Schreier}}, \bibinfo {author}
  {\bibfnamefont {L.}~\bibnamefont {Frunzio}}, \bibinfo {author} {\bibfnamefont
  {D.~I.}\ \bibnamefont {Schuster}}, \bibinfo {author} {\bibfnamefont {A.~A.}\
  \bibnamefont {Houck}}, \bibinfo {author} {\bibfnamefont {A.}~\bibnamefont
  {Wallraff}}, \bibinfo {author} {\bibfnamefont {A.}~\bibnamefont {Blais}},
  \bibinfo {author} {\bibfnamefont {M.~H.}\ \bibnamefont {Devoret}}, \bibinfo
  {author} {\bibfnamefont {S.~M.}\ \bibnamefont {Girvin}}, \ and\ \bibinfo
  {author} {\bibfnamefont {R.~J.}\ \bibnamefont {Schoelkopf}},\ }\href@noop {}
  {\bibfield  {journal} {\bibinfo  {journal} {Nature}\ }\textbf {\bibinfo
  {volume} {449}},\ \bibinfo {pages} {443} (\bibinfo {year}
  {2007})}\BibitemShut {NoStop}%
\bibitem [{\citenamefont {DiCarlo}\ \emph {et~al.}(2010)\citenamefont
  {DiCarlo}, \citenamefont {Reed}, \citenamefont {Sun}, \citenamefont
  {Johnson}, \citenamefont {Chow}, \citenamefont {Gambetta}, \citenamefont
  {Frunzio}, \citenamefont {Girvin}, \citenamefont {Devoret},\ and\
  \citenamefont {Schoelkopf}}]{DiCarlo2010N}%
  \BibitemOpen
  \bibfield  {author} {\bibinfo {author} {\bibfnamefont {L.}~\bibnamefont
  {DiCarlo}}, \bibinfo {author} {\bibfnamefont {M.~D.}\ \bibnamefont {Reed}},
  \bibinfo {author} {\bibfnamefont {L.}~\bibnamefont {Sun}}, \bibinfo {author}
  {\bibfnamefont {B.~R.}\ \bibnamefont {Johnson}}, \bibinfo {author}
  {\bibfnamefont {J.~M.}\ \bibnamefont {Chow}}, \bibinfo {author}
  {\bibfnamefont {J.~M.}\ \bibnamefont {Gambetta}}, \bibinfo {author}
  {\bibfnamefont {L.}~\bibnamefont {Frunzio}}, \bibinfo {author} {\bibfnamefont
  {S.~M.}\ \bibnamefont {Girvin}}, \bibinfo {author} {\bibfnamefont {M.~H.}\
  \bibnamefont {Devoret}}, \ and\ \bibinfo {author} {\bibfnamefont {R.~J.}\
  \bibnamefont {Schoelkopf}},\ }\href@noop {} {\bibfield  {journal} {\bibinfo
  {journal} {Nature}\ }\textbf {\bibinfo {volume} {467}},\ \bibinfo {pages}
  {574} (\bibinfo {year} {2010})}\BibitemShut {NoStop}%
\bibitem [{\citenamefont {Vijay}\ \emph {et~al.}(2012)\citenamefont {Vijay},
  \citenamefont {Macklin}, \citenamefont {Slichter}, \citenamefont {Weber},
  \citenamefont {Murch}, \citenamefont {Naik}, \citenamefont {Korotkov},\ and\
  \citenamefont {Siddiqi}}]{KorotkovSiddiqi2012N}%
  \BibitemOpen
  \bibfield  {author} {\bibinfo {author} {\bibfnamefont {R.}~\bibnamefont
  {Vijay}}, \bibinfo {author} {\bibfnamefont {C.}~\bibnamefont {Macklin}},
  \bibinfo {author} {\bibfnamefont {D.~H.}\ \bibnamefont {Slichter}}, \bibinfo
  {author} {\bibfnamefont {S.~J.}\ \bibnamefont {Weber}}, \bibinfo {author}
  {\bibfnamefont {K.~W.}\ \bibnamefont {Murch}}, \bibinfo {author}
  {\bibfnamefont {R.}~\bibnamefont {Naik}}, \bibinfo {author} {\bibfnamefont
  {A.~N.}\ \bibnamefont {Korotkov}}, \ and\ \bibinfo {author} {\bibfnamefont
  {I.}~\bibnamefont {Siddiqi}},\ }\href@noop {} {\bibfield  {journal} {\bibinfo
   {journal} {Nature}\ }\textbf {\bibinfo {volume} {490}},\ \bibinfo {pages}
  {77} (\bibinfo {year} {2012})}\BibitemShut {NoStop}%
\bibitem [{\citenamefont {Rist\`{e}}\ \emph {et~al.}(2013)\citenamefont
  {Rist\`{e}}, \citenamefont {Dukalski}, \citenamefont {Watson}, \citenamefont
  {de~Lange}, \citenamefont {Tiggelman}, \citenamefont {Blanter}, \citenamefont
  {Lehnert}, \citenamefont {Schouten},\ and\ \citenamefont
  {DiCarlo}}]{RisteDiCarlo2013N}%
  \BibitemOpen
  \bibfield  {author} {\bibinfo {author} {\bibfnamefont {D.}~\bibnamefont
  {Rist\`{e}}}, \bibinfo {author} {\bibfnamefont {M.}~\bibnamefont {Dukalski}},
  \bibinfo {author} {\bibfnamefont {C.~A.}\ \bibnamefont {Watson}}, \bibinfo
  {author} {\bibfnamefont {G.}~\bibnamefont {de~Lange}}, \bibinfo {author}
  {\bibfnamefont {M.~J.}\ \bibnamefont {Tiggelman}}, \bibinfo {author}
  {\bibfnamefont {Y.~M.}\ \bibnamefont {Blanter}}, \bibinfo {author}
  {\bibfnamefont {K.~W.}\ \bibnamefont {Lehnert}}, \bibinfo {author}
  {\bibfnamefont {R.~N.}\ \bibnamefont {Schouten}}, \ and\ \bibinfo {author}
  {\bibfnamefont {L.}~\bibnamefont {DiCarlo}},\ }\href@noop {} {\bibfield
  {journal} {\bibinfo  {journal} {Nature}\ }\textbf {\bibinfo {volume} {502}},\
  \bibinfo {pages} {350} (\bibinfo {year} {2013})}\BibitemShut {NoStop}%
\bibitem [{\citenamefont {Roch}\ \emph {et~al.}(2014)\citenamefont {Roch},
  \citenamefont {Schwartz}, \citenamefont {Motzoi}, \citenamefont {Macklin},
  \citenamefont {Vijay}, \citenamefont {Eddins}, \citenamefont {Korotkov},
  \citenamefont {Whaley}, \citenamefont {Sarovar},\ and\ \citenamefont
  {Siddiqi}}]{KorotkovSiddiqi2014PRL}%
  \BibitemOpen
  \bibfield  {author} {\bibinfo {author} {\bibfnamefont {N.}~\bibnamefont
  {Roch}}, \bibinfo {author} {\bibfnamefont {M.~E.}\ \bibnamefont {Schwartz}},
  \bibinfo {author} {\bibfnamefont {F.}~\bibnamefont {Motzoi}}, \bibinfo
  {author} {\bibfnamefont {C.}~\bibnamefont {Macklin}}, \bibinfo {author}
  {\bibfnamefont {R.}~\bibnamefont {Vijay}}, \bibinfo {author} {\bibfnamefont
  {A.~W.}\ \bibnamefont {Eddins}}, \bibinfo {author} {\bibfnamefont {A.~N.}\
  \bibnamefont {Korotkov}}, \bibinfo {author} {\bibfnamefont {K.~B.}\
  \bibnamefont {Whaley}}, \bibinfo {author} {\bibfnamefont {M.}~\bibnamefont
  {Sarovar}}, \ and\ \bibinfo {author} {\bibfnamefont {I.}~\bibnamefont
  {Siddiqi}},\ }\href@noop {} {\bibfield  {journal} {\bibinfo  {journal} {Phys.
  Rev. Lett.}\ }\textbf {\bibinfo {volume} {112}},\ \bibinfo {pages} {170501}
  (\bibinfo {year} {2014})}\BibitemShut {NoStop}%
\bibitem [{\citenamefont {Srinivasa}\ \emph {et~al.}(2016)\citenamefont
  {Srinivasa}, \citenamefont {Taylor},\ and\ \citenamefont
  {Tahan}}]{Srinivasa2016PRB}%
  \BibitemOpen
  \bibfield  {author} {\bibinfo {author} {\bibfnamefont {V.}~\bibnamefont
  {Srinivasa}}, \bibinfo {author} {\bibfnamefont {J.~M.}\ \bibnamefont
  {Taylor}}, \ and\ \bibinfo {author} {\bibfnamefont {C.}~\bibnamefont
  {Tahan}},\ }\href@noop {} {\bibfield  {journal} {\bibinfo  {journal} {Phys.
  Rev. B}\ }\textbf {\bibinfo {volume} {94}},\ \bibinfo {pages} {205421}
  (\bibinfo {year} {2016})}\BibitemShut {NoStop}%
\bibitem [{\citenamefont {Childress}\ \emph {et~al.}(2004)\citenamefont
  {Childress}, \citenamefont {S{\o}rensen},\ and\ \citenamefont
  {Lukin}}]{Childress2004}%
  \BibitemOpen
  \bibfield  {author} {\bibinfo {author} {\bibfnamefont {L.}~\bibnamefont
  {Childress}}, \bibinfo {author} {\bibfnamefont {A.~S.}\ \bibnamefont
  {S{\o}rensen}}, \ and\ \bibinfo {author} {\bibfnamefont {M.~D.}\ \bibnamefont
  {Lukin}},\ }\href@noop {} {\bibfield  {journal} {\bibinfo  {journal} {Phys.
  Rev. A}\ }\textbf {\bibinfo {volume} {69}},\ \bibinfo {pages} {042302}
  (\bibinfo {year} {2004})}\BibitemShut {NoStop}%
\bibitem [{\citenamefont {Jin}\ \emph {et~al.}(2012)\citenamefont {Jin},
  \citenamefont {Marthaler}, \citenamefont {Shnirman},\ and\ \citenamefont
  {Sch\"{o}n}}]{ShnirmanSchon2012PRL}%
  \BibitemOpen
  \bibfield  {author} {\bibinfo {author} {\bibfnamefont {P.-Q.}\ \bibnamefont
  {Jin}}, \bibinfo {author} {\bibfnamefont {M.}~\bibnamefont {Marthaler}},
  \bibinfo {author} {\bibfnamefont {A.}~\bibnamefont {Shnirman}}, \ and\
  \bibinfo {author} {\bibfnamefont {G.}~\bibnamefont {Sch\"{o}n}},\ }\href@noop
  {} {\bibfield  {journal} {\bibinfo  {journal} {Phys. Rev. Lett.}\ }\textbf
  {\bibinfo {volume} {108}},\ \bibinfo {pages} {190506} (\bibinfo {year}
  {2012})}\BibitemShut {NoStop}%
\bibitem [{\citenamefont {Hu}\ \emph {et~al.}(2012)\citenamefont {Hu},
  \citenamefont {Liu},\ and\ \citenamefont {Nori}}]{HuLiuNori2012PRB}%
  \BibitemOpen
  \bibfield  {author} {\bibinfo {author} {\bibfnamefont {X.}~\bibnamefont
  {Hu}}, \bibinfo {author} {\bibfnamefont {Y.-x.}\ \bibnamefont {Liu}}, \ and\
  \bibinfo {author} {\bibfnamefont {F.}~\bibnamefont {Nori}},\ }\href@noop {}
  {\bibfield  {journal} {\bibinfo  {journal} {Phys. Rev. B}\ }\textbf {\bibinfo
  {volume} {86}},\ \bibinfo {pages} {035314(R)} (\bibinfo {year}
  {2012})}\BibitemShut {NoStop}%
\bibitem [{\citenamefont {Samkharadze}\ \emph {et~al.}(2017)\citenamefont
  {Samkharadze}, \citenamefont {Zheng}, \citenamefont {Kalhor}, \citenamefont
  {Brousse}, \citenamefont {Sammak}, \citenamefont {Mendes}, \citenamefont
  {Blais}, \citenamefont {Scappucci},\ and\ \citenamefont
  {Vandersypen}}]{Samkharadze-Vandersypen2017preprint}%
  \BibitemOpen
  \bibfield  {author} {\bibinfo {author} {\bibfnamefont {N.}~\bibnamefont
  {Samkharadze}}, \bibinfo {author} {\bibfnamefont {G.}~\bibnamefont {Zheng}},
  \bibinfo {author} {\bibfnamefont {N.}~\bibnamefont {Kalhor}}, \bibinfo
  {author} {\bibfnamefont {D.}~\bibnamefont {Brousse}}, \bibinfo {author}
  {\bibfnamefont {A.}~\bibnamefont {Sammak}}, \bibinfo {author} {\bibfnamefont
  {U.~C.}\ \bibnamefont {Mendes}}, \bibinfo {author} {\bibfnamefont
  {A.}~\bibnamefont {Blais}}, \bibinfo {author} {\bibfnamefont
  {G.}~\bibnamefont {Scappucci}}, \ and\ \bibinfo {author} {\bibfnamefont
  {L.~M.~K.}\ \bibnamefont {Vandersypen}},\ }\href@noop {} {\enquote {\bibinfo
  {title} {{S}trong spin-photon coupling in silicon},}\ } (\bibinfo {year}
  {2017}),\ \Eprint {http://arxiv.org/abs/1711.02040} {arXiv:1711.02040
  [cond-mat]} \BibitemShut {NoStop}%
\bibitem [{\citenamefont {Cross}\ and\ \citenamefont
  {Gambetta}(2015)}]{CrossGambetta2015PRA}%
  \BibitemOpen
  \bibfield  {author} {\bibinfo {author} {\bibfnamefont {A.~W.}\ \bibnamefont
  {Cross}}\ and\ \bibinfo {author} {\bibfnamefont {J.~M.}\ \bibnamefont
  {Gambetta}},\ }\href@noop {} {\bibfield  {journal} {\bibinfo  {journal}
  {Phys. Rev. A}\ }\textbf {\bibinfo {volume} {91}},\ \bibinfo {pages} {032325}
  (\bibinfo {year} {2015})}\BibitemShut {NoStop}%
\bibitem [{\citenamefont {Paik}\ \emph {et~al.}(2016)\citenamefont {Paik},
  \citenamefont {Mezzacapo}, \citenamefont {Sandberg}, \citenamefont {McClure},
  \citenamefont {Abdo}, \citenamefont {Corcoles}, \citenamefont {Dial},
  \citenamefont {Bogorin}, \citenamefont {Plourde}, \citenamefont {Steffen},
  \citenamefont {Cross}, \citenamefont {Gambetta},\ and\ \citenamefont
  {Chow}}]{PaikChow2016PRL}%
  \BibitemOpen
  \bibfield  {author} {\bibinfo {author} {\bibfnamefont {H.}~\bibnamefont
  {Paik}}, \bibinfo {author} {\bibfnamefont {A.}~\bibnamefont {Mezzacapo}},
  \bibinfo {author} {\bibfnamefont {M.}~\bibnamefont {Sandberg}}, \bibinfo
  {author} {\bibfnamefont {D.~T.}\ \bibnamefont {McClure}}, \bibinfo {author}
  {\bibfnamefont {B.}~\bibnamefont {Abdo}}, \bibinfo {author} {\bibfnamefont
  {A.~D.}\ \bibnamefont {Corcoles}}, \bibinfo {author} {\bibfnamefont
  {O.}~\bibnamefont {Dial}}, \bibinfo {author} {\bibfnamefont {D.~F.}\
  \bibnamefont {Bogorin}}, \bibinfo {author} {\bibfnamefont {B.~L.~T.}\
  \bibnamefont {Plourde}}, \bibinfo {author} {\bibfnamefont {M.}~\bibnamefont
  {Steffen}}, \bibinfo {author} {\bibfnamefont {A.~W.}\ \bibnamefont {Cross}},
  \bibinfo {author} {\bibfnamefont {J.~M.}\ \bibnamefont {Gambetta}}, \ and\
  \bibinfo {author} {\bibfnamefont {J.~M.}\ \bibnamefont {Chow}},\ }\href@noop
  {} {\bibfield  {journal} {\bibinfo  {journal} {Phys. Rev. Lett.}\ }\textbf
  {\bibinfo {volume} {117}},\ \bibinfo {pages} {250502} (\bibinfo {year}
  {2016})}\BibitemShut {NoStop}%
\bibitem [{\citenamefont {Blais}\ \emph {et~al.}(2007)\citenamefont {Blais},
  \citenamefont {Gambetta}, \citenamefont {Wallraff}, \citenamefont {Schuster},
  \citenamefont {Girvin}, \citenamefont {Devoret},\ and\ \citenamefont
  {Schoelkopf}}]{Blais2007PRA}%
  \BibitemOpen
  \bibfield  {author} {\bibinfo {author} {\bibfnamefont {A.}~\bibnamefont
  {Blais}}, \bibinfo {author} {\bibfnamefont {J.~M.}\ \bibnamefont {Gambetta}},
  \bibinfo {author} {\bibfnamefont {A.}~\bibnamefont {Wallraff}}, \bibinfo
  {author} {\bibfnamefont {D.~I.}\ \bibnamefont {Schuster}}, \bibinfo {author}
  {\bibfnamefont {S.~M.}\ \bibnamefont {Girvin}}, \bibinfo {author}
  {\bibfnamefont {M.~H.}\ \bibnamefont {Devoret}}, \ and\ \bibinfo {author}
  {\bibfnamefont {R.~J.}\ \bibnamefont {Schoelkopf}},\ }\href@noop {}
  {\bibfield  {journal} {\bibinfo  {journal} {Phys. Rev. A}\ }\textbf {\bibinfo
  {volume} {75}},\ \bibinfo {pages} {032329} (\bibinfo {year}
  {2007})}\BibitemShut {NoStop}%
\bibitem [{\citenamefont {Wallraff}\ \emph {et~al.}(2007)\citenamefont
  {Wallraff}, \citenamefont {Schuster}, \citenamefont {Blais}, \citenamefont
  {Gambetta}, \citenamefont {Schreier}, \citenamefont {Frunzio}, \citenamefont
  {Devoret}, \citenamefont {Girvin},\ and\ \citenamefont
  {Schoelkopf}}]{Wallraff2007PRL}%
  \BibitemOpen
  \bibfield  {author} {\bibinfo {author} {\bibfnamefont {A.}~\bibnamefont
  {Wallraff}}, \bibinfo {author} {\bibfnamefont {D.~I.}\ \bibnamefont
  {Schuster}}, \bibinfo {author} {\bibfnamefont {A.}~\bibnamefont {Blais}},
  \bibinfo {author} {\bibfnamefont {J.~M.}\ \bibnamefont {Gambetta}}, \bibinfo
  {author} {\bibfnamefont {J.}~\bibnamefont {Schreier}}, \bibinfo {author}
  {\bibfnamefont {L.}~\bibnamefont {Frunzio}}, \bibinfo {author} {\bibfnamefont
  {M.~H.}\ \bibnamefont {Devoret}}, \bibinfo {author} {\bibfnamefont {S.~M.}\
  \bibnamefont {Girvin}}, \ and\ \bibinfo {author} {\bibfnamefont {R.~J.}\
  \bibnamefont {Schoelkopf}},\ }\href@noop {} {\bibfield  {journal} {\bibinfo
  {journal} {Phys. Rev. Lett.}\ }\textbf {\bibinfo {volume} {99}},\ \bibinfo
  {pages} {050501} (\bibinfo {year} {2007})}\BibitemShut {NoStop}%
\bibitem [{\citenamefont {Leek}\ \emph {et~al.}(2009)\citenamefont {Leek},
  \citenamefont {Filipp}, \citenamefont {Maurer}, \citenamefont {Baur},
  \citenamefont {Bianchetti}, \citenamefont {Fink}, \citenamefont {G\"{o}ppl},
  \citenamefont {Steffen},\ and\ \citenamefont {Wallraff}}]{Leek2009PRB}%
  \BibitemOpen
  \bibfield  {author} {\bibinfo {author} {\bibfnamefont {P.~J.}\ \bibnamefont
  {Leek}}, \bibinfo {author} {\bibfnamefont {S.}~\bibnamefont {Filipp}},
  \bibinfo {author} {\bibfnamefont {P.}~\bibnamefont {Maurer}}, \bibinfo
  {author} {\bibfnamefont {M.}~\bibnamefont {Baur}}, \bibinfo {author}
  {\bibfnamefont {R.}~\bibnamefont {Bianchetti}}, \bibinfo {author}
  {\bibfnamefont {J.~M.}\ \bibnamefont {Fink}}, \bibinfo {author}
  {\bibfnamefont {M.}~\bibnamefont {G\"{o}ppl}}, \bibinfo {author}
  {\bibfnamefont {L.}~\bibnamefont {Steffen}}, \ and\ \bibinfo {author}
  {\bibfnamefont {A.}~\bibnamefont {Wallraff}},\ }\href@noop {} {\bibfield
  {journal} {\bibinfo  {journal} {Phys. Rev. B}\ }\textbf {\bibinfo {volume}
  {79}},\ \bibinfo {pages} {180511} (\bibinfo {year} {2009})}\BibitemShut
  {NoStop}%
\bibitem [{\citenamefont {You}\ \emph {et~al.}(2007)\citenamefont {You},
  \citenamefont {Hu}, \citenamefont {Ashhab},\ and\ \citenamefont
  {Nori}}]{YouHuAshhabNori2007PRB}%
  \BibitemOpen
  \bibfield  {author} {\bibinfo {author} {\bibfnamefont {J.~Q.}\ \bibnamefont
  {You}}, \bibinfo {author} {\bibfnamefont {X.}~\bibnamefont {Hu}}, \bibinfo
  {author} {\bibfnamefont {S.}~\bibnamefont {Ashhab}}, \ and\ \bibinfo {author}
  {\bibfnamefont {F.}~\bibnamefont {Nori}},\ }\href@noop {} {\bibfield
  {journal} {\bibinfo  {journal} {Phys. Rev. B}\ }\textbf {\bibinfo {volume}
  {75}},\ \bibinfo {pages} {140515(R)} (\bibinfo {year} {2007})}\BibitemShut
  {NoStop}%
\bibitem [{\citenamefont {Kerman}(2010)}]{Kerman2010PRL}%
  \BibitemOpen
  \bibfield  {author} {\bibinfo {author} {\bibfnamefont {A.~J.}\ \bibnamefont
  {Kerman}},\ }\href@noop {} {\bibfield  {journal} {\bibinfo  {journal} {Phys.
  Rev. Lett.}\ }\textbf {\bibinfo {volume} {104}},\ \bibinfo {pages} {027002}
  (\bibinfo {year} {2010})}\BibitemShut {NoStop}%
\bibitem [{\citenamefont {Khezri}\ \emph {et~al.}(2015)\citenamefont {Khezri},
  \citenamefont {Dressel},\ and\ \citenamefont
  {Korotkov}}]{KhezriDresselKorotkov2015PRA}%
  \BibitemOpen
  \bibfield  {author} {\bibinfo {author} {\bibfnamefont {M.}~\bibnamefont
  {Khezri}}, \bibinfo {author} {\bibfnamefont {J.}~\bibnamefont {Dressel}}, \
  and\ \bibinfo {author} {\bibfnamefont {A.~N.}\ \bibnamefont {Korotkov}},\
  }\href@noop {} {\bibfield  {journal} {\bibinfo  {journal} {Phys. Rev. A}\
  }\textbf {\bibinfo {volume} {92}},\ \bibinfo {pages} {052306} (\bibinfo
  {year} {2015})}\BibitemShut {NoStop}%
\bibitem [{\citenamefont {Sank}\ \emph {et~al.}(2016)\citenamefont {Sank},
  \citenamefont {Chen}, \citenamefont {Khezri}, \citenamefont {Kelly},
  \citenamefont {Barends}, \citenamefont {Campbell}, \citenamefont {Chen},
  \citenamefont {Chiaro}, \citenamefont {Dunsworth}, \citenamefont {Fowler},
  \citenamefont {Jeffrey}, \citenamefont {Lucero}, \citenamefont {Megrant},
  \citenamefont {Mutus}, \citenamefont {Neeley}, \citenamefont {Neill},
  \citenamefont {O'Malley}, \citenamefont {Quintana}, \citenamefont {Roushan},
  \citenamefont {Vainsencher}, \citenamefont {White}, \citenamefont {Wenner},
  \citenamefont {Korotkov},\ and\ \citenamefont
  {Martinis}}]{SankKorotkov2016PRL}%
  \BibitemOpen
  \bibfield  {author} {\bibinfo {author} {\bibfnamefont {D.}~\bibnamefont
  {Sank}}, \bibinfo {author} {\bibfnamefont {Z.}~\bibnamefont {Chen}}, \bibinfo
  {author} {\bibfnamefont {M.}~\bibnamefont {Khezri}}, \bibinfo {author}
  {\bibfnamefont {J.}~\bibnamefont {Kelly}}, \bibinfo {author} {\bibfnamefont
  {R.}~\bibnamefont {Barends}}, \bibinfo {author} {\bibfnamefont
  {B.}~\bibnamefont {Campbell}}, \bibinfo {author} {\bibfnamefont
  {Y.}~\bibnamefont {Chen}}, \bibinfo {author} {\bibfnamefont {B.}~\bibnamefont
  {Chiaro}}, \bibinfo {author} {\bibfnamefont {A.}~\bibnamefont {Dunsworth}},
  \bibinfo {author} {\bibfnamefont {A.}~\bibnamefont {Fowler}}, \bibinfo
  {author} {\bibfnamefont {E.}~\bibnamefont {Jeffrey}}, \bibinfo {author}
  {\bibfnamefont {E.}~\bibnamefont {Lucero}}, \bibinfo {author} {\bibfnamefont
  {A.}~\bibnamefont {Megrant}}, \bibinfo {author} {\bibfnamefont
  {J.}~\bibnamefont {Mutus}}, \bibinfo {author} {\bibfnamefont
  {M.}~\bibnamefont {Neeley}}, \bibinfo {author} {\bibfnamefont
  {C.}~\bibnamefont {Neill}}, \bibinfo {author} {\bibfnamefont {P.~J.~J.}\
  \bibnamefont {O'Malley}}, \bibinfo {author} {\bibfnamefont {C.}~\bibnamefont
  {Quintana}}, \bibinfo {author} {\bibfnamefont {P.}~\bibnamefont {Roushan}},
  \bibinfo {author} {\bibfnamefont {A.}~\bibnamefont {Vainsencher}}, \bibinfo
  {author} {\bibfnamefont {T.}~\bibnamefont {White}}, \bibinfo {author}
  {\bibfnamefont {J.}~\bibnamefont {Wenner}}, \bibinfo {author} {\bibfnamefont
  {A.~N.}\ \bibnamefont {Korotkov}}, \ and\ \bibinfo {author} {\bibfnamefont
  {J.~M.}\ \bibnamefont {Martinis}},\ }\href@noop {} {\bibfield  {journal}
  {\bibinfo  {journal} {Phys. Rev. Lett.}\ }\textbf {\bibinfo {volume} {117}},\
  \bibinfo {pages} {190503} (\bibinfo {year} {2016})}\BibitemShut {NoStop}%
\bibitem [{\citenamefont {Kerman}(2013)}]{Kerman2013}%
  \BibitemOpen
  \bibfield  {author} {\bibinfo {author} {\bibfnamefont {A.~J.}\ \bibnamefont
  {Kerman}},\ }\href@noop {} {\bibfield  {journal} {\bibinfo  {journal} {New
  Journal of Physics}\ }\textbf {\bibinfo {volume} {15}},\ \bibinfo {pages}
  {123011} (\bibinfo {year} {2013})}\BibitemShut {NoStop}%
\bibitem [{\citenamefont {Ruskov}\ and\ \citenamefont
  {Tahan}(2019)}]{RuskovTahan2019PRB99}%
  \BibitemOpen
  \bibfield  {author} {\bibinfo {author} {\bibfnamefont {R.}~\bibnamefont
  {Ruskov}}\ and\ \bibinfo {author} {\bibfnamefont {C.}~\bibnamefont {Tahan}},\
  }\href@noop {} {\bibfield  {journal} {\bibinfo  {journal} {Phys. Rev. B}\
  }\textbf {\bibinfo {volume} {99}},\ \bibinfo {pages} {245306} (\bibinfo
  {year} {2019})}\BibitemShut {NoStop}%
\bibitem [{\citenamefont {Ruskov}\ and\ \citenamefont
  {Tahan}(2017)}]{RuskovTahan2017preprint1}%
  \BibitemOpen
  \bibfield  {author} {\bibinfo {author} {\bibfnamefont {R.}~\bibnamefont
  {Ruskov}}\ and\ \bibinfo {author} {\bibfnamefont {C.}~\bibnamefont {Tahan}},\
  }\href@noop {} {\bibfield  {journal} {\bibinfo  {journal}
  {arXiv:cond-mat.mes-hall/1704.05876}\ } (\bibinfo {year} {2017})}\BibitemShut
  {NoStop}%
\bibitem [{Sta()}]{StaticLongitudinal}%
  \BibitemOpen
  \href@noop {} {}\Eprint {http://arxiv.org/abs/A static longitudinal coupling,
  $\propto \left[ g_{\parallel}^{\rm st}\, \sigma_z + g_0^{\rm st} \right]
  (\hat{a} + \hat{a}^{\dagger} )$, would lead to an always on entangling
  interaction: ${\cal H}_{\rm 2Qbs} = \frac{2 g^{\rm st}_{1,\|} g^{\rm
  st}_{2,\|}}{\omega_r} \,\, \sigma_z^{(1)}\!\!\!\otimes\sigma_z^{(2)}$,
  derived via a Lang-Firsov transformation (see, e.g.,
  Ref.~\onlinecite{Billangeon2015PRB}). This coupling is, however,
  parametrically small ($g^{\rm st}_{i,\|} \ll \omega_r$)} {A static
  longitudinal coupling, $\propto \left[ g_{\parallel}^{\rm st}\, \sigma_z +
  g_0^{\rm st} \right] (\hat{a} + \hat{a}^{\dagger} )$, would lead to an always
  on entangling interaction: ${\cal H}_{\rm 2Qbs} = \frac{2 g^{\rm st}_{1,\|}
  g^{\rm st}_{2,\|}}{\omega_r} \,\, \sigma_z^{(1)}\!\!\!\otimes\sigma_z^{(2)}$,
  derived via a Lang-Firsov transformation (see, e.g.,
  Ref.~\onlinecite{Billangeon2015PRB}). This coupling is, however,
  parametrically small ($g^{\rm st}_{i,\|} \ll \omega_r$)} \BibitemShut
  {NoStop}%
\bibitem [{\citenamefont {Averin}\ \emph {et~al.}(1985)\citenamefont {Averin},
  \citenamefont {Zorin},\ and\ \citenamefont
  {Likharev}}]{AverinZorinLikharevZhETF1985}%
  \BibitemOpen
  \bibfield  {author} {\bibinfo {author} {\bibfnamefont {D.~V.}\ \bibnamefont
  {Averin}}, \bibinfo {author} {\bibfnamefont {A.~B.}\ \bibnamefont {Zorin}}, \
  and\ \bibinfo {author} {\bibfnamefont {K.~K.}\ \bibnamefont {Likharev}},\
  }\href@noop {} {\bibfield  {journal} {\bibinfo  {journal} {Sov. Phys. JETP}\
  }\textbf {\bibinfo {volume} {61}},\ \bibinfo {pages} {407} (\bibinfo {year}
  {1985})}\BibitemShut {NoStop}%
\bibitem [{\citenamefont {Averin}\ and\ \citenamefont
  {Bruder}(2003)}]{AverinBruderPRL2003}%
  \BibitemOpen
  \bibfield  {author} {\bibinfo {author} {\bibfnamefont {D.~V.}\ \bibnamefont
  {Averin}}\ and\ \bibinfo {author} {\bibfnamefont {C.}~\bibnamefont
  {Bruder}},\ }\href@noop {} {\bibfield  {journal} {\bibinfo  {journal} {Phys.
  Rev. Lett.}\ }\textbf {\bibinfo {volume} {91}},\ \bibinfo {pages} {057003}
  (\bibinfo {year} {2003})}\BibitemShut {NoStop}%
\bibitem [{\citenamefont {Sillanp\"{a}\"{a}}\ \emph {et~al.}(2005)\citenamefont
  {Sillanp\"{a}\"{a}}, \citenamefont {Lehtinen}, \citenamefont {Paila},
  \citenamefont {Makhlin}, \citenamefont {Roschier},\ and\ \citenamefont
  {Hakonen}}]{SillanpaaHakonenPRL2005}%
  \BibitemOpen
  \bibfield  {author} {\bibinfo {author} {\bibfnamefont {M.~A.}\ \bibnamefont
  {Sillanp\"{a}\"{a}}}, \bibinfo {author} {\bibfnamefont {T.}~\bibnamefont
  {Lehtinen}}, \bibinfo {author} {\bibfnamefont {A.}~\bibnamefont {Paila}},
  \bibinfo {author} {\bibfnamefont {Y.}~\bibnamefont {Makhlin}}, \bibinfo
  {author} {\bibfnamefont {L.}~\bibnamefont {Roschier}}, \ and\ \bibinfo
  {author} {\bibfnamefont {P.~J.}\ \bibnamefont {Hakonen}},\ }\href@noop {}
  {\bibfield  {journal} {\bibinfo  {journal} {Phys. Rev. Lett.}\ }\textbf
  {\bibinfo {volume} {95}},\ \bibinfo {pages} {206806} (\bibinfo {year}
  {2005})}\BibitemShut {NoStop}%
\bibitem [{\citenamefont {Duty}\ \emph {et~al.}(2005)\citenamefont {Duty},
  \citenamefont {Johansson}, \citenamefont {Bladh}, \citenamefont {Gunnarsson},
  \citenamefont {Wilson},\ and\ \citenamefont {Delsing}}]{DutyDelsingPRL2005}%
  \BibitemOpen
  \bibfield  {author} {\bibinfo {author} {\bibfnamefont {T.}~\bibnamefont
  {Duty}}, \bibinfo {author} {\bibfnamefont {G.}~\bibnamefont {Johansson}},
  \bibinfo {author} {\bibfnamefont {K.}~\bibnamefont {Bladh}}, \bibinfo
  {author} {\bibfnamefont {D.}~\bibnamefont {Gunnarsson}}, \bibinfo {author}
  {\bibfnamefont {C.}~\bibnamefont {Wilson}}, \ and\ \bibinfo {author}
  {\bibfnamefont {P.}~\bibnamefont {Delsing}},\ }\href@noop {} {\bibfield
  {journal} {\bibinfo  {journal} {Phys. Rev. Lett.}\ }\textbf {\bibinfo
  {volume} {95}},\ \bibinfo {pages} {206807} (\bibinfo {year}
  {2005})}\BibitemShut {NoStop}%
\bibitem [{\citenamefont {Samkharadze}\ \emph {et~al.}(2016)\citenamefont
  {Samkharadze}, \citenamefont {Bruno}, \citenamefont {Scarlino}, \citenamefont
  {Zheng}, \citenamefont {DiVincenzo}, \citenamefont {DiCarlo},\ and\
  \citenamefont {Vandersypen}}]{Samkharadze2016}%
  \BibitemOpen
  \bibfield  {author} {\bibinfo {author} {\bibfnamefont {N.}~\bibnamefont
  {Samkharadze}}, \bibinfo {author} {\bibfnamefont {A.}~\bibnamefont {Bruno}},
  \bibinfo {author} {\bibfnamefont {P.}~\bibnamefont {Scarlino}}, \bibinfo
  {author} {\bibfnamefont {G.}~\bibnamefont {Zheng}}, \bibinfo {author}
  {\bibfnamefont {D.~P.}\ \bibnamefont {DiVincenzo}}, \bibinfo {author}
  {\bibfnamefont {L.}~\bibnamefont {DiCarlo}}, \ and\ \bibinfo {author}
  {\bibfnamefont {L.~M.~K.}\ \bibnamefont {Vandersypen}},\ }\href@noop {}
  {\bibfield  {journal} {\bibinfo  {journal} {Phys. Rev. Applied}\ }\textbf
  {\bibinfo {volume} {5}},\ \bibinfo {pages} {044004} (\bibinfo {year}
  {2016})}\BibitemShut {NoStop}%
\bibitem [{\citenamefont {Niepce}\ \emph {et~al.}(2019)\citenamefont {Niepce},
  \citenamefont {Burnett},\ and\ \citenamefont {Bylander}}]{Bylander2018}%
  \BibitemOpen
  \bibfield  {author} {\bibinfo {author} {\bibfnamefont {D.}~\bibnamefont
  {Niepce}}, \bibinfo {author} {\bibfnamefont {J.}~\bibnamefont {Burnett}}, \
  and\ \bibinfo {author} {\bibfnamefont {J.}~\bibnamefont {Bylander}},\
  }\href@noop {} {\bibfield  {journal} {\bibinfo  {journal} {Phys. Rev.
  Applied}\ }\textbf {\bibinfo {volume} {11}},\ \bibinfo {pages} {044014}
  (\bibinfo {year} {2019})}\BibitemShut {NoStop}%
\bibitem [{\citenamefont {M{\o}lmer}\ and\ \citenamefont
  {S{\o}rensen}(1999)}]{Molmer1999PRL}%
  \BibitemOpen
  \bibfield  {author} {\bibinfo {author} {\bibfnamefont {K.}~\bibnamefont
  {M{\o}lmer}}\ and\ \bibinfo {author} {\bibfnamefont {A.}~\bibnamefont
  {S{\o}rensen}},\ }\href@noop {} {\bibfield  {journal} {\bibinfo  {journal}
  {Phys. Rev. Lett.}\ }\textbf {\bibinfo {volume} {82}},\ \bibinfo {pages}
  {1835} (\bibinfo {year} {1999})}\BibitemShut {NoStop}%
\bibitem [{\citenamefont {Milburn}\ \emph {et~al.}(2000)\citenamefont
  {Milburn}, \citenamefont {Schneider},\ and\ \citenamefont
  {James}}]{Milburn2000F}%
  \BibitemOpen
  \bibfield  {author} {\bibinfo {author} {\bibfnamefont {G.~J.}\ \bibnamefont
  {Milburn}}, \bibinfo {author} {\bibfnamefont {S.}~\bibnamefont {Schneider}},
  \ and\ \bibinfo {author} {\bibfnamefont {D.~F.~V.}\ \bibnamefont {James}},\
  }\href@noop {} {\bibfield  {journal} {\bibinfo  {journal} {Fortschr. Phys.}\
  }\textbf {\bibinfo {volume} {48}},\ \bibinfo {pages} {801} (\bibinfo {year}
  {2000})}\BibitemShut {NoStop}%
\bibitem [{\citenamefont {Leibfried}\ \emph {et~al.}(2003)\citenamefont
  {Leibfried}, \citenamefont {DeMarco}, \citenamefont {Meyer}, \citenamefont
  {Lucas}, \citenamefont {Barrett}, \citenamefont {Britton}, \citenamefont
  {Itano}, \citenamefont {Jelenkovi\'{c}}, \citenamefont {Langer},
  \citenamefont {Rosenband},\ and\ \citenamefont {Wineland}}]{Leibried2003N}%
  \BibitemOpen
  \bibfield  {author} {\bibinfo {author} {\bibfnamefont {D.}~\bibnamefont
  {Leibfried}}, \bibinfo {author} {\bibfnamefont {B.}~\bibnamefont {DeMarco}},
  \bibinfo {author} {\bibfnamefont {V.}~\bibnamefont {Meyer}}, \bibinfo
  {author} {\bibfnamefont {D.}~\bibnamefont {Lucas}}, \bibinfo {author}
  {\bibfnamefont {M.}~\bibnamefont {Barrett}}, \bibinfo {author} {\bibfnamefont
  {J.}~\bibnamefont {Britton}}, \bibinfo {author} {\bibfnamefont {W.~M.}\
  \bibnamefont {Itano}}, \bibinfo {author} {\bibfnamefont {B.}~\bibnamefont
  {Jelenkovi\'{c}}}, \bibinfo {author} {\bibfnamefont {C.}~\bibnamefont
  {Langer}}, \bibinfo {author} {\bibfnamefont {T.}~\bibnamefont {Rosenband}}, \
  and\ \bibinfo {author} {\bibfnamefont {D.~J.}\ \bibnamefont {Wineland}},\
  }\href@noop {} {\bibfield  {journal} {\bibinfo  {journal} {Nature}\ }\textbf
  {\bibinfo {volume} {422}},\ \bibinfo {pages} {412} (\bibinfo {year}
  {2003})}\BibitemShut {NoStop}%
\bibitem [{\citenamefont {Haljan}\ \emph {et~al.}(2005)\citenamefont {Haljan},
  \citenamefont {Brickman}, \citenamefont {Deslauriers}, \citenamefont {Lee},\
  and\ \citenamefont {Monroe}}]{HaljanMonroe2005PRL}%
  \BibitemOpen
  \bibfield  {author} {\bibinfo {author} {\bibfnamefont {P.~C.}\ \bibnamefont
  {Haljan}}, \bibinfo {author} {\bibfnamefont {K.-A.}\ \bibnamefont
  {Brickman}}, \bibinfo {author} {\bibfnamefont {L.}~\bibnamefont
  {Deslauriers}}, \bibinfo {author} {\bibfnamefont {P.~J.}\ \bibnamefont
  {Lee}}, \ and\ \bibinfo {author} {\bibfnamefont {C.}~\bibnamefont {Monroe}},\
  }\href@noop {} {\bibfield  {journal} {\bibinfo  {journal} {Phys. Rev. Lett.}\
  }\textbf {\bibinfo {volume} {94}},\ \bibinfo {pages} {153602} (\bibinfo
  {year} {2005})}\BibitemShut {NoStop}%
\bibitem [{\citenamefont {Billangeon}\ \emph {et~al.}(2015)\citenamefont
  {Billangeon}, \citenamefont {Tsai},\ and\ \citenamefont
  {Nakamura}}]{Billangeon2015PRB}%
  \BibitemOpen
  \bibfield  {author} {\bibinfo {author} {\bibfnamefont {P.~M.}\ \bibnamefont
  {Billangeon}}, \bibinfo {author} {\bibfnamefont {J.~S.}\ \bibnamefont
  {Tsai}}, \ and\ \bibinfo {author} {\bibfnamefont {Y.}~\bibnamefont
  {Nakamura}},\ }\href@noop {} {\bibfield  {journal} {\bibinfo  {journal}
  {Phys. Rev. B}\ }\textbf {\bibinfo {volume} {91}},\ \bibinfo {pages} {094517}
  (\bibinfo {year} {2015})}\BibitemShut {NoStop}%
\bibitem [{\citenamefont {Didier}\ \emph {et~al.}(2015)\citenamefont {Didier},
  \citenamefont {Bourassa},\ and\ \citenamefont {Blais}}]{Didier2015PRL}%
  \BibitemOpen
  \bibfield  {author} {\bibinfo {author} {\bibfnamefont {N.}~\bibnamefont
  {Didier}}, \bibinfo {author} {\bibfnamefont {J.}~\bibnamefont {Bourassa}}, \
  and\ \bibinfo {author} {\bibfnamefont {A.}~\bibnamefont {Blais}},\
  }\href@noop {} {\bibfield  {journal} {\bibinfo  {journal} {Phys. Rev. Lett.}\
  }\textbf {\bibinfo {volume} {115}},\ \bibinfo {pages} {203601} (\bibinfo
  {year} {2015})}\BibitemShut {NoStop}%
\bibitem [{\citenamefont {Richer}\ and\ \citenamefont
  {DiVincenzo}(2016)}]{RicherDiVincenzo2016PRB}%
  \BibitemOpen
  \bibfield  {author} {\bibinfo {author} {\bibfnamefont {S.}~\bibnamefont
  {Richer}}\ and\ \bibinfo {author} {\bibfnamefont {D.~P.}\ \bibnamefont
  {DiVincenzo}},\ }\href@noop {} {\bibfield  {journal} {\bibinfo  {journal}
  {Phys. Rev. B}\ }\textbf {\bibinfo {volume} {93}},\ \bibinfo {pages} {134501}
  (\bibinfo {year} {2016})}\BibitemShut {NoStop}%
\bibitem [{\citenamefont {Stockklauser}\ \emph {et~al.}(2017)\citenamefont
  {Stockklauser}, \citenamefont {Scarlino}, \citenamefont {Koski},
  \citenamefont {Gasparinetti}, \citenamefont {Andersen}, \citenamefont
  {Reichl}, \citenamefont {Wegscheider}, \citenamefont {Ihn}, \citenamefont
  {Ensslin},\ and\ \citenamefont
  {Wallraff}}]{Stockklauser-Ihn-Ensslin-Wallraff2017PRX}%
  \BibitemOpen
  \bibfield  {author} {\bibinfo {author} {\bibfnamefont {A.}~\bibnamefont
  {Stockklauser}}, \bibinfo {author} {\bibfnamefont {P.}~\bibnamefont
  {Scarlino}}, \bibinfo {author} {\bibfnamefont {J.~V.}\ \bibnamefont {Koski}},
  \bibinfo {author} {\bibfnamefont {S.}~\bibnamefont {Gasparinetti}}, \bibinfo
  {author} {\bibfnamefont {C.~K.}\ \bibnamefont {Andersen}}, \bibinfo {author}
  {\bibfnamefont {C.}~\bibnamefont {Reichl}}, \bibinfo {author} {\bibfnamefont
  {W.}~\bibnamefont {Wegscheider}}, \bibinfo {author} {\bibfnamefont
  {T.}~\bibnamefont {Ihn}}, \bibinfo {author} {\bibfnamefont {K.}~\bibnamefont
  {Ensslin}}, \ and\ \bibinfo {author} {\bibfnamefont {A.}~\bibnamefont
  {Wallraff}},\ }\href@noop {} {\bibfield  {journal} {\bibinfo  {journal}
  {Phys. Rev. X}\ }\textbf {\bibinfo {volume} {7}},\ \bibinfo {pages} {011030}
  (\bibinfo {year} {2017})}\BibitemShut {NoStop}%
\bibitem [{\citenamefont {Caldeira}\ and\ \citenamefont
  {Leggett}(1983{\natexlab{a}})}]{CaldeiraLeggett1983a}%
  \BibitemOpen
  \bibfield  {author} {\bibinfo {author} {\bibfnamefont {A.~O.}\ \bibnamefont
  {Caldeira}}\ and\ \bibinfo {author} {\bibfnamefont {A.~J.}\ \bibnamefont
  {Leggett}},\ }\href@noop {} {\bibfield  {journal} {\bibinfo  {journal} {Ann.
  Phys. N.Y.}\ }\textbf {\bibinfo {volume} {149}},\ \bibinfo {pages} {374}
  (\bibinfo {year} {1983}{\natexlab{a}})}\BibitemShut {NoStop}%
\bibitem [{\citenamefont {Caldeira}\ \emph {et~al.}(1989)\citenamefont
  {Caldeira}, \citenamefont {Cerdeira},\ and\ \citenamefont
  {Ramaswamy}}]{CaldeiraCerdeira1989}%
  \BibitemOpen
  \bibfield  {author} {\bibinfo {author} {\bibfnamefont {A.~O.}\ \bibnamefont
  {Caldeira}}, \bibinfo {author} {\bibfnamefont {H.~A.}\ \bibnamefont
  {Cerdeira}}, \ and\ \bibinfo {author} {\bibfnamefont {R.}~\bibnamefont
  {Ramaswamy}},\ }\href@noop {} {\bibfield  {journal} {\bibinfo  {journal}
  {Phys. Rev. A}\ }\textbf {\bibinfo {volume} {40}},\ \bibinfo {pages} {3438}
  (\bibinfo {year} {1989})}\BibitemShut {NoStop}%
\bibitem [{RLC()}]{RLC-position-momentum}%
  \BibitemOpen
  \href@noop {} {}\Eprint {http://arxiv.org/abs/For the purposes of this paper
  one can use the RLC-equation for the resonator: $L d^2 Q(t)/dt^2 + R dQ(t)/dt
  + Q(t)/C = 0$ and interpret inductance $L$ as a mechanical mass and charge
  $Q(t)$ as position variable. This allows us, in the style of Devoret's Les
  Houches lectures \cite{Devoret1997-Les_Houches}, to interpret $Q(t)$ as a
  loop charge, see Ref.~\onlinecite{RuskovTahan2019PRB99}} {For the purposes of
  this paper one can use the RLC-equation for the resonator: $L d^2 Q(t)/dt^2 +
  R dQ(t)/dt + Q(t)/C = 0$ and interpret inductance $L$ as a mechanical mass
  and charge $Q(t)$ as position variable. This allows us, in the style of
  Devoret's Les Houches lectures \cite{Devoret1997-Les_Houches}, to interpret
  $Q(t)$ as a loop charge, see Ref.~\onlinecite{RuskovTahan2019PRB99}}
  \BibitemShut {NoStop}%
\bibitem [{Mar()}]{Markovian}%
  \BibitemOpen
  \href@noop {} {}\Eprint {http://arxiv.org/abs/We assume a Markovian
  environment, so that the typical time scale $\hbar/k_B T_r$ is much shorter
  than the system time scales} {We assume a Markovian environment, so that the
  typical time scale $\hbar/k_B T_r$ is much shorter than the system time
  scales} \BibitemShut {NoStop}%
\bibitem [{\citenamefont {Walls}\ and\ \citenamefont
  {Milburn}(2008)}]{MilburnWalls-book2008}%
  \BibitemOpen
  \bibfield  {author} {\bibinfo {author} {\bibfnamefont {D.~F.}\ \bibnamefont
  {Walls}}\ and\ \bibinfo {author} {\bibfnamefont {G.~J.}\ \bibnamefont
  {Milburn}},\ }\href@noop {} {\emph {\bibinfo {title} {{Q}uantum optics}}}\
  (\bibinfo  {publisher} {Berlin Heidelberg, DE: Springer-Verlag},\ \bibinfo
  {year} {2008})\BibitemShut {NoStop}%
\bibitem [{\citenamefont {Wiseman}\ and\ \citenamefont
  {Milburn}(2010)}]{WisemanMilburn-book2010}%
  \BibitemOpen
  \bibfield  {author} {\bibinfo {author} {\bibfnamefont {H.~M.}\ \bibnamefont
  {Wiseman}}\ and\ \bibinfo {author} {\bibfnamefont {G.~J.}\ \bibnamefont
  {Milburn}},\ }\href@noop {} {\emph {\bibinfo {title} {{Q}uantum measurement
  and control}}}\ (\bibinfo  {publisher} {Cambridge, UK: Cambridge University
  Press},\ \bibinfo {year} {2010})\BibitemShut {NoStop}%
\bibitem [{\citenamefont {Ruskov}\ \emph {et~al.}(2005)\citenamefont {Ruskov},
  \citenamefont {Schwab},\ and\ \citenamefont
  {Korotkov}}]{RuskovSchwabKorotkov2005}%
  \BibitemOpen
  \bibfield  {author} {\bibinfo {author} {\bibfnamefont {R.}~\bibnamefont
  {Ruskov}}, \bibinfo {author} {\bibfnamefont {K.}~\bibnamefont {Schwab}}, \
  and\ \bibinfo {author} {\bibfnamefont {A.~N.}\ \bibnamefont {Korotkov}},\
  }\href@noop {} {\bibfield  {journal} {\bibinfo  {journal} {Phys. Rev. B}\
  }\textbf {\bibinfo {volume} {71}},\ \bibinfo {pages} {235407} (\bibinfo
  {year} {2005})}\BibitemShut {NoStop}%
\bibitem [{\citenamefont {Hayes}\ \emph {et~al.}(2012)\citenamefont {Hayes},
  \citenamefont {Clark}, \citenamefont {Debnath}, \citenamefont {Hucul},
  \citenamefont {Inlek}, \citenamefont {Lee}, \citenamefont {Quraishi},\ and\
  \citenamefont {Monroe}}]{HayesMonroe2012PRL}%
  \BibitemOpen
  \bibfield  {author} {\bibinfo {author} {\bibfnamefont {D.}~\bibnamefont
  {Hayes}}, \bibinfo {author} {\bibfnamefont {S.}~\bibnamefont {Clark}},
  \bibinfo {author} {\bibfnamefont {S.}~\bibnamefont {Debnath}}, \bibinfo
  {author} {\bibfnamefont {D.}~\bibnamefont {Hucul}}, \bibinfo {author}
  {\bibfnamefont {I.}~\bibnamefont {Inlek}}, \bibinfo {author} {\bibfnamefont
  {K.}~\bibnamefont {Lee}}, \bibinfo {author} {\bibfnamefont {Q.}~\bibnamefont
  {Quraishi}}, \ and\ \bibinfo {author} {\bibfnamefont {C.}~\bibnamefont
  {Monroe}},\ }\href@noop {} {\bibfield  {journal} {\bibinfo  {journal} {Phys.
  Rev. Lett.}\ }\textbf {\bibinfo {volume} {109}},\ \bibinfo {pages} {020503}
  (\bibinfo {year} {2012})}\BibitemShut {NoStop}%
\bibitem [{Wal()}]{Walsh}%
  \BibitemOpen
  \href@noop {} {}\Eprint {http://arxiv.org/abs/We note that more sophisticated
  coupling sign flips via the so-called Walsh
  functions\cite{HayesMonroe2012PRL} does not help either} {We note that more
  sophisticated coupling sign flips via the so-called Walsh
  functions\cite{HayesMonroe2012PRL} does not help either} \BibitemShut
  {NoStop}%
\bibitem [{\citenamefont {Clerk}\ and\ \citenamefont
  {Utami}(2007)}]{ClerkUtami2012PRA}%
  \BibitemOpen
  \bibfield  {author} {\bibinfo {author} {\bibfnamefont {A.~A.}\ \bibnamefont
  {Clerk}}\ and\ \bibinfo {author} {\bibfnamefont {D.~W.}\ \bibnamefont
  {Utami}},\ }\href@noop {} {\bibfield  {journal} {\bibinfo  {journal} {Phys.
  Rev. A}\ }\textbf {\bibinfo {volume} {75}},\ \bibinfo {pages} {042302}
  (\bibinfo {year} {2007})}\BibitemShut {NoStop}%
\bibitem [{Mat()}]{MatrixPhase}%
  \BibitemOpen
  \href@noop {} {}\Eprint {http://arxiv.org/abs/This ``matrix phase factor'' is
  a member of a more general non-Abelian phase transformation of an arbitrary
  two-qubit state\cite{Mosseri2001} that leaves the two-qubit density matrix
  intact} {This ``matrix phase factor'' is a member of a more general
  non-Abelian phase transformation of an arbitrary two-qubit
  state\cite{Mosseri2001} that leaves the two-qubit density matrix intact}
  \BibitemShut {NoStop}%
\bibitem [{\citenamefont {Royer}\ \emph {et~al.}(2017)\citenamefont {Royer},
  \citenamefont {Grimsmo}, \citenamefont {Didier},\ and\ \citenamefont
  {Blais}}]{Royer2017Quantum}%
  \BibitemOpen
  \bibfield  {author} {\bibinfo {author} {\bibfnamefont {B.}~\bibnamefont
  {Royer}}, \bibinfo {author} {\bibfnamefont {A.~L.}\ \bibnamefont {Grimsmo}},
  \bibinfo {author} {\bibfnamefont {N.}~\bibnamefont {Didier}}, \ and\ \bibinfo
  {author} {\bibfnamefont {A.}~\bibnamefont {Blais}},\ }\href@noop {}
  {\bibfield  {journal} {\bibinfo  {journal} {Quantum}\ }\textbf {\bibinfo
  {volume} {1}} (\bibinfo {year} {2017})}\BibitemShut {NoStop}%
\bibitem [{\citenamefont {Harvey}\ \emph {et~al.}(2018)\citenamefont {Harvey},
  \citenamefont {B{\o}ttcher}, \citenamefont {Orona}, \citenamefont {Bartlett},
  \citenamefont {Doherty},\ and\ \citenamefont {Yacoby}}]{HarveyYacoby2018PRB}%
  \BibitemOpen
  \bibfield  {author} {\bibinfo {author} {\bibfnamefont {S.~P.}\ \bibnamefont
  {Harvey}}, \bibinfo {author} {\bibfnamefont {C.~G.~L.}\ \bibnamefont
  {B{\o}ttcher}}, \bibinfo {author} {\bibfnamefont {L.~A.}\ \bibnamefont
  {Orona}}, \bibinfo {author} {\bibfnamefont {S.~D.}\ \bibnamefont {Bartlett}},
  \bibinfo {author} {\bibfnamefont {A.~C.}\ \bibnamefont {Doherty}}, \ and\
  \bibinfo {author} {\bibfnamefont {A.}~\bibnamefont {Yacoby}},\ }\href@noop {}
  {\bibfield  {journal} {\bibinfo  {journal} {Phys. Rev. B}\ }\textbf {\bibinfo
  {volume} {97}},\ \bibinfo {pages} {235409} (\bibinfo {year}
  {2018})}\BibitemShut {NoStop}%
\bibitem [{\citenamefont {Ruskov}(2016)}]{Delft2016-Ruskov-poster}%
  \BibitemOpen
  \bibfield  {author} {\bibinfo {author} {\bibfnamefont {R.}~\bibnamefont
  {Ruskov}},\ }\href@noop {} {\enquote {\bibinfo {title} {{\it Spin-Cavity
  longitudinal coupling for two-qubit gates and measurement}, {P}oster at the
  {S}ilicon quantum electronics conference, {D}elft, {T}he {N}etherlands,
  {J}une 13-14},}\ } (\bibinfo {year} {2016})\BibitemShut {NoStop}%
\bibitem [{\citenamefont {Tahan}(2016)}]{QCPR2016-Alexandria-Tahan}%
  \BibitemOpen
  \bibfield  {author} {\bibinfo {author} {\bibfnamefont {C.}~\bibnamefont
  {Tahan}},\ }\href@noop {} {\enquote {\bibinfo {title} {Talk at the {Q}uantum
  {C}omputing {P}rogram {R}eview, {A}lexandria, {VA}, {J}uly 18-21},}\ }
  (\bibinfo {year} {2016})\BibitemShut {NoStop}%
\bibitem [{\citenamefont {Hao}\ \emph {et~al.}(2014)\citenamefont {Hao},
  \citenamefont {Ruskov}, \citenamefont {Xiao}, \citenamefont {Tahan},\ and\
  \citenamefont {Jiang}}]{Xiao2014NC}%
  \BibitemOpen
  \bibfield  {author} {\bibinfo {author} {\bibfnamefont {X.}~\bibnamefont
  {Hao}}, \bibinfo {author} {\bibfnamefont {R.}~\bibnamefont {Ruskov}},
  \bibinfo {author} {\bibfnamefont {M.}~\bibnamefont {Xiao}}, \bibinfo {author}
  {\bibfnamefont {C.}~\bibnamefont {Tahan}}, \ and\ \bibinfo {author}
  {\bibfnamefont {H.}~\bibnamefont {Jiang}},\ }\href@noop {} {\bibfield
  {journal} {\bibinfo  {journal} {Nature Communications}\ }\textbf {\bibinfo
  {volume} {5}},\ \bibinfo {pages} {3860} (\bibinfo {year} {2014})}\BibitemShut
  {NoStop}%
\bibitem [{\citenamefont {Mi}\ \emph {et~al.}(2017{\natexlab{b}})\citenamefont
  {Mi}, \citenamefont {Cady}, \citenamefont {Zajac}, \citenamefont {Stehlik},
  \citenamefont {Edge},\ and\ \citenamefont {Petta}}]{MiCadyPetta2017SiCQED}%
  \BibitemOpen
  \bibfield  {author} {\bibinfo {author} {\bibfnamefont {X.}~\bibnamefont
  {Mi}}, \bibinfo {author} {\bibfnamefont {J.~V.}\ \bibnamefont {Cady}},
  \bibinfo {author} {\bibfnamefont {D.~M.}\ \bibnamefont {Zajac}}, \bibinfo
  {author} {\bibfnamefont {J.}~\bibnamefont {Stehlik}}, \bibinfo {author}
  {\bibfnamefont {L.~F.}\ \bibnamefont {Edge}}, \ and\ \bibinfo {author}
  {\bibfnamefont {J.~R.}\ \bibnamefont {Petta}},\ }\href@noop {} {\bibfield
  {journal} {\bibinfo  {journal} {Applied Physics Letters}\ }\textbf {\bibinfo
  {volume} {110}},\ \bibinfo {pages} {043502} (\bibinfo {year}
  {2017}{\natexlab{b}})}\BibitemShut {NoStop}%
\bibitem [{\citenamefont {West}\ \emph {et~al.}(2019)\citenamefont {West},
  \citenamefont {Hensen}, \citenamefont {Jouan}, \citenamefont {Tanttu},
  \citenamefont {Yang}, \citenamefont {Rossi}, \citenamefont {Gonzalez-Zalba},
  \citenamefont {Hudson}, \citenamefont {Morello}, \citenamefont {Reilly},\
  and\ \citenamefont {Dzurak}}]{WestDzurak2019NNano}%
  \BibitemOpen
  \bibfield  {author} {\bibinfo {author} {\bibfnamefont {A.}~\bibnamefont
  {West}}, \bibinfo {author} {\bibfnamefont {B.}~\bibnamefont {Hensen}},
  \bibinfo {author} {\bibfnamefont {A.}~\bibnamefont {Jouan}}, \bibinfo
  {author} {\bibfnamefont {T.}~\bibnamefont {Tanttu}}, \bibinfo {author}
  {\bibfnamefont {C.}~\bibnamefont {Yang}}, \bibinfo {author} {\bibfnamefont
  {A.}~\bibnamefont {Rossi}}, \bibinfo {author} {\bibfnamefont
  {M.}~\bibnamefont {Gonzalez-Zalba}}, \bibinfo {author} {\bibfnamefont
  {F.}~\bibnamefont {Hudson}}, \bibinfo {author} {\bibfnamefont
  {A.}~\bibnamefont {Morello}}, \bibinfo {author} {\bibfnamefont
  {D.}~\bibnamefont {Reilly}}, \ and\ \bibinfo {author} {\bibfnamefont
  {A.}~\bibnamefont {Dzurak}},\ }\href@noop {} {\bibfield  {journal} {\bibinfo
  {journal} {Nat Nanotechnol.}\ }\textbf {\bibinfo {volume} {14}},\ \bibinfo
  {pages} {447} (\bibinfo {year} {2019})}\BibitemShut {NoStop}%
\bibitem [{\citenamefont {Freeman}\ \emph {et~al.}(2016)\citenamefont
  {Freeman}, \citenamefont {Schoenfield},\ and\ \citenamefont
  {Jiang}}]{FreemanJiang2016APL}%
  \BibitemOpen
  \bibfield  {author} {\bibinfo {author} {\bibfnamefont {B.~M.}\ \bibnamefont
  {Freeman}}, \bibinfo {author} {\bibfnamefont {J.~S.}\ \bibnamefont
  {Schoenfield}}, \ and\ \bibinfo {author} {\bibfnamefont {H.-W.}\ \bibnamefont
  {Jiang}},\ }\href@noop {} {\bibfield  {journal} {\bibinfo  {journal} {Appl.
  Phys. Lett.}\ }\textbf {\bibinfo {volume} {108}},\ \bibinfo {eid} {032103}
  (\bibinfo {year} {2016})}\BibitemShut {NoStop}%
\bibitem [{\citenamefont {Connors}\ \emph {et~al.}(2019)\citenamefont
  {Connors}, \citenamefont {Nelson}, \citenamefont {Qiao}, \citenamefont
  {Edge},\ and\ \citenamefont {Nichol}}]{ConnorsNichol2019PRB}%
  \BibitemOpen
  \bibfield  {author} {\bibinfo {author} {\bibfnamefont {E.~J.}\ \bibnamefont
  {Connors}}, \bibinfo {author} {\bibfnamefont {J.}~\bibnamefont {Nelson}},
  \bibinfo {author} {\bibfnamefont {H.}~\bibnamefont {Qiao}}, \bibinfo {author}
  {\bibfnamefont {L.~F.}\ \bibnamefont {Edge}}, \ and\ \bibinfo {author}
  {\bibfnamefont {J.~M.}\ \bibnamefont {Nichol}},\ }\href@noop {} {\bibfield
  {journal} {\bibinfo  {journal} {Phys. Rev. B}\ }\textbf {\bibinfo {volume}
  {100}},\ \bibinfo {pages} {165305} (\bibinfo {year} {2019})}\BibitemShut
  {NoStop}%
\bibitem [{\citenamefont {Culcer}\ \emph {et~al.}(2010)\citenamefont {Culcer},
  \citenamefont {Cywi\ifmmode~\acute{n}\else \'{n}\fi{}ski}, \citenamefont
  {Li}, \citenamefont {Hu},\ and\ \citenamefont {Das~Sarma}}]{Culcer2010PRB}%
  \BibitemOpen
  \bibfield  {author} {\bibinfo {author} {\bibfnamefont {D.}~\bibnamefont
  {Culcer}}, \bibinfo {author} {\bibfnamefont {L.}~\bibnamefont
  {Cywi\ifmmode~\acute{n}\else \'{n}\fi{}ski}}, \bibinfo {author}
  {\bibfnamefont {Q.}~\bibnamefont {Li}}, \bibinfo {author} {\bibfnamefont
  {X.}~\bibnamefont {Hu}}, \ and\ \bibinfo {author} {\bibfnamefont
  {S.}~\bibnamefont {Das~Sarma}},\ }\href@noop {} {\bibfield  {journal}
  {\bibinfo  {journal} {Phys. Rev. B}\ }\textbf {\bibinfo {volume} {82}},\
  \bibinfo {pages} {155312} (\bibinfo {year} {2010})}\BibitemShut {NoStop}%
\bibitem [{\citenamefont {Lodari}\ \emph {et~al.}(2020)\citenamefont {Lodari},
  \citenamefont {Hendrickx}, \citenamefont {Lawrie}, \citenamefont {Hsiao},
  \citenamefont {Vandersypen}, \citenamefont {Sammak}, \citenamefont
  {Veldhorst},\ and\ \citenamefont
  {Scappucci}}]{ScapucciVeldhorst2020-preprint}%
  \BibitemOpen
  \bibfield  {author} {\bibinfo {author} {\bibfnamefont {M.}~\bibnamefont
  {Lodari}}, \bibinfo {author} {\bibfnamefont {N.~W.}\ \bibnamefont
  {Hendrickx}}, \bibinfo {author} {\bibfnamefont {W.~I.~L.}\ \bibnamefont
  {Lawrie}}, \bibinfo {author} {\bibfnamefont {T.~K.}\ \bibnamefont {Hsiao}},
  \bibinfo {author} {\bibfnamefont {L.~M.~K.}\ \bibnamefont {Vandersypen}},
  \bibinfo {author} {\bibfnamefont {A.}~\bibnamefont {Sammak}}, \bibinfo
  {author} {\bibfnamefont {M.}~\bibnamefont {Veldhorst}}, \ and\ \bibinfo
  {author} {\bibfnamefont {G.}~\bibnamefont {Scappucci}},\ }\href@noop {}
  {\enquote {\bibinfo {title} {{L}ow percolation density and charge noise with
  holes in germanium},}\ } (\bibinfo {year} {2020}),\ \Eprint
  {http://arxiv.org/abs/2007.06328v1} {arXiv:2007.06328v1 [cond-mat]}
  \BibitemShut {NoStop}%
\bibitem [{\citenamefont {Ruskov}\ and\ \citenamefont
  {Korotkov}(2003)}]{RuskovKorotkov2003PRB67}%
  \BibitemOpen
  \bibfield  {author} {\bibinfo {author} {\bibfnamefont {R.}~\bibnamefont
  {Ruskov}}\ and\ \bibinfo {author} {\bibfnamefont {A.~N.}\ \bibnamefont
  {Korotkov}},\ }\href@noop {} {\bibfield  {journal} {\bibinfo  {journal}
  {Phys. Rev. B}\ }\textbf {\bibinfo {volume} {67}},\ \bibinfo {pages}
  {241305(R)} (\bibinfo {year} {2003})}\BibitemShut {NoStop}%
\bibitem [{\citenamefont {Ruskov}\ \emph {et~al.}(2006)\citenamefont {Ruskov},
  \citenamefont {Korotkov},\ and\ \citenamefont
  {Mizel}}]{RuskovKorotkovMizel2006PRB72}%
  \BibitemOpen
  \bibfield  {author} {\bibinfo {author} {\bibfnamefont {R.}~\bibnamefont
  {Ruskov}}, \bibinfo {author} {\bibfnamefont {A.~N.}\ \bibnamefont
  {Korotkov}}, \ and\ \bibinfo {author} {\bibfnamefont {A.}~\bibnamefont
  {Mizel}},\ }\href@noop {} {\bibfield  {journal} {\bibinfo  {journal} {Phys.
  Rev. B}\ }\textbf {\bibinfo {volume} {73}},\ \bibinfo {pages} {085317}
  (\bibinfo {year} {2006})}\BibitemShut {NoStop}%
\bibitem [{\citenamefont {Lalumiere}\ \emph {et~al.}(2010)\citenamefont
  {Lalumiere}, \citenamefont {Gambetta},\ and\ \citenamefont
  {Blais}}]{Lalumiere2010}%
  \BibitemOpen
  \bibfield  {author} {\bibinfo {author} {\bibfnamefont {K.}~\bibnamefont
  {Lalumiere}}, \bibinfo {author} {\bibfnamefont {J.~M.}\ \bibnamefont
  {Gambetta}}, \ and\ \bibinfo {author} {\bibfnamefont {A.}~\bibnamefont
  {Blais}},\ }\href@noop {} {\bibfield  {journal} {\bibinfo  {journal} {Phys.
  Rev. A}\ }\textbf {\bibinfo {volume} {81}},\ \bibinfo {pages} {040301(R)}
  (\bibinfo {year} {2010})}\BibitemShut {NoStop}%
\bibitem [{\citenamefont {Roos}(2008)}]{Roos2008NJP}%
  \BibitemOpen
  \bibfield  {author} {\bibinfo {author} {\bibfnamefont {C.~F.}\ \bibnamefont
  {Roos}},\ }\href@noop {} {\bibfield  {journal} {\bibinfo  {journal} {New
  Journal of Physics}\ }\textbf {\bibinfo {volume} {10}},\ \bibinfo {pages}
  {013002} (\bibinfo {year} {2008})}\BibitemShut {NoStop}%
\bibitem [{\citenamefont {Caldeira}\ and\ \citenamefont
  {Leggett}(1983{\natexlab{b}})}]{CaldeiraLeggett1983b}%
  \BibitemOpen
  \bibfield  {author} {\bibinfo {author} {\bibfnamefont {A.~O.}\ \bibnamefont
  {Caldeira}}\ and\ \bibinfo {author} {\bibfnamefont {A.~J.}\ \bibnamefont
  {Leggett}},\ }\href@noop {} {\bibfield  {journal} {\bibinfo  {journal}
  {Physica A}\ }\textbf {\bibinfo {volume} {121}},\ \bibinfo {pages} {587}
  (\bibinfo {year} {1983}{\natexlab{b}})}\BibitemShut {NoStop}%
\bibitem [{\citenamefont {Gardiner}\ and\ \citenamefont
  {Zoller}(2000)}]{GardinerZoller-book2000}%
  \BibitemOpen
  \bibfield  {author} {\bibinfo {author} {\bibfnamefont {C.~W.}\ \bibnamefont
  {Gardiner}}\ and\ \bibinfo {author} {\bibfnamefont {P.}~\bibnamefont
  {Zoller}},\ }\href@noop {} {\emph {\bibinfo {title} {{Q}uantum noise}}}\
  (\bibinfo  {publisher} {Berlin, DE: Springer},\ \bibinfo {year}
  {2000})\BibitemShut {NoStop}%
\bibitem [{\citenamefont {Gambetta}\ \emph {et~al.}(2008)\citenamefont
  {Gambetta}, \citenamefont {Blais}, \citenamefont {Boissonneault},
  \citenamefont {Houck}, \citenamefont {Schuster},\ and\ \citenamefont
  {Girvin}}]{Gambetta2008PRA}%
  \BibitemOpen
  \bibfield  {author} {\bibinfo {author} {\bibfnamefont {J.~M.}\ \bibnamefont
  {Gambetta}}, \bibinfo {author} {\bibfnamefont {A.}~\bibnamefont {Blais}},
  \bibinfo {author} {\bibfnamefont {M.}~\bibnamefont {Boissonneault}}, \bibinfo
  {author} {\bibfnamefont {A.~A.}\ \bibnamefont {Houck}}, \bibinfo {author}
  {\bibfnamefont {D.~I.}\ \bibnamefont {Schuster}}, \ and\ \bibinfo {author}
  {\bibfnamefont {S.~M.}\ \bibnamefont {Girvin}},\ }\href@noop {} {\bibfield
  {journal} {\bibinfo  {journal} {Phys. Rev. A}\ }\textbf {\bibinfo {volume}
  {77}},\ \bibinfo {pages} {012112} (\bibinfo {year} {2008})}\BibitemShut
  {NoStop}%
\bibitem [{\citenamefont {Keane}\ and\ \citenamefont
  {Korotkov}(2012)}]{KeaneKorotkov2012PRA}%
  \BibitemOpen
  \bibfield  {author} {\bibinfo {author} {\bibfnamefont {K.}~\bibnamefont
  {Keane}}\ and\ \bibinfo {author} {\bibfnamefont {A.~N.}\ \bibnamefont
  {Korotkov}},\ }\href@noop {} {\bibfield  {journal} {\bibinfo  {journal}
  {Phys. Rev. A}\ }\textbf {\bibinfo {volume} {86}},\ \bibinfo {pages} {012333}
  (\bibinfo {year} {2012})}\BibitemShut {NoStop}%
\bibitem [{\citenamefont {Mosseri}\ and\ \citenamefont
  {Dandoloff}(2001)}]{Mosseri2001}%
  \BibitemOpen
  \bibfield  {author} {\bibinfo {author} {\bibfnamefont {R.}~\bibnamefont
  {Mosseri}}\ and\ \bibinfo {author} {\bibfnamefont {R.}~\bibnamefont
  {Dandoloff}},\ }\href@noop {} {\bibfield  {journal} {\bibinfo  {journal}
  {Journal of Physics A-Mathematical and General}\ }\textbf {\bibinfo {volume}
  {34}},\ \bibinfo {pages} {10243} (\bibinfo {year} {2001})}\BibitemShut
  {NoStop}%
\bibitem [{\citenamefont {Cabrera}\ and\ \citenamefont
  {Baylis}(2007)}]{CabreraBaylis2007PL}%
  \BibitemOpen
  \bibfield  {author} {\bibinfo {author} {\bibfnamefont {R.}~\bibnamefont
  {Cabrera}}\ and\ \bibinfo {author} {\bibfnamefont {W.~E.}\ \bibnamefont
  {Baylis}},\ }\href@noop {} {\bibfield  {journal} {\bibinfo  {journal}
  {Physics Letters A}\ }\textbf {\bibinfo {volume} {368}},\ \bibinfo {pages}
  {25} (\bibinfo {year} {2007})}\BibitemShut {NoStop}%
\bibitem [{\citenamefont {Wiseman}\ and\ \citenamefont
  {Milburn}(1994)}]{WisemanMilburn1994PRA}%
  \BibitemOpen
  \bibfield  {author} {\bibinfo {author} {\bibfnamefont {H.~M.}\ \bibnamefont
  {Wiseman}}\ and\ \bibinfo {author} {\bibfnamefont {G.~J.}\ \bibnamefont
  {Milburn}},\ }\href@noop {} {\bibfield  {journal} {\bibinfo  {journal} {Phys.
  Rev. A}\ }\textbf {\bibinfo {volume} {49}},\ \bibinfo {pages} {1350}
  (\bibinfo {year} {1994})}\BibitemShut {NoStop}%
\bibitem [{\citenamefont {{\O}ksendal}(1998)}]{Oksendal-book1998}%
  \BibitemOpen
  \bibfield  {author} {\bibinfo {author} {\bibfnamefont {B.}~\bibnamefont
  {{\O}ksendal}},\ }\href@noop {} {\emph {\bibinfo {title} {{S}tochastic
  differential equations}}}\ (\bibinfo  {publisher} {Springer, Berlin, 1998},\
  \bibinfo {year} {1998})\BibitemShut {NoStop}%
\bibitem [{\citenamefont {Korotkov}(2001)}]{KorotkovPRB2001}%
  \BibitemOpen
  \bibfield  {author} {\bibinfo {author} {\bibfnamefont {A.~N.}\ \bibnamefont
  {Korotkov}},\ }\href@noop {} {\bibfield  {journal} {\bibinfo  {journal}
  {Phys. Rev. B}\ }\textbf {\bibinfo {volume} {63}},\ \bibinfo {pages} {115403}
  (\bibinfo {year} {2001})}\BibitemShut {NoStop}%
\bibitem [{\citenamefont {Makhlin}\ and\ \citenamefont
  {Shnirman}(2004)}]{MakhlinShnirman2004PRL}%
  \BibitemOpen
  \bibfield  {author} {\bibinfo {author} {\bibfnamefont {Y.}~\bibnamefont
  {Makhlin}}\ and\ \bibinfo {author} {\bibfnamefont {A.}~\bibnamefont
  {Shnirman}},\ }\href@noop {} {\bibfield  {journal} {\bibinfo  {journal}
  {Phys. Rev. Lett.}\ }\textbf {\bibinfo {volume} {92}},\ \bibinfo {pages}
  {178301} (\bibinfo {year} {2004})}\BibitemShut {NoStop}%
\bibitem [{\citenamefont {Ithier}\ \emph {et~al.}(2005)\citenamefont {Ithier},
  \citenamefont {Collin}, \citenamefont {Joyez}, \citenamefont {Meeson},
  \citenamefont {Vion}, \citenamefont {Esteve}, \citenamefont {Chiarello},
  \citenamefont {Shnirman}, \citenamefont {Makhlin}, \citenamefont {Schriefl},\
  and\ \citenamefont {Sch\"{o}n}}]{Ithier2005PRB}%
  \BibitemOpen
  \bibfield  {author} {\bibinfo {author} {\bibfnamefont {G.}~\bibnamefont
  {Ithier}}, \bibinfo {author} {\bibfnamefont {E.}~\bibnamefont {Collin}},
  \bibinfo {author} {\bibfnamefont {P.}~\bibnamefont {Joyez}}, \bibinfo
  {author} {\bibfnamefont {P.~J.}\ \bibnamefont {Meeson}}, \bibinfo {author}
  {\bibfnamefont {D.}~\bibnamefont {Vion}}, \bibinfo {author} {\bibfnamefont
  {D.}~\bibnamefont {Esteve}}, \bibinfo {author} {\bibfnamefont
  {F.}~\bibnamefont {Chiarello}}, \bibinfo {author} {\bibfnamefont
  {A.}~\bibnamefont {Shnirman}}, \bibinfo {author} {\bibfnamefont
  {Y.}~\bibnamefont {Makhlin}}, \bibinfo {author} {\bibfnamefont
  {J.}~\bibnamefont {Schriefl}}, \ and\ \bibinfo {author} {\bibfnamefont
  {G.}~\bibnamefont {Sch\"{o}n}},\ }\href@noop {} {\bibfield  {journal}
  {\bibinfo  {journal} {Phys. Rev. B}\ }\textbf {\bibinfo {volume} {72}},\
  \bibinfo {pages} {134519} (\bibinfo {year} {2005})}\BibitemShut {NoStop}%
\bibitem [{\citenamefont {Kratochwil}\ \emph {et~al.}(2020)\citenamefont
  {Kratochwil}, \citenamefont {Koski}, \citenamefont {Landig}, \citenamefont
  {Scarlino}, \citenamefont {Abadillo-Uriel}, \citenamefont {Reichl},
  \citenamefont {Coppersmith}, \citenamefont {Wegscheider}, \citenamefont
  {Friesen}, \citenamefont {Wallraff}, \citenamefont {Ihn},\ and\ \citenamefont
  {Ensslin}}]{KratochwilEnslin2020-preprint}%
  \BibitemOpen
  \bibfield  {author} {\bibinfo {author} {\bibfnamefont {B.}~\bibnamefont
  {Kratochwil}}, \bibinfo {author} {\bibfnamefont {J.~V.}\ \bibnamefont
  {Koski}}, \bibinfo {author} {\bibfnamefont {A.~J.}\ \bibnamefont {Landig}},
  \bibinfo {author} {\bibfnamefont {P.}~\bibnamefont {Scarlino}}, \bibinfo
  {author} {\bibfnamefont {J.~C.}\ \bibnamefont {Abadillo-Uriel}}, \bibinfo
  {author} {\bibfnamefont {C.}~\bibnamefont {Reichl}}, \bibinfo {author}
  {\bibfnamefont {S.~N.}\ \bibnamefont {Coppersmith}}, \bibinfo {author}
  {\bibfnamefont {W.}~\bibnamefont {Wegscheider}}, \bibinfo {author}
  {\bibfnamefont {M.}~\bibnamefont {Friesen}}, \bibinfo {author} {\bibfnamefont
  {A.}~\bibnamefont {Wallraff}}, \bibinfo {author} {\bibfnamefont
  {T.}~\bibnamefont {Ihn}}, \ and\ \bibinfo {author} {\bibfnamefont
  {K.}~\bibnamefont {Ensslin}},\ }\href@noop {} {\enquote {\bibinfo {title}
  {{R}ealization of a {CQ3} qubit: energy spectroscopy and coherence},}\ }
  (\bibinfo {year} {2020}),\ \Eprint {http://arxiv.org/abs/2006.05883}
  {arXiv:2006.05883 [cond-mat]} \BibitemShut {NoStop}%
\bibitem [{\citenamefont {Devoret}(1997)}]{Devoret1997-Les_Houches}%
  \BibitemOpen
  \bibfield  {author} {\bibinfo {author} {\bibfnamefont {M.~H.}\ \bibnamefont
  {Devoret}},\ }\href@noop {} {\bibfield  {journal} {\bibinfo  {journal} {in
  Quantum Fluctuations, Lecture notes of July 1995 Les Houches LXIII, edited by
  S. Reynaud, E. Giacobino, and J. Zinn-Justin (Amsterdam: Elsevier 1997)}\ ,\
  \bibinfo {pages} {351}} (\bibinfo {year} {1997})}\BibitemShut {NoStop}%
\end{thebibliography}


%


\end{document}